\documentclass[fleqn,usenatbib]{mnras}

\usepackage{newtxtext,newtxmath}

\usepackage[T1]{fontenc}
\usepackage{ae,aecompl}

\newcommand{\taucl}{\tau_{\rm cl}}

\usepackage{graphicx}	
\usepackage{amsmath}	
\usepackage{wasysym}    
\usepackage{xspace}
\makeatletter 
  \patchcmd{\NAT@citex}
    {\@citea\NAT@hyper@{%
      \NAT@nmfmt{\NAT@nm}%
      \hyper@natlinkbreak{\NAT@aysep\NAT@spacechar}{\@citeb\@extra@b@citeb}%
      \NAT@date}}
    {\@citea\NAT@nmfmt{\NAT@nm}%
    \NAT@aysep\NAT@spacechar\NAT@hyper@{\NAT@date}}{}{}

  \patchcmd{\NAT@citex}
    {\@citea\NAT@hyper@{%
      \NAT@nmfmt{\NAT@nm}%
      \hyper@natlinkbreak{\NAT@spacechar\NAT@@open\if*#1*\else#1\NAT@spacechar\fi}%
        {\@citeb\@extra@b@citeb}%
      \NAT@date}}
    {\@citea\NAT@nmfmt{\NAT@nm}%
    \NAT@spacechar\NAT@@open\if*#1*\else#1\NAT@spacechar\fi\NAT@hyper@{\NAT@date}}
    {}{}
\makeatother

\usepackage{mathtools}
\DeclarePairedDelimiter\bra{\langle}{\rvert}
\DeclarePairedDelimiter\ket{\lvert}{\rangle}
\DeclarePairedDelimiterX\braket[2]{\langle}{\rangle}{#1\,\delimsize\vert\,\mathopen{}#2}

\newcommand\Msun{\text{M}_{\astrosun}} 
\newcommand\Lsun{\text{L}_{\astrosun}} 
\newcommand\HI{\ion{H}{I}\xspace} 
\newcommand\HII{\ion{H}{II}\xspace} 
\newcommand\HeI{\ion{He}{I}\xspace} 
\newcommand\OI{\ion{O}{I}\xspace} 

\newcommand\orcid[1]{\href{http://orcid.org/#1}{{\includegraphics[height=12pt]{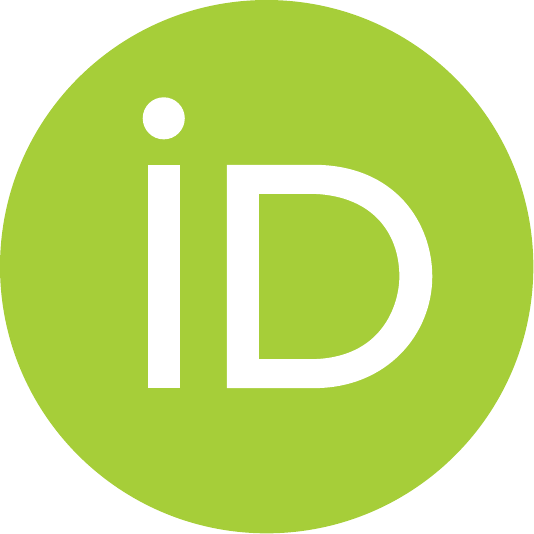}}}}

\title[Lyman-$\alpha$ radiation pressure feedback]{Lyman-$\boldsymbol{\alpha}$ feedback prevails at Cosmic Dawn: \\ Implications for the first galaxies, stars, and star clusters}

\author[O. Nebrin et al.]{
\parbox[t]{\textwidth}{
Olof Nebrin\orcid{0000-0003-3877-360X},$^1$\thanks{Email: \href{mailto:olof.nebrin@astro.su.se}{olof.nebrin@astro.su.se}} Aaron Smith\orcid{0000-0002-2838-9033},$^2$ Kevin Lorinc\orcid{0009-0005-3827-8774},$^2$ Johan Hörnquist\orcid{0000-0002-4291-2636},$^3$ Åsa Larson\orcid{0000-0003-2182-7165},$^3$ \\ Garrelt Mellema\orcid{0000-0002-2512-6748},$^1$ and Sambit K. Giri\orcid{0000-0002-2560-536X}$^4$}
\vspace*{6pt} \\
$^1$ Department of Astronomy \& Oskar Klein Centre for Cosmoparticle Physics, AlbaNova, Stockholm University, SE-106 91 Stockholm, Sweden \\ $^2$ Department of Physics, The University of Texas at Dallas, Richardson, Texas 75080, USA \\ $^3$ Department of Physics, AlbaNova, Stockholm University, SE-106 91 Stockholm, Sweden \\ $^4$ Nordita, KTH Royal Institute of Technology and Stockholm University, Hannes Alfvéns väg 12, SE-106 91 Stockholm, Sweden}

\date{Accepted XXX. Received YYY; in original form ZZZ}

\pubyear{2024}

\begin{document}
\label{firstpage}
\pagerange{\pageref{firstpage}--\pageref{lastpage}}
\maketitle

\begin{abstract}
Radiation pressure from Lyman-$\alpha$ (Ly$\alpha$) scattering is a potentially dominant form of early stellar feedback, capable of injecting up to $\sim 100 \, \times$ more momentum into the interstellar medium (ISM) than UV continuum radiation pressure and stellar winds. Ly$\alpha$ feedback is particularly strong in dust-poor environments and is thus especially important during the formation of the first stars and galaxies. As upcoming galaxy formation simulations incorporate Ly$\alpha$ feedback, it is crucial to consider processes that can limit it to avoid placing $\Lambda$CDM in apparent tension with recent \textit{JWST} observations indicating efficient star formation at Cosmic Dawn. We study Ly$\alpha$ feedback using a novel analytical Ly$\alpha$ radiative transfer solution that includes the effects of continuum absorption, gas velocity gradients, Ly$\alpha$ destruction (e.g. by $2p \rightarrow 2s$ transitions), ISM turbulence, and atomic recoil. We verify our solution for uniform clouds using extensive Monte Carlo radiative transfer (MCRT) tests, and resolve a previous discrepancy between analytical and MCRT predictions. We then study the sensitivity of Ly$\alpha$ feedback to the aforementioned effects. While these can dampen Ly$\alpha$ feedback by a factor $\lesssim \textrm{few} \times 10$, we find it remains $\gtrsim 5 - 100 \, \times$ stronger than direct radiation pressure and therefore cannot be neglected. We provide an accurate fit for the Ly$\alpha$ force multiplier $M_{\rm F}$, suitable for implementation in subgrid models for galaxy formation simulations. Our findings highlight the critical role of Ly$\alpha$ feedback in regulating star formation at Cosmic Dawn, and underscore the necessity of incorporating it into simulations to accurately model early galaxy evolution.

\end{abstract}

\begin{keywords}
galaxies: formation -- dark ages, reionization, first stars -- radiative transfer -- atomic processes -- atomic data
\end{keywords}



\section{Introduction}
\label{Intro}

\begin{figure*}
\includegraphics[width=1.00\textwidth]{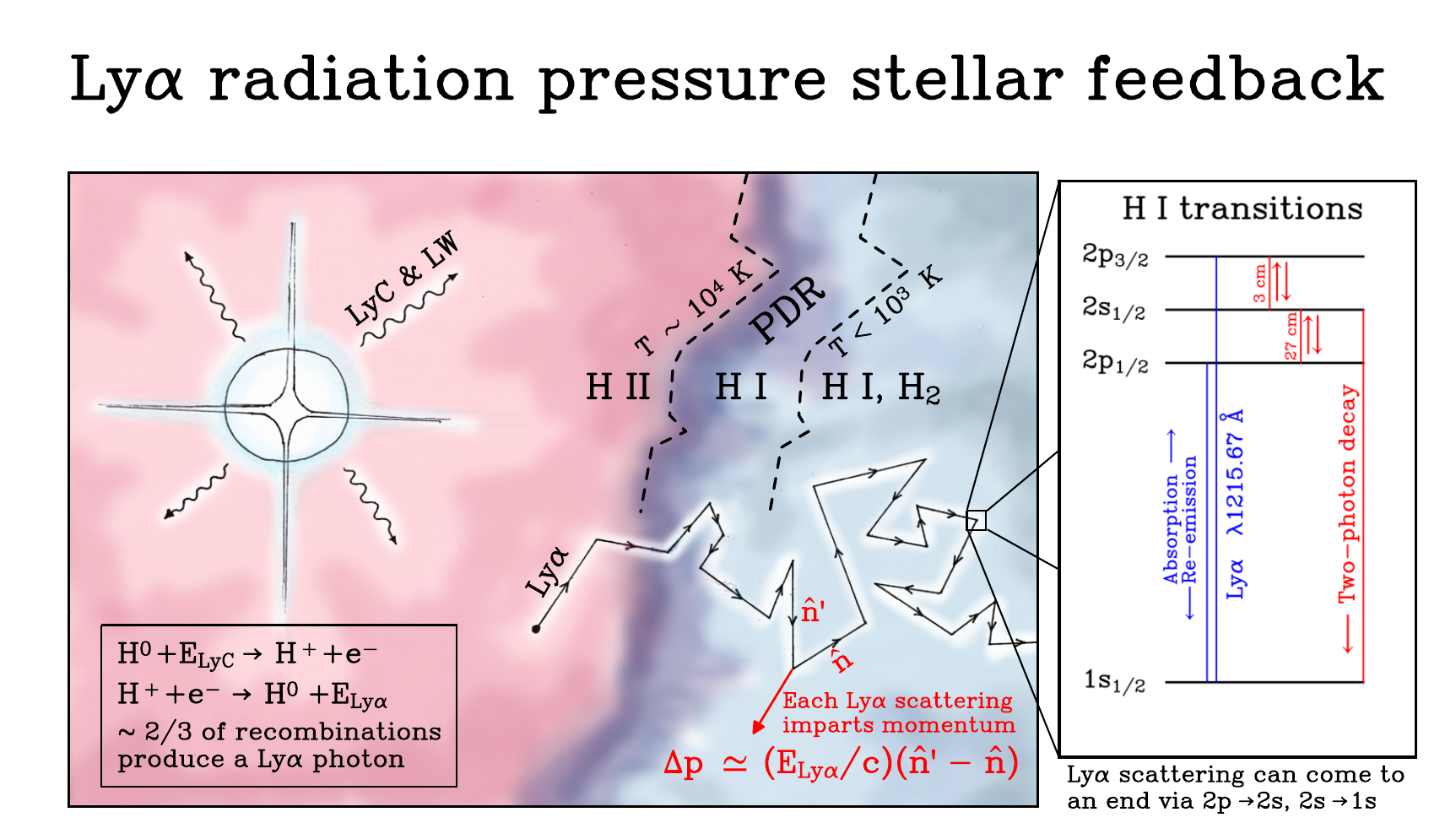}
\caption{A schematic overview of Ly$\alpha$ stellar feedback in a star-forming cloud. \textbf{Left panel}: In \HII regions around young massive stars, $\sim 2/3$ of recombinations produce a Ly$\alpha$ photon. If the Ly$\alpha$ photon escapes absorption by any dust present in the \HII region, it may eventually scatter into the surrounding photodissociation region \citep[PDR, see e.g.][]{Hollenbach1999, Draine2011}, and eventually into a partially or fully molecular H$_2$ region. In metal-poor star-forming clouds, these regions are typically extremely optically thick to Ly$\alpha$, with a line centre optical depth $\tau_{0} \gtrsim 10^9$. As a result, the Ly$\alpha$ photon will undergo many scattering events until it can either escape the cloud or be destroyed. Each scattering event imparts some momentum $\Delta \boldsymbol{p}$ into the gas. \textbf{Right panel}: The relevant energy levels in \HI. Usually, Ly$\alpha$ absorption to one of the $2p$ levels is almost immediately ($A_{\rm Ly\alpha}^{-1} \sim 10^{-9} \, \rm s$) followed by decay to the ground state ($1s$), and re-emission of a Ly$\alpha$ photon. However, there is also a small probability $p_{\rm d,HI}$ of a $2p \rightarrow 2s$ transition, followed by two-photon decay to $1s$. In this case, the original Ly$\alpha$ photon is destroyed and Ly$\alpha$ scattering comes to an end.}
\label{LyafeedbackSketch}
\end{figure*}

Radiation from stars, stellar winds, and supernovae (SNe) inject copious amounts of momentum and energy into the interstellar medium (ISM), capable of destroying star-forming clouds. These stellar feedback processes are crucial for understanding why star formation in the local Universe is, on average, observed to be inefficient, with only $\sim 1 \%$ of the total (stellar and gas) mass of giant molecular clouds (GMCs) in our Galaxy being in the form of stars \citep[e.g.][]{Lee2016, Grudic2019_SFE}. A key goal for theoretical astrophysics is to predict and understand the star formation efficiency from first principles, which requires detailed modelling of these feedback processes. 

This issue has become increasingly important with the recent launch of the \textit{JWST} and subsequent discoveries of surprisingly massive galaxies \citep[e.g.][]{Labbe2023, Chemerynska2024, Xiao2024} and very dense star clusters \citep{Vanzella2023, Adamo2024} during the Epoch of Reionization and Cosmic Dawn, when the first stars and galaxies formed (at redshifts of $z \sim 5$--$30$) \citep[e.g.][]{Bromm2011review, Klessen2023}. These discoveries of highly efficient star formation at high redshifts could indicate that our understanding of stellar feedback in the early Universe is incomplete, or that the standard model of cosmology ($\Lambda$CDM) needs modification \citep[e.g.][]{BoylanKolchin2023}. Addressing the former (among other things) is necessary before we can conclude the latter. Still, the recent observations highlight potential discrepancies between theoretical models and inferred stellar masses, constraining the efficacy of feedback in these primeval galaxies.

To address the nature of feedback at high redshifts, it is prudent to consider the unique environments of star-forming clouds during Cosmic Dawn, which are expected to be qualitatively different from typical local GMCs. In particular, they exhibit the following characteristics:
\begin{enumerate}
    \item \textit{Very dense}: High-resolution cosmological simulations and models have consistently predicted large gas densities ($\sim 10^4$--$10^9\,\rm cm^{-3}$) and surface densities ($\sim 10^3$--$10^6\,\Msun ~ \rm pc^{-2}$) to be common in star-forming clouds at Cosmic Dawn \citep[e.g.][]{Oh2002, Safranek2014, Kimm2016_GC, Ricotti2016, Latif2022, Nebrin2022, Garcia2023}. This prediction is broadly in line with recent detections of bound star clusters with stellar surface densities $\sim 10^5\,\Msun ~ \rm pc^{-2}$ at $z = 6$--$10$ \citep{Vanzella2023, Adamo2024}. Similarly, a significant fraction of the so-called `Little Red Dots' may be extremely compact galaxies with stellar densities $\sim 10^4 - 10^8\,\Msun ~ \rm pc^{-3}$ \citep{Audric2024}.
    \item \textit{Dust-poor}: The ISM of the first galaxies has not undergone many episodes of metal enrichment by SNe. Consequently, the gas is expected to be dust-poor. Additionally, dust may be driven out of the first galaxies by radiation pressure \citep{Fukushima2018, Hirashita2019_radiationpressuredust}. For the very first stars, there is no dust at all. 
    \item \textit{Abundant atomic hydrogen}: The formation of H$_2$ becomes relatively inefficient in dust-poor environments \citep[e.g.][]{Cazaux2009, Nebrin2023cooling, Sternberg2021}. Strong Lyman-Werner (LW) feedback from young stars can also suppress H$_2$-formation in star-forming clouds \citep{Sugimura2024}. As a result, even dense and cool clouds can have substantial \HI column densities. 
\end{enumerate}
These characteristics have significant implications for the nature of stellar feedback at high redshifts. Due to their relatively high densities, many if not most of the star-forming clouds at Cosmic Dawn will have short free-fall timescales $t_{\rm ff} < \textrm{few Myrs}$. As a result, massive stars do not have time to explode as SNe before the cloud is converted into stars, and star formation `overshoots' and becomes bursty \citep[e.g.][]{Kimm2016_GC, Faucher2018_bursty, Nebrin2022}. Thus, only so-called `early' (pre-SNe) stellar feedback can regulate the star formation efficiency in such starbursts. Indeed, this holds true even for many local GMCs, which have estimated dispersal timescales of $< 3 ~ \textrm{Myrs}$, consistent with disruption by early stellar feedback \citep{Kruijssen2019, Chevance2022, Chevance2023}. 

Early stellar feedback from photoionization, stellar winds, and radiation pressure on dust has been extensively explored in the literature and incorporated into many state-of-the-art cloud-scale star formation simulations \citep[e.g.][]{Raskutti2016, Grudic2018, Kim2018_RadPressure, Grudic2022_Starforge, Menon2022_IR, Menon2023_directRP, Menon2024}, and galaxy formation simulations \citep[e.g.][]{Kimm2017minihaloes, Hopkins2018_FIRE2, Agertz2020, Lahen2020, Hopkins2023}. At Cosmic Dawn, radiation pressure on dust becomes increasingly irrelevant because of the lack of dust. Photoionization feedback is also expected to be weaker since very dense clouds, with escape velocities $\gtrsim 10\, \rm km ~ s^{-1}$, can avoid disruption after being photoheated to $T \sim 10^4\,\rm K$ \citep{Dale2012, Dale2013, Kimm2016_GC, Grudic2018}. Similarly, line-driven stellar winds become less effective in metal-poor stars \citep[e.g.][]{Krtivcka2006_winds, Smith2014_winds}. Thus, while important in certain contexts, these processes likely play a smaller role in regulating star formation in the early Universe.

However, as traditional feedback processes become increasingly ineffective, the effect of Lyman-$\alpha$ (Ly$\alpha$) radiation pressure is expected to become increasingly significant. Because of (i), (ii), and (iii), combined with the large cross-section of the Ly$\alpha$ line ($1215.67$ Å), Ly$\alpha$ photons produced within \HII regions of young massive stars will scatter many times before escaping or being absorbed (see Fig. \ref{LyafeedbackSketch}). In this process, they exert substantial radiation pressure on the gas. Although less explored compared to many other feedback mechanisms, several studies have investigated early stellar feedback from Ly$\alpha$ radiation pressure \citep[e.g.][]{George1973, Cox1985, Birthell1990, McKee2008, Dijkstra2008, Yajima2014, Smith2016, Smith2017, Abe2018, Kimm2018, Kimm2019, Tomaselli2021, Han2022, Thomson2024}.

The resonant scattering of Ly$\alpha$ photons can inject significant momentum into the gas, potentially exceeding the momentum input from other early feedback processes by a factor of $\sim 10$--$1000$ \citep[e.g.][]{Kimm2018, Tomaselli2021}, and in some cases even being comparable to the cumulative momentum input from SNe \citep[][]{Kimm2018}. Despite its potential dominance, Ly$\alpha$ feedback is not incorporated in current cosmological galaxy formation simulations, largely due to the high computational demands of having self-consistent on-the-fly Ly$\alpha$ radiative transfer (RT). This has motivated the development of subgrid models for Ly$\alpha$ feedback. 

By fitting Monte Carlo Ly$\alpha$ RT experiments performed with the \textsc{rasacas} \citep{RASCAS} code, \cite{Kimm2018} incorporated a subgrid model for Ly$\alpha$ feedback in a high-resolution simulation of a single low-metallicity dwarf galaxy \citep[see also][]{Kimm2019, Han2022}. These studies found that the inclusion of Ly$\alpha$ feedback (in addition to other forms of feedback) suppressed the number of star clusters formed by a factor of $\sim 5$ compared to simulations without Ly$\alpha$ feedback. This finding is in qualitative agreement with \cite{Abe2018}, who used analytic estimates to argue that Ly$\alpha$ feedback is critically important in the formation of metal-poor globular clusters (GCs).

\cite{Tomaselli2021} studied Ly$\alpha$ feedback analytically \citep[see also][]{Lao2020}, and also concluded that it is often very strong, and cannot be ignored when modelling high-redshift galaxies. \cite{Nebrin2022} implemented their Ly$\alpha$ feedback model in an early version of \textsc{Anaxagoras}, a detailed semi-analytical model of starbursts in low-mass haloes at Cosmic Dawn \citep[][Nebrin, in prep.]{Nebrin2023cooling}. There it was found that Ly$\alpha$ feedback strongly regulated both the formation of Population III (Pop III) stars in minihaloes (dark matter mass $M_\text{vir} \sim 10^5$--$10^7\,\Msun$), and the formation of bound star clusters and Ultra-Faint Dwarf galaxies in atomic-cooling haloes ($M_\text{vir} \sim 10^7$--$10^9\,\Msun$). Including Ly$\alpha$ feedback may also affect the spectral energy distributions (SEDs) from unresolved star-forming regions in hydrodynamical simulations \citep{Kapoor2023}.

Recently, \cite{Jaura2022} simulated the formation of Pop III stars and found that \HII regions typically remained confined in the dense protostellar discs \citep[see also][]{Sharda2024}. To properly model the breakout of \HII regions, and hence accurately predict the IMF of Pop III stars, \cite{Jaura2022} suggested that Ly$\alpha$ feedback must be included in future simulations.\footnote{The role of Ly$\alpha$ radiation pressure feedback during the formation of the first stars has also been discussed earlier by \cite{Doroshkevich1976}, \cite{Omukai2002}, \cite{McKee2008}, and \cite{Stacy2012, Stacy2016}.} Additionally, trapped Ly$\alpha$ photons can affect the fragmentation and chemistry of metal-free $T \sim 10^4\,\rm K$ gas, and thereby play a role in determining the mass scale of direct-collapse black holes, the possible progenitors of supermassive black holes \citep[][]{Spaans2006, Milosavljevic2009, Latif2011, Johnson2017, Smith2017_DCBH, Ge2017}.

On paper, the above studies strongly suggest the importance of Ly$\alpha$ feedback. However, as upcoming simulations and models start to incorporate Ly$\alpha$ feedback in the pursuit of greater realism, we anticipate a looming problem. Namely, the \textit{risk of severely underpredicting the star formation efficiency at Cosmic Dawn compared to recent \textit{JWST} observations}. As alluded to earlier, this is a potential issue already facing current simulations that do not incorporate Ly$\alpha$ feedback, although models often employ arbitrary momentum boosting calibrations to account for uncertain factors such as the high-mass
end of the IMF, numerical overcooling, and missing feedback processes \citep[e.g.][]{Rosdahl2018}. Adding an additional dominant feedback process could exacerbate this problem, potentially sparking a new `small-scale challenge' for $\Lambda$CDM \citep{Bullock2017}. We also note that Ly$\alpha$ feedback is immune to the arguments of \cite{Dekel2023} against the importance of feedback from SNe, winds, and radiation pressure on dust at Cosmic Dawn, which we largely agree with in a qualitative sense \citep[see the above discussion and][]{Nebrin2022}.

This motivates a careful study of Ly$\alpha$ feedback and ways it could be dampened. It is well-known that dust absorption of Ly$\alpha$ photons can set an upper limit to Ly$\alpha$ radiation pressure \cite[see e.g. discussions and results in][]{Cox1985, Birthell1990, Henney1998, Oh2002, Wise2012radpressure, Kimm2018, Tomaselli2021}. However, even with Solar-like dust abundances, momentum input from Ly$\alpha$ radiation pressure can exceed the momentum input from direct radiation pressure \citep[][]{Kimm2018, Tomaselli2021}. For lower dust abundances, most relevant to Cosmic Dawn, dust alone cannot prevent Ly$\alpha$ feedback from dominating over other forms of early feedback.

A less explored suppression mechanism for Ly$\alpha$ feedback comes from $2p \rightarrow 2s$ transitions in hydrogen (Fig. \ref{LyafeedbackSketch}), which can destroy Ly$\alpha$ photons if followed by two-photon emission to the ground state. Collisions with electrons and protons can induce such transitions \citep[e.g.][]{Pengelly1964, Guzman2017}, but the rates are typically too small to have a significant effect in largely neutral star-forming clouds. As a result, most studies have ignored the impact of $2p \rightarrow 2s$ transitions on Ly$\alpha$ feedback \citep[with rare exceptions including collisional de-excitation by charged particles, see e.g.][]{Chiu1998, McKee2008}. However, collisions with neutral atoms and molecules (e.g. \HI, \HeI, H$_2$) can also induce $2p \rightleftharpoons 2s$ transitions \citep[e.g.][]{Slocomb1971,Ryan1977, Vassilev1990}, the rates of which were not available or known during early discussions of $2p \rightleftharpoons 2s$ transitions in astrophysical contexts \citep[e.g.][]{Spitzer1951, Pengelly1964}. The importance of this suppression mechanism has not been quantified.

Furthermore, most studies of Ly$\alpha$ feedback have assumed uniform and/or static clouds \citep[e.g.][]{Dijkstra2008, Smith2017, Kimm2018, Lao2020, Tomaselli2021}. Ly$\alpha$ photons can escape more easily from clouds with velocity gradients \citep{Bonilha1979, Seon2020}. Similarly, cloud inhomogeneity and anisotropy can create low-density channels through which Ly$\alpha$ photons can escape \citep[e.g.][]{Behrens2014, Zheng2014, Smith2019, Kakiichi2021, Monter2024, Yuan2024}. Whether these factors can significantly suppress Ly$\alpha$ feedback remains unknown but is crucial to assess in detail \citep[][]{Thomson2024}.

Finally, there is also an unexplained factor $\sim 2$--$3$ discrepancy between the estimated Ly$\alpha$ momentum input using analytical solutions \citep{Lao2020, Tomaselli2021} compared to Monte Carlo methods \citep{Kimm2018}. \cite{Tomaselli2021} suggested that this could be due to the inclusion of atomic recoil by the latter, which slowly drains energy from Ly$\alpha$ photons, pushing them to the red wing of the Ly$\alpha$ line and facilitating escape \citep{Adams1971, Smith2017}. Unfortunately, there are no analytical solutions with recoil,\footnote{In the case of an infinite expanding medium, appropriate for the intergalactic medium, there are a few known analytical solutions with recoil \citep[e.g.][]{Grachev1989, Chuzhoy2006, Furlanetto2006_FokkerPlanck}. These solutions are unfortunately inapplicable to clouds of finite extent in the ISM.} nor any published comparisons of Ly$\alpha$ feedback with and without recoil.

In this paper we study Ly$\alpha$ feedback with the goal of quantifying the importance of various suppression mechanisms, and derive an accurate estimate for the Ly$\alpha$ force multiplier $M_{\rm F}$, for use in subgrid models in galaxy and star formation simulations. The paper is structured as follows. In Sec. \ref{Solution}, we derive a novel analytical Ly$\alpha$ RT solution, generalizing earlier solutions for spherical clouds to include Ly$\alpha$ photon destruction (e.g. by $2p \rightarrow 2s$ transitions and dust absorption), velocity gradients, atomic recoil, and density fluctuations from supersonic turbulence. In Sec. \ref{Force multiplier} we use the solution to study Ly$\alpha$ radiation pressure and its sensitivity to the aforementioned effects. In Sec. \ref{Lya destruction} we carefully examine processes that can destroy Ly$\alpha$ photons. In the process of doing so we perform a new \textit{ab initio} calculation of the rate of $2p \rightarrow 2s$ transitions in collisions with hydrogen (see Appendix \ref{Appendix H(2s) cross section} for details), and later show that this can suppress Ly$\alpha$ feedback around the first stars. We discuss the implications of our results for star and galaxy formation at Cosmic Dawn in Sec. \ref{Discussion section}. Finally, we summarize our results in Sec. \ref{Summary and conclusion}. Several additional details and derivations can be found in the extensive Appendix.

\section{Modelling Lyman-\texorpdfstring{\boldmath$\alpha$}{alpha} feedback}
\subsection{Problem setup and background}
\label{Problem setup and background}

\begin{table}
\caption{Approximate emission rates of ionizing (LyC) photons per $\Msun$ of stars formed, for different stellar populations and masses.}
\begin{tabular}{l c}
\hline
\hline
Stellar Population & $\Dot{Q}_{\rm LyC}$ ($\textrm{s}^{-1} ~ \Msun^{-1}$)  \\

\hline
\hline

\\
\vspace{8 pt}

Pop II, Kroupa IMF$^\dagger$ & $5 \times 10^{46}$  \\

\vspace{8 pt}

Pop III, $9 ~ \Msun$ stars$^{\diamond}$ & $5.2 \times 10^{46}$  \\

\vspace{8 pt}

Pop III, $20 ~ \Msun$ stars$^{\diamond}$ & $2.4 \times 10^{47}$   \\

\vspace{8 pt}

Pop III, $40 ~ \Msun$ stars$^{\diamond}$ & $6.1 \times 10^{47}$  \\

\vspace{8 pt}

Pop III, $60 ~ \Msun$ stars$^{\diamond}$ & $8.7 \times 10^{47}$   \\

\vspace{8 pt}

Pop III, $85 ~ \Msun$ stars$^{\diamond}$ & $1.1 \times 10^{48}$  \\

\vspace{8 pt}

Pop III, $120 ~ \Msun$ stars$^{\diamond}$ & $1.3 \times 10^{48}$  

\\

\hline
\hline
\end{tabular}
\vspace{1 pt}\\
$^\dagger$: The value of $\Dot{Q}_{\rm LyC}$ is taken as representative of $Z/Z_{\odot} = 0.05$ Pop II stars with ages $\lesssim 3.5 ~ \rm Myrs$ \citep[see e.g.][]{Ma2016, Rosdahl2018}. Variation in metallicity (for a fixed IMF) can lead to small ($\sim \pm 0.1 ~ \rm dex$) changes around this value \citep[e.g.][]{Stanway2019, Deng2024}.

$^{\diamond}$: These are lifetime-average values, taken from \cite{Klessen2023}. The stellar lifetime goes from $20 ~ \rm Myrs$ for $9 ~ \Msun$, down to $3 ~ \rm Myrs$ for $120 ~ \Msun$.
\label{Qdot}
\end{table}

For simplicity we consider a uniform, spherical Ly$\alpha$ source of effective radius $R_{\rm s}$ situated at the centre of a uniform spherical cloud of radius $R_{\rm cl} \geq R_{\rm s}$. If we are studying a star cluster in formation, then $R_{\rm s}$ would represent the radius of the cluster, or their combined \HII region(s) where the Ly$\alpha$ photons are produced (see Fig. \ref{LyafeedbackSketch} for a schematic sketch). If, on the other hand, we are studying a single star, then $R_{\rm s}$ would represent the \HII region of that star, which, at least initially, could satisfy $R_{\rm s} \ll R_{\rm cl}$. Thus, the entire parameter range $0 \leq R_{\rm s} \leq R_{\rm cl}$ has applications. We allow for isotropic Hubble-like cloud expansion/contraction, $u(r) = (\Dot{R}_{\rm cl}/R_{\rm cl})r$, with $\Dot{R}_{\rm cl}$ being the cloud expansion/contraction rate at the cloud radius. 

The Ly$\alpha$ emissivity (in erg cm$^{-3}$ s$^{-1}$ Hz$^{-1}$ sr$^{-1}$) in the comoving frame is then:
\begin{equation}
    j_{\rm s}(r < R_{\rm s}, \nu) = \dfrac{3 L_{\rm Ly\alpha} \phi_{\rm s}(\nu)}{(4 \pi)^2 R_{\rm s}^3} \hspace{1 pt} \label{js},
\end{equation}
where $L_{\rm Ly\alpha}$ is the total Ly$\alpha$ luminosity of the source(s) (erg s$^{-1}$), and $\phi_{\rm s}(\nu)$ is the Ly$\alpha$ emission line profile. The latter is normalized so that $\int_{0}^{\infty} \textrm{d}\nu \hspace{1 pt} \phi_{\rm s} = 1$. We will only consider cases where $\phi_{\rm s}$ is narrow and centered on $\nu_{\rm Ly\alpha}$, which is typically true. With the above caveats, our modelling is agnostic with respect to the nature of the Ly$\alpha$ source --- it could be the \HII region around a young star cluster, the \HII region around a single star, collisionally excited \HI gas, etc.

In the scenario most relevant for Ly$\alpha$ stellar feedback, Ly$\alpha$ photons are mainly produced in the \HII region(s) of young massive stars. Typically, a fraction $f_{\rm rec, Ly\alpha} \simeq 2/3$ of recombinations in an \HII region produce a Ly$\alpha$ photon \citep[e.g.][]{Draine2011, Dijkstra2014}. Assuming photoionization equilibrium, and a star cluster (or single star) of mass $M_{\star}$, we then have 
\begin{align}
    L_{\rm Ly\alpha} ~&=~ (1 - f_{\rm esc,LyC})f_{\rm rec, Ly\alpha} \Dot{Q}_{\rm LyC} M_{\star} E_{\rm Ly\alpha} \nonumber \\ ~&=~ 5.4 \times 10^{38} \hspace{1 pt} (1 - f_{\rm esc,LyC}) \left( \dfrac{f_{\rm rec, Ly\alpha}}{2/3} \right) \nonumber \\ ~&\times~ \left( \dfrac{\Dot{Q}_{\rm LyC}}{5 \times 10^{46}~ \textrm{s}^{-1} ~ \Msun^{-1}} \right) \left( \dfrac{M_{\star}}{10^3 ~\Msun} \right) ~ \textrm{erg s}^{-1} \hspace{1 pt} \label{Lya luminosity},
\end{align}
where $f_{\rm esc,LyC}$ is the escape fraction of ionizing (LyC) photons, $\Dot{Q}_{\rm LyC}$ is the LyC photon emission rate per $\Msun$ of stars formed, and $E_{\rm Ly\alpha} = h\nu_{\rm Ly\alpha}$ is the energy of a Ly$\alpha$ photon (Table \ref{TableSymbols}). $\Dot{Q}_{\rm LyC}$ is typically fairly constant before the death of the first massive stars after $\sim 3 - 20 ~ \rm Myrs$, depending on the IMF. Typical values of $\Dot{Q}_{\rm LyC}$ are given in Table \ref{Qdot} for Population II (Pop II) stars with a Kroupa IMF \citep{Kroupa2001}, and some Pop III stars of different masses \citep[adopted from][]{Klessen2023}. 

The goal of this paper is to compute the Ly$\alpha$ pressure on the cloud for a given source distribution. We are primarily interested in the \textit{relative} suppression of Ly$\alpha$ feedback when accounting for various suppression mechanisms, most of which have been previously overlooked. For this reason, the precise value of $L_{\rm Ly\alpha}$ will be of less importance to us here, except in the discussion in Sec. \ref{Discussion section}. To compute the radiation pressure, we need to determine the Ly$\alpha$ flux within the cloud, which in turn requires us to consider Ly$\alpha$ RT. We do this in the next subsection.

\begin{table*}
\caption{Frequently used variables in this paper, including their symbols, definitions, meanings, and references for further discussion.}
\begin{tabular}{l l l l }
\hline
\hline
Symbol(s) & Definition(s) or value(s) & Meaning   & Discussion   \\

\hline
\hline

\\

 $J$, $\boldsymbol{H}$, $\mathbfss{K}$, $\mathbfss{L}$ & Eq.~(\ref{A1moments})  & The zeroth, first, second, and third angular moments of the Ly$\alpha$ &  Appendix \ref{AppendixRTeq}, Sec. \ref{Solution} \\ 

 &    &specific intensity $I$, respectively. $J$ is the Ly$\alpha$ mean specific  &  \\ 

 &    &intensity (in erg cm$^{-2}$ s$^{-1}$ Hz$^{-1}$ sr$^{-1}$).   &  \\ 

 \\

 $\boldsymbol{F}$, $F$ & $= 4 \pi \boldsymbol{H}$, $4\pi \lvert \boldsymbol{H} \rvert$   & The Ly$\alpha$ specific flux (in erg cm$^{-2}$ s$^{-1}$ Hz$^{-1}$). & Appendix \ref{AppendixRTeq}, Sec. \ref{Solution}, \ref{Force multiplier} \\ 

  \\
 
   $\nu_{\rm Ly\alpha}$, $E_{\rm Ly\alpha}$ &  $= 2.466 \times 10^{15} ~ \rm Hz$, $10.20 ~ \rm eV$  & Frequency and energy of Ly$\alpha$ photons at line centre, respectively. &  \\ 

  \\
 
   $b$ &  $= \sqrt{ 2k_{\rm B}T/m_{\rm H}}$  & The thermal velocity, optionally including microturbulence. &  \\ 

  \\

  $\Delta \nu_{\rm D}$ &  $= (b/c) \nu_{\rm Ly\alpha}$  & The Doppler width of the Ly$\alpha$ line. &  \\ 

  \\
 
   $x$ &  $\equiv (\nu - \nu_{\rm Ly\alpha})/\Delta \nu_{\rm D}$  & Frequency displacement from Ly$\alpha$ line centre in units of $\Delta \nu_{\rm D}$. &  \\ 
 
   \\
 
   $\sigma_0$ &  Eq.~(\ref{sigma0 definition})  & The Ly$\alpha$ cross-section at line centre. &  \\ 

   \\
 
   $\tau(r)$, $\taucl$, $\tau_{\rm s}$ &  $\equiv n_{\rm HI} \sigma_0 r$, $n_{\rm HI} \sigma_0 R_{\rm cl}$, 
 $n_{\rm HI} \sigma_0 R_{\rm s}$  & The Ly$\alpha$ line centre optical depth to radius $r$, $r = R_{\rm cl}$ &  \\ 

   &    &(the cloud radius), and $r = R_{\rm s}$ (emission region radius),   &  \\ 

   &    &respectively.   &  \\ 

   \\
 
  $a_{\rm v}$ &  $= 4.7 \times 10^{-3} \hspace{1 pt} (T/100\,\rm K)^{-1/2}$  & The Voigt parameter for the Ly$\alpha$ line. &  \\

 \\
 
  $\mathcal{H}(x)$ &  Eq.  (\ref{Voigt definition})  & The dimensionless Voigt profile for the Ly$\alpha$ line. &  \\ 

 \\

   $\Bar{x}$ &  $= \dfrac{h \Delta \nu_{\rm D}}{2 k_{\rm B}T}$,  Eq.~(\ref{xbar numerical})  & Atomic recoil parameter. &  Appendix \ref{AppendixRTeq}, Sec. \ref{Solution},  \\ 

    &   & &   Sec. \ref{effect of recoil on Lya feedback} \\ 

 \\
 
  $\epsilon$ &  $\simeq \tau_{\rm c,abs}/\taucl$  & Continuum absorption parameter. Approximately the ratio between & Appendix \ref{AppendixRTeq}, Sec. \ref{Solution}, \ref{Sec. M_F continuum absorption},  \\

   &  & the cloud continuum absorption optical depth $\tau_{\rm c,abs}$, and $\taucl$. &  \\ 

    \\

  $\Dot{R}_{\rm cl}$ & $=  \boldsymbol{u}(r = R_{\rm cl}) \cdot \boldsymbol{\hat{r}}$   & Radial expansion/contraction velocity at the cloud edge. &  \\

 \\

   $\Tilde{x}_n$, $\Tilde{\gamma}_n$, $\Tilde{\epsilon}_n$  & Eqs.~(\ref{xtilde expression})--(\ref{epsilontilde expression}) (uniform clouds),  & Convenient dimensionless parameters for recoil, velocity gradients, & Appendix \ref{AppendixRTfreq}, Sec. \ref{uniform cloud solution} \\ 

      & Eqs.~(\ref{xtilde expression turb})--(\ref{epsilontilde expression turb}) (turbulent clouds) & and continuum absorption, respectively. &  \\ 

 \\

  $\mathcal{M}$, $b_s$ &  Eq.~(\ref{turbulence variance})  & The 3D turbulent Mach number, and the turbulence & Sec. \ref{turbulent cloud solution section}, \ref{effect of turbulence on M_F section} \\ 

   &   & driving parameter, respectively. &  \\

 \\

  $p_{\rm d}$ &  Eq.~(\ref{pd final expression})  & The Ly$\alpha$ destruction probability per scattering event. & Sec. \ref{Lya destruction} \\ 

  \\

  $\mathcal{P}$ &  $\equiv \sqrt{6 \pi} p_{\rm d} \tau_{\rm cl}/2 $  & Convenient variable. $\mathcal{P} \gtrsim \rm few$ for Ly$\alpha$ feedback suppression. & Sec. \ref{Solution}, \ref{Lyadestruction general effect} \\ 

  \\

  $L_{\rm Ly\alpha}$ &  Eq.~(\ref{Lya luminosity})  & The total Ly$\alpha$ luminosity from sources in the cloud (erg s$^{-1}$). & Sec. \ref{Problem setup and background} \\ 

  \\

  $\Dot{p}_{\rm Ly\alpha}$ &  Eq.~(\ref{Pdot general})  &  The net radial force on the cloud from Ly$\alpha$ radiation pressure. & Sec. \ref{Force multiplier} \\ 

  \\

  $M_{\rm F}$ &  $\equiv \Dot{p}_{\rm Ly\alpha}/(L_{\rm Ly\alpha}/c)$, & The Ly$\alpha$ force multiplier. $M_{\rm F} \gg 1$ for strong Ly$\alpha$ feedback. & Sec. \ref{Force multiplier} \\ 

   &  Eq.~(\ref{M_F general}) \& Eq.~(\ref{M_F general turbulence}) & . & \\

 \\
 
\hline
\hline
\end{tabular}

\label{TableSymbols}
\end{table*}

\subsection{Radiative transfer solution}
\label{Solution}

Ly$\alpha$ radiative transfer is notoriously complicated, and therefore only a few analytical solutions of the radiative transfer equation exist. \cite{Harrington1973} \citep[building on work by][]{Unno1955} and \cite{Neufeld1990} were the first to find analytical solutions of resonant Ly$\alpha$ scattering in slab geometries. These authors allowed for the potential destruction of Ly$\alpha$ photons by e.g. $2p \rightarrow 2s$ transitions, and dust absorption. Later on, \cite{Dijkstra2006} and \cite{Tasitsiomi2006} found the corresponding solution for uniform static spherical and cubic clouds, respectively, with a central point source of Ly$\alpha$ photons (e.g. a star), but ignoring Ly$\alpha$ destruction \citep[see also e.g.][]{Seon2020, McClellan2022}. More recently, \cite{Lao2020} and \cite{Tomaselli2021} generalized the earlier solutions for spherical clouds to consider density gradients and different source distributions. However, these studies neglected all destruction mechanisms besides dust absorption, and they did not consider the effects of velocity gradients, atomic recoil, or turbulent density fluctuations.

In this paper we derive a novel analytical solution describing resonant Ly$\alpha$ scattering in a spherical, potentially turbulent, cloud. We then use this solution to estimate the momentum input from Ly$\alpha$ radiation pressure, and its sensitivity to various effects (e.g. $2p \rightarrow 2s$ transitions, cloud expansion). New ingredients in this solution are:
\begin{enumerate}
    \item Generalizing the solutions of \cite{Dijkstra2006} and \cite{Lao2020} for a spherical geometry to include a finite Ly$\alpha$ destruction probability (e.g. by $2p \rightarrow 2s$ transitions, Sec. \ref{Lya destruction}). This effect had only been considered earlier by \cite{Harrington1973} and \cite{Neufeld1990} for a slab geometry. \\
    
    \item For the first time we allow for velocity gradients, as long as they are constant on cloud scales, as in, e.g., a cloud undergoing Hubble-like expansion or contraction. This generalizes the only known previous analytical solution by \cite{Loeb1999} of a point source in an infinite expanding medium that had assumed zero gas temperature \citep[see also][]{Higgins2012}. We explore the effect of velocity gradients on Ly$\alpha$ RT (e.g. spectra) in greater detail in a separate paper \citep{Smith2025}. \\
    
    \item We include atomic recoil, i.e. the (small) energy loss during each Ly$\alpha$ scattering event, which can help push Ly$\alpha$ photons to the red side of line centre, and hence make cloud escape easier. Recoil leads, on average, to a frequency jump of $-(h \Delta\nu_{\rm D}/2k_{\rm B}T)\Delta\nu_{\rm D} \equiv -\Bar{x} \Delta \nu_{\rm D}$ \citep[see e.g. ][ and Appendix \ref{AppendixRTeq}]{Chuzhoy2006, Rybicki2006}. Numerically we get:
    \begin{equation}
        \Bar{x} = 2.54 \times 10^{-3} \left( \dfrac{T}{100\,\rm K} \right)^{-1/2} \hspace{1 pt} \label{xbar numerical}.
    \end{equation}
    
    \item We include continuum absorption/scattering, allowing us to model the effects of dust absorption/scattering or other continuum absorption effects (e.g. Raman scattering by H$_2$). Dust absorption and isotropic scattering of Ly$\alpha$ photons had only been considered analytically before for a slab geometry by \cite{Neufeld1990}. As discussed in Appendix \ref{AppendixRTeq}, we allow for anisotropic scattering using the $\delta$-Eddington approximation \citep{Joseph1976, Wiscombe1977}, often employed in climate modelling \citep[e.g.][]{NCAR1992}. However, we find that there is no dependence on the scattering asymmetry as long as the Ly$\alpha$ opacity is much larger than the continuum opacity. \\

    \item We generalize the solution further in Sec. \ref{turbulent cloud solution section} to include an approximate treatment of (potentially large) turbulent density fluctuations. This is achieved using the theory of asymptotic homogenization \citep[e.g.][]{Zwillinger1992_Diffeqs, Holmes2013_Perturbation}, which has been used in several other non-astrophysical contexts to derive effective diffusion coefficients in media with small-scale fluctuations. This allows us to study, analytically, the sensitivity of Ly$\alpha$ feedback to Ly$\alpha$ escape along low-density channels in turbulent media \citep{Kimm2019, Kakiichi2021}. \\ 
    
    \item Finally, while our main focus is on applying the solution to study Ly$\alpha$ radiation pressure, we also derive analytical expressions for other quantities of interest -- e.g. the predicted emergent spectrum, and the escape fraction of Ly$\alpha$ photons. We will explore these applications of the solution in more detail in future work. 
\end{enumerate}

\subsubsection{Solution for a uniform spherical \HI cloud}
\label{uniform cloud solution}

\begin{figure*}
\includegraphics[width=0.85\textwidth]{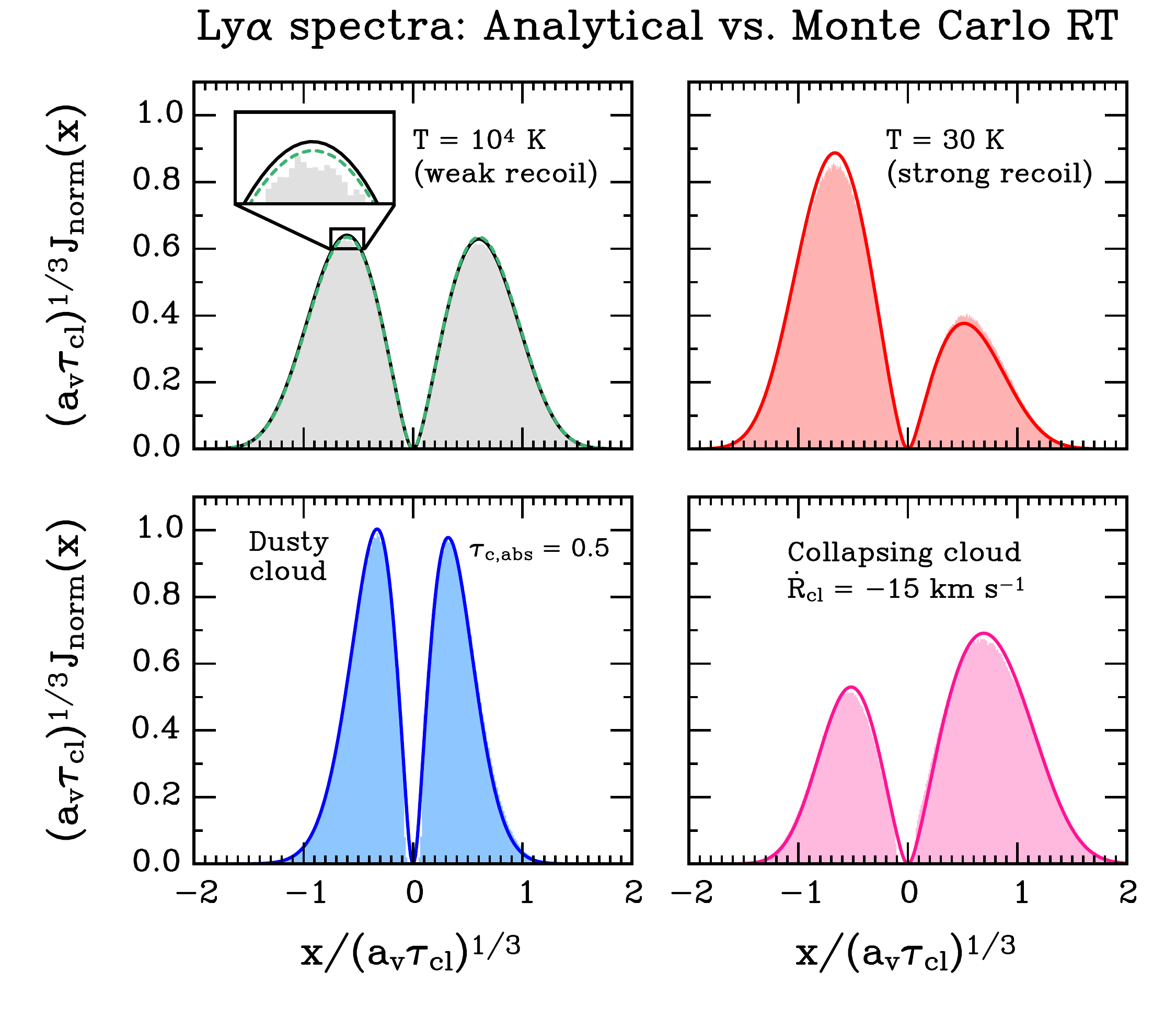}
\caption{The analytically predicted normalized emergent Ly$\alpha$ spectra (solid lines, Eq.~\ref{spectrum main text}), $J_{\rm norm}(x) \equiv J(x)/\int_{-\infty}^{\infty}\textrm{d}x \hspace{1 pt} J(x)$, compared to MCRT results using \textsc{colt} (shaded histograms), for a few example \HI clouds. The MCRT spectra were computed with $5\times 10^6$ photons and dynamical core-skipping. We expect the agreement between analytical and MCRT predictions to become even better with a more careful treatment of boundary conditions \citep{McClellan2022}. Each cloud has $\taucl = 10^8$ and uniform emission ($\Bar{\tau}_{\rm s} = R_{\rm s}/R_{\rm cl} = 1$). Recoil is taken into account for all spectra, but we have ignored the Ly$\alpha$ destruction probability ($p_{\rm d} = 0$). \textbf{Upper left panel}: A static dust-free cloud with $T = 10^4\,\rm K$. Also shown is the solution of \citet{Lao2020} for the same setup (dashed green line). The tiny deviation between the solid and dashed lines is due to including recoil in this paper. \textbf{Upper right panel}: Same as the upper left panel, but for $T = 30\,\rm K$, to better showcase the effect of recoil. \textbf{Lower left panel}: The same as the upper left panel, but including continuum absorption (e.g. from dust) with $\tau_{\rm c,abs} = 0.5$. \textbf{Lower right panel}: Same as the upper left panel, but for a collapsing cloud with $\Dot{R}_{\rm cl} = -15\,\rm km ~ s^{-1}$.  }
\label{spectrum}
\end{figure*}
\begin{figure}
    \includegraphics[trim={0.0cm 0.0cm 0.0cm 0.0cm},clip,width = \columnwidth]{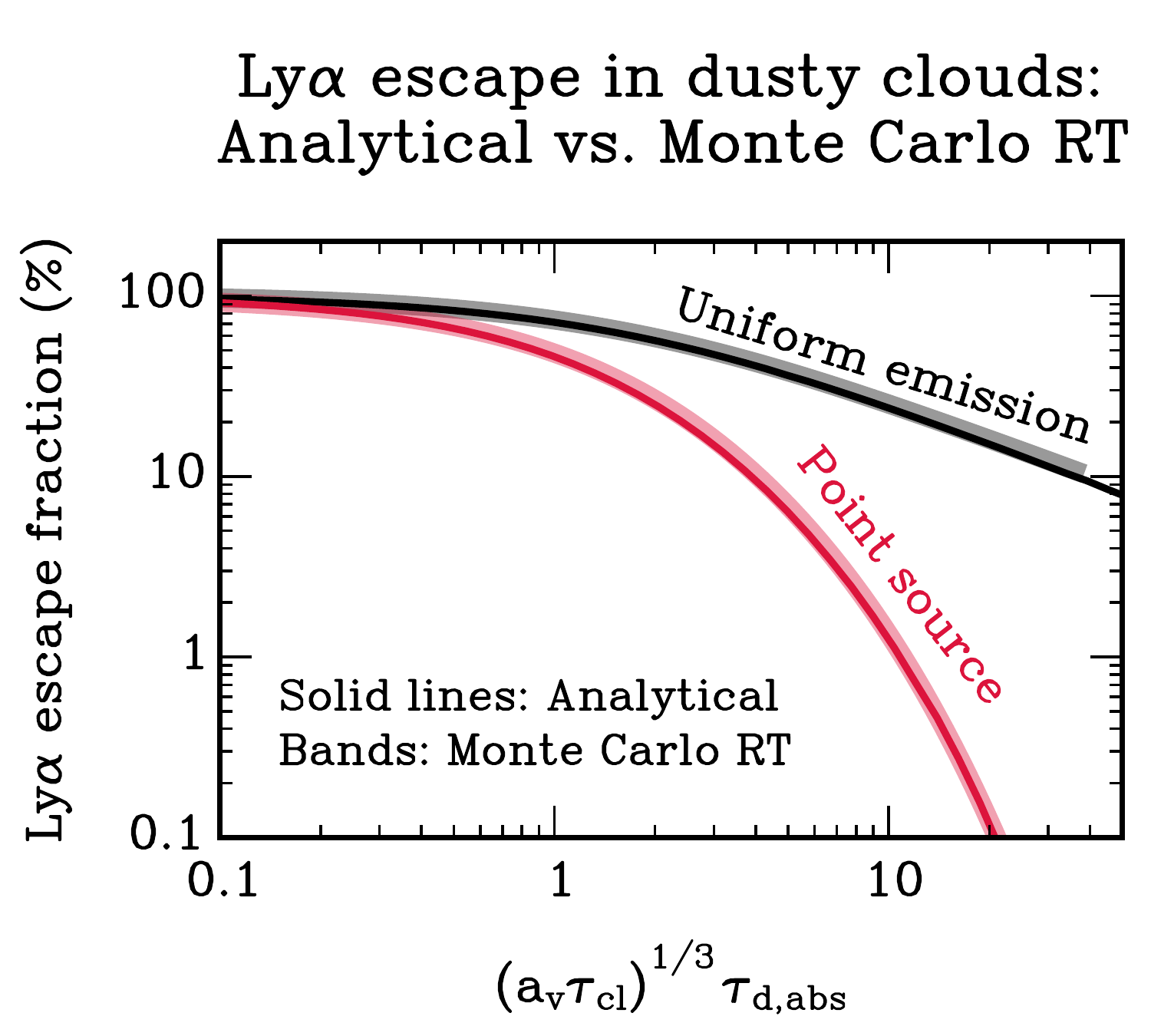}
    \caption{The predicted Ly$\alpha$ escape fraction in dusty static uniform clouds from the analytical solution of this paper (lines, Eq.~\ref{fesc main text}), compared to MCRT results (bands) for the same setup using \textsc{colt}. We have ignored recoil for simplicity. For the MCRT runs we have used $10^5$ photons, static core-skipping ($x_{\rm crit} = 2$), and fixed $T = 10^4 \, \rm K$ and $\taucl = 10^8$, so that $a_{\rm v} \taucl = 4.7 \times 10^4$. We are therefore in the regime where the analytical solution is expected to be valid, i.e. $a_{\rm v} \taucl \gg 1000$. We expect the agreement to become even better for larger $a_{\rm v} \taucl$.
}
    \label{fesc Lya prediction vs Monte Carlo}
\end{figure}

We begin by solving the radiative transfer equation for a uniform cloud. As shown in Appendix \ref{AppendixRTeq}, the Ly$\alpha$ radiative equation for an optically thick spherical uniform cloud in the comoving frame reads:
\begin{align}
    \underbrace{\dfrac{\partial^2 J}{\partial \tau^2} + \dfrac{2}{\tau} \dfrac{\partial J}{\partial \tau}}_{\textrm{Spatial diffusion}} + \underbrace{\dfrac{3\Dot{R}_{\rm cl}}{b \taucl } \mathcal{H} \dfrac{\partial J}{\partial x}}_{\textrm{Doppler shift}} ~=~ &-\underbrace{\dfrac{3 \mathcal{H} j_{\rm s}}{n_{\rm HI} \sigma_0}}_{\textrm{Emission}} + \underbrace{3 (p_{\rm d} \mathcal{H}^2 + \epsilon \mathcal{H}) J}_{\textrm{Ly}\alpha\textrm{ destruction}} \nonumber \\ ~&-~ \underbrace{\dfrac{3}{2} \mathcal{H} \dfrac{\partial}{\partial x} \left( \mathcal{H} \dfrac{\partial J}{\partial x} + 2\Bar{x} \mathcal{H} J \right)}_{\textrm{Frequency diffusion \& recoil}} \hspace{1 pt} \label{radtransfereq}.
\end{align}
The definitions of the variables that enter this equation (and others) can be found in Table \ref{TableSymbols}. Here $J(x,\tau)$ is the mean specific intensity (erg cm$^{-2}$ s$^{-1}$ Hz$^{-1}$ sr$^{-1}$), $x \equiv (\nu - \nu_{\textrm{Ly}\alpha})/\Delta \nu_{\rm D}$ is the frequency displacement from line centre in units of the Doppler width, and $\tau = n_{\rm HI} \sigma_0 r$ is the Ly$\alpha$ optical depth at line centre from cloud centre, with $\taucl \equiv  n_{\rm HI} \sigma_0 R_{\rm cl}$ being the cloud optical depth. $\mathcal{H}(x)$ is the dimensionless Voigt profile, and $p_{\rm d}$ is the probability of Ly$\alpha$ photon destruction per scattering event (i.e. it is one minus the single-scattering albedo), assumed here to be $\ll 1$ (as will be shown to be the case in Sec. \ref{Lya destruction}). A non-zero $p_{\rm d}$ can arise from $2p \rightarrow 2s$ transitions as we will see in detail later. For future reference, the Ly$\alpha$ cross-section at line centre $\sigma_0$ and the Voigt profile are given by \citep[e.g.][]{Dijkstra2014}:\footnote{Far from line centre, the profile is modified because of Rayleigh scattering. This lowers the cross-section by a factor proportional to $(\nu/\nu_{\rm Ly\alpha})^4$ on the red side \citep[e.g.][]{Peebles1993}. \cite{McKee2008} took this effect into account in their modelling of Ly$\alpha$ feedback from Pop III protostars, arguing that it could be important at very high column densities $\gtrsim 3 \times 10^{24} ~ \rm cm^{-2}$. We will ignore this effect because more accurate calculations of the effect of Rayleigh scattering, taking many different states of hydrogen into account \citep[in contrast to][]{Peebles1993}, have shown that the deviation from a Lorentzian profile (assumed to get the Voigt profile) is negligible for $\lvert \Delta \nu \rvert/\nu_{\rm Ly\alpha} \lesssim 0.2$ \citep[see e.g.][and discussion therein]{Mortlock2016, Kokubo2024}. }
\begin{align}
    \sigma_0 &=~ 5.88 \times 10^{-13} \hspace{1 pt} T_{100}^{-1/2} \hspace{1 pt} \rm cm^2 \hspace{1 pt} \label{sigma0 definition}, \\ \mathcal{H}(x) &= \dfrac{a_{\rm v}}{\pi} \int_{-\infty}^{\infty} \textrm{d}u \hspace{1 pt} \dfrac{e^{-u^2}}{(x-u)^2 + a_{\rm v}^2} \hspace{ 1 pt} , \label{Voigt definition}
\end{align}
where $a_{\rm v} = 4.7 \times 10^{-3} \hspace{1 pt} T_{100}^{-1/2}$ is the Voigt parameter, and $T_{\rm x} \equiv T/\rm x ~ \rm K$. The Voigt profile has the normalization $\int_{-\infty}^{\infty} \textrm{d}x \hspace{1 pt} \mathcal{H}(x) = \sqrt{\pi}$, and the following approximate behaviour in the core and in the wings:
\begin{align}
    \mathcal{H}(x) \simeq \begin{cases}
        e^{-x^2} & \quad \lvert x \rvert < x_{\rm crit} \\ \dfrac{a_{\rm v}}{\sqrt{\pi} x^2} & \quad \lvert x \rvert > x_{\rm crit} 
    \end{cases} \label{VoigtApprox},
\end{align}
with $x_{\rm crit} \simeq 2-4$ depending on the temperature \citep[see e.g. fig. 1 in][]{Smith2015}. $\Dot{R}_{\rm cl}$ is the cloud expansion or contraction rate, and $b = \sqrt{2 k_{\rm B}T/m_{\rm H}}$ is the thermal velocity. 

Equation (\ref{radtransfereq}) describes the emission and spatial diffusion of Ly$\alpha$ photons in the cloud, as well as their red/blueshift (for non-zero $\Dot{R}_{\rm cl}$), and frequency diffusion using a Fokker-Planck approximation to the frequency redistribution. For a static cloud ($\Dot{R}_{\rm cl} = 0$) and no Ly$\alpha$ destruction ($p_{\rm d} = \epsilon = 0$), Eq.~(\ref{radtransfereq}) reduces to the equation solved by several previous authors \citep[e.g.][]{Dijkstra2006, Lao2020, Seon2020, Tomaselli2021, McClellan2022}. We can solve Eq.~(\ref{radtransfereq}) following earlier work using an eigenfunction expansion Ansatz:
\begin{equation}
    J(x,\tau) = \sum_{n=1}^{\infty} T_n(\tau)f_n(x) \hspace{1 pt} \label{variable separataion eq}.
\end{equation}
It is also convenient to follow \cite{Harrington1973} and introduce a new frequency variable $y(x)$:
\begin{equation}
    y(x) \equiv \sqrt{\frac{2}{3}} \int_0^{x} \dfrac{\textrm{d}u}{\mathcal{H}(u)} \simeq \begin{cases}
        \sqrt{\dfrac{2}{3}} x  & \quad \lvert x \rvert < x_{\rm crit} \\ \sqrt{\dfrac{2 \pi}{3}} \dfrac{x^3}{3 a_{\rm v}} & \quad \lvert x \rvert > x_{\rm crit}
    \end{cases} \label{y(x) def}.
\end{equation}
Using the above Ansatz and variable change in Eq.~(\ref{radtransfereq}) yields 
\begin{align}
     \sum_{n} \left( \dfrac{\textrm{d}^2 T_n}{\textrm{d} \tau^2} + \dfrac{2}{\tau} \dfrac{\textrm{d} T_n}{\textrm{d} \tau} \right) f_n =~ &- \dfrac{3 \mathcal{H} j_{\rm s}}{ n_{\rm HI} \sigma_0} +  \sum_{n} 3 (p_{\rm d} \mathcal{H}^2 + \mathcal{H} \epsilon) f_n T_n \nonumber  \\ &-~ \sum_{n} \left[\gamma \dfrac{\textrm{d}f_n}{\textrm{d}y} +   \dfrac{\textrm{d}^2 f_n}{\textrm{d}y^2} \right]  T_n \nonumber \\  &-~ \sum_{n} \sqrt{6} \Bar{x} \left[ \dfrac{\textrm{d}}{\textrm{d}y} \left( \mathcal{H} f_n \right) \right] T_n  \hspace{1 pt} \label{before integrating},
\end{align}
where $\gamma \equiv \sqrt{6} \Dot{R}_{\rm cl}/b \taucl$. We can choose the eigenfunctions $T_n(\tau)$ such that 
\begin{equation}
    \dfrac{\textrm{d}^2 T_n}{\textrm{d} \tau^2} + \frac{2}{\tau} \dfrac{\textrm{d} T_n}{\textrm{d} \tau} = - \lambda_n^2 T_n \hspace{1 pt} \label{Tn equation},
\end{equation}
for some eigenvalue $\lambda_n$ to be determined. The solution which is finite at the cloud centre is then $T_n(\tau) = C_n \sin(\lambda_n \tau)/\tau$, where $C_n$ is a constant. We can determine the eigenvalues $\lambda_n$ from the boundary condition that there is no incoming Ly$\alpha$ intensity at the cloud edge, $\tau = \tau_{\rm cl}$. Using the two-stream approximation \citep[e.g.][]{Rybicki1986, Lao2020, Nebrin2023}, the incoming intensity is $I^- = J - \sqrt{3}H$, where $H \equiv \frac{1}{2} \int_{-1}^{1} \textrm{d}\mu \hspace{1 pt} \mu I$. Since we have assumed the Eddington approximation, we also have $H = - (1/3 \mathcal{H}) (\partial J/\partial\tau)$. Using this and the boundary condition $I^-(\tau_{\rm cl}) = 0$ then leads to
\begin{equation}
    \dfrac{1}{\sqrt{3} \mathcal{H}(x)} \dfrac{\textrm{d}T_n}{\textrm{d}\tau}\bigg|_{\taucl} + T_n(\tau_{\rm cl}) = 0 \hspace{1 pt} \label{BCcloudedge}.
\end{equation}
This leads to $\lambda_n \simeq n \pi /\tau_{\rm cl}$ for large $\tau_{\rm cl}$. We therefore find that the eigenfunctions are orhogonal, i.e. $\int_0^{\tau_{\rm cl}} \textrm{d}\tau \hspace{1 pt} \tau^2 T_n(\tau) T_{n'}(\tau) \propto \delta_{n',n}$. We can normalize the eigenfunctions such that $4 \pi \int_0^{\tau_{\rm cl}} \textrm{d}\tau \hspace{1 pt} \tau^2 T_n^2(\tau) = 1$, giving us
\begin{equation}
    T_n(\tau) = \dfrac{\sin(\lambda_n \tau)}{\sqrt{2 \pi \tau_{\rm cl}} \tau} \hspace{1 pt} \label{Tn sol}.
\end{equation}
Using Eq.~(\ref{Tn equation}) in Eq.~(\ref{before integrating}), integrating both sides over $4 \pi \int_0^{\tau_{\rm cl}} \tau^2 T_n$, and rearranging then yields an equation for the frequency dependence:
\begin{align}
    \dfrac{\textrm{d}^2 f_n}{\textrm{d}y^2} + \gamma \dfrac{\textrm{d} f_n}{\textrm{d}y} + \sqrt{6} \Bar{x} \dfrac{\textrm{d}}{\textrm{d}y} \left( \mathcal{H} f_n \right)  &=~ \lambda_n^2 f_n + 3 (p_{\rm d} \mathcal{H}^2 + \epsilon \mathcal{H}) f_n \nonumber \\ &-~ 4\pi \int_0^{\tau_{\rm cl}} \textrm{d}\tau \hspace{1 pt} \tau^2  \dfrac{3 \mathcal{H} j_{\rm s} T_n}{ n_{\rm HI} \sigma_0}\hspace{1 pt}. \label{freqeq1}
\end{align}
This differential equation is hard to solve exactly, not in the least because of the $\mathcal{H}^2$ term on the right-hand side. However, \cite{Harrington1973} pointed out that such terms are sharply peaked around $y = 0$, and hence can be approximated as Dirac delta functions. In particular, since $\int_{-\infty}^{\infty} \textrm{d}y \hspace{1 pt} \mathcal{H}^2(y) = \sqrt{2/3} \int_{-\infty}^{\infty} \textrm{d}x \hspace{1 pt} \mathcal{H}(x) = \sqrt{2\pi/3}$ we can simply make the approximate replacement $\mathcal{H}^2 \rightarrow \sqrt{2 \pi/3} \hspace{1 pt} \delta_{\rm D}(y)$ in Eq.~(\ref{freqeq1}), which renders the differential equation analytically solvable. Similarly, since the source term $\propto j_{\rm s}$ is also sharply peaked around $y = 0$, we can take it as proportional to $\delta_{\rm D}(y)$ too. We then end up with:
\begin{align}
    \dfrac{\textrm{d}^2 f_n}{\textrm{d}y^2} + \gamma \dfrac{\textrm{d} f_n}{\textrm{d}y} + \sqrt{6} \Bar{x} \dfrac{\textrm{d}}{\textrm{d}y} \left( \mathcal{H} f_n \right)  ~&=~ (\lambda_n^2 + 3 \epsilon \mathcal{H} )f_n \nonumber \\~&+~ \left[ \sqrt{6 \pi} p_{\rm d} f_n - \mathcal{I}_{\textrm{s},n} \right] \delta_{\rm D}(y) \hspace{1 pt} , \label{freqeq2}
\end{align}
where $\mathcal{I}_{\textrm{s}, n}$ is a constant dependent on the spatial distribution of sources:
\begin{align}
    \mathcal{I}_{\textrm{s}, n} &\equiv~ \int_{-\infty}^{\infty} \textrm{d}y \left[ 4\pi \int_0^{\tau_{\rm cl}} \textrm{d}\tau \hspace{1 pt} \tau^2  \dfrac{3 \mathcal{H}(y) j_{\rm s}(\tau, y) T_n(\tau)}{ n_{\rm HI} \sigma_0} \right] \nonumber \\
    &=~ \dfrac{4 \sqrt{6} \pi}{n_{\rm HI} \sigma_0 \Delta \nu_{\rm D}} \int_0^{\infty} \textrm{d}\nu \int_0^{\taucl}  \textrm{d}\tau \hspace{1 pt}  \dfrac{\sin(n \pi \tau/\taucl) \tau}{\sqrt{2 \pi \taucl}} j_{\rm s}(\tau,\nu) \hspace{1 pt} .
\end{align}
Eqs.~(\ref{y(x) def}) and (\ref{Tn sol}) have been used to get the second line. With Eq.~(\ref{js}) for the emissivity $j_{\rm s}$, this can be evaluated, giving us
\begin{equation}
    \mathcal{I}_{\textrm{s}, n} = \dfrac{3\sqrt{6} L_{\rm Ly\alpha} \taucl}{4 \pi \Delta \nu_{\rm D} R_{\rm cl}^2} \dfrac{[\sin(n\pi \Bar{\tau}_{\rm s}) - n\pi \Bar{\tau}_{\rm s} \cos( n\pi \Bar{\tau}_{\rm s})]}{ \sqrt{2 \pi \taucl} \Bar{\tau}_{\rm s}^3 (n \pi)^2} \hspace{1 pt} \label{Isn},
\end{equation}
where $\Bar{\tau} \equiv \tau/\taucl$ (so $\Bar{\tau}_{\rm s} = n_{\rm HI} \sigma_0 R_{\rm s}/\taucl$).

In the absence of recoil and continuum absorption (i.e. $\Bar{x} = \epsilon = 0$) and away from line centre, Eq.~(\ref{freqeq2}) has simple exponentially growing and decaying solutions. To get the physically relevant solution we impose the boundary condition $f_n(\lvert y \rvert = \infty) = 0$, since there should be no intensity sufficiently far from line centre. The normalization can be determined by the jump in derivative across $y = 0$, with the following result \textit{if we ignore recoil and continuum absorption}:
\begin{equation}
     f_n(y) = \dfrac{\mathcal{I}_{\textrm{s}, n} \tau_{\rm cl}}{2(\mathcal{P} + n \pi \Bar{\mathcal{D}}_n)} \exp\left[  - \dfrac{\sqrt{6} (\Dot{R}_{\rm cl}/b) y}{2\taucl} - \dfrac{n \pi \Bar{\mathcal{D}}_n \lvert y \rvert}{\tau_{\rm cl}} \right] \hspace{1 pt} \label{Solution f_n no recoil or dust},
\end{equation}
where $\mathcal{P} \equiv \sqrt{6 \pi} p_{\rm d} \tau_{\rm cl}/2$, and
\begin{equation}
    \Bar{\mathcal{D}}_n \equiv  \sqrt{1 +   \dfrac{3}{2 (n \pi)^2} \left( \dfrac{\Dot{R}_{\rm cl}}{b} \right)^2 } \hspace{1 pt} \label{Overbar D_n definition}.
\end{equation}
In the more general and realistic case with recoil and continuum absorption, Eq.~(\ref{freqeq2}) becomes significantly harder to solve. We derive a series solution in Appendix \ref{AppendixRTfreq}, generalizing the solution found by \cite{Neufeld1990} to also include recoil and (approximately) cloud expansion/contraction. The solution can be written in the form:
\begin{equation}
    f_n(z) = \dfrac{\mathcal{I}_{\textrm{s}, n} \tau_{\rm cl}}{2(\mathcal{P} + n \pi\mathcal{D}_n)} \mathcal{F}_n(z) \hspace{1 pt} \label{f_n expression general},
\end{equation}
where $x = z \beta_n$, with $\beta_n \equiv (3 a_{\rm v}^2/ 2\pi \lambda_n^2)^{1/6}$, and $\mathcal{F}_n(z)$ and $\mathcal{D}_n$ depends on recoil, velocity gradients, and continuum absorption via the dimensionless parameters $\Tilde{x}_n \equiv 2 \beta_n \Bar{x}$, $\Tilde{\gamma}_n \equiv \gamma \sqrt{2\pi/ 3} \beta_n^3 /a_{\rm v}$, and $\Tilde{\epsilon}_n \equiv 2 \sqrt{\pi} \beta_n^4 \epsilon/a_{\rm v}$, respectively. For the dimensionless recoil parameter we find:
\begin{align}
    \Tilde{x}_n ~&=~ \left( \dfrac{3}{2 \pi^3} \right)^{1/6} \dfrac{(a_{\rm v} \taucl)^{1/3}}{n^{1/3}} \dfrac{h \Delta\nu_{\rm D}}{k_{\rm B} T} \nonumber \\ ~&=~ \dfrac{1.10}{n^{1/3}} \left(\dfrac{\taucl}{10^{10}}\right)^{1/3} T_{100}^{-2/3} \hspace{1 pt} \label{xtilde expression}.
\end{align}
For the dimensionless velocity gradient parameter:
\begin{align}
     \Tilde{\gamma}_n ~&=~ \dfrac{\sqrt{6}\Dot{R}_{\rm cl}}{n \pi b }  \nonumber \\  ~&=~ \dfrac{6.07}{n} \left( \dfrac{\Dot{R}_{\rm cl}}{10\,\rm km ~ s^{-1}} \right) T_{100}^{-1/2} \hspace{1 pt} \label{gammatilde expression}.
\end{align}
And for the dimensionless continuum absorption parameter:
\begin{align}
    \Tilde{\epsilon}_n ~&=~ \dfrac{2\sqrt{\pi}}{n^{4/3}} \left( \dfrac{3}{2 \pi^3} \right)^{2/3}   (a_{\rm v}\taucl)^{1/3} \tau_{\rm c, abs}  \nonumber \\ ~&=~ \dfrac{170}{n^{4/3}} \left(\dfrac{\taucl}{10^{10}}\right)^{1/3} \tau_{\rm c, abs} T_{100}^{-1/6} \hspace{1 pt} \label{epsilontilde expression}.
\end{align}
The full series solution for $\mathcal{F}_n(z)$ can be found in Eq.~(\ref{f_n series solution final}). Unfortunately, there is no simple expression for $\mathcal{F}_n$ or $\mathcal{D}_n$ in the general case, since the series solution has coefficients that are determined recursively \citep[as in the series solution found by][]{Neufeld1990}. However, we note that for Ly$\alpha$ feedback (and some other quantities like the Ly$\alpha$ escape fraction and trapping time) we will only need  $\mathcal{D}_n$ and integrals of $\mathcal{F}_n(z)$, which are only functions of $\Tilde{x}_n$, $\Tilde{\gamma}_n$, and $\Tilde{\epsilon}_n$. 

With Eq.~(\ref{f_n expression general}), we can write the full series solution for the Ly$\alpha$ mean specific intensity:
\begin{equation}
    J(x, \Bar{\tau}) = \sum_{n=1}^{\infty} \dfrac{\mathcal{I}_{\textrm{s}, n}}{\sqrt{8 \pi \taucl}(\mathcal{P} + n \pi\mathcal{D}_n)}  \dfrac{\sin(n \pi \Bar{\tau})}{\Bar{\tau}}  \mathcal{F}_n(z)  \hspace{1 pt} ,
\end{equation}
or written out in full using Eq.~(\ref{Isn}):
\begin{align}
    J(x, \Bar{\tau}) &=~ \dfrac{3 \sqrt{6} L_{\rm Ly\alpha}}{(4\pi)^2 \Delta\nu_{\rm D} R_{\rm cl}^2 } \nonumber \\ &\times~ \sum_{n = 1}^{\infty} \dfrac{[\sin(n \pi \Bar{\tau}_{\rm s}) - n\pi \Bar{\tau}_{\rm s}\cos(n \pi \Bar{\tau}_{\rm s})]}{\Bar{\tau}_{\rm s}^3 (n\pi)^2 (\mathcal{P} + n \pi \mathcal{D}_n)} \dfrac{\sin(n \pi \Bar{\tau})}{\Bar{\tau}} \mathcal{F}_n(z) \hspace{1 pt} \label{series solution}.
\end{align}
With this solution we can derive the corresponding emergent spectrum $J(x)$ at the cloud edge, and the Ly$\alpha$ escape fraction $f_{\rm esc, Ly\alpha}$. This is done in Appendix \ref{AppendixSpectrum} and \ref{Escape fraction appendix}, respectively. For the emergent spectrum we find: 
\begin{align}
    J(x) ~&=~ -\dfrac{\sqrt{18} L_{\rm Ly\alpha}}{(4\pi)^2  \taucl \Delta \nu_{\rm D} R_{\rm cl}^2 \mathcal{H}(x)} \nonumber \\ ~&\times ~\sum_{n=1}^{\infty} \dfrac{ [\sin(n\pi \Bar{\tau}_{\rm s}) - n\pi \Bar{\tau}_{\rm s} \cos( n\pi \Bar{\tau}_{\rm s})] }{\Bar{\tau}_{\rm s}^3 (n \pi) (\mathcal{P} + n\pi \mathcal{D}_n)} \hspace{1 pt}  (-1)^n \mathcal{F}_n(z) \hspace{1 pt} \, \label{spectrum main text},
\end{align}
and for the escape fraction:\footnote{This series converges very slowly for a point source. We therefore used Wynn's Epsilon method to speed up convergence.}
\begin{align}
    f_{\rm esc, Ly\alpha} ~&=~ -3 \sum_{n = 1}^{\infty} (-1)^n \dfrac{[\sin(n \pi \Bar{\tau}_{\rm s}) - n\pi \Bar{\tau}_{\rm s}\cos(n \pi \Bar{\tau}_{\rm s})]}{\Bar{\tau}_{\rm s}^3 (n\pi)^2 (\mathcal{P} + n \pi \mathcal{D}_n)} \nonumber\\ ~&\times~ \int_{-\infty}^{\infty} \textrm{d}z \hspace{1 pt} z^2  \mathcal{F}_{n}(z) \hspace{1 pt} \label{fesc main text}.
\end{align}
We plot a few example spectra in Fig. \ref{spectrum}, and the escape fraction in Fig. \ref{fesc Lya prediction vs Monte Carlo}. Also shown in these plots are our results from Monte Carlo RT (MCRT) experiments for the same setups, using \textsc{colt} \citep[the Cosmic Lyman-$\alpha$ Transfer code,][]{Smith2015}.\footnote{\cite{ForeroRomero2011} and \cite{Garavito2014} have also estimated the Ly$\alpha$ escape fraction for spherical clouds using MCRT simulations, but for $a_{\rm v} \taucl \sim 4700$, near the border where the diffusion approximation is expected to break down. These authors therefore find slightly lower values of $f_{\rm esc, Ly\alpha}$ than we do, but still qualitatively consistent with our results.}
We find excellent agreement between the analytical solution and the MCRT results, both for the spectra and the escape fraction. We perform additional tests in Sec. \ref{Force multiplier} for the predicted Ly$\alpha$ radiation pressure, finding good agreement with MCRT simulations. These tests show that the new analytical solution accurately captures Ly$\alpha$ scattering in optically thick clouds. A companion study of Ly$\alpha$ RT in moving media, including predicting spectra, is presented in \cite{Smith2025}.

\subsubsection{Approximate solution for a spherical turbulent \HI cloud}
\label{turbulent cloud solution section}

\begin{figure}
    \includegraphics[trim={0.0cm 0.0cm 0.0cm 0.0cm},clip,width = \columnwidth]{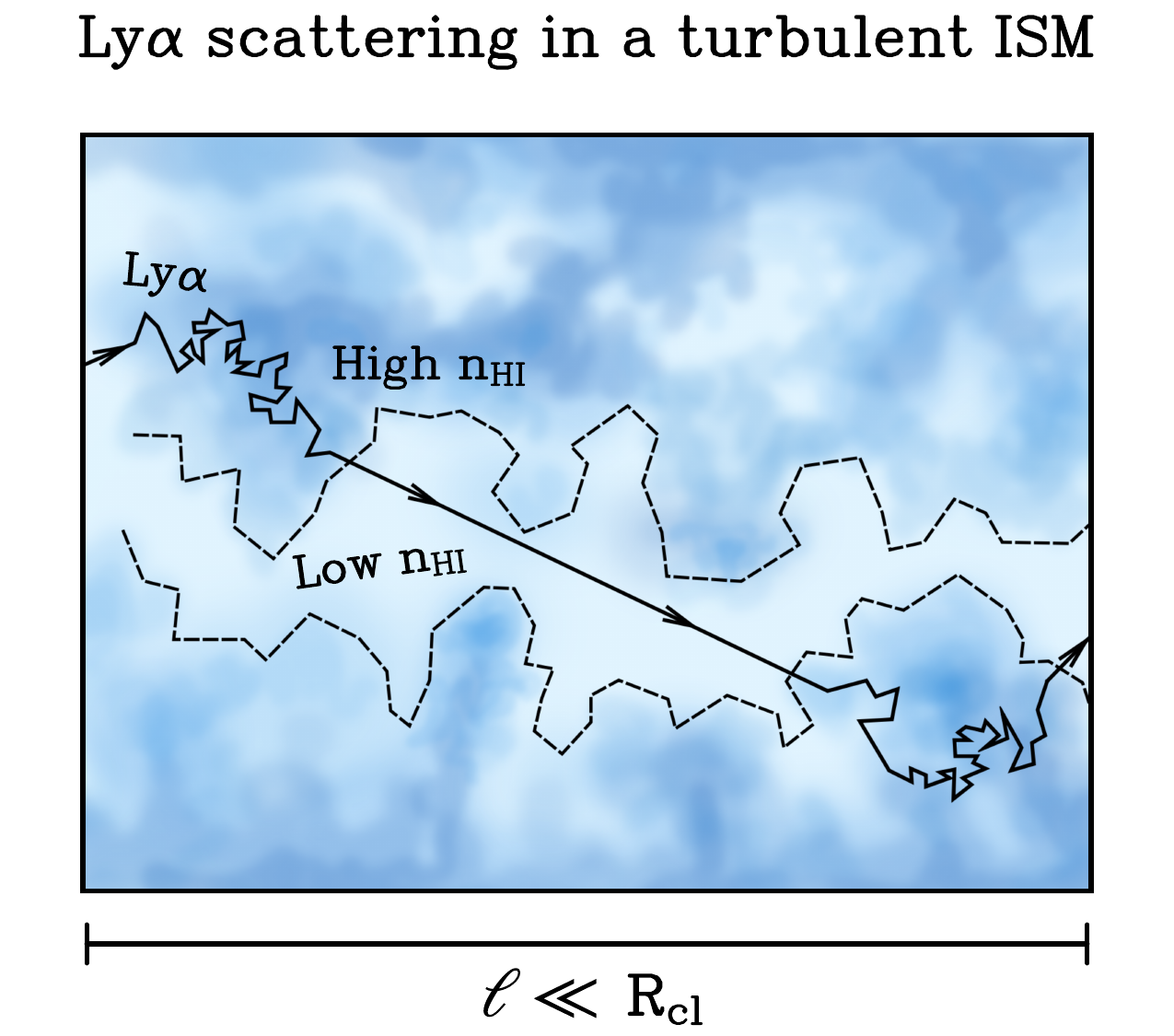}
    \caption{A schematic sketch illustrating the impact of small-scale ($\ell \ll R_{\rm cl}$) turbulent density fluctuations on Ly$\alpha$ scattering. Dark (light) regions illustrate dense (tenuous) regions. A scattering Ly$\alpha$ photon (solid line) can preferentially escape along low-density channels. The goal of asymptotic homogenization theory (HT), as applied to Ly$\alpha$ RT, is to find the \textit{effective} spatial diffusion term on scales $\gtrsim \ell$, by averaging over the small-scale ($\lesssim \ell$) fluctuations in a rigorous manner.
}
    \label{Schematic turbulent media}
\end{figure}

Real clouds in the ISM are not perfectly uniform. Supersonic turbulence, for instance, can generate large fluctuations in volume densities, column densities, and velocity gradients \citep[e.g.][]{Federrath2010, Federrath2013, Thompson2016, Buck2022}. As illustrated in Fig. \ref{Schematic turbulent media}, Ly$\alpha$ photons are expected to escape more easily via low-density channels, and spatially fluctuating velocity gradients could induce large Doppler shifts of Ly$\alpha$ photons and further aid in their escape \citep{Munirov2023}. To assess the corresponding impact on Ly$\alpha$ feedback, we have to consider Ly$\alpha$ RT in non-uniform clouds.
If we allow for spatial dependence of the gas density and velocity, the Ly$\alpha$ RT equation is generalized to (Eq.~\ref{Appendix Lya RT equation final general result}):
\begin{align}
    \underbrace{ \dfrac{1}{\alpha_0} \boldsymbol{\nabla} \boldsymbol{\cdot} \left( \dfrac{1}{\alpha_0}  \boldsymbol{\nabla} J  \right)}_{\textrm{Spatial diffusion}} + \underbrace{\dfrac{(\boldsymbol{\nabla \cdot u})}{b \alpha_0 } \mathcal{H} \dfrac{\partial J}{\partial x}}_{\textrm{Doppler shift}} ~&=~ -\underbrace{\dfrac{3 \mathcal{H} j_{\rm s}}{\alpha_0}}_{\textrm{Emission}} + \underbrace{3 (p_{\rm d} \mathcal{H}^2 + \epsilon \mathcal{H}) J}_{\textrm{Ly}\alpha\textrm{ destruction}} \nonumber \\ ~-&~ \underbrace{\dfrac{3}{2} \mathcal{H} \dfrac{\partial}{\partial x} \left( \mathcal{H} \dfrac{\partial J}{\partial x} + 2\Bar{x} \mathcal{H} J \right)}_{\textrm{Frequency diffusion \& recoil}} \, , \label{Lya RT equation general main text} 
\end{align}
where $\alpha_0(\boldsymbol{r}) \equiv n_{\rm HI} (\boldsymbol{r}) \sigma_0$, and $\boldsymbol{u}(\boldsymbol{r})$ is the gas velocity. There is no general exact analytical solution of Eq.~(\ref{Lya RT equation general main text}).\footnote{Indeed, similar problems of radiative diffusion in inhomogeneous clouds arise in atmospheric physics, where only limited analytical progress has been made \citep[for a review, see][]{Davis2001}.} However, we can render Eq.~(\ref{Lya RT equation general main text}) analytically tractable by utilizing asymptotic homogenization theory (HT) \citep[for good overviews, see e.g.][]{Zwillinger1992_Diffeqs, Holmes2013_Perturbation}. HT is a powerful technique that can approximately solve diffusion equations with diffusion coefficients that exhibit large spatial fluctuations, provided these fluctuations occur on relatively small scales. With HT applied to Ly$\alpha$ RT (Eq.~\ref{Lya RT equation general main text}), we make the following assumptions:
\begin{enumerate}
    \item We assume that the gas density and properties ($\alpha_0$, $p_{\rm d}$, and $\epsilon$), and the gas velocity $\boldsymbol{u}$ fluctuate on some formally small scale $\ell \ll R_{\rm cl}$ (see Fig. \ref{Schematic turbulent media} for a schematic sketch). Most often it is assumed that the fluctuations are periodic over `cells' of volume $\sim \ell^3$, which can make the derivation more convenient \citep[e.g.][]{Zwillinger1992_Diffeqs, Holmes2013_Perturbation}. However, HT can also be applied for random fluctuations \citep[e.g.][]{Kozlow1993_centrallimit, Hanasoge2013}. Here, we model the gas density as randomly fluctuating on small scales. \\

    \item Because of the first assumption, we assume there is a separation of scales, such that we can generally write $\alpha_0 = \alpha_0(\boldsymbol{r}, \boldsymbol{\xi})$, and similarly for $\boldsymbol{u}$, $p_{\rm d}$, and $\epsilon$. Here $\boldsymbol{\xi} \equiv \boldsymbol{r}/\varepsilon$ is the `fast' spatial variable, characterizing fluctuations on small scales $\ll R_{\rm cl}$. For simplicity, we assume that the cloud is approximately uniform on large scales, so that $\alpha_0$, $p_{\rm d}$, and $\epsilon$ are only functions of $\boldsymbol{\xi}$. 
\end{enumerate}
The goal is now to find the \textit{effective} Ly$\alpha$ RT equation, valid on scales $> \ell$. To do so, we expand $J(x,\boldsymbol{r}, \boldsymbol{\xi})$ as a power series in $\varepsilon \equiv \ell/R_{\rm cl}$, and use the chain rule for the gradient $\boldsymbol{\nabla}$:
\begin{align}
    J(x,\boldsymbol{r}, \boldsymbol{\xi}) &=~ J_{(0)}(x,\boldsymbol{r}) + \varepsilon J_{(1)}(x,\boldsymbol{r}, \boldsymbol{\xi}) + \varepsilon^2 J_{(2)}(x,\boldsymbol{r}, \boldsymbol{\xi}) + \mathcal{O}(\varepsilon^3) \, , \nonumber \\ \boldsymbol{\nabla} &=~ \boldsymbol{\nabla}_{\boldsymbol{r}} + \varepsilon^{-1} \boldsymbol{\nabla}_{\boldsymbol{\xi}} \, ,
\end{align}
where $\boldsymbol{\nabla}_{\boldsymbol{r}} \equiv \partial/\partial\boldsymbol{r}$ and $\boldsymbol{\nabla}_{\boldsymbol{\xi}} \equiv \partial/\partial\boldsymbol{\xi}$ are the gradients for the `slow' and `fast' variations, respectively. The assumption that $J_{(0)}$ is independent of $\boldsymbol{\xi}$ can be understood intuitively to capture the fact that $J$ should look uniform on scales $\gg \ell$ (but smaller than $R_{\rm cl}$), which can also be derived self-consistently in HT \citep[e.g.][]{Zwillinger1992_Diffeqs, Holmes2013_Perturbation}. Finally, we note that we do not expand $\boldsymbol{\nabla} \boldsymbol{\cdot} \boldsymbol{u}$, but rather treat it as a given function of $\boldsymbol{r}$ and $\boldsymbol{\xi}$, as done for $\alpha_0$. We now use the above expansions in Eq.~(\ref{Lya RT equation general main text}). The spatial diffusion term becomes:
\begin{align}
    \boldsymbol{\nabla} \boldsymbol{\cdot} \left( \dfrac{1}{\alpha_0}  \boldsymbol{\nabla} J  \right) &=~ \left[ \boldsymbol{\nabla}_{\boldsymbol{\xi}} \left( \dfrac{1}{\alpha_0} \right)  \boldsymbol{\cdot}  \boldsymbol{\nabla}_{\boldsymbol{r}} J_{(0)} + \boldsymbol{\nabla}_{\boldsymbol{\xi}} \boldsymbol{\cdot} \left( \dfrac{1}{\alpha_0} \boldsymbol{\nabla}_{\boldsymbol{\xi}} J_{(1)}\right) \right] \varepsilon^{-1} \nonumber \\ &+~  \dfrac{1}{\alpha_0} \nabla_{\boldsymbol{r}}^{2}J_{(0)} + \boldsymbol{\nabla}_{\boldsymbol{\xi}} \boldsymbol{\cdot} \left( \dfrac{1}{\alpha_0}   \boldsymbol{\nabla}_{\boldsymbol{r}}J_{(1)} \right) \label{diffusion operator 1} \\ &+~ \dfrac{1}{\alpha_0} \boldsymbol{\nabla}_{\boldsymbol{r}} \boldsymbol{\cdot} \boldsymbol{\nabla}_{\boldsymbol{\xi}}  J_{(1)} + \boldsymbol{\nabla}_{\boldsymbol{\xi}} \boldsymbol{\cdot} \left( \dfrac{1}{\alpha_0} \boldsymbol{\nabla}_{\boldsymbol{\xi}} J_{(2)} \right) + \mathcal{O}(\varepsilon) \, \nonumber.
\end{align}
If we insert this result into Eq.~(\ref{Lya RT equation general main text}), expand the rest of the terms, and collect terms by order in $\varepsilon$, we find that to order $\mathcal{O}(\varepsilon^{-1})$:
\begin{equation}
    0 = \boldsymbol{\nabla}_{\boldsymbol{\xi}} \left( \dfrac{1}{\alpha_0} \right)  \boldsymbol{\cdot}  \boldsymbol{\nabla}_{\boldsymbol{r}} J_{(0)} + \boldsymbol{\nabla}_{\boldsymbol{\xi}} \boldsymbol{\cdot} \left( \dfrac{1}{\alpha_0} \boldsymbol{\nabla}_{\boldsymbol{\xi}} J_{(1)}\right) \, \label{J_1 equation O(l^-1)},
\end{equation}
This can be solved using separation of variables. One finds that $J_{(1)} = \boldsymbol{\Lambda} \boldsymbol{\cdot} \boldsymbol{\nabla}_{\boldsymbol{r}} J_{(0)}$, where the components of the vector $\boldsymbol{\Lambda}$ satisfy \citep{Zwillinger1992_Diffeqs, Holmes2013_Perturbation}:
\begin{equation}
    \dfrac{\partial}{\partial \xi_j} \left( \dfrac{1}{\alpha_0} \dfrac{\partial \Lambda_i}{\partial \xi_j} \right) = - \dfrac{\partial}{\partial \xi_i} \left( \dfrac{1}{\alpha_0} \right) \, \label{cell equation},
\end{equation}
where we use the Einstein summation convention (both here and in the rest of the text). For periodic fluctuations, Eq.~(\ref{cell equation}) is to be solved for a given cell, assuming periodic boundary conditions: $\boldsymbol{\Lambda}(\boldsymbol{\xi}) = \boldsymbol{\Lambda}(\boldsymbol{\xi}+\boldsymbol{\xi}_{\rm p})$. In contrast, in the case of random fluctuations, Eq.~(\ref{cell equation}) is to be solved for $\boldsymbol{\xi} \in \mathbb{R}^3$, to then find the expectation value of $\boldsymbol{\Lambda}$. We can volume-average Eq.~(\ref{diffusion operator 1}), first noting that with periodic boundary conditions we have:\footnote{While this terminology is most clear for the case of periodic fluctuations within `cells', it carries over to random fluctuations. As pointed out by \cite{Hanasoge2013}, random media in this context can be viewed as periodic media with an infinite periodicity length scale.} 
\begin{equation}
    \left\langle \boldsymbol{\nabla}_{\boldsymbol{\xi}} \boldsymbol{\cdot} \left( \dfrac{1}{\alpha_0}   \boldsymbol{\nabla}_{\boldsymbol{r}}J_{(1)} \right) + \boldsymbol{\nabla}_{\boldsymbol{\xi}} \boldsymbol{\cdot} \left( \dfrac{1}{\alpha_0} \boldsymbol{\nabla}_{\boldsymbol{\xi}} J_{(2)} \right) \right\rangle = 0 \, ,
\end{equation}
where $\langle Q \rangle (\boldsymbol{r}) \equiv (1/\mathcal{V}) \int_{\mathcal{V}} \textrm{d}^3 \boldsymbol{\xi} \, Q(\boldsymbol{r}, \boldsymbol{\xi})$. Thus:
\begin{align}
    \left\langle \boldsymbol{\nabla} \boldsymbol{\cdot} \left( \dfrac{1}{\alpha_0}  \boldsymbol{\nabla} J  \right) \right\rangle &=~ \left\langle \dfrac{1}{\alpha_0} \right\rangle \nabla_{\boldsymbol{r}}^2 J_{(0)} + \left\langle \dfrac{1}{\alpha_0} \boldsymbol{\nabla}_{\boldsymbol{\xi}} \boldsymbol{\Lambda}  \right\rangle \boldsymbol{:} \boldsymbol{\nabla}_{\boldsymbol{r}} \boldsymbol{\nabla}_{\boldsymbol{r}} J_{(0)} \nonumber \\ &+~ \mathcal{O}(\varepsilon) \, \label{effective diffusion term 1},
\end{align}
with $\mathbfss{A} \boldsymbol{:} \mathbfss{B} \equiv \mathbfss{A}_{ij} \mathbfss{B}_{ij}$. This is the general expression for the effective Ly$\alpha$ spatial diffusion term. What remains to be determined is $\boldsymbol{\Lambda}$, governed by Eq.~(\ref{cell equation}). To solve it, we need to make additional assumptions about the \HI density fluctuations. 

Isothermal turbulence is well-known to produce an approximately log-normal distribution of densities \citep[e.g.][]{Passot1998, Federrath2010, Krumholz2014review}. Most of the observed distributions of column densities in dense clouds are consistent with this prediction \citep[e.g.][]{Kainulainen2013, Burkhart2015, Schneider2022_turb}, with some deviations, especially at very high densities, that could be the result of, e.g., self-gravity and stellar feedback \citep[][]{Burkhart2017_PLT, Appel2022, Schneider2022_turb}. As a simple first approximation, we adopt a log-normal probability density function (PDF) for the gas density:
\begin{equation}
    p_s(s) \, \textrm{d}s = \dfrac{1}{\sqrt{2 \pi \sigma_s^2}} \exp\left[ - \dfrac{(s - \mu_s)^2}{2 \sigma_s^2} \right] \, \textrm{d}s \, \label{turbulence PDF},
\end{equation}
where $s \equiv \ln(\rho/\langle \rho \rangle)$, and $\langle \rho \rangle$ is the volume-averaged gas density of the cloud. The mean $\mu_s$ and variance $\sigma_s^2$ of $s$ is determined by the 3D turbulent Mach number $\mathcal{M}$:
\begin{equation}
    \sigma_s^2 \simeq \ln\left( 1 + b_s^2 \mathcal{M}^2 \right) \, , \hspace{10 pt} \mu_s = -\dfrac{1}{2} \sigma_s^2 \label{turbulence variance}. 
\end{equation}
The parameter $b_s$ is $1$, $1/3$, and $\simeq 0.4$ for turbulence driven by purely compressive (curl-free) modes, purely solenoidal (divergence-free) modes, or a `natural' mix of the two, respectively \citep[e.g.][]{Federrath2008, Federrath2010, Krumholz2014review}. The latter is often adopted in analytical/subgrid models \citep[e.g.][]{Hopkins2012_GMCs, Faucher2013_feedbackregulated, Kimm2017minihaloes}, but observations and high-resolution simulations are consistent with a range of values, depending on the environment \citep[e.g.][]{Orkisz2017, Menon2020_HIIexpansionturbulence, Sharda2022_turbulenceobs, Gerrard2023}. For this reason we leave $b_s$ unspecified. For future reference, we provide some useful expectation values of the log-normal PDF in Table \ref{turbulent density fluctuations}.

\begin{table}
\caption{Useful volume-weighted expectation values in a turbulent ISM, assuming a log-normal distribution of densities (Eq.~\ref{turbulence PDF}).}
\begin{tabular}{l c}
\hline
\hline
Quantity & Meaning  \\

\hline
\hline

\\
\vspace{8 pt}

$\langle \rho^n \rangle =  \langle \rho \rangle^n\exp[n(n - 1) \sigma_s^2 /2 ] $ & Mean of $\rho^n$  \\

\vspace{8 pt}

$\exp[\langle \ln \rho \rangle] =  \langle \rho \rangle \exp(\mu_s) $ & Geometric mean of $\rho$  

\\

\hline
\hline
\end{tabular}
\vspace{1 pt}\\.
\label{turbulent density fluctuations}
\end{table}

We now approximately solve Eq.~(\ref{cell equation}) for $\boldsymbol{\Lambda}$, assuming that the \HI density has the PDF of Eq.~(\ref{turbulence PDF}). We follow the approach of \cite{Dagan1993}, who considered a nearly identical problem in the study of the effective conductivity in media with random log-normal fluctuations. First, we expand and re-write Eq.~(\ref{cell equation}) as follows:
\begin{equation}
    \nabla^2_{\boldsymbol{\xi}} \Lambda_i - \boldsymbol{\nabla}_{\boldsymbol{\xi}} s \boldsymbol{\cdot} \boldsymbol{\nabla}_{\boldsymbol{\xi}} \Lambda_i = \partial_{\xi_i} s \, , \hspace{10 pt} s = \ln(\alpha_0/\langle \alpha_0 \rangle) \,. \label{Gamma_i PDE exact}
\end{equation}
Next, we define $s - \mu_s \equiv \sigma_s \Tilde{s}$ and make a perturbative expansion in $\sigma_s$, i.e. $\Lambda_i = \sum_{k = 0}^\infty \Lambda_{(k), i} \sigma_s^k$, but keeping $\mu_s$ fixed. Inserting this expansion into Eq.~(\ref{Gamma_i PDE exact}) and collecting terms by order in $\sigma_s$ yields: 
\begin{align}
     \nabla^2_{\boldsymbol{\xi}} \Lambda_{(0),i} &=~ 0 \, , \label{Lambda0 eq}\\ \nabla^2_{\boldsymbol{\xi}} \Lambda_{(1),i} - \boldsymbol{\nabla}_{\boldsymbol{\xi}} \Tilde{s} \boldsymbol{\cdot} \boldsymbol{\nabla}_{\boldsymbol{\xi}} \Lambda_{(0),i} &=~ \partial_{\xi_i} \Tilde{s} \, , \label{Lambda1 eq}\\ \nabla^2_{\boldsymbol{\xi}} \Lambda_{(n),i} - \boldsymbol{\nabla}_{\boldsymbol{\xi}} \Tilde{s} \boldsymbol{\cdot} \boldsymbol{\nabla}_{\boldsymbol{\xi}} \Lambda_{(n-1),i} &=~ 0 \, , \hspace{10 pt} \textrm{for } n \geq 2 \, \label{Lambda2 eq}.
\end{align}
\cite{Dagan1993} considered $n \leq 3$, but in the interest of space we will only consider $n \leq 1$ here, and later comment on how the result is generalized by retaining higher-order terms. First we note that the finite solution of Eq.~(\ref{Lambda0 eq}), over $\boldsymbol{\xi} \in \mathbb{R}^3$, is simply $\Lambda_{(0),i} = 0$. With this result, Eq.~(\ref{Lambda1 eq}) can be solved for $\Lambda_{(1),i}$ using the Green's function $\mathcal{G}(\boldsymbol{\xi},\boldsymbol{\xi}')$ of the Laplacian, satisfying $\boldsymbol{\nabla}_{\boldsymbol{\xi}}^2 \mathcal{G} = - \delta_{\rm D}^{(3)}(\boldsymbol{\xi} - \boldsymbol{\xi}')$ \citep[e.g.][]{Arfken2013}. The first-order result for $\Lambda_i$ is then:
\begin{equation}
    \Lambda_{i}(\boldsymbol{\xi}) = -\int \textrm{d}^3 \boldsymbol{\xi}' \, \mathcal{G}(\boldsymbol{\xi},\boldsymbol{\xi}') \dfrac{\partial s}{\partial \xi_{i}'} + \mathcal{O}(\sigma_s^2) \hspace{1 pt} . \label{Lambda_i solution}
\end{equation}
If we also expand $1/\alpha_0 = \exp(-\mu_s-\sigma_s \Tilde{s})/\langle \alpha_0 \rangle$ in powers of $\sigma_s$, the $\boldsymbol{\Lambda}$-term in Eq.~(\ref{effective diffusion term 1}) becomes:
\begin{equation}
    \left\langle \dfrac{1}{\alpha_0}  \dfrac{\partial \Lambda_i}{\partial \xi_j}   \right\rangle = \dfrac{e^{-\mu_s}}{\langle \alpha_0 \rangle} \left\langle \left[1 - \sigma_s \Tilde{s} + \mathcal{O}(\sigma_s^2) \right]  \dfrac{\partial \Lambda_i}{\partial \xi_j}   \right\rangle  \, \label{Lambda_i average 1st order}.
\end{equation}
Using Eqs.~(\ref{Lambda_i solution})--(\ref{Lambda_i average 1st order}), and only retaining the lowest-order terms, yields:
\begin{align}
    \left\langle \dfrac{1}{\alpha_0}  \dfrac{\partial \Lambda_i}{\partial \xi_j}   \right\rangle &\simeq~   -\dfrac{e^{-\mu_s}}{\langle \alpha_0 \rangle} \left\langle [1- \sigma_s\Tilde{s}(\boldsymbol{\xi})]   \int \textrm{d}^3 \boldsymbol{\xi}' \,  \dfrac{\partial \mathcal{G}}{\partial \xi_j}  \dfrac{\partial s}{\partial \xi_{i}'}   \right\rangle  \nonumber \\ &=~ \dfrac{e^{-\mu_s}}{\langle \alpha_0 \rangle}     \left\langle [1- \sigma_s\Tilde{s}(\boldsymbol{\xi})] \int \textrm{d}^3 \boldsymbol{\xi}' \,  \dfrac{\partial^2 \mathcal{G}}{\partial \xi_i' \partial \xi_j }     s(\boldsymbol{\xi}') \right\rangle \, \nonumber ,
\end{align}
where we have used integration by parts on the second line. Next we note that $\partial^2 \mathcal{G}/\partial \xi_i' \partial \xi_j = -\partial^2 \mathcal{G}/\partial \xi_i' \partial \xi_j'$, and that
\begin{equation}
    \int \textrm{d}^3 \boldsymbol{\xi}' \,  \dfrac{\partial^2 \mathcal{G}}{\partial \xi_i' \partial \xi_j' }  s(\boldsymbol{\xi}') = \begin{cases}
        -\dfrac{1}{3} s(\boldsymbol{\xi}) & \quad \textrm{for } i=j \\ 0 & \quad \textrm{for }  i \neq j
    \end{cases} \, .
\end{equation}
We also have $\langle \Tilde{s}(\boldsymbol{\xi}) \rangle = 0$ and $\langle \Tilde{s}(\boldsymbol{\xi})\Tilde{s}(\boldsymbol{\xi}) \rangle = 1$ from Table \ref{turbulent density fluctuations}, recalling that $\Tilde{s} \equiv (s-\mu_s)/\sigma_s$. In summary, we find that
\begin{equation}
    \left\langle \dfrac{1}{\alpha_0}  \dfrac{\partial \Lambda_i}{\partial \xi_j}   \right\rangle =  -\dfrac{e^{-\mu_s}}{\langle \alpha_0 \rangle} \dfrac{1}{3} \sigma_s^2 \delta_{ij} + \textrm{higher-order terms} \, \label{1st order correction lambda deriv}.
\end{equation}
We also note that $\langle 1/\alpha_0 \rangle = e^{-\mu_s} \langle e^{-\sigma_s \Tilde{s}} \rangle/\langle \alpha_0 \rangle$. Taylor expanding $e^{-\sigma_s \Tilde{s}}$ yields:
\begin{equation}
    \left\langle \dfrac{1}{\alpha_0} \right\rangle = \dfrac{e^{-\mu_s}}{\langle \alpha_0 \rangle} \left(1 + \dfrac{1}{2} \sigma_s^2 \right) + \textrm{higher-order terms} \, \label{1st order correction 1/alpha0}.
\end{equation}
We can now combine Eqs.~(\ref{1st order correction lambda deriv})--(\ref{1st order correction 1/alpha0}) to determine the lowest-order effect of supersonic turbulence on the effective Ly$\alpha$ spatial diffusion term (Eq.~\ref{effective diffusion term 1}):
\begin{align}
    \left\langle \boldsymbol{\nabla} \boldsymbol{\cdot} \left( \dfrac{1}{\alpha_0}  \boldsymbol{\nabla} J  \right) \right\rangle &=~  \dfrac{e^{-\mu_s}}{\langle \alpha_0 \rangle}\left( 1 + \dfrac{1}{6} \sigma_s^2 \right) \nabla_{\boldsymbol{r}}^2 J_{(0)} \label{effective diffusion term 1st order} \\ &+~ \textrm{higher-order terms} \nonumber \, .
\end{align}
By retaining the next couple of higher-order terms in Eq.~(\ref{Lambda2 eq}), \cite{Dagan1993} showed that the expression in brackets is consistent with a Taylor expansion of $\exp( \sigma_s^2 /6)$ up to $\mathcal{O}(\sigma_s^4)$. The idea that the latter is the exact result, to all orders, is known as the Landau–Lifschitz–Matheron (LLM) conjecture \citep[after][]{Landau1960, Matheron1967}. The LLM conjecture has been discussed and tested by many other authors in different contexts \citep[e.g.][]{Dykaar1992, Neuman1993, Noetinger1994, Abramovich1995, DeWit1995, Renard1997, Sanchez2006, Jankovic2017}. 

\cite{Abramovich1995} and \cite{DeWit1995} have shown that the LLM conjecture becomes inexact at $\mathcal{O}(\sigma_s^6)$, i.e. at higher orders than those considered by \cite{Dagan1993}. Despite this, numerical simulations show that the LLM conjectured relation is a good approximation for $\sigma_s^2 \lesssim 7-8$ \citep{Dykaar1992, Neuman1993, Jankovic2017},\footnote{The deviation is on the order of $5\%$--$10\%$ for $\sigma_s^2 = 7-8$ in the simulations of \cite{Neuman1993} and \cite{Jankovic2017}, which is insignificant for the purposes of this paper. Although we are not aware of any data at higher $\sigma_s^2$, it seems likely that the LLM formula can be extrapolated up to at least $\sigma_s^2 \sim 10$, corresponding to $\mathcal{M} \sim 150/b_s$, and still give correct results within a few $10\%$, as long as the assumptions underlying HT still hold.} which in our context (Eq.~\ref{turbulence variance}) corresponds to a turbulent ISM with $\mathcal{M} \lesssim 55/b_s$.  We therefore adopt the LLM conjecture as an approximate non-linear extension of the first-order result in Eq.~(\ref{effective diffusion term 1st order}), and hence write the effective Ly$\alpha$ spatial diffusion term in turbulent media as
\begin{align}
     \left\langle \boldsymbol{\nabla} \boldsymbol{\cdot} \left( \dfrac{1}{\alpha_0}  \boldsymbol{\nabla} J  \right) \right\rangle &\simeq~ \dfrac{e^{2 \sigma_s^2 /3}}{\langle \alpha_0 \rangle}  \nabla_{\boldsymbol{r}}^2 J_{(0)} + \mathcal{O}(\varepsilon) \label{effective diffusion term final} \\ &\equiv~  \dfrac{1}{\Tilde{\alpha}_0} \nabla_{\boldsymbol{r}}^2 J_{(0)} + \mathcal{O}(\varepsilon) \, \nonumber ,
\end{align}
where we have used $\mu_s = - \sigma_s^2 /2$ (Eq.~\ref{turbulence variance}). We can now finally write down the effective, `homogenized', Ly$\alpha$ RT equation for $J_{(0)}$. Dropping subscripts, i.e. $J_{(0)} \rightarrow J$ and $\boldsymbol{\nabla}_{\boldsymbol{r}} \rightarrow \boldsymbol{\nabla}$, yields:
\begin{align}
     \dfrac{\nabla^2 J}{\Tilde{\alpha}_0 \langle \alpha_0 \rangle }   + \dfrac{\langle\boldsymbol{\nabla \cdot u}\rangle}{b \langle \alpha_0 \rangle } \mathcal{H} \dfrac{\partial J}{\partial x} &=~ -\dfrac{3 \mathcal{H} j_{\rm s}}{\langle \alpha_0 \rangle} + 3 [ \langle p_{\rm d} \rangle_{m} \mathcal{H}^2 + \langle \epsilon \rangle_{m} \mathcal{H}] J \nonumber \\ &-~ \dfrac{3}{2} \mathcal{H}  \dfrac{\partial}{\partial x} \left(  \mathcal{H} \dfrac{\partial J}{\partial x} +  2 \Bar{x} \mathcal{H} J \right) \, \label{homogenized Lya RT eq}, 
\end{align}
where we have introduced the small-scale \HI mass-weighted average $\langle Q \rangle_{m}$ of a function $Q(\boldsymbol{\xi})$:
\begin{equation}
    \langle Q \rangle_{m} \equiv \dfrac{\langle \alpha_0 Q \rangle}{\langle \alpha_0 \rangle} \, .
\end{equation}
We can solve Eq.~(\ref{homogenized Lya RT eq}) in the same way we did for a uniform cloud in Sec. \ref{uniform cloud solution}. While having small-scale turbulent density fluctuations, on large scales we assume that the cloud is approximately spherically symmetric and uniform \citep[for a similar setup in simulations of star formation in turbulent GMCs, see][]{Lane2022}. Next we use an eigenfunction expansion, $J(x,\boldsymbol{r}) = \sum_n T_n(\boldsymbol{r}) f_n(x)$, and find that the spatial eigenfunctions satisfy (c.f. Eq.~\ref{Tn equation}):
\begin{equation}
    \dfrac{\textrm{d}^2 T_n}{\textrm{d} r^2} + \dfrac{2}{r} \dfrac{\textrm{d} T_n}{\textrm{d} r} = - \lambda_n^2 \langle \alpha_0 \rangle T_n \, , \quad \int_{\mathcal{V}_{\rm cl}} \textrm{d}^3 \boldsymbol{r} \, \langle \alpha_0 \rangle  T_{n'} T_n = \delta_{n'n} \, ,
\end{equation}
where $\lambda_n$ is the eigenvalue, and where the last equation is a convenient normalization for the orthogonal eigenfunctions. The eigenfunctions and eigenvalues, satisfying the boundary condition of no incoming radiation at $r = R_{\rm cl}$, are then:
\begin{equation}
    T_n(r) = \dfrac{\sin(n \pi r/R_{\rm cl})}{ \sqrt{2 \pi \langle \taucl \rangle} r} \, , \quad \lambda_n \simeq \dfrac{n \pi}{\taucl^{\star}} \, ,
\end{equation}
where we have defined $\langle \taucl \rangle$ and $\taucl^{\star}$, respectively, as:
\begin{equation}
    \langle \taucl \rangle \equiv \langle \alpha_0 \rangle R_{\rm cl} \, , \quad \taucl^{\star} \equiv \sqrt{\langle \alpha_0 \rangle \Tilde{\alpha}_0} R_{\rm cl} = e^{- \sigma_s^2/3} \langle \taucl \rangle \, .
\end{equation}
Using $J(x,r) = \sum_{n'} T_{n'}(r) f_{n'}(x)$ in Eq.~(\ref{homogenized Lya RT eq}), multiplying by $\langle \alpha_0 \rangle T_n(r)$ and integrating over volume then yields
\begin{align}
    \dfrac{\textrm{d}^2 f_n}{\textrm{d}y^2} + \gamma \dfrac{\textrm{d} f_n}{\textrm{d}y} + \sqrt{6} \Bar{x} \dfrac{\textrm{d}}{\textrm{d}y} \left( \mathcal{H} f_n \right)  ~&=~ (\lambda_n^2 + 3 \langle \epsilon \rangle_{ m} \mathcal{H} )f_n  \label{freqeq non uniform} \\~&+~ \left[ \sqrt{6 \pi} \langle p_{\rm d} \rangle_{m} f_n - \mathcal{I}_{\textrm{s},n} \right] \delta_{\rm D}(y) \hspace{1 pt} , \nonumber
\end{align}
where $\gamma \equiv \sqrt{2/3} \langle \boldsymbol{\nabla} \boldsymbol{\cdot} \boldsymbol{u} \rangle/b\langle \alpha_0 \rangle$  (assumed constant), and $\mathcal{I}_{\textrm{s},n}$ is now
\begin{align}
    \mathcal{I}_{\textrm{s}, n} &\equiv~ \int_{-\infty}^{\infty} \textrm{d}y \, \int_{\mathcal{V}_{\rm cl}} \textrm{d}^3 \boldsymbol{r} \, 3 \mathcal{H} j_{\rm s} T_n \nonumber \\
    &=~ \dfrac{3\sqrt{6} L_{\rm Ly\alpha} }{4 \pi \Delta \nu_{\rm D} R_{\rm cl}} \dfrac{[\sin(n\pi \Bar{\tau}_{\rm s}) - n\pi \Bar{\tau}_{\rm s} \cos( n\pi \Bar{\tau}_{\rm s})]}{ \sqrt{2 \pi \langle \taucl \rangle} \Bar{\tau}_{\rm s}^3 (n \pi)^2} \, ,
\end{align}
where, as before, $\Bar{\tau}_{\rm s} = R_{\rm s}/R_{\rm cl}$. The solution for $f_n$ is equivalent to Eq.~(\ref{f_n expression general}) but with re-scaled variables that take turbulent fluctuations into account. The end result for the mean specific intensity becomes:
\begin{align}
    J(x, \Bar{\tau}) &=~ \dfrac{3 \sqrt{6} L_{\rm Ly\alpha}}{(4\pi)^2 \Delta\nu_{\rm D} R_{\rm cl}^2 } \dfrac{\taucl^\star}{\langle \taucl \rangle} \nonumber \\ &\times~ \sum_{n = 1}^{\infty} \dfrac{[\sin(n \pi \Bar{\tau}_{\rm s}) - n\pi \Bar{\tau}_{\rm s}\cos(n \pi \Bar{\tau}_{\rm s})]}{\Bar{\tau}_{\rm s}^3 (n\pi)^2 (\mathcal{P} + n \pi \mathcal{D}_n)} \dfrac{\sin(n \pi \Bar{\tau})}{\Bar{\tau}} \mathcal{F}_n(z) \hspace{1 pt} \label{series solution turbulent cloud},
\end{align}
where $\mathcal{P} = \sqrt{6 \pi} \langle p_{\rm d} \rangle_{ m} \taucl^{\star}$, $\Bar{\tau} = r/R_{\rm cl}$, and the dimensionless variables that determine $\mathcal{F}_n$ and $\mathcal{D}_n$ are now generalized as (c.f. Eqs. \ref{xtilde expression}--\ref{epsilontilde expression}):
\begin{align}
    \Tilde{x}_n &\equiv~ 2 \beta_n \Bar{x} \, = \, \left( \dfrac{3}{2 \pi^3} \right)^{1/6} \eta_{\rm t}^{1/6} \dfrac{[a_{\rm v} \langle \taucl \rangle]^{1/3}}{n^{1/3}} \dfrac{h \Delta\nu_{\rm D}}{k_{\rm B} T} \, ,  \\ \Tilde{\gamma}_n &\equiv~ \sqrt{\dfrac{2\pi}{3}} \dfrac{\gamma \beta_n^3}{a_{\rm v}} \, = \, \sqrt{\dfrac{2}{3}} \eta_{\rm t}^{1/2} \dfrac{\langle \boldsymbol{\nabla} \boldsymbol{\cdot} \boldsymbol{u} \rangle R_{\rm cl}}{n \pi b } \, , \\ \Tilde{\epsilon}_n &\equiv~ 2 \sqrt{\pi} \beta_n^4 \dfrac{\langle \epsilon \rangle_{m}}{a_{\rm v}}  \, = \, \dfrac{2 \sqrt{\pi} \langle \epsilon \rangle_{ m}}{n^{4/3} a_{\rm v}} \left( \dfrac{3}{2 \pi^3} \right)^{2/3} \eta_{\rm t}^{2/3} [a_{\rm v} \langle \taucl \rangle]^{4/3}  \, ,
\end{align}
where $\eta_{\rm t} \equiv \Tilde{\alpha}_0/\langle \alpha_0 \rangle$. We can simplify these expressions further under the following assumptions. First, that the turbulence is statistically isotropic entails that the small-scale turbulent contribution to $\langle \boldsymbol{\nabla} \boldsymbol{\cdot} \boldsymbol{u} \rangle$ averages out to $\sim$ zero \citep{Pan2018}. This conclusion is qualitatively consistent with Ly$\alpha$ MCRT simulations in turbulent clouds, which find a very small, if any, effect on the the ratio of the red and blue peaks in the emergent spectra \citep[][]{Kimm2019, Kakiichi2021}.\footnote{\cite{Kimm2019} simulated Ly$\alpha$ RT in turbulent star-forming GMCs. Despite initial turbulent Mach numbers of $\sim \rm few-10$ in the cold ($\sim 100 ~ \rm K$) gas, the spectra in their simulations are very symmetric with respect to the red and blue peaks (see their fig. 13). } Instead, the major contribution to $\boldsymbol{\nabla} \boldsymbol{\cdot} \boldsymbol{u}$ is expected to come from cloud-scale bulk gas motion (at radial expansion/contraction rate $\Dot{R}_{\rm cl}$).  

Finally, we also neglect fluctuations in $\epsilon \equiv \alpha_{\rm c,abs}/\alpha_0$ (e.g. due to variation in the dust-to-gas ratio within the cloud), so that $\langle \epsilon \rangle_{m} = \langle \tau_{\rm c,abs} \rangle/\langle \taucl \rangle$. With these simplifications the dimensionless parameters can be written as:
\begin{align}
    \Tilde{x}_n &=~  \left( \dfrac{3}{2 \pi^3} \right)^{1/6}  \dfrac{[a_{\rm v} \langle \taucl \rangle]^{1/3}}{n^{1/3} [1 + (b_s \mathcal{M})^2]^{1/9}} \dfrac{h \Delta\nu_{\rm D}}{k_{\rm B} T} \, \label{xtilde expression turb},  \\ \Tilde{\gamma}_n  &=~   \dfrac{\sqrt{6}\Dot{R}_{\rm cl}}{n \pi b [1 + (b_s \mathcal{M})^2]^{1/3}} \, \label{gammatilde expression turb}, \\ \Tilde{\epsilon}_n  &=~  \dfrac{2 \sqrt{\pi}}{n^{4/3} } \left( \dfrac{3}{2 \pi^3} \right)^{2/3}  \dfrac{[a_{\rm v} \langle \taucl \rangle]^{1/3} \langle \tau_{\rm c,abs} \rangle}{[1 + (b_s \mathcal{M})^2]^{4/9}}\, \label{epsilontilde expression turb},
\end{align}
where we have used $\eta_{\rm t} = e^{-2 \sigma_s^2 /3} = [1 + (b_s \mathcal{M})^2]^{-2/3}$ (from Eqs. \ref{turbulence variance} and \ref{effective diffusion term final}). These equations reduce to Eqs.  (\ref{xtilde expression})--(\ref{epsilontilde expression}) for a non-turbulent/uniform medium. The strongest effect we see is on the continuum absorption variable, which scales as $\Tilde{\epsilon}_n \propto \mathcal{M}^{-8/9}$ for $\mathcal{M} \gg 1$. Since $\Tilde{\epsilon}_n$ controls the Ly$\alpha$ escape fraction (see Appendix \ref{Escape fraction appendix}), this shows that turbulence can boost Ly$\alpha$ escape by creating low-density channels, thereby lowering the chance of dust absorption before cloud escape. This is qualitatively consistent with the findings of Ly$\alpha$ MCRT simulations in dusty turbulent clouds \citep{Kimm2019, Kimm2022}.\footnote{A rough order-of-magnitude comparison can be made in this case to the simulations of \cite{Kimm2019}. For a non-turbulent cloud with a central source and Ly$\alpha$ escape fraction $f_{\rm esc, Ly\alpha} \sim 0.1 \%$, we expect that when considering turbulence with $\mathcal{M} \sim 10$ should boost the escape fraction to $f_{\rm esc,Ly\alpha} \sim 10\%$ (judging from Fig. \ref{fesc Lya prediction vs Monte Carlo}). This is consistent with what is seen in their fig. 15 for similar conditions.} Another immediate consequence of Eq.~(\ref{series solution turbulent cloud}) is that the emergent spectrum, which peaks at $x \sim (a_{\rm v} \taucl^{\star})^{1/3} \propto  \mathcal{M}^{-2/9}$ for $\mathcal{M} \gg 1$, should be narrower in turbulent media, which is also seen in MCRT simulations \citep{Kimm2019, Kakiichi2021}. A more detailed quantitative comparison of the predictions of the approximate solution of this section requires detailed Ly$\alpha$ MCRT simulations in turbulent clouds, which is beyond the scope of this paper, but will be explored in future work. 

\subsection{The Ly\texorpdfstring{\boldmath$\alpha$}{alpha} radiation pressure force multiplier}
\label{Force multiplier}

\begin{figure*}
\includegraphics[width=0.90\textwidth]{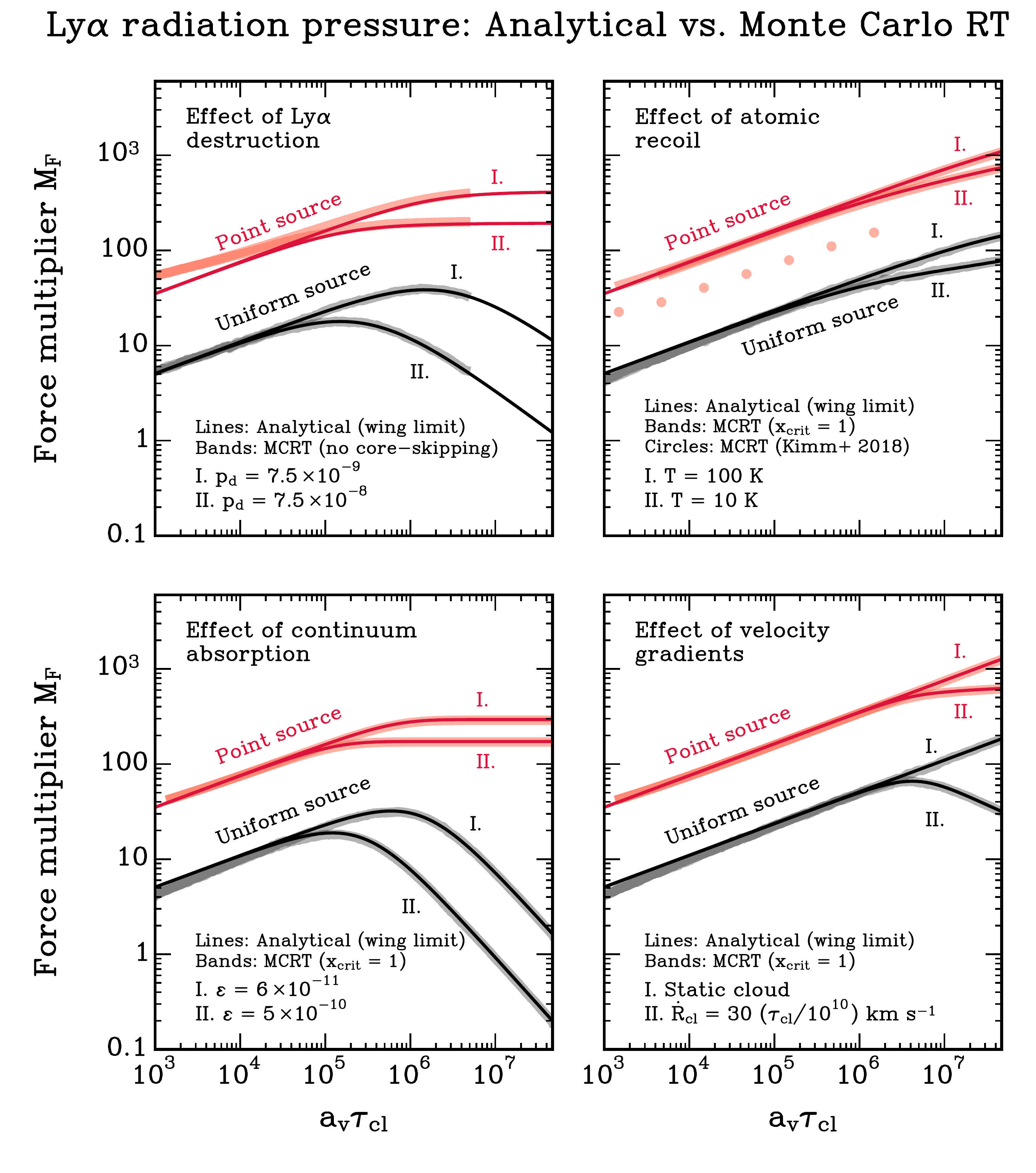}
\caption{Comparison of the predicted Ly$\alpha$ force multiplier $M_{\rm F}$ from the analytical solution (lines, Eqs. \ref{M_F general}--\ref{M_F velocity gradients exact}) and MCRT experiments using \textsc{colt} (bands). Every MCRT simulation (both for a point source and a uniform source), excluding the upper left panel, was run with $10^5$ photons. A relatively small amount of static core-skipping was used when the photon had a frequency of $x \leq 1$ (i.e. $x_{\rm crit} = 1$) for all panels except the upper left one, since we found that the effect of Ly$\alpha$ destruction was sensitive to core-skipping. \textbf{Upper left panel}: The effect of a finite Ly$\alpha$ destruction probability, for $T = 10 \, \rm K$, and ignoring recoil. The predictions for two $p_{\rm d}$ values are plotted: $p_{\rm d} = 7.5 \times 10^{-9}$, and  $p_{\rm d} = 7.5 \times 10^{-8}$. The uniform source MCRT tests were run with $1.5 \times 10^5$ photons to aide convergence. The discrepancy at low $a_{\rm v} \taucl$ for the point source is because of the wing approximation in deriving $M_{\rm F}$ analytically. With a more accurate treatment in this regime, following \citet{Tomaselli2021}, we expect the agreement to become better. Note that better agreement at low $a_{\rm v} \taucl$ in the other panels is a consequence of core-skipping. Without core-skipping, the MCRT results would resemble this panel in the low $a_{\rm v} \taucl$ limit for both the point source and uniform source. \textbf{Upper right panel}: The effect on $M_{\rm F}$ of including atomic recoil. Predictions are shown for $T = 100 \, \rm K$, and $T = 10 \, \rm K$. The circles show the MCRT results by \citet{Kimm2018} for a point source in a $T=100 \, \rm K$ uniform spherical cloud. It is evident that their MCRT results, obtained using \textsc{rascas}, disagree with both the analytical predictions, and MCRT results from \textsc{colt} of this paper by a factor $\sim 2 - 3$.   \textbf{Lower left panel}: The effect of continuum absorption, for $T = 100 \, \rm K$, and ignoring recoil. Results are shown for two different values of $\epsilon \equiv \tau_{\rm c,abs}/\taucl$:  $\epsilon = 6 \times 10^{-11}$, and $\epsilon = 5 \times 10^{-10}$. These correspond approximately to dust-to-gas ratios of $\mathscr{D}/\mathscr{D}_{\odot} \simeq  0.04$ and $\mathscr{D}/\mathscr{D}_{\odot} \simeq  0.3$, respectively \citep[see e.g. the Astrodust+PAH dust model of][listed in Table \ref{Dust models}]{Hensley2023}. \textbf{Lower right panel}: The effect of velocity gradients, for $T = 100 \, \rm K$, and ignoring recoil. Results are shown for the static case, and for a Hubble-like expansion $\Dot{R}_{\rm cl} = 30 \, (\taucl/10^{10}) \, \rm km \, s^{-1}$. The latter corresponds to $\Dot{R}_{\rm cl}/b = 23.3 \, (a_{\rm v} \taucl/4.7\times10^7)$, i.e. supersonic motion for the largest shown values of $a_{\rm v} \taucl$. }
\label{M_F analytical vs MCRT}
\end{figure*}

Having derived a new Ly$\alpha$ RT solution in the previous section, we can now estimate the Ly$\alpha$ radiation pressure in the cloud, and study its sensitivity to various effects (e.g. Ly$\alpha$ destruction, velocity gradients, and turbulence). As shown in Appendix \ref{AppendixRadPressure}, the acceleration due to Ly$\alpha$ radiation pressure is $\boldsymbol{a}_{\rm Ly\alpha} = (1/c\rho) \int_{0}^{\infty}\textrm{d}\nu \hspace{1 pt} n_{\rm HI} \sigma_0 \mathcal{H} \boldsymbol{F}$, where the flux and cross-section are evaluated in the \textit{comoving} frame. We will focus on non-turbulent (i.e. uniform) spherical clouds first, and discuss the effect of turbulence in Sec. \ref{effect of turbulence on M_F section}.

Because of the assumed spherical symmetry, the flux, and hence the acceleration, is radially outward. The net radiation pressure force $\Dot{p}_{\textrm{Ly}\alpha}$ on the cloud is then
\begin{align}
    \Dot{p}_{\textrm{Ly}\alpha} &=~ \dfrac{4 \pi}{c} \int_{0}^{\infty} \textrm{d}\nu \hspace{1 pt} \int_{0}^{R_{\rm cl}} \textrm{d}r \hspace{1 pt} r^2 n_{\rm HI} \sigma_0 \mathcal{H} F \nonumber \\ &\simeq~ -\dfrac{(4 \pi)^2 \Delta \nu_{\rm D}}{3c( n_{\rm HI} \sigma_0)^2} \int_{-\infty}^{\infty} \textrm{d}x \hspace{1 pt} \int_{0}^{\tau_{\rm cl}} \textrm{d}\tau \hspace{1 pt} \tau^2  \dfrac{\partial J}{\partial \tau} \nonumber \\  &=~ \dfrac{2(4 \pi)^2 \Delta \nu_{\rm D} \tau_{\rm cl}^2}{3c( n_{\rm HI} \sigma_0)^2} \int_{-\infty}^{\infty} \textrm{d}x \hspace{1 pt}  \int_{0}^{1} \textrm{d}\Bar{\tau} \hspace{1 pt} \Bar{\tau} J   \hspace{1 pt} \label{Pdot general}.
\end{align}
On the second line the Ly$\alpha$ flux has been expressed as $F = 4 \pi H = - (4 \pi/3 \mathcal{H})(\partial J/\partial \tau)$ (from the Eddington approximation), and the third line reflects an expansion using integration by parts, along with adopting the variable $\Bar{\tau} \equiv \tau/\tau_{\rm cl}$, and the fact that $[\tau^2 J]_{0}^{\tau_{\rm cl}} = 0$.\footnote{As noted by \cite{McClellan2022}, the series solution has $J = 0$ at the cloud edge, which stems from our approximate treatment of the boundary condition there. These authors propose a more complicated but rigorous solution to enforce the boundary condition at the cloud edge. With this they found that for $(a_{\rm v} \tau_{\rm cl})^{1/3} \gg 1$ the series solution is fairly accurate, with the the flux $\propto \partial J/\partial \tau$ even being accurate at the cloud edge. Thus, no large errors are expected from using the series solution in the cases of interest to us.} 

As a measure of the strength of Ly$\alpha$ feedback, it is customary to define the dimensionless Ly$\alpha$ force multiplier, $M_{\rm F} \equiv \Dot{p}_{\rm Ly\alpha}/(L_{\rm Ly\alpha}/c)$ \citep[e.g.][]{Dijkstra2008, Kimm2018, Smith2019, Tomaselli2021}. A value of $M_{\rm F} \gg 1$ corresponds to strong Ly$\alpha$ feedback, whereas $M_{\rm F} \sim 1$ is akin to the single-scattering limit characteristic of radiation pressure on dust. Physically, $M_{\rm F}$ is expected to be proportional to the time $t_{\rm trap}$ that Ly$\alpha$ photons remain trapped in the cloud, implying $M_{\rm F} \sim t_{\rm trap}/(R_{\rm cl}/c)$. In the case of static uniform clouds and ignoring Ly$\alpha$ destruction, simple physical arguments and more detailed analytical and MCRT calculations predict that $t_{\rm trap} \sim (a_{\rm v} \taucl)^{1/3} (R_{\rm cl}/c)$ \citep[e.g.][]{Adams1975, Bonilha1979, Spaans2006, Smith2018_discretediffusion, Lao2020}. Thus, we expect in this limit that $M_{\rm F} \sim (a_{\rm v} \taucl)^{1/3} \sim 100$--$10^3$, and therefore that Ly$\alpha$ feedback is very strong. Below we calculate $M_{\rm F}$ and its dependence on various effects using the full Ly$\alpha$ RT solution.  

Using the series solution (Eq.~\ref{series solution}) in Eq.~(\ref{Pdot general}) and simplifying gives us an expression for the force multiplier:
\begin{align}
    M_{\rm F} &=~  2 \sqrt{6} \sum_{n=1}^{\infty}  \dfrac{[\sin(n \pi \Bar{\tau}_{\rm s}) - n\pi \Bar{\tau}_{\rm s}\cos(n \pi \Bar{\tau}_{\rm s})]}{\Bar{\tau}_{\rm s}^3 (n\pi)^2 (\mathcal{P} + n \pi \mathcal{D}_n)} \nonumber \\ &\times~ \int_{0}^{1} \textrm{d}\Bar{\tau} \hspace{1 pt} \sin(n \pi \Bar{\tau}) \beta_n \int_{-\infty}^{\infty} \textrm{d}z \hspace{1 pt} \mathcal{F}_n(z) \hspace{1 pt} \label{M_F general, before integration}.
\end{align}
The spatial integral can be evaluated: $\int_{0}^{1} \textrm{d}\Bar{\tau} \hspace{1 pt} \sin(n\pi\Bar{\tau}) = [1 - (-1)^n]/(n\pi)$. Using this and $\beta_n = (3 a_{\rm v}^2/ 2\pi \lambda_n^2)^{1/6}$ then yields
\begin{align}
    \dfrac{M_{\rm F}}{(a_{\rm v}\taucl)^{1/3}} &=~ 4\sqrt{6} \left( \dfrac{3}{2 \pi^3} \right)^{1/6} \sum_{n = 0}^{\infty} \dfrac{[\sin(n_{\rm o} \pi \Bar{\tau}_{\rm s}) - n_{\rm o}\pi \Bar{\tau}_{\rm s}\cos(n_{\rm o} \pi \Bar{\tau}_{\rm s})]}{\Bar{\tau}_{\rm s}^3 (n_{\rm o}\pi)^3 n_{\rm o}^{1/3} (\mathcal{P} + n_{\rm o} \pi \mathcal{D}_{n_{\rm o}})} \nonumber \\ &\times~ \int_{-\infty}^{\infty} \textrm{d}z \hspace{1 pt} \mathcal{F}_{n_{\rm o}}(z) \hspace{1 pt} \label{M_F general},
\end{align}
where $n_{\rm o} = 2n+1$. For reference, if we neglect recoil and continuum absorption, we can write a more explicit solution for $M_{\rm F}$ using Eq.~(\ref{Solution f_n no recoil or dust}) in Eq.~(\ref{M_F general, before integration}):
\begin{align}
    \dfrac{M_{\rm F}}{(a_{\rm v}\taucl)^{1/3}} &=~ 4\sqrt{6} \Gamma\left( \dfrac{4}{3} \right) \sum_{n = 0}^{\infty} \dfrac{[\sin(n_{\rm o} \pi \Bar{\tau}_{\rm s}) - n_{\rm o}\pi \Bar{\tau}_{\rm s}\cos(n_{\rm o} \pi \Bar{\tau}_{\rm s})]}{\Bar{\tau}_{\rm s}^3 (n_{\rm o}\pi)^3 (\mathcal{P} + n_{\rm o} \pi \Bar{\mathcal{D}}_{n_{\rm o}})} \nonumber \\ &\times~ ( \mathcal{K}_{n_{\rm o},+} + \mathcal{K}_{n_{\rm o},-} ) \hspace{1 pt} \label{M_F velocity gradients exact},
\end{align}
where $\Bar{\mathcal{D}}_n$ is given in Eq.~(\ref{Overbar D_n definition}), $\Gamma(4/3) \simeq 0.89298$, and 
\begin{equation}
    \mathcal{K}_{n_{\rm o},\pm} \equiv  \left[\sqrt{\dfrac{2\pi}{3}} \dfrac{n_{\rm o} \pi \Bar{\mathcal{D}}_{n_{\rm o}}}{3} \pm \dfrac{\sqrt{\pi}}{3} \dfrac{\Dot{R}_{\rm cl}}{b} \right]^{-1/3} \, .
\end{equation}
We plot the predicted $M_{\rm F}$ from Eqs.~(\ref{M_F general})--(\ref{M_F velocity gradients exact}) in Fig. \ref{M_F analytical vs MCRT} and compare it to the results of MCRT experiments from \textsc{colt}. In each panel we show the impact of different effects -- Ly$\alpha$ destruction, recoil, continuum absorption, and velocity gradients -- and we find that the analytical solution is in excellent agreement with our MCRT results. Also shown in the upper right panel are earlier \textsc{rascas} MCRT results from \cite{Kimm2018}, which yields $M_{\rm F}$ values lower than ours by a factor $\sim 2-3$. \cite{Tomaselli2021} speculated that the discrepancy between their analytical results and the MCRT results of \cite{Kimm2018} was due to the implementation of recoil by the latter. We have confirmed that \cite{Kimm2018} underestimated $M_{\rm F}$, owing to a coordinate transformation problem (Taysun Kimm, priv. comm. 2024). As we show in Fig. \ref{M_F analytical vs MCRT}, even with recoil taken into account, the new analytical solution and the \textsc{colt} MCRT results agree and give larger $M_{\rm F}$-values than those of \cite{Kimm2018}. If we ignore recoil and consider a static cloud with a point source (top red line and band in the lower right panel), we find that the original predictions in \cite{Lao2020} and \cite{Tomaselli2021} agree with our \textsc{colt} MCRT results to within $1.4\%$ ($2\%$) at $a_{\rm v}\taucl = 4.7 \times 10^7$ ($a_{\rm v}\taucl = 4.7 \times 10^6$).

These MCRT tests for $M_{\rm F}$, in addition to the tests for the spectra and the Ly$\alpha$ escape fraction, show that the analytical solution is remarkably accurate. In the following subsections we will discuss the impact of different effects on $M_{\rm F}$ in more detail, and in the process develop a fit for subgrid modelling. This fit is significantly faster and more convenient to evaluate than the general expression in Eq.~(\ref{M_F general}). We leave the study of physical processes that determine $p_{\rm d}$ and $\epsilon$ to Sec. \ref{Lya destruction} and \ref{Lya absorption processes}, respectively.

\subsubsection{Effect of the spatial source distribution on Ly$\alpha$ feedback}
\label{spatial source distribution section}

\begin{figure}
    \includegraphics[trim={0.1cm 0.1cm 0.1cm 0.1cm},clip,width = 0.94\columnwidth]{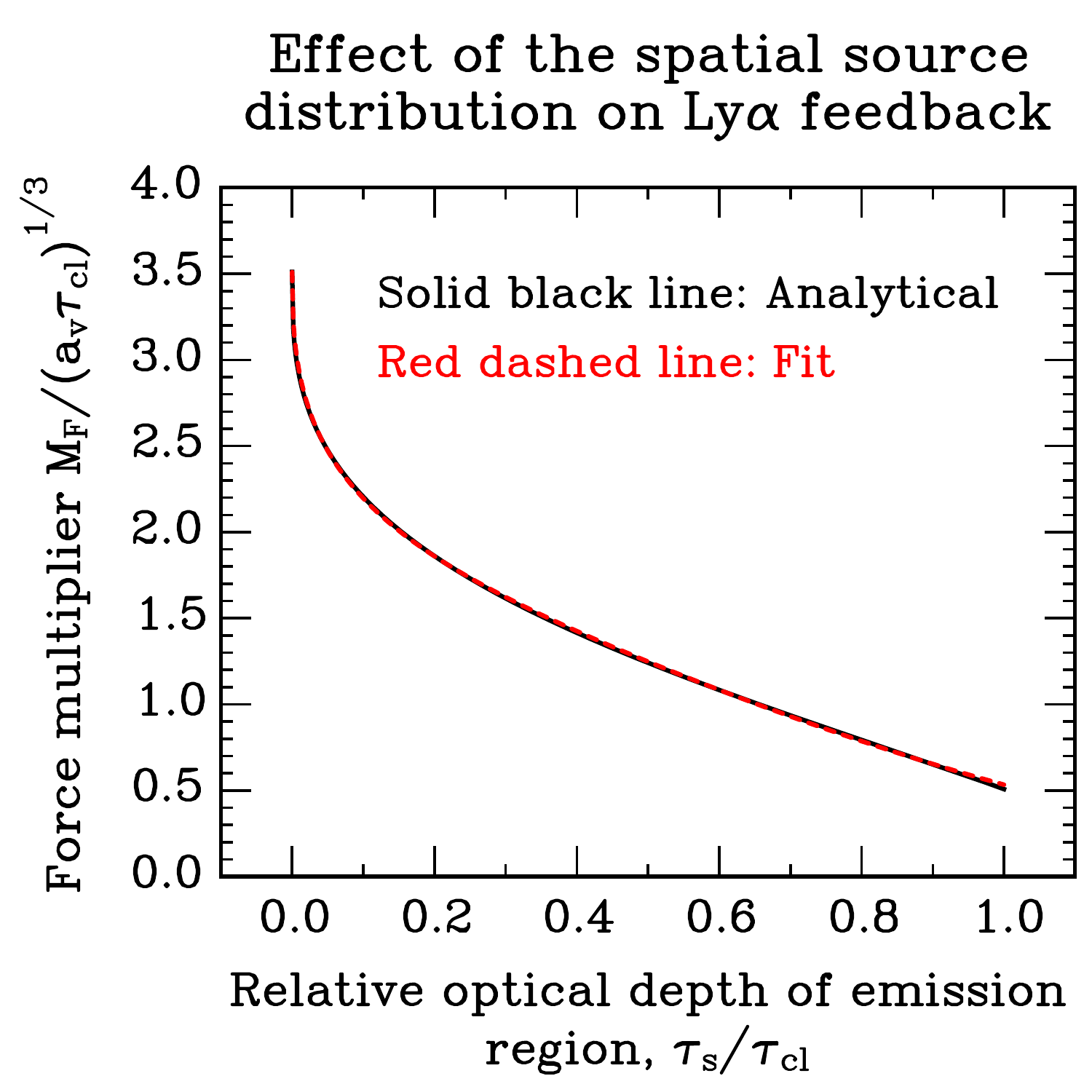}
    \caption{The effect of the spatial source distribution on the Ly$\alpha$ force multiplier in a uniform and static cloud, ignoring Ly$\alpha$ destruction and recoil. For a point source ($\tau_{\rm s} \rightarrow 0$), we find $M_{\rm F} \simeq 3.51 (a_{\rm v} \taucl)^{1/3}$. In the case of uniform emission ($\tau_{\rm s}/\taucl = 1$), we find $M_{\rm F} \simeq 0.51 (a_{\rm v} \taucl)^{1/3}$. These results are in agreement with the calculations of \citet{Lao2020} and \citet{Tomaselli2021}. For intermediate cases, $M_{\rm F}$ interpolates between the extremes, which seem to be well-fitted by Eq.~(\ref{M_F fit source distribution}) (red dashed curve in the figure).
}
    \label{M_F source distribution}
\end{figure}

We see from Eq.~(\ref{M_F general}) that the Ly$\alpha$ force multiplier depends on the source distribution in the cloud, via the parameter $\Bar{\tau}_{\rm s} = \tau_{\rm s}/\tau_{\rm cl}$. For the uniform \HI cloud assumed in Sec. \ref{uniform cloud solution}, this would give $\Bar{\tau}_{\rm s} = R_{\rm s}/R_{\rm cl}$. More generally, if we consider a radial dependence for $n_{\rm HI}$, we argue that $\Bar{\tau}_{\rm s}$ should be interpreted as the relative \textit{optical depth} (rather than physical size) of the emission region. This conclusion can be motivated as follows:
\begin{enumerate}
    \item In the case of a slab geometry, the general solution for an arbitrary \HI vertical density profile can be derived \citep{Lao2020}. This solution is only a function of the optical depth, so that the source distribution only enters through the relative optical depth, $\Bar{\tau}_{\rm s} = \tau_{\rm s}/\taucl$.
    
    \item A spherical cloud with uniform Ly$\alpha$ emission from an \HII region of size $R_{\rm s} \simeq R_{\rm cl}$ should be approximately described by a shell/slab solution, in which case it follows that $\Bar{\tau}_{\rm s} = \tau_{\rm s}/\taucl$. The Ly$\alpha$ optical depth of the \HII region is typically $\tau_{\rm s} \sim 10^4 - 10^5$ \citep[e.g.][]{Draine2011}, which is negligible compared to the Ly$\alpha$ optical depth of the outer \HI shell, $\taucl = 5.88 \times 10^8 \, T_{100}^{-1/2} (N_{\rm HI}/10^{21} \, \textrm{cm}^{-2})$, where we have normalized to a typical \HI shell/cloud column density (see Fig. \ref{discussion_overview} and its caption). Thus, $\Bar{\tau}_{\rm s} \ll 1$, even if the \HII region is of comparable size to the cloud. 

    \item In the opposite case of a small \HII region, $R_{\rm s} \ll R_{\rm cl}$, we must have $\Bar{\tau}_{\rm s} \ll 1$, regardless of whether it is defined in physical or optical depth space. 
\end{enumerate}
Thus, we conclude that $\Bar{\tau}_{\rm s} \simeq \tau_{\rm s}/\taucl$, i.e. the relative optical depth of the emission region should approximately describe the dependence on the spatial source distribution. In future work we will generalize to consider a radially dependent \HI density in a self-consistent manner. If we consider the computed $M_{\rm F}$ for a slab, we expect this to change $M_{\rm F}$ by a factor $\lesssim \rm few \times 10\%$ compared to simply making the identification $\Bar{\tau}_{\rm s} = \tau_{\rm s}/\taucl$.\footnote{In the case of Ly$\alpha$ emission from a physically small \HII region we expect the spherical solution with $\Bar{\tau}_{\rm s}$ to be accurate. In the opposite case of emission from a physically large \HII region, the solution approaches that of a slab/shell with a `central' source (i.e. at $\tau_{\rm s} \rightarrow 0$). This solution has $M_{\rm F} = 2.2 (a_{\rm v} \taucl)^{1/3}$ \citep{Lao2020}, compared to $M_{\rm F} = 3.51 (a_{\rm v} \taucl)^{1/3}$ for a point source in a spherical cloud (see Fig. \ref{M_F source distribution}). This yields a maximum error of $\sim 60\%$.}

In Fig. \ref{M_F source distribution} we plot $M_{\rm F}$ as a function of $\Bar{\tau}_{\rm s}$ for a static and uniform cloud, ignoring Ly$\alpha$ destruction and recoil. For an effective central point source ($\Bar{\tau}_{\rm s} \rightarrow 0$) we find $M_{\rm F} \simeq 3.51 (a_{\rm v} \taucl)^{1/3}$, and in the case of uniform emission ($\Bar{\tau}_{\rm s} = 1$) we find $M_{\rm F} \simeq 0.51 (a_{\rm v} \taucl)^{1/3}$, These results were derived earlier by \cite{Lao2020} and \cite{Tomaselli2021}. The lower radiation pressure in the case of extended sources ($\Bar{\tau}_{\rm s} > 0$) is mainly due to flux cancellation, which lowers the net radial force on the cloud.\footnote{The fact that the average trapping time is lower in the case of uniform emission also plays a role, but not as much. For example, going from a central point source to uniform emission reduce the trapping time by less than a factor $2$ \citep[see fig. 1 in][]{Lao2020}. In comparison, the force multiplier drops by a factor $\sim 6.9$, as per Fig. \ref{M_F source distribution}.} The same phenomenon is observed in simulations of star formation with radiation pressure on dust, where the momentum injection rate drops from the idealized prediction $\Dot{p}_{\rm dust} \simeq L_{\rm UV}/c$ (point source) to $\Dot{p}_{\rm dust} \sim 0.1 \hspace{1 pt} L_{\rm UV}/c$ because stars can form throughout the cloud \citep{Kim2018_RadPressure, Menon2023_directRP}.
We see in Fig. \ref{M_F source distribution} that the general case $0 \leq \Bar{\tau}_{\rm s} \leq 1$ is well-fit by:\footnote{It would be inappropriate to extrapolate this fit to $\Bar{\tau}_{\rm s} > 1$, which represents a source larger than the cloud, a case not covered in the solution of this paper.}
\begin{align}
    M_{\rm F}^{\rm fit} &=~ \eta_{\rm s} (a_{\rm v} \taucl)^{1/3} \hspace{1 pt} , \nonumber \\ 
    \eta_{\rm s} &=~ 3.51 \hspace{1 pt} \exp(- 0.674 \hspace{1 pt} \Bar{\tau}_{\rm s}^{2.42} - 1.21 \hspace{1 pt} \Bar{\tau}_{\rm s}^{0.414}) \, .  \label{M_F fit source distribution} 
\end{align}
We will now generalize this fit further to include additional effects. 

\subsubsection{Effect of Ly$\alpha$ photon destruction on Ly$\alpha$ feedback}
\label{Lyadestruction general effect}
\begin{figure}
    \includegraphics[trim={0.1cm 0.1cm 0.1cm 0.1cm},clip,width = \columnwidth]{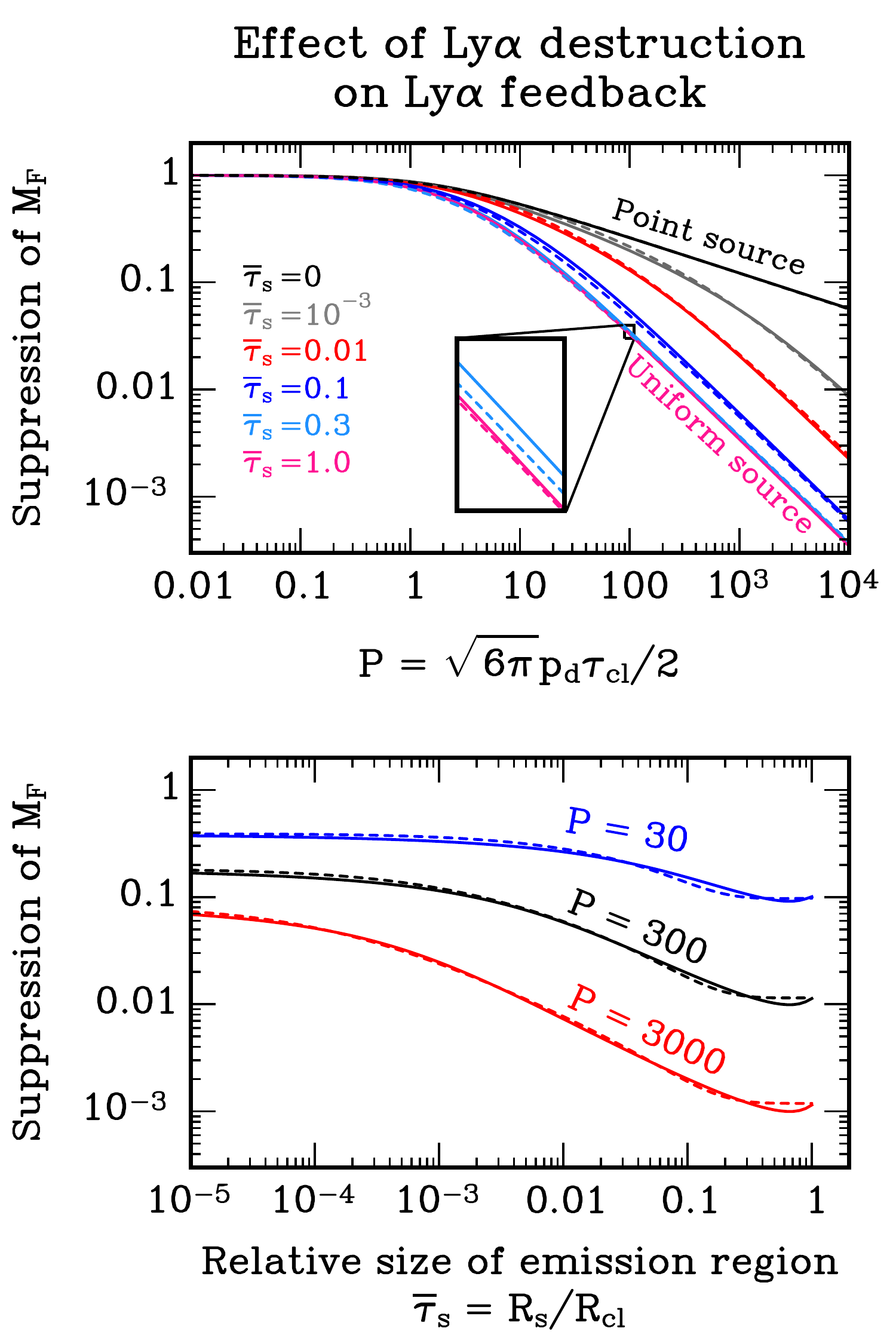}
    \caption{The effect of a non-zero Ly$\alpha$ destruction probability ($p_{\rm d} > 0$) on the Ly$\alpha$ force multiplier $M_{\rm F}$. Recoil, velocity gradients, and continuum absorption are ignored here. Solid lines correspond to the prediction from the series solution, while dashed lines show the fit from Eq.~(\ref{M_F fit p_d}). \textbf{Upper panel}: A plot of the relative suppression in $M_{\rm F}$ as a function of $\mathcal{P}$ for clouds with fixed $\Bar{\tau}_{\rm s}$. \textbf{Lower panel}: Also a plot of the relative suppression in $M_{\rm F}$, but as a function of $\Bar{\tau}_{\rm s}$ for fixed values of $\mathcal{P}$.
}
    \label{M_F p_d effect}
\end{figure}
What is the general impact on Ly$\alpha$ feedback of introducing a non-zero Ly$\alpha$ destruction probability ($p_{\rm d} > 0$)? In Fig. \ref{M_F p_d effect} we plot the relative suppression in $M_{\rm F}$ as a function of $\mathcal{P} \equiv \sqrt{6 \pi}p_{\rm d}\taucl/2$ in a static cloud for different source distributions, ignoring recoil and continuum absorption. We also compare the analytical predictions to MCRT experiments with \textsc{colt} in the upper left panel of Fig. \ref{M_F analytical vs MCRT}, and find excellent agreement in the optically thick limit. 

The following fit, which generalizes Eq.~(\ref{M_F fit source distribution}), is seen to approximately capture the suppression for $0 \leq R_{\rm s}/R_{\rm cl} \leq 1$ and all $\mathcal{P}$:
\begin{equation}
    M_{\rm F}^{\rm fit} =  \dfrac{\eta_{\rm s} (a_{\rm v}\taucl)^{1/3}}{(1 + \mathcal{P}/\mathcal{P}_{\rm crit,point})^{1/3} + (\mathcal{P}/\mathcal{P}_{\rm crit,ext})} \hspace{1 pt} , \label{M_F fit p_d}
\end{equation}
where
\begin{align}
    \mathcal{P}_{\rm crit, point} &=~ 1.8 \, \exp(\Bar{\tau}_{\rm s}^{1/4})  \, , \\  \mathcal{P}_{\rm crit, ext} &=~ 1.5 \, \Bar{\tau}_{\rm s}^{-0.6}\exp(-10 \, \Bar{\tau}_{\rm s}) + 3.6  \, ,
\end{align}
and $\eta_{\rm s}$ the same as in Eq.~(\ref{M_F fit source distribution}). As is evident from both Fig. \ref{M_F p_d effect} and  Eq.~(\ref{M_F fit p_d}), Ly$\alpha$ feedback can be significantly suppressed if $\mathcal{P} \gtrsim \rm few$. Or equivalently, when $p_{\rm d} \gtrsim 1/\taucl$. This result has a simple physical interpretation. The mean number of scatterings before escape is $\sim \taucl$ \citep[e.g.][]{Adams1972, Harrington1973, Lao2020}. A Ly$\alpha$ photon is therefore likely to be destroyed if $p_{\rm d} \gtrsim 1/\taucl$, above which we expect Ly$\alpha$ scattering and feedback to be suppressed.

\subsubsection{Effect of atomic recoil on Ly$\alpha$ feedback}
\label{effect of recoil on Lya feedback}

\begin{figure}
   \includegraphics[trim={0.1cm 0.1cm 0.1cm 0.1cm},clip,width = 0.94\columnwidth]{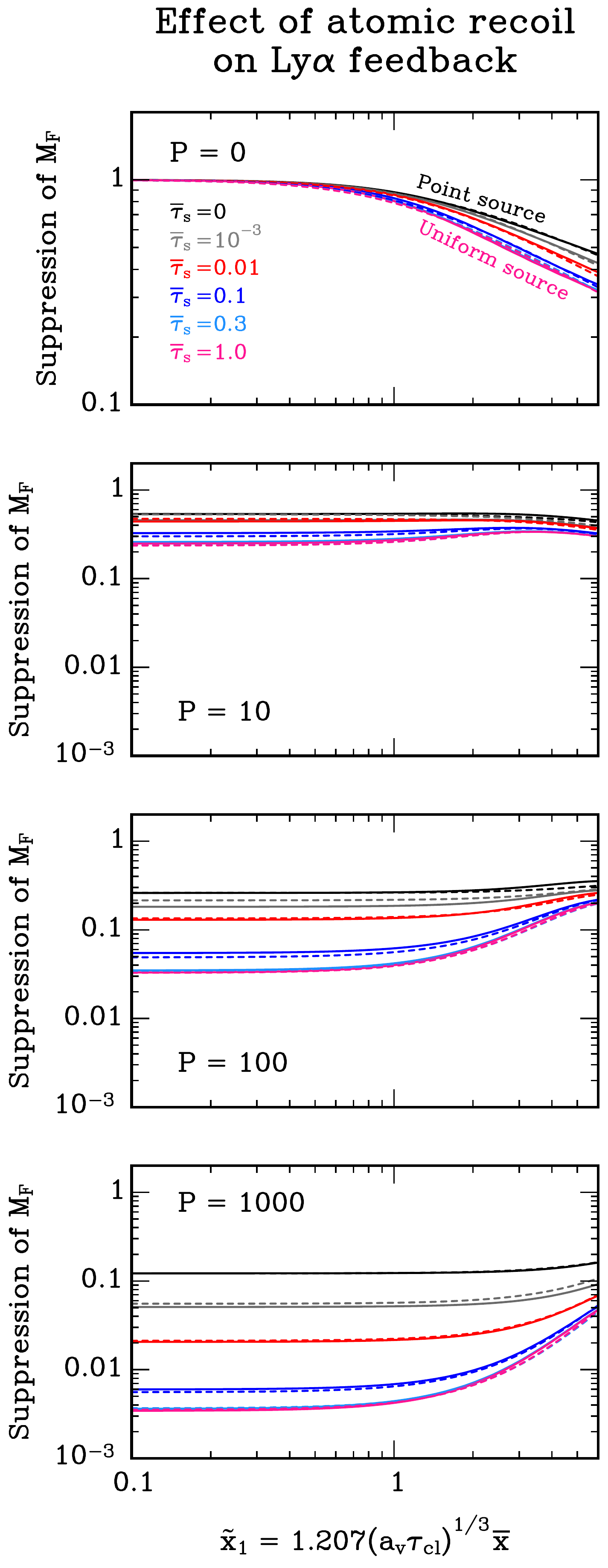}
    \caption{The effect of recoil on the Ly$\alpha$ force multiplier $M_{\rm F}$ in static clouds, ignoring continuum absorption. Solid lines, coloured by the relative size of the emission region as in Fig. \ref{M_F p_d effect}, show the analytical predictions (Eq.~\ref{M_F general}), and dashed lines show the fit in Eq.~(\ref{M_F fit recoil}). In the top panel, Ly$\alpha$ destruction is ignored ($\mathcal{P} = 0$). The lower panels show the combined effect of recoil and Ly$\alpha$ destruction ($\mathcal{P} > 0$).
}
    \label{M_F_recoil}
\end{figure}

Atomic recoil slowly pushes Ly$\alpha$ photons to the red side of the wings where they can more easily escape \citep{Adams1971, Smith2017}. Does this have a significant impact on Ly$\alpha$ feedback? In Fig. \ref{M_F_recoil}, we plot the relative suppression of the force multiplier $M_{\rm F}$ by recoil. We also compare the analytical predictions for $M_{\rm F}$ to MCRT results from \textsc{colt} in Fig. \ref{M_F analytical vs MCRT}, showing excellent agreement. The following generalization of the fit in Eq.~(\ref{M_F fit p_d}) captures the effect of recoil:
\begin{equation}
    M_{\rm F}^{\rm fit} =  \dfrac{\eta_{\rm s} (a_{\rm v}\taucl)^{1/3}}{\mathcal{A}^{1/3} + (\mathcal{P}/\mathcal{G}_{\rm rec}\mathcal{P}_{\rm crit,ext})} \, , \label{M_F fit recoil}
\end{equation}
where:
\begin{align}
    \mathcal{A} &=~ 1 + (\mathcal{P}/\mathcal{G}_{\rm rec} \mathcal{P}_{\rm crit,point}) + (\Tilde{x}_1/\Tilde{x}_{\rm 1,crit})^{3c_{\rm rec}} \, , \\ \Tilde{x}_{\rm 1,crit} &=~ 0.55 \exp(-5.8 \Bar{\tau}_{\rm s}^{1/2}) + 0.98 \, , \\ c_{\rm rec} &=~ 0.54 \exp(-10^3 \Bar{\tau}_{\rm s}) + 0.63[1 - \exp(-10^3 \Bar{\tau}_{\rm s})] \, , \\ \mathcal{G}_{\rm rec} &=~ 1 + \eta_{\rm rec}[0.2 (\Tilde{x}_1/\Tilde{x}_{\rm 1,crit})^2 + 0.005(\Tilde{x}_1/\Tilde{x}_{\rm 1,crit})^4] \, , \\ \eta_{\rm rec} &=~ 0.35 + \dfrac{0.65 \, \Bar{\tau}_{\rm s}^{3/2}}{\Bar{\tau}_{\rm s}^{3/2} + 0.00395} \, .
\end{align}
We find that recoil can dampen Ly$\alpha$ feedback by a factor $\sim \rm few$ when $\tilde{x}_1 \gtrsim 1$, corresponding to optical depths $\taucl > 7.4 \times 10^{9} ~ T_{100}^2$. This threshold is in agreement with expectations from simple physical arguments \citep{Adams1971, Smith2017}. In particular, on average recoil push Ly$\alpha$ photons to the red wing by an amount $\Bar{x} = 2.54 \times 10^{-3} \hspace{1 pt} T_{100}^{-1/2}$ per scattering event (Eq.~\ref{xbar numerical}). In comparison, on average Ly$\alpha$ photons are pushed back to the line centre by an amount $-1/\lvert x \rvert$. Thus, recoil starts to dominate at frequencies $\lvert x_{\rm rec} \rvert \sim 1/\Bar{x}$. If $\lvert x_{\rm rec} \rvert$ is smaller than the frequency for cloud escape, $x_{\rm esc} \sim (a_{\rm v} \taucl)^{1/3}$, recoil will aid in the escape of the photon. This gives the condition $\taucl \gtrsim 10^{10} ~ T_{100}^2$, which is consistent with the results of the analytical solution and the MCRT experiments.

We also see from the lower panels in Fig. \ref{M_F_recoil} that recoil can dampen the suppression from Ly$\alpha$ destruction ($\mathcal{P} > 0$) when $\Tilde{x}_1 \gtrsim 1$. This interference effect can be understood as follows. Ly$\alpha$ destruction suppress $M_{\rm F}$ as long as the photons can scatter in the core of the Voigt profile enough times to make Ly$\alpha$ destruction probable. However, strong recoil pushes the photons to the red wing before this happens, thereby making suppression of $M_{\rm F}$ by Ly$\alpha$ destruction less efficient.

\subsubsection{Effect of continuum absorption on Ly$\alpha$ feedback}
\label{Sec. M_F continuum absorption}

\begin{figure}
    \includegraphics[trim={0.1cm 0.1cm 0.1cm 0.1cm},clip,width = 0.94\columnwidth]{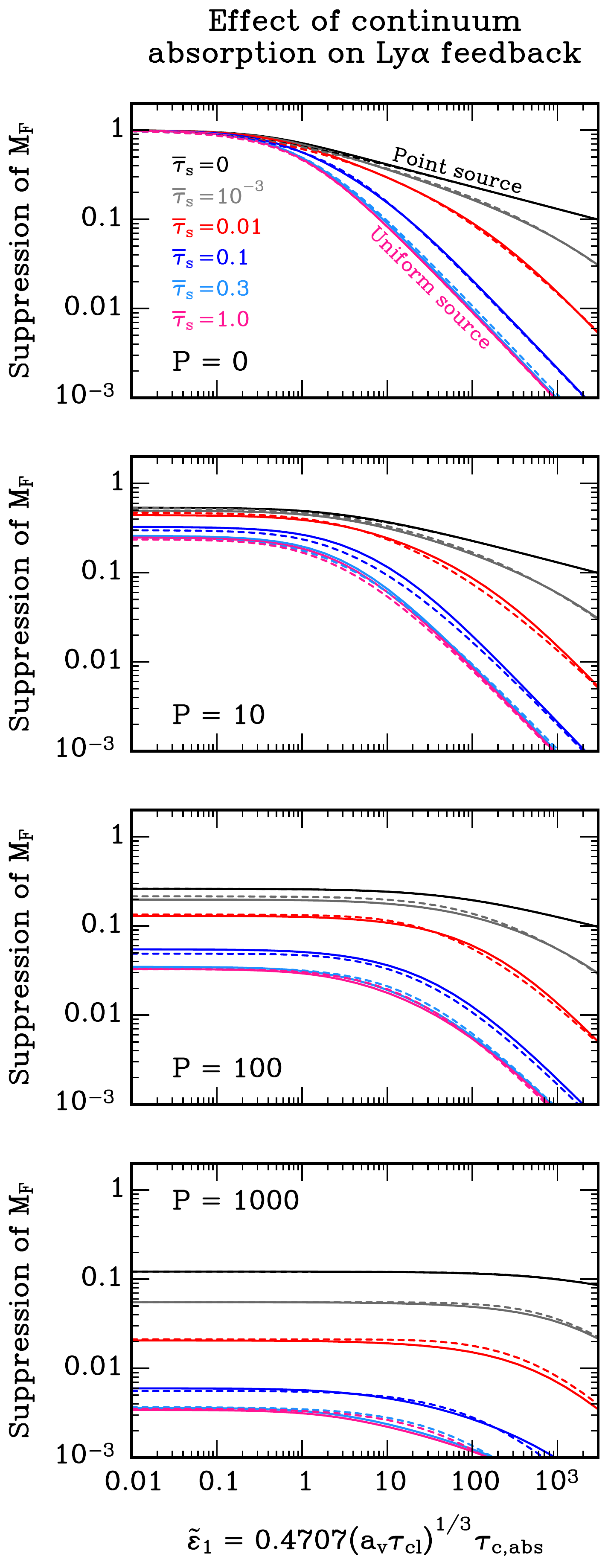}
    \caption{The effect of continuum absorption (e.g. by dust) on the Ly$\alpha$ force multiplier in static clouds, ignoring recoil. Solid lines, coloured by the relative size of the emission region as in Fig. \ref{M_F p_d effect}, show the analytical predictions (Eq.~\ref{M_F general}), and dashed lines show the fit in Eq.~(\ref{M_F fit continuum absorption}).  In the top panel, the Ly$\alpha$ destruction probability is set to zero (so $\mathcal{P} = 0$). The lower panels show the combined effect of continuum absorption and Ly$\alpha$ destruction ($\mathcal{P} > 0$).
}
    \label{M_F dust}
\end{figure}

It is well-known that Ly$\alpha$ photons can be absorbed by dust, and that this can suppress Ly$\alpha$ feedback \citep[e.g.][]{Birthell1990, Henney1998, Krumholz2009, Abe2018, Kimm2018, Tomaselli2021}. Recently, \cite{Tomaselli2021} and \cite{Kimm2018} found that continuum absorption of Ly$\alpha$ photons around a central point source leads to a constant plateau in $M_{\rm F}$ as $\taucl \rightarrow \infty$. Here, we extend their analysis to consider more general source distributions. 

In Fig. \ref{M_F analytical vs MCRT} we compare the analytical predictions for $M_{\rm F}$ to MCRT results from \textsc{colt}. We find excellent agreement between the two. In Fig. \ref{M_F dust} we plot the relative suppression of $M_{\rm F}$ as a function of $\Tilde{\epsilon}_1 \equiv 0.4707(a_{\rm v} \taucl)^{1/3} \tau_{\rm c, abs}$, where $\tau_{\rm c,abs}$ is the cloud continuum absorption optical depth. We remind the reader that the effect of continuum absorption in the optically thick limit only enters through this combination of variables, as shown in Appendix \ref{AppendixRTfreq} \citep[see also][]{Neufeld1990}. The result can be fitted by the following generalization of Eq.~(\ref{M_F fit recoil}):
\begin{equation}
    M_{\rm F}^{\rm fit} =  \dfrac{\eta_{\rm s} (a_{\rm v}\taucl)^{1/3}}{\mathcal{A}^{1/3} + (\mathcal{P}/\mathcal{G}_{\rm rec}\mathcal{P}_{\rm crit,ext})} \hspace{1 pt} , \label{M_F fit continuum absorption}
\end{equation}
where:
\begin{align}
    \mathcal{A} &=~ 1 + ( \mathcal{P}/\mathcal{G}_{\rm rec}\mathcal{P}_{\rm crit,point}) + (\Tilde{x}_1/\Tilde{x}_{\rm 1,crit})^{3c_{\rm rec}} + \mathcal{F}_{\rm abs} \, , \\ \mathcal{F}_{\rm abs} &=~ [(3.4 \Tilde{\epsilon}_1)^{c_{\rm abs}} + (\eta_{\rm abs} \Tilde{\epsilon}_1)^{4c_{\rm abs}}]^{3/4c_{\rm abs}} \, , \\ \eta_{\rm abs} &=~ 1.1 \, [1 - \exp(-6.54 \Bar{\tau}_{\rm s}^{1.06})]\, , \\ c_{\rm abs} &=~ \dfrac{[1 - \exp(-20 \Bar{\tau}_{\rm s})]}{3[1 + (\mathcal{P}/50)^{1/4} + (\mathcal{P}/300)^{3/4}]} + 0.17 \exp(-20 \Bar{\tau}_{\rm s}) \, .
\end{align}
In the top panel of Fig. \ref{M_F dust}, where we ignore Ly$\alpha$ destruction at line centre (i.e. $\mathcal{P} = 0$), we find that for a point source, $M_{\rm F} \propto 1/\Tilde{\epsilon}_1^{1/4}$, and so reach a plateau for constant $\epsilon = \tau_{\rm c,abs}/\taucl$ (see also Fig. \ref{M_F analytical vs MCRT}). In particular, we find that $M_{\rm F}(\taucl \rightarrow \infty) \simeq 3.12 (a_{\rm v}/\epsilon)^{1/4}$. We therefore confirm the analytical results of \cite{Tomaselli2021}, who derived the constraint $M_{\rm F}(\taucl \rightarrow \infty) \simeq (3.06 \pm 0.3) \times (a_{\rm v}/\epsilon)^{1/4}$ for a point source, by a different method.\footnote{Rather than directly solving the Ly$\alpha$ RT equation for a dusty cloud in detail as we have done in this paper, \cite{Tomaselli2021} used mathematical and numerical techniques to derive tight constraints on the limiting analytical value of $M_{\rm F}$. The two methods give results that are in excellent agreement, but the detailed solution in this paper is more accurate, and also provides additional information of interest -- such as how continuum absorption interacts with other effects (e.g. $p_{\rm d} > 0$), as well as the emergent spectrum and Ly$\alpha$ escape fraction.}   

In the lower panels of Fig. \ref{M_F dust}, we study the impact of adding a finite Ly$\alpha$ destruction probability ($\mathcal{P}>0$), on top of Ly$\alpha$ destruction by continuum absorption. The two effects do not interfere -- instead, the former can potentially suppress $M_{\rm F}$ even further in dense, dust-enriched clouds. Because of this, one cannot simply replace $\taucl$ in $M_{\rm F}$ by the critical optical depth for dust absorption, which can be $\ll \tau_{\rm cl}$, as had been suggested by \cite{Tomaselli2021}.\footnote{\cite{Tomaselli2021} show that the plateau for $M_{\rm F}$ in the case of a point source can be understood to occur when the probability of dust absorption becomes $\sim 1$ \citep[see also][]{Kimm2018}. Using a random-walk argument, they show that this occurs when $\taucl \sim a_{\rm v}^{-1/4} \epsilon^{-3/4}$. To get the effect of dust on $M_{\rm F}$ they therefore suggest an analytical prescription where one makes the replacement $\taucl \rightarrow \min[\taucl, \, a_{\rm v}^{-1/4} \epsilon^{-3/4}]$ (ignoring order unity factors in their more exact analysis). However, such a prescription can severely underestimate $\mathcal{P}$ in dust-enriched clouds, since in many circumstances $\taucl \gg a_{\rm v}^{-1/4} \epsilon^{-3/4}$.} Doing so would greatly underestimate the effect of Ly$\alpha$ destruction, since $\mathcal{P} \propto p_{\rm d} \taucl$. 

Another effect we see from Fig. \ref{M_F dust} is that continuum absorption can suppress $M_{\rm F}$ much more severely in clouds with extended sources. However, in practice, we do not expect this to be very important for Ly$\alpha$ stellar feedback, since as discussed in Sec. \ref{spatial source distribution section}, $\Bar{\tau}_{\rm s} \simeq \tau_{\rm s}/\taucl \ll 1$ if the Ly$\alpha$ photons originate from a central but extended \HII region, trapped in a larger \HI cloud. This effect could, however, conceivably suppress Ly$\alpha$ pressure in other contexts where Ly$\alpha$ photons are mainly produced by Ly$\alpha$ collisional excitation in \HI gas.

Finally, we note that our detailed analytical and MCRT results, as well as those of \cite{Kimm2018} and \cite{Tomaselli2021}, find significantly weaker suppression of Ly$\alpha$ feedback by continuum absorption than order-of-magnitude estimates by \cite{Henney1998}. The latter had motivated the neglect of Ly$\alpha$ feedback in dust-enriched environments in several later works \citep[e.g.][]{Krumholz2009, Draine2011_radpressure, Wise2012radpressure, Jeffreson2021, Olivier2021}, but overestimate the suppression by dust by a factor of order $\sim 50$ \citep[as recently pointed out by][]{Thomson2024}. Our results therefore suggest that Ly$\alpha$ feedback may even be important in dust-enriched/low-redshift galaxies (as also seen in Fig. \ref{discussion_overview}).

\subsubsection{Effect of velocity gradients on Ly$\alpha$ feedback}

\begin{figure}
    \includegraphics[trim={0.1cm 0.1cm 0.1cm 0.1cm},clip,width = 0.94\columnwidth]{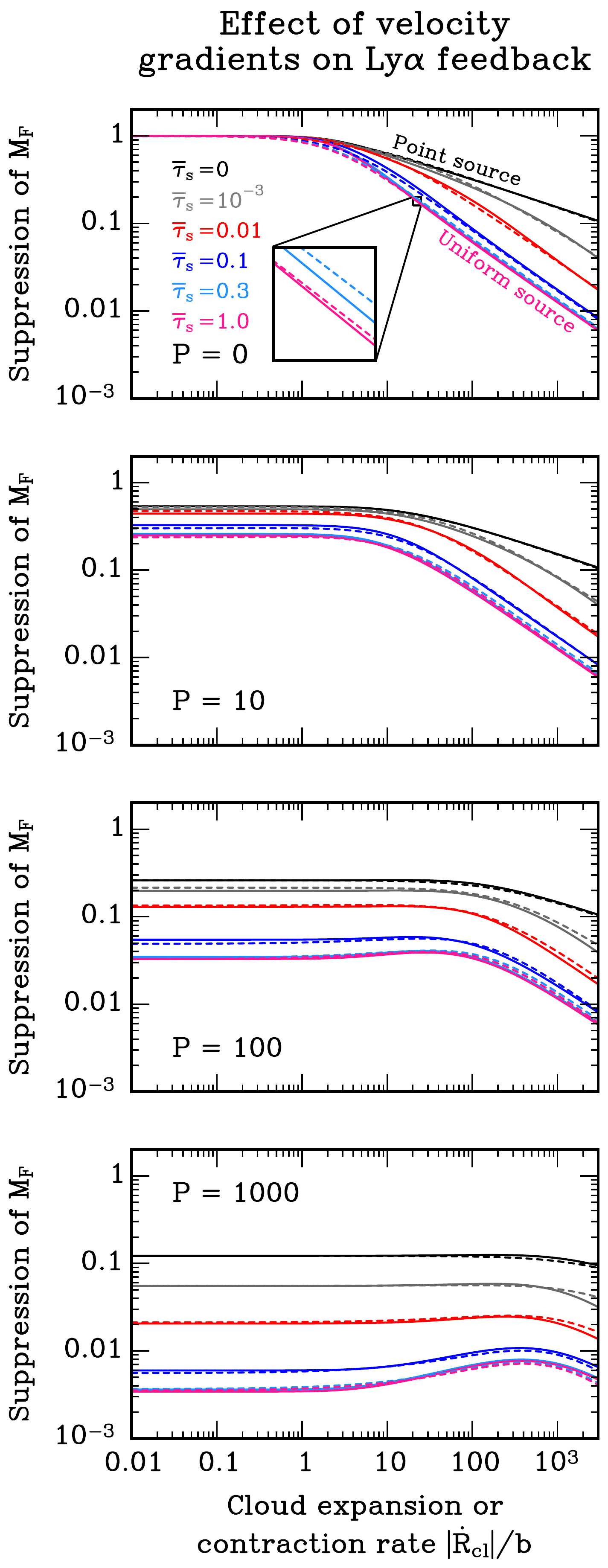}
    \caption{The relative suppression of the Ly$\alpha$ force multiplier $M_{\rm F}$ as a function of the cloud expansion/contraction rate, $\lvert \Dot{R}_{\rm cl} \rvert/b$, ignoring recoil and continuum absorption. Solid lines show the predictions of the analytical solution (Eq.~\ref{M_F velocity gradients exact}), whereas the dashed lines show the fit in Eq.~(\ref{M_F fit velocity}). All lines are coloured by the spatial source distribution $\Bar{\tau}_{\rm s}$. In the top panel the Ly$\alpha$ destruction probability is set to zero ($\mathcal{P} = 0$), while it is non-zero in the lower panels. 
}
    \label{M_f velocity gradients plot}
\end{figure}

Velocity gradients could shift the Ly$\alpha$ photons to the wings, facilitating cloud escape \citep[e.g.][]{Bonilha1979, Braun1989, McKee2008, Seon2020}. What is the resulting impact on Ly$\alpha$ feedback? In Fig. \ref{M_f velocity gradients plot} we plot the relative suppression of $M_{\rm F}$ as a function of $\lvert \Dot{R}_{\rm cl} \rvert/b$, for different values of $\mathcal{P}$ and spatial source distributions, as predicted from Eq.~(\ref{M_F velocity gradients exact}). We also compare the analytical predictions for $M_{\rm F}$ to MCRT results from \textsc{colt} in the lower right panel of Fig. \ref{M_F analytical vs MCRT}, showing excellent agreement. 

In the absence of Ly$\alpha$ destruction, and for $\lvert \Dot{R}_{\rm cl} \rvert/b \gtrsim \rm few$, we find that the force multiplier scales as $M_{\rm F} \propto (\lvert \Dot{R}_{\rm cl} \rvert/b)^{-2/3}$ and $(\lvert \Dot{R}_{\rm cl} \rvert/b)^{-1/3}$ for extended sources and point sources, respectively. As in the case of atomic recoil (Fig. \ref{M_F_recoil}), we find that velocity gradients can slightly dampen the effect of Ly$\alpha$ destruction, by Doppler-shifting the photons out of the core. However, this effect is only visible for extended sources and clouds with $\mathcal{P} \gtrsim 100$ and $\lvert \Dot{R}_{\rm cl} \rvert/b \sim 10 - 10^3$ (see the two lower panels of Fig. \ref{M_f velocity gradients plot}).

To incorporate velocity gradients in our fit for $M_{\rm F}$, we generalize Eq.~(\ref{M_F fit continuum absorption}) as follows:
\begin{equation}
    M_{\rm F}^{\rm fit} =  \dfrac{\eta_{\rm s} (a_{\rm v}\taucl)^{1/3}}{\mathcal{A}^{1/3} + [\mathcal{P}/(\mathcal{G}_{\rm rec} + \mathcal{G}_{\rm vel})\mathcal{P}_{\rm crit,ext}] + \mathcal{F}_{\rm vel}} \hspace{1 pt} , \label{M_F fit velocity}
\end{equation}
where:
\begin{align}
    \mathcal{A} &=~ 1 + ( \mathcal{P}/\mathcal{G}_{\rm rec}\mathcal{P}_{\rm crit,point}) + (\Tilde{x}_1/\Tilde{x}_{\rm 1,crit})^{3c_{\rm rec}} \\ &+~ \mathcal{F}_{\rm abs} + 0.29 \, \Dot{\mathcal{R}}_{\rm cl} \, , \nonumber \\
    \mathcal{F}_{\rm vel} &=~ \left[\left(1 + \dfrac{3}{4} \eta_{\rm vel} \Dot{\mathcal{R}}_{\rm cl} \right)^{2c_{\rm vel}/3} - 1 \right]^{1/c_{\rm vel}} \, , \\
    \eta_{\rm vel} &=~ 1 - \exp[ - (\Bar{\tau_{\rm s}}/0.1)^{0.7}] \, , \\ c_{\rm vel} &=~ \dfrac{0.6}{1 + (\mathcal{P}/10^3)^{1/2}} \, , \\
    \mathcal{G}_{\rm vel} &=~ 0.058 \, \Dot{\mathcal{R}}_{\rm cl}^{0.65} [1 - \exp(- 9.5\Bar{\tau}_{\rm s}^{3/4})] \exp(- 0.02 \, \Dot{\mathcal{R}}_{\rm cl}^{0.6}) \, ,
\end{align}
and $\Dot{\mathcal{R}}_{\rm cl} \equiv \lvert \Dot{R}_{\rm cl} \rvert/b$. The predictions of this fit are plotted in Fig. \ref{M_f velocity gradients plot}, and seen to be in good agreement with the full analytical solution from Eq.~(\ref{M_F velocity gradients exact}). In gravitationally bound star-forming clouds, we expect that $\lvert \Dot{R}_{\rm cl} \rvert/b \lesssim 10$, and so that velocity gradients during this phase can suppress Ly$\alpha$ feedback by a factor $\lesssim \textrm{few}$. If the cloud is disrupted and the gas accelerated to much higher velocities --- by Ly$\alpha$ feedback or some other mechanism --- velocity gradients will suppress Ly$\alpha$ feedback further, along with the reduced \HI column densities.

Earlier work has also pointed out the potential dampening of Ly$\alpha$ radiation pressure by velocity gradients \citep[][]{Braun1989, Birthell1990, Oh2002, Dijkstra2008, McKee2008, Tomaselli2021}. \cite{Birthell1990} and \cite{Oh2002} studied the effect of velocity gradients using the fit to the trapping time in a slab from \cite{Bonilha1979}. They found a marginal impact on Ly$\alpha$ radiation pressure in bound clouds, consistent with our results. We note, however, that in this paper we have, for the first time, calculated $M_{\rm F}$ directly, rather than relying on the trapping time as a proxy for $M_{\rm F}$. As a result, we find a weaker scaling of $M_{\rm F}$ with $\lvert \Dot{R}_{\rm cl} \rvert/b$ than \cite{Bonilha1979} found for the trapping time.\footnote{In fact, we have also computed the trapping time and found $t_{\rm trap}/(R_{\rm cl}/c) \propto (\lvert \Dot{R}_{\rm cl} \rvert/b)^{-2/3}$, regardless of $\Bar{\tau}_{\rm s}$ \citep[presented in][]{Smith2025}. This scaling is less steep than that of \cite{Bonilha1979}, who found $t_{\rm trap}/(R_{\rm cl}/c) \propto (\lvert \Dot{R}_{\rm cl} \rvert/b)^{-3/2}$. The discrepancy may simply be due to advancements in MCRT simulations (backed up by accurate analytical calculations here) since the late 70's. } 

\subsubsection{Effect of turbulent density fluctuations on Ly$\alpha$ feedback}
\label{effect of turbulence on M_F section}

\begin{figure}
    \includegraphics[trim={0.1cm 0.1cm 0.1cm 0.1cm},clip,width = 0.94\columnwidth]{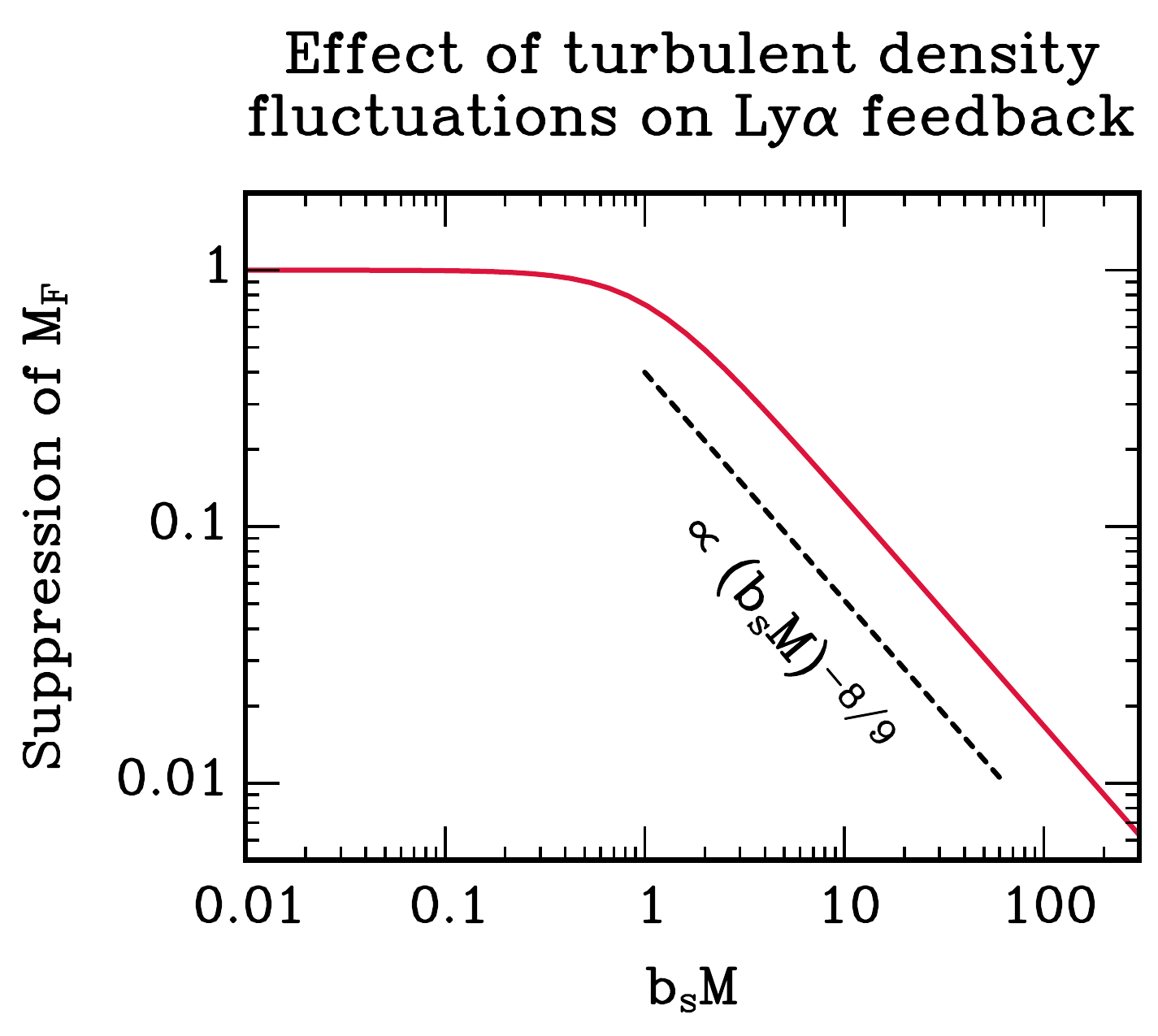}
    \caption{The relative suppression of $M_{\rm F}$ due to turbulent density fluctuations (red solid line), as a function of $b_s \mathcal{M}$ (as predicted by Eq.~\ref{M_F general turbulence}), where $\mathcal{M}$ is the turbulent Mach number, and $b_s$ the turbulent driving parameter. Ly$\alpha$ destruction, large-scale velocity gradients, and recoil have been ignored in this plot. The suppression scales as $M_{\rm F} \propto (b_s \mathcal{M})^{-8/9}$ for $b_s \mathcal{M} \gtrsim 1$.
}
    \label{M_F turbulence}
\end{figure}

So far we have focused on Ly$\alpha$ feedback in uniform clouds. How is Ly$\alpha$ feedback affected if we introduce large turbulent density fluctuations? We can address this using the approximate solution in Eq.~(\ref{series solution turbulent cloud}) for a turbulent cloud. The force multiplier can be derived following the same steps as in the non-turbulent case. The end result is:\footnote{We find that $M_{\rm F} \propto [\taucl^{\star}/\langle \taucl \rangle](a_{\rm v} \taucl^{\star})^{1/3}$. Since $\taucl^{\star} = e^{-\sigma_s^2/3} \langle \taucl \rangle$, and $\sigma_s^2 = \ln[1 + (b_s \mathcal{M})^2]$, this yields the final scaling with the Mach number. }
\begin{align}
    \dfrac{M_{\rm F}}{[a_{\rm v} \langle \taucl \rangle]^{1/3}} &=~ 4\sqrt{6} \left( \dfrac{3}{2 \pi^3} \right)^{1/6} \dfrac{1}{[1 + (b_s \mathcal{M})^2]^{4/9}}  \nonumber \\ &\times~  \sum_{n = 0}^{\infty} \dfrac{[\sin(n_{\rm o} \pi \Bar{\tau}_{\rm s}) - n_{\rm o}\pi \Bar{\tau}_{\rm s}\cos(n_{\rm o} \pi \Bar{\tau}_{\rm s})]}{\Bar{\tau}_{\rm s}^3 (n_{\rm o}\pi)^3 n_{\rm o}^{1/3} (\mathcal{P} + n_{\rm o} \pi \mathcal{D}_{n_{\rm o}})} \label{M_F general turbulence} \\ &\times~ \int_{-\infty}^{\infty} \textrm{d}z \hspace{1 pt} \mathcal{F}_{n_{\rm o}}(z) \,  \nonumber,
\end{align}
where we remind the reader that $\mathcal{M}$ is the 3D turbulent Mach number, and $1/3 < b_s < 1$ is the turbulent driving parameter ($b_s \simeq 0.4$ for a `natural' mix of solenoidal and compressive modes). Thus, for a cloud that is (on average) static and dust-free, we expect that strong turbulence ($\mathcal{M} \gg 1$) will suppress the force multiplier as $M_{\rm F} \propto \mathcal{M}^{-8/9}$, as shown in Fig. \ref{M_F turbulence}. This suppression is due to the creation of low-density channels in the cloud, which hasten Ly$\alpha$ escape (Fig. \ref{Schematic turbulent media}). 

Typical GMCs in the Milky Way have $\mathcal{M} \lesssim 10$ \citep{Heyer2015, Syed2020, Wang2020}, which suggest that turbulence can suppress Ly$\alpha$ feedback by a factor $\sim \rm few - 10$. This is significant and cannot be ignored, but it is far from enough to render Ly$\alpha$ feedback unimportant --- even in a turbulent $\mathcal{M} \sim 10$ cloud, the force multiplier can still easily be $M_{\rm F} \gtrsim \textrm{few} \times 10$ (see Fig. \ref{discussion_overview}). More detailed tests of this prediction would have to await systematic Ly$\alpha$ MCRT experiments in turbulent clouds. We note, however, that our results --- that turbulence can dampen Ly$\alpha$ feedback by a factor $\lesssim 10$ --- are consistent with some earlier preliminary simulations and order-of-magnitude estimates. For example, \cite{Smith2019} calculated $M_{\rm F}$ in a simulated metal-poor and turbulent reionization-era galaxy, using post-processing Ly$\alpha$ MCRT. Despite significant turbulence, many regions still had $M_{\rm F} \sim 100$. 

In one of the early pioneering works on Ly$\alpha$ feedback, \cite{Cox1985} used simple order-of-magnitude estimates to argue that low-density channels in the ISM, in the form of `cracks' and `crevices', should only lead to moderate suppression of Ly$\alpha$ trapping. Our more detailed calculation in this paper supports this conclusion. Finally, we note that \cite{Skinner2015} and \cite{Menon2022_IR} have found that turbulence can dampen multiple-scattering infrared radiation pressure on dust, which, albeit somewhat different from Ly$\alpha$ radiation pressure, share similar features (e.g. preferential escape along low-density channels).\footnote{UV and optical photons can be absorbed by dust and re-radiated in the IR. If the medium is optically thick in the IR ($\tau_{\rm IR} \gg 1$), the photons can scatter many times. The boost in radiation pressure (the equivalent to $M_{\rm F}$ for Ly$\alpha$ feedback) is $\simeq \tau_{\rm IR}$ for a uniform medium with a central source \citep[see e.g. Appendix 6 in][]{Nebrin2022}. As in the case of Ly$\alpha$ feedback, low-density channels and flux cancellation can dampen the momentum input \citep[e.g.][]{Krumholz2009, Krumholz2012, Krumholz2013, Skinner2015, Menon2022_IR}.} These authors find a suppression by a factor $\sim \rm few - 10$ for $\mathcal{M} \sim 10$, similar to what we find for Ly$\alpha$ feedback.

It is straightforward to generalize our fit in Eq.~(\ref{M_F fit velocity}) to incorporate turbulent density fluctuations. From Eqs.~(\ref{M_F general turbulence}) and (\ref{xtilde expression turb})--(\ref{epsilontilde expression turb}) we get 
\begin{equation}
    M_{\rm F}^{\rm fit} =  \dfrac{\eta_{\rm s} [a_{\rm v}\langle \taucl \rangle ]^{1/3} \, [1 + (b_s \mathcal{M})^2]^{-4/9}}{\mathcal{A}^{1/3} + [\mathcal{P}/(\mathcal{G}_{\rm rec} + \mathcal{G}_{\rm vel})\mathcal{P}_{\rm crit,ext}] + \mathcal{F}_{\rm vel}} \hspace{1 pt} , \label{M_F fit turbulence, final}
\end{equation}
where the dimensionless parameters that enter this expression are now generalized as follows:
\begin{align}
    \mathcal{P} &=~ \dfrac{\sqrt{6 \pi} \langle p_{\rm d} \rangle_{m} \langle \taucl \rangle}{2[1+ (b_s \mathcal{M})^2]^{1/3}} \,, \\ \Tilde{x}_1 &=~ \dfrac{1.207 \, [a_{\rm v} \langle \taucl \rangle]^{1/3} \Bar{x}}{[1 + (b_s \mathcal{M})^2]^{1/9}}   \,, \\ \Tilde{\epsilon}_1 &=~ \dfrac{0.4707 \, [a_{\rm v} \langle \taucl \rangle]^{1/3} \langle \tau_{\rm c,abs} \rangle }{[1 + (b_s \mathcal{M})^2]^{4/9}}  \,,  \\ \Dot{\mathcal{R}}_{\rm cl} &=~  \dfrac{\lvert \Dot{R}_{\rm cl} \rvert/b}{ [1 + (b_s \mathcal{M})^2]^{1/3}} \, .
\end{align}
Eq.~(\ref{M_F fit turbulence, final}) is the final version of the fit $M_{\rm F}^{\rm fit}$ for the Ly$\alpha$ force multiplier. For convenience, we provide a simple \textsc{Python} implementation of the fit at the Github repository associated with this project.\footnote{\url{https://github.com/olofnebrin/Lyman-alpha-feedback}.} In the next section we study the physical processes that determine $\langle p_{\rm d} \rangle_{ m}$.

\subsection{Destruction probability of Lyman-\texorpdfstring{\boldmath$\alpha$}{alpha} photons}
\label{Lya destruction}

In Sec. \ref{Lyadestruction general effect}, we saw that the Ly$\alpha$ radiation pressure force multiplier drops as $M_{\rm F} \propto 1/(p_{\rm d} \tau_{\rm cl})$ for $p_{\rm d} \tau_{\rm cl} \gtrsim 1$. Thus, even a tiny destruction probability $p_{\rm d} \gtrsim 10/\taucl \sim 10^{-11} - 10^{-9}$ can significantly limit Ly$\alpha$ feedback in star-forming clouds with \HI surface densities $\Sigma_{\rm H I} \sim 10^2 - 10^4 ~ \Msun \hspace{1 pt} \rm pc^{-2}$. In this section we estimate the destruction probability $p_{\rm d}$.

\subsubsection{Ly$\alpha$ destruction via $2p \rightarrow 2s$ transitions}

If Ly$\alpha$ absorption is followed by a $2p\rightarrow 2s$ transition (with probability $p_{2p,2s}$), and then a $2s \rightarrow 1s$ two-photon decay transition (with probability $p_{2 \gamma}$), the original Ly$\alpha$ photon is effectively destroyed (see Fig. \ref{LyafeedbackSketch}). The destruction probability $p_{\rm d, HI}$ associated with this process is given by
\begin{align}
    p_{\rm d, HI} &=~ p_{2p,2s} p_{2\gamma} + p_{2p,2s}(p_{2s,2p}p_{2p,2s}) p_{2\gamma} \nonumber \\ &+~  p_{2p,2s}(p_{2s,2p}p_{2p,2s}p_{2s,2p}p_{2p,2s}) p_{2\gamma} + ... \nonumber \\ &=~ p_{2p,2s} p_{2\gamma} \sum_{k=0}^{\infty} (p_{2s,2p}p_{2p,2s})^k \hspace{1 pt}.
\end{align}
The terms with parentheses allow for multiple transitions between the $2p$ and $2s$ levels before the final decay to $1s$ via two-photon emission, with $p_{2s,2p}$ being the probability of a $2s \rightarrow 2p$ transition. The infinite series can be evaluated, giving us a closed-form expression for the Ly$\alpha$ destruction probability: 
\begin{equation}
    p_{\rm d, HI} = \frac{p_{2p,2s} p_{2\gamma}}{1 - p_{2s,2p}p_{2p,2s}} \hspace{1 pt} \label{p_d first version}.
\end{equation}
The probabilities that enter this expression are:
\begin{align}
    p_{2p,2s} &=~ \frac{\mathcal{R}_{2p,2s}}{A_{\rm Ly\alpha}+ \mathcal{R}_{2p,2s}} \hspace{1 pt} \label{p_2p2s}, \\ p_{2s,2p} &=~ \frac{\mathcal{R}_{2s,2p}}{A_{2\gamma} + \mathcal{R}_{2s,2p}} \hspace{1 pt} , \label{p_2s2p} \\ p_{2 \gamma} &=~ \frac{A_{2\gamma}}{A_{2\gamma} + \mathcal{R}_{2s,2p}} \hspace{1 pt} \label{p_2phot} .
\end{align}
Here $\mathcal{R}_{ij}$ is the rate of $i \rightarrow j$ transitions (s$^{-1}$), and the Einstein-A coefficients are $A_{\rm Ly\alpha} = 6.265 \times 10^{8} ~ \rm s^{-1}$, and $A_{2\gamma} = 8.2206 ~ \rm s^{-1}$ for the Ly$\alpha$ and two-photon transitions, respectively \citep[e.g.][and references therein]{Draine2011, Chluba2008}. With Eqs.~(\ref{p_2p2s})--(\ref{p_2phot}), we rewrite the destruction probability in Eq.~(\ref{p_d first version}) as:\footnote{If we ignore stimulated and spontaneous transitions between $2s$ and $2p$, and use $\mathcal{R}_{2p,2s} \ll (A_{\rm Ly\alpha}/A_{2\gamma})(A_{2\gamma} + \mathcal{R}_{2s,2p})$, and detailed balance ($\mathcal{C}_{2p,2s} \simeq \mathcal{C}_{2s,2p}/3$), we get:
\begin{equation}
    p_{\rm d,HI} \simeq \dfrac{A_{2\gamma} \mathcal{C}_{2s,2p}}{3A_{\rm Ly\alpha}(A_{2\gamma} +  \mathcal{C}_{2s,2p})} = \dfrac{A_{2\gamma}}{3 A_{\rm Ly\alpha}} \dfrac{1}{(A_{2\gamma}/\mathcal{C}_{2s,2p} + 1)} \, \nonumber.
\end{equation}
This is identical to eq.~(21) of \cite{Spitzer1951}, and eq.~(B17) of \cite{McKee2008}, if we note that $A_{2\gamma}/3A_{\rm Ly\alpha} \simeq 4.4 \times 10^{-9}$. }
\begin{equation}
    p_{\rm d, HI} = \frac{\mathcal{R}_{2p,2s}}{(A_{\rm Ly \alpha}/A_{2 \gamma})( A_{2 \gamma} + \mathcal{R}_{2s,2p}) + \mathcal{R}_{2p,2s}} \hspace{1 pt} \label{p_d general}.
\end{equation}
Considering collisional excitation/de-excitation ($\mathcal{C}_{ij}$), CMB-stimulated transitions ($\Gamma_{ij}^{\rm CMB}$), and spontaneous emission ($A_{ij}$), the rates $\mathcal{R}_{ij}$ are:
\begin{align}
    \mathcal{R}_{2p,2s} &=~ \dfrac{2}{3} A_{2p,2s} + \mathcal{C}_{2p,2s} + \Gamma_{2p,2s}^{\rm CMB} \hspace{1 pt} \label{R2p2s}, \\  \mathcal{R}_{2s,2p} &=~ A_{2s,2p} + \mathcal{C}_{2s,2p} + \Gamma_{2s,2p}^{\rm CMB} \hspace{1 pt} \label{R2s2p} .
\end{align}
The terms entering these equations are discussed in detail below.

\subsubsection{Spontaneous emission rates}

$A_{2p,2s} = 8.78 \times 10^{-7} ~ \rm s^{-1}$ and $A_{2s,2p} = 1.597 \times 10^{-9} ~ \rm s^{-1}$ are the Einstein-A coefficients for the $2p_{3/2} \rightarrow 2s_{1/2}$ and $2s_{1/2} \rightarrow 2p_{1/2}$ transitions, respectively \citep{Dennison2005}. The factor $2/3$ in Eq.~(\ref{R2p2s}) comes from the assumption that the two $2p$ levels ($2p_{3/2}$ and $2p_{1/2}$) are populated according to their statistical weights, so that $n_{2p_{3/2}} = (2/3)n_{2p}$ and $n_{2p_{1/2}} = (1/3)n_{2p}$. As noted by \cite{Dijkstra2016}, this is a safe assumption because of the Wouthuysen-Field (WF) effect \citep{Wouthuysen1952, Field1958, Field1959}, wherein the colour temperature of the Ly$\alpha$ spectrum is tightly coupled to the gas temperature as a result of atomic recoil from Ly$\alpha$ scattering --- indeed, we saw this explicitly in Fig. \ref{spectrum}. Using Ly$\alpha$ MCRT experiments, \cite{Seon2020} have shown that the WF effect is strong in clouds with Ly$\alpha$ optical depths $\tau_{\rm cl} \gtrsim 100$, which is easily satisfied for star-forming clouds of interest to us. Since the energy difference between the two $2p$ levels is only $\Delta E/k_{\rm B} \simeq 0.53\,\rm K$ \citep{Dennison2005}, and the gas temperature limited to $T > 27.3 [(1+z)/10]\,\rm K$ by the CMB, we conclude that WF effect is expected to populate the $2p$ levels according to their statistical weights.\footnote{Collisional mixing at very high gas densities can conceivably also lead to a statistical distribution for the $2p$ levels. }

\subsubsection{Stimulated transition rates}

\cite{Hirata2006} and \cite{Dijkstra2016} have calculated the rate of stimulated $2p \rightleftharpoons 2s$ transitions by the CMB at redshift $z$ to be:
\begin{align}
    \Gamma_{2p,2s}^{\rm CMB} ~&=~  3.1 \times 10^{-6} \, (1+z) \hspace{2 pt} \rm s^{-1} \, \label{Gamm CMB 2p2s}, \\ 
    \Gamma_{2s,2p}^{\rm CMB} ~&=~ 3 \Gamma_{2p,2s}^{\rm CMB} = 9.3  \times 10^{-6} \, (1+z) \hspace{2 pt} \rm s^{-1} \, \label{Gamm CMB 2s2p}.
\end{align}
With these rates, we find that the CMB alone can supply a destruction probability $p_{\rm d,HI} \sim 10^{-13} \, [(1+z)/20]$, which is small at typical redshifts of interest ($z \lesssim 30$). However, we keep the rates in our modelling for a few reasons. Firstly, these rates are easily implemented, and provide a floor to $p_{\rm d}$ at a given redshift. Secondly, it makes it easier to generalize the model further in the future to consider, e.g., maser amplification of the CMB \citep[see][and the discussion in Sec. \ref{neglected processes for pd}]{Dijkstra2016}. Finally, there is the possibility that the very first stars formed at redshifts $z \sim 100-400$ if the primordial power spectrum is blue-tilted on small scales \citep{Hirano2015_bluetilt, Hirano2024, Ito2024}. In such scenarios, the impact of the CMB on Ly$\alpha$ stellar feedback from Pop III stars cannot be ignored. For these reasons, we include the stimulated transition rates in Eqs.~(\ref{Gamm CMB 2p2s})--(\ref{Gamm CMB 2s2p}).

\subsubsection{Collisional excitation/de-excitation rates}
\label{Subsection CE/CDE rates}

\begin{table}
\caption{Collisional excitation/de-excitation processes affecting Ly$\alpha$ feedback in star-forming clouds.}
\begin{tabular}{l l l}
\hline
\hline
Reaction & Rate & Relative importance \\

\hline
\hline

\vspace{3 pt}

$\text{H}(2s) + \text{H}^{+} \rightleftharpoons \text{H}(2p) + \text{H}^{+}$ & Eq.~(\ref{H+rate}) & Negligible or minor \\

\vspace{3 pt}

$\text{H}(2s) + e^{-} \rightleftharpoons \text{H}(2p) + e^{-}$ & Eq.~(\ref{e-rate}) & Negligible or minor  \\

\vspace{3 pt}

$\text{H}(2s) + \text{He}^0 \rightleftharpoons \text{H}(2p) + \text{He}^0$ & Eq.~(\ref{HeRate}) & Major  \\

\vspace{3 pt}

$\text{H}(2s) + \text{H}_2 \rightleftharpoons \text{H}(2p) + \text{H}_2$ & Eq.~(\ref{H2Rate}) & Minor/major$^\dagger$ \\

\vspace{3 pt}

$\text{H}(2s) + \text{H}(1s) \rightleftharpoons \text{H}(2p) + \text{H}(1s)$ & Eq.~(\ref{HIrate}) & Major/minor$^\dagger$ \\

\vspace{3 pt}

$\text{H}(2s) + \text{H}(2s) \rightleftharpoons \text{H}(2p) + \text{H}(2p)$ &  & Negligible$^{\diamond}$  \\

\hline
\hline
\end{tabular}
\vspace{1 pt}\\
$^\dagger$: If hydrogen is mostly atomic, collisions with H($1s$) dominates over collisions with H$_2$. If hydrogen is mostly molecular, the reverse is the case.
$^{\diamond}$: See the discussion in Sec. \ref{neglected processes for pd}.
\label{CEtable}
\end{table}

\begin{figure}
    \includegraphics[trim={0.1cm 0.1cm 0.1cm 0.1cm},clip,width = \columnwidth]{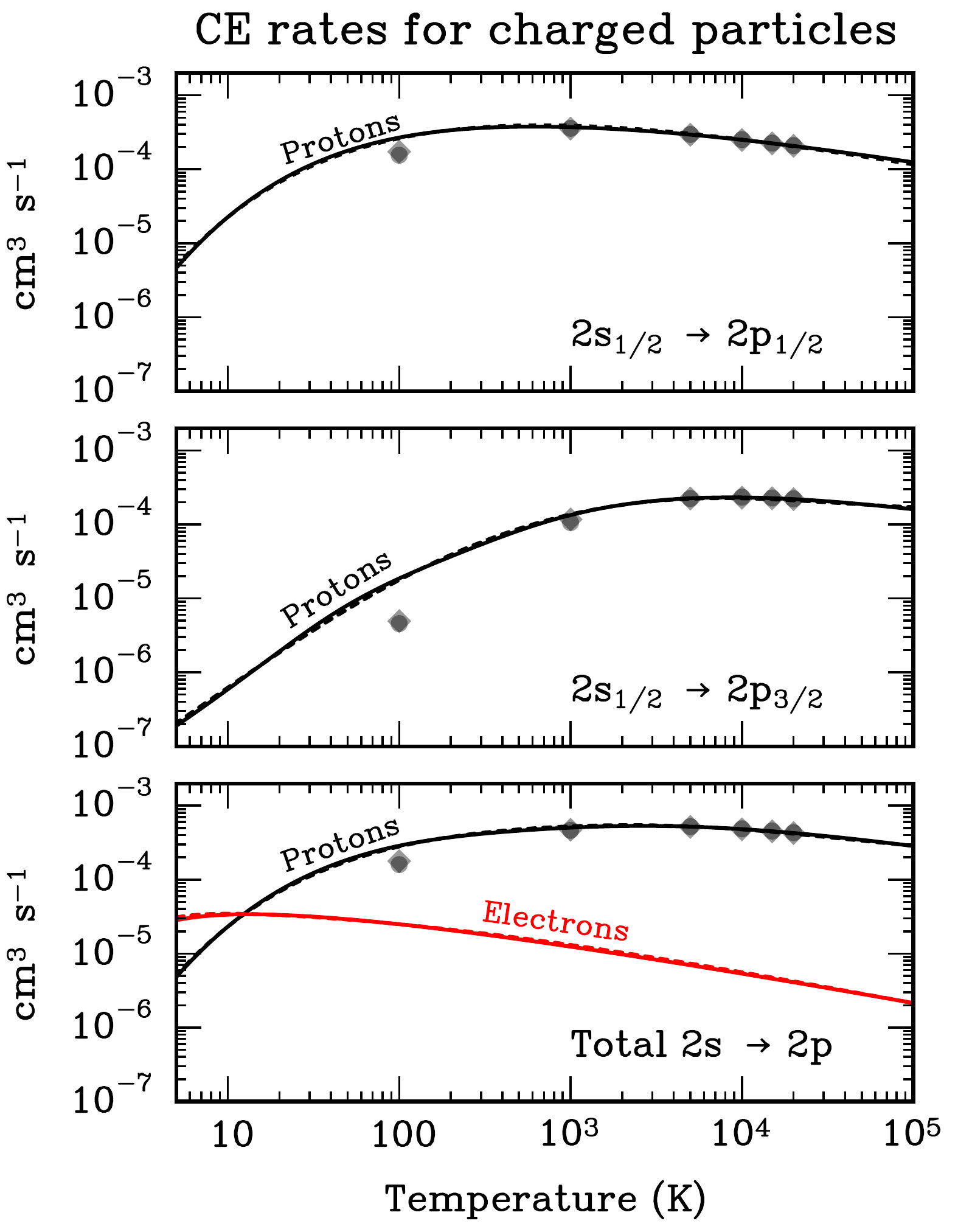}
    \caption{Plots showing the CE rate coefficients for protons (black lines), and electrons (red line). Solid lines show numerical integration using cross-sections by \citet{Struensee1988} and \citet{Chibisov1969} for protons and electrons, respectively. Fits are shown as dashed lines, and are given in Eqs.~(\ref{H+rate})--(\ref{e-rate}). Grey symbols are results for protons from other calculations \citep{Vrinceaunu2001, Guzman2017_PSM}, as compiled in \citet{Guzman2017}. The upper and middle panels show the CE rate coefficients for $2s \rightarrow 2p_{1/2}$ and $2s \rightarrow 2p_{3/2}$, respectively, while the lower panel panel shows the total $2s \rightarrow 2p$ CE rate coefficient. 
}
    \label{C2s2p_charged}
\end{figure}

$\mathcal{C}_{2p,2s}$ ($\mathcal{C}_{2s,2p}$) is the $2p \rightarrow 2s$ collisional de-excitation (CDE) ($2s \rightarrow 2p$ collisional excitation, CE) rate. A summary of the CE/CDE processes considered in this paper are listed in Table \ref{CEtable} and discussed in order below. For all relevant temperatures they are related via $\mathcal{C}_{2s,2p} \simeq 3\mathcal{C}_{2p,2s}$ by detailed balance. 

It is well-known that protons and electrons contribute to this $l$-changing CDE rate \citep[e.g.][]{Spitzer1951, Purcell1952, Seaton1955, Pengelly1964, Chibisov1969, Struensee1988, Guzman2016, Guzman2017, Aggarwal2018, Badnell2021}. As noted by \cite{Guzman2017}, there are unfortunately only theoretical estimates for these rates, with differing results depending on the method used. Using the quantum mechanically computed cross-section for collisions with protons by \cite{Struensee1988} yields a CE rate that is well-fitted by:
\begin{align}
    \mathcal{C}^{\rm HII}_{2s,2p} &\simeq ~ \dfrac{6.98 \times 10^{-3}}{T^{0.5}} \left(1 + 0.993 \log_{10} T \right) e^{-(601/T)^{0.406}} n_{\rm HII}  \nonumber \\ &+~ 10^{-7.69 + 1.21 y + 0.455 y^2 - 0.199 y^3 + 0.0182 y^4} n_{\rm HII} \label{H+rate} \hspace{1 pt} ,
\end{align}
where $y \equiv \log_{10} T$. The first and second lines in Eq.~(\ref{H+rate}) represent the contributions from transitions to $2p_{1/2}$ and $2p_{3/2}$, respectively. We plot these rates in Fig. \ref{C2s2p_charged} (black lines), and also show results from the quantum mechanical calculation by \cite{Vrinceaunu2001} and the Bethe--Borne calculation of \cite{Guzman2017_PSM}, as compiled in \cite{Guzman2017}. All rates are very similar for $T \gtrsim 10^3\,\rm K$, but the estimate of \cite{Struensee1988} is higher by a factor $\sim 2-4$ at $T = 100\,\rm K$, highlighting theoretical uncertainties.

\cite{Chibisov1969} has estimated the CE cross-section for collisions with electrons \citep[see also][p. 22]{Janev1987}. The corresponding CE rate can be fitted by:
\begin{equation}
     \mathcal{C}^{ e^-}_{2s,2p} \simeq \dfrac{5 \times 10^{-5} \hspace{1 pt} n_{\rm e}}{1 + T_{100}^{0.45} } \hspace{1 pt} e^{-(2.5/T)^2}  \label{e-rate} \hspace{1 pt} .
\end{equation}
The numerically integrated CE rate and the fit in Eq.~(\ref{e-rate}) are shown in Fig. \ref{C2s2p_charged} as red solid and dashed lines, respectively. We also note that more recently, \cite{Aggarwal2018} has estimated the rate of this process, finding a rate that is larger than the one by \cite{Chibisov1969} by a factor $\sim 10$. Regardless of which estimate one adopts, however, collisions with protons dominate the CE rate for charged particles, and so we neglect these uncertainties here. 

Since star-forming regions are cold and self-shielding, the ionization fraction is expected to be very small (see discussion below). Thus, the CDE rate by charged particles is typically small. In the astrophysical literature, the small rate of CDE by \textit{charged} particles has sometimes been taken to imply that CDE as a whole has a negligible effect on Ly$\alpha$ scattering and feedback \citep[e.g.][]{McKee2008}. However, neutral atoms and molecules also contribute to the CDE rate, and in fact are likely to dominate in star-forming clouds. 

Cross-sections for quenching of metastable hydrogen (i.e. $2s \rightarrow 2p$) by \HeI and H$_2$ have been estimated both theoretically and experimentally \citep[e.g.][]{Byron1971, Slocomb1971, Dose1974, Dose1975, Ryan1977, Vassilev1990, Terazawa1993}. \cite{Ryan1977} determined the quenching cross-sections for H($2s$)--He and H($2s$)--H$_2$ collisions for velocities $> 3 ~ \textrm{km s}^{-1}$. \cite{Vassilev1990} revisited the H($2s$)--H$_2$ cross-section for a similar velocity range, finding similar results to \cite{Ryan1977}, and good agreement with simple theoretical predictions. 

Using the quenching cross-sections from \cite{Ryan1977} and \cite{Vassilev1990} for H($2s$)--He and H($2s$)--H$_2$, respectively, we find approximate CE rates of:\footnote{The power law fits/models of \cite{Ryan1977} and \cite{Vassilev1990} for the cross-sections were used to give simple analytical expressions for the CE rates. For a power law cross-section $\sigma_{ij}(u) = \sigma_{ij,0} (u/u_0)^{-\beta}$ the Maxwellian-averaged rate coefficient $\langle \sigma_{ij} u \rangle$ can be evaluated analytically:
\begin{equation}
    \langle \sigma_{ij} u \rangle = \sigma_{ij,0} u_{0}^{\beta} \left( \dfrac{2 k_{\rm B} T}{\mu} \right)^{(1-\beta)/2} \dfrac{2}{\sqrt{\pi}}\Gamma\left( 2 - \dfrac{\beta}{2} \right) \hspace{1 pt} \nonumber,
\end{equation}
where $\mu$ is the reduced mass of the target-projectile system, and $\Gamma$ the gamma function. \cite{Ryan1977} and \cite{Vassilev1990} give $(\sigma_{2s,2p,0},\beta) = (8.5\times 10^{-15} ~ \textrm{cm}^2, \hspace{1 pt} 0.37)$ and $(\sigma_{2s,2p,0},\beta) = (9.9\times 10^{-15} ~ \textrm{cm}^2, \hspace{1 pt} 2/5)$ for \HeI and H$_2$, respectively, with $u_0 = 10^6 ~ \textrm{cm s}^{-1}$.} 
\begin{align}
    \mathcal{C}_{2s,2p}^{\rm HeI} &\simeq~ 2.6 \times 10^{-9} \hspace{1 pt} T_{100}^{0.315} \hspace{1 pt} n_{\rm HeI} \label{HeRate} \hspace{1 pt} , \\
    \mathcal{C}_{2s,2p}^{\rm H_2} &\simeq~ 1.7 \times 10^{-9} \hspace{1 pt} T_{100}^{0.3} \hspace{1 pt} n_{\rm H_2} \label{H2Rate} \hspace{1 pt} .
\end{align}
It should be cautioned that \cite{Ryan1977} and \cite{Vassilev1990} (and others) measured cross-sections for relative velocities $> 3  ~ \textrm{km }\textrm{s}^{-1}$. Thus, the estimates of Eqs.~(\ref{HeRate}) and (\ref{H2Rate}) are extrapolations for temperatures $T \lesssim \textrm{few} \times 100\,\rm K$. However, the velocity dependence of the cross-section is predicted to be weak theoretically, and it therefore seems unlikely that the rates are off by more than a factor $\sim 2$--$3$ down to $T \sim 10\,\rm K$. More experimental and theoretical estimates in the low-temperature regime would clearly be useful to model Ly$\alpha$ feedback. 

Finally, metastable hydrogen could also be quenched in collisions with neutral hydrogen. Given the great abundance of atomic hydrogen in early star-forming clouds, such reactions could be very important, if not dominant. Apart from a theoretical estimate by \cite{Flannery1971} at very high energies ($> 2.25\,\rm keV$), we are not aware of any published cross-section for $\textrm{H}(2s)+\textrm{H}(1s) \rightarrow \textrm{H}(2p)+\textrm{H}(1s)$.\footnote{In the context of modelling Ly$\alpha$ emission from hot Jupiters, \cite{Menager2013} too identified this process as potentially important, but they could not find any published data and therefore did not include it. \cite{Laursen2010} argued that the cross-section for this process should be of order $n^2 a_0^2 \sim 10^{-16} ~ \rm cm^2$ for energy level $n$, which is a factor $\sim \mathcal{O}(10)$ smaller than the result of our detailed calculation, and does not capture the complicated energy dependence of the cross-section that we find. \cite{Spitzer1951} claimed that CE/CDE by neutral hydrogen can be neglected if $n_{\rm HI} < 10^3 n_{\rm e}$, but provide no reference or calculation in support of this result. In the context of self-shielded gas clouds, we \textit{do} have $n_{\rm HI} > 10^3 n_{\rm e}$, which sparked our study of this process.} Because of this, we perform an \textit{ab initio} calculation of the cross-section over a wide range of collision energies, $5 \times 10^{-6} ~ \textrm{eV} < E < 100 ~ \textrm{eV}$. 

The cross-section is calculated using a molecular close-coupling approach, where all degrees of freedom are treated quantum mechanically \citep[][]{stenrup2009}. In this calculation, we use the adiabatic potential energy curves and non-adiabatic couplings for states of $^1\Sigma^+_{g/u}$ and $^3\Sigma^+_{u}$ that were previously calculated \textit{ab initio} by some of the present authors \citep[][]{hornquist2022,hornquist2023}. For states of $^3\Sigma^+_g$ symmetry, new structure calculations are performed. The radial non-adiabatic couplings are utilized to transform the coupled Schrödinger equation into a strict diabatic representation \citep[][]{mead1982}. A numerical solution of the diabatic coupled Schrödinger equation is then obtained by using the log-derivative method developed by \cite{johnson1973}. The reader is referred to Appendix \ref{Appendix H(2s) cross section} for a more detailed description of the structure calculations and the methodology used to calculate the cross-section.

\begin{figure}
    \includegraphics[trim={0.1cm 0.1cm 0.1cm 0.1cm},clip,width = \columnwidth]{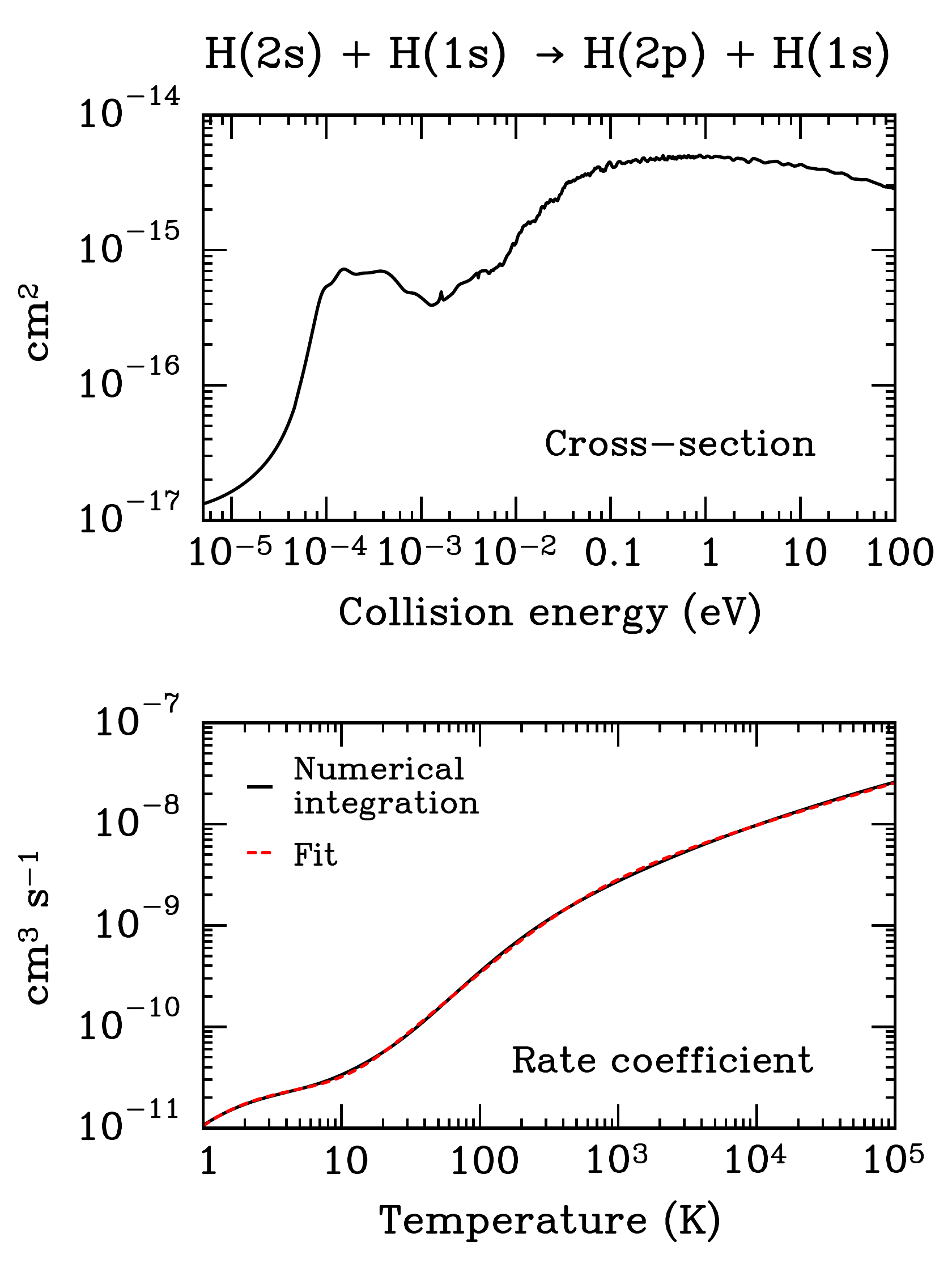}
    \caption{ \textbf{Upper panel}: The calculated cross-section as a function of collision energy for the collisional excitation process $\textrm{H}(2s)+\textrm{H}(1s) \rightarrow \textrm{H}(2p)+\textrm{H}(1s)$. See the main text and Appendix \ref{Appendix H(2s) cross section} for details on the calculation of the cross-section. The cross-section drops rapidly below $E \sim 10^{-4} ~ \rm eV$ since the collision energy approach the threshold for excitation. \textbf{Lower panel}: The corresponding rate coefficient. The black solid line show the rate coefficient obtained by numerical integration of the cross-section from the upper panel. The red dashed line show a fit (Eq.~\ref{HIrate}) to the numerical result for the rate coefficient.
}
    \label{C2s2p_atomicH}
\end{figure}

The cross-section and the corresponding rate coefficient are shown in Fig. \ref{C2s2p_atomicH}.\footnote{The data for the calculcated cross-section, seen in the upper panel of Fig. \ref{C2s2p_atomicH}, is available at the Github repository associated with this project: \url{https://github.com/olofnebrin/Lyman-alpha-feedback}.} As seen in the lower panel of the figure, the computed rate for $\textrm{H}(2s)+\textrm{H}(1s) \rightarrow \textrm{H}(2p)+\textrm{H}(1s)$ can be fitted by:
\begin{align}
    \mathcal{C}_{2s,2p}^{\rm HI} ~&\simeq~ 4.1 \times 10^{-11} \exp[-(1.6/T)^{3/4} - 0.15 \hspace{1 pt} T] \hspace{1 pt} n_{\rm HI} \nonumber \\ ~&+~ \dfrac{(1.6\times10^{-10} + 1.57 \times 10^{-12} \hspace{1 pt} T^{1.1})}{( 8.9 \hspace{1 pt} T^{-0.55} +  0.078 \hspace{1 pt} T^{0.35})^{2}} \hspace{1 pt} n_{\rm HI} \label{HIrate} \hspace{1 pt} ,
\end{align}
where we have used $n_{1s} \simeq n_{\rm HI}$ (see justification in Sec. \ref{neglected processes for pd}). The maximum relative error in the fit to the rate coefficient is $\sim 4.6\%$ for $1 ~ \textrm{K} < T < 10^5 ~ \textrm{K}$.

\begin{figure}
    \includegraphics[trim={0.1cm 0.1cm 0.1cm 0.1cm},clip,width = \columnwidth]{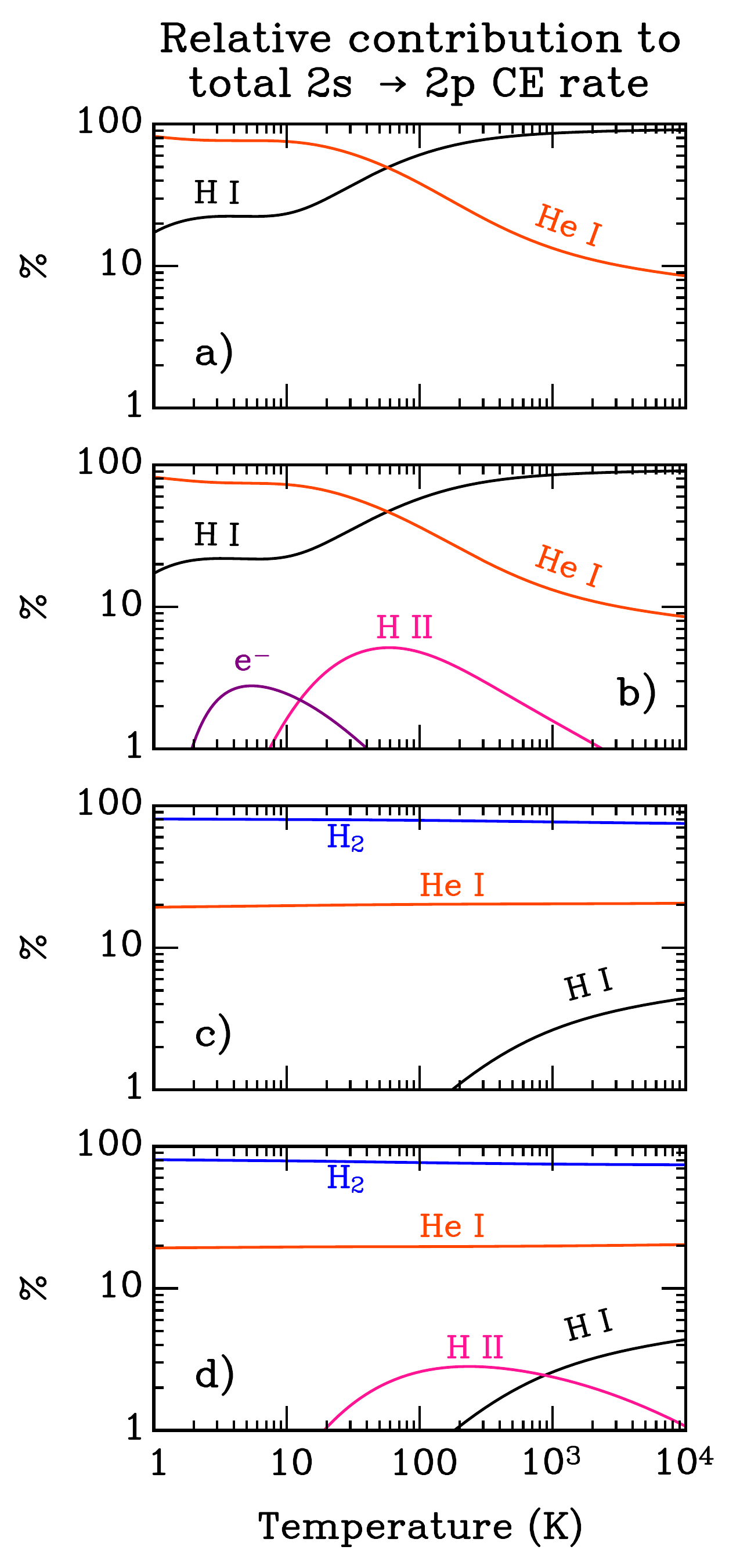}
    \caption{The relative contribution (in $\%$) to the total CE rate $\mathcal{C}_{2s,2p}$, for four different combinations of ionization fractions $x \equiv n_{\rm HII}/n_{\rm H} = n_{\rm e}/n_{\rm H}$ and molecular hydrogen fractions $f_{\rm H_2} \equiv n_{\rm H_2}/n_{\rm H}$. \textbf{Panel a)}: $x = 10^{-8}$ and $f_{\rm H_2} = 10^{-3}$ (typical values for dense gas in a primordial minihalo). \textbf{Panel b)}: $x = 10^{-6}$ and $f_{\rm H_2} = 10^{-3}$ (e.g. dense gas in a primordial minihalo that is unusually ionized). \textbf{Panel c)}: $x = 10^{-8}$ and $f_{\rm H_2} = 0.49$ (e.g. a dense molecular cloud which is efficiently self-shielded). \textbf{Panel d)}: $x = 10^{-6}$ and $f_{\rm H_2} = 0.49$ (e.g. a dense molecular cloud which is less self-shielded). In each scenario it is seen that collisions with charged particles contribute only a relatively small fraction to the total CE rate.
}
    \label{RelativeCErates}
\end{figure}

In Fig. \ref{RelativeCErates} the relative contributions to the total CE rate from all the considered processes are shown, for different combinations of ionization fractions $x = (10^{-8}, 10^{-6})$ and molecular hydrogen fractions $f_{\rm H_2} = (10^{-3}, 0.49)$. These values span a range of typical values for gas in present-day GMCs, as well as primordial gas clouds. GMCs can self-shield efficiently against photoionization of \HI and photodissociation of H$_2$ by the interstellar radiation field. Within them, H$_2$ can then form efficiently on dust grains. As a result, GMCs are almost fully molecular ($f_{\rm H_2} \sim 0.5$), and only a very small ionization fraction $x \sim 10^{-8} - 10^{-6}$ can be maintained by cosmic-ray ionization \citep{Caselli1998, Draine2011, Pineda2024}. In chemically pristine dark matter minihaloes and atomic-cooling haloes, wherein the first stars and galaxies are expected to have formed, gas-phase formation of H$_2$ leads to a typical H$_2$ abundance $f_{\rm H_2} \sim 10^{-3}$ \citep[e.g.][]{Tegmark1997, Oh2002, Gao2007,  Nebrin2023cooling, Prole2023}. Because of rapid recombinations in the dense and cool ($T \sim  \textrm{few} \times 100 - 10^3\,\rm K$) star-forming gas within these haloes, only a tiny ionization fraction $x \sim 10^{-8} - 10^{-7}$ is typically maintained \citep[e.g.][]{Prole2023, Gurian2024}. 

In panels b) and d) of Fig. \ref{RelativeCErates}, the relative CE rates for $x = 10^{-6}$ is shown for H$_2$ fractions $f_{\rm H_2} = 10^{-3}$ and $f_{\rm H_2} = 0.49$, respectively. In either case it is seen that CE by charged particles only contribute $< 10\%$ of the total CE rate. For a lower ionization fraction of $x = 10^{-8}$ -- adopted in panels a) and c) -- the contribution from collisions with charged particles is $< 1 \%$.

Given the above calculated rates and considerations we conclude that collisions with \HI, \HeI, and H$_2$ are the most important collisional processes with respect to Ly$\alpha$ feedback. Collisions with charged particles are never dominant in those dense --- and therefore self-shielded --- clouds where suppression of Ly$\alpha$ feedback is potentially important in the first place. Thus, to keep subgrid modelling as simple as possible while still retaining the essential physics, we will neglect collisions with charged particles. We will only include the CE rates in Eqs.~(\ref{HeRate})--(\ref{HIrate}), so that:
\begin{equation}
    \mathcal{C}_{2s,2p} \simeq \mathcal{C}_{2s,2p}^{\rm HeI} + \mathcal{C}_{2s,2p}^{\rm H_2} + \mathcal{C}_{2s,2p}^{\rm HI} \hspace{1 pt} . \label{C_2s2p summary}
\end{equation}
In this case the collisional de-excitation rate is $\mathcal{C}_{2p,2s} = \mathcal{C}_{2s,2p}/3$.

\subsubsection{Ly$\alpha$ destruction via H$_2$ line absorption}

\begin{figure}
    \includegraphics[trim={0.1cm 0.1cm 0.1cm 0.1cm},clip,width = 0.9\columnwidth]{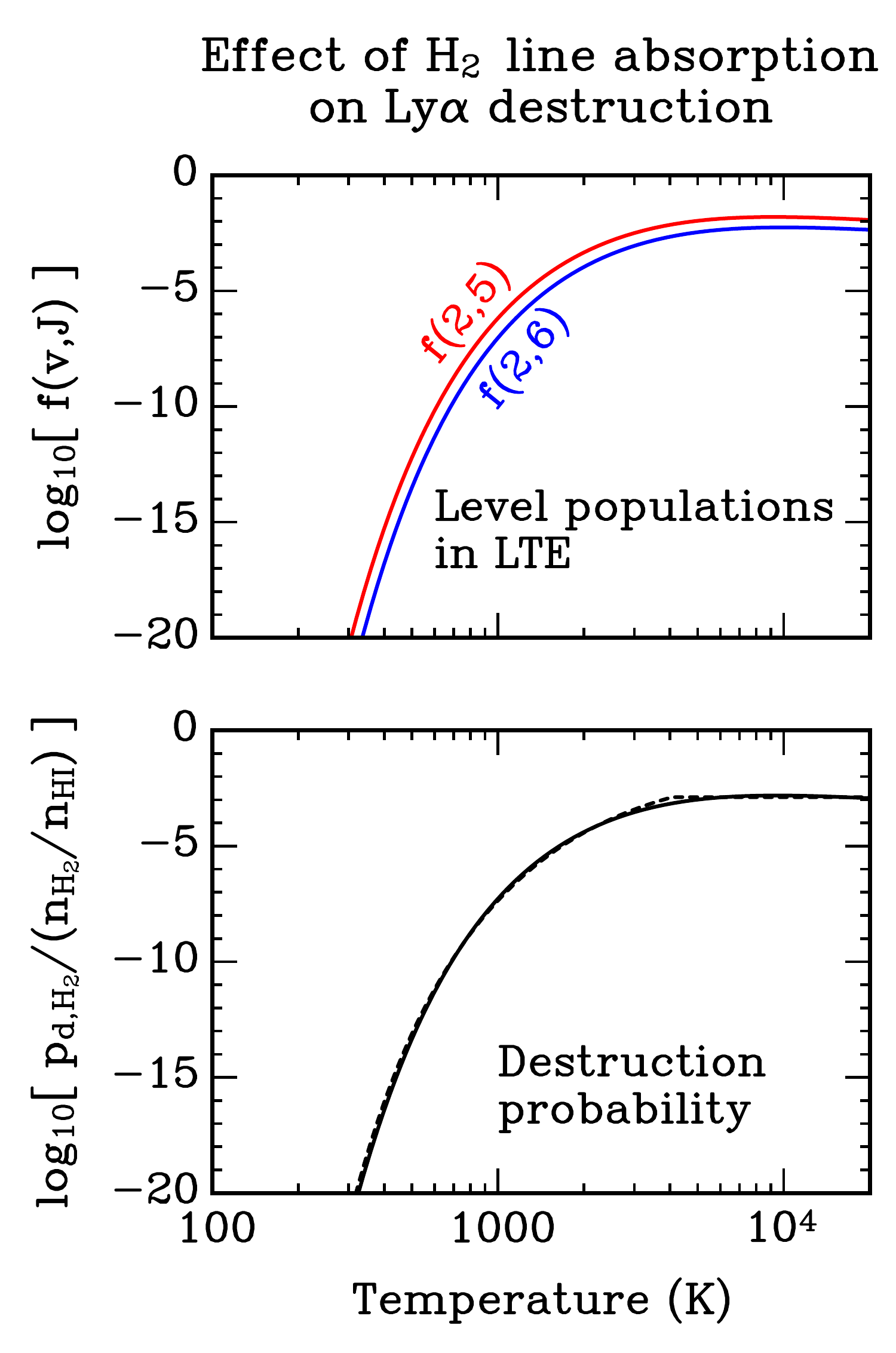}
    \caption{\textbf{Upper panel}: A plot of the relevant fractional level populations $f(v,J)$ in the ro-vibrational states $(v,J)=(2,5)$, $(2,6)$, assuming LTE. We have assumed an ortho-to-para ratio of 3:1. This plot is consistent with fig. 19 of \citet{Neufeld1990}, but also shows the behaviour for $T < 1000 \,\rm K$. \textbf{Lower panel}: The corresponding contribution $p_{\rm d,H_2}$ to the Ly$\alpha$ destruction probability (solid line). The dashed line shows the fit in Eq.~(\ref{pd from H2}).  
}
    \label{H2 line absorption pd plot}
\end{figure}

Absorption lines that neighbour Ly$\alpha$ line centre can also contribute to the destruction probability.\footnote{As shown by \cite{Neufeld1990}, an absorption line is considered `close' if its frequency satisfies $|x| \lesssim (a_{\rm v} \taucl)^{1/3}$, indicating significant overlap between the absorption and Ly$\alpha$ lines. This condition is always met for the B-X 1-2 $R(6)$ $1215.73~ \textrm{Å}$ line. Although the B-X 1-2 $P(5)$ $1216.07~ \textrm{Å}$ line is further away from Ly$\alpha$ line centre, it still has sufficient overlap in dense clouds to be treated similarly \citep[for similar modelling, see][]{Chiu1998}. } In particular, \cite{Neufeld1990} has pointed out that Ly$\alpha$ scattering can be suppressed if Ly$\alpha$ photons are absorbed by the B-X 1-2 $R(6)$ $1215.73~ \textrm{Å}$ and B-X 1-2 $P(5)$ $1216.07~ \textrm{Å}$ lines of H$_2$ \citep[e.g.][]{Shull1978, Lupu2006}. In local thermodynamic equilibrium (LTE),\footnote{\cite{Wolcott2019} show that LTE is expected to hold for the relevant levels $(v,J) = (2,5)$ and $(2,6)$ for gas densities $n \gtrsim 10^5 - 10^6 ~ \rm cm^{-3}$.} \cite{Neufeld1990} find that the relevant fractional level populations $f(v,J)$ drop exponentially below $T \sim 2000\,\rm K$ (see his fig. 19), but do not show the predicted population for $T < 1000\,\rm K$. Here we extend his plot to lower temperatures. The predicted fractional level population in LTE is shown in the upper panel of Fig. \ref{H2 line absorption pd plot}. In the lower panel we plot the corresponding destruction probability \citep[see eqs. 5.7--5.8 in][]{Neufeld1990}:
\begin{align}
    p_{\textrm{d}, \textrm{H}_2} &\simeq~ [0.082 \hspace{1 pt} f(2,6)  + 0.069 \hspace{1 pt} f(2,5) ] \hspace{1 pt} \dfrac{n_{\rm H_2}}{n_{\rm HI}} \label{pd from H2} \\ &\simeq~ \min \left[ \dfrac{33 \, e^{-14540/T} + 13 \, e^{-15564/T}}{46 + 10^3 \, e^{-(10^3/T)^{3/2}}} \, , \, 1.3 \times 10^{-3} \right]  \dfrac{n_{\rm H_2}}{n_{\rm HI}} \nonumber \, ,
\end{align}
where the second line is a fit, shown in the lower panel of Fig. \ref{H2 line absorption pd plot} as the dashed line. We see that $p_{\rm d,H_2}/(n_{\rm H_2}/n_{\rm HI})$ has a strong temperature dependence, jumping from $\sim 10^{-20}$ at $T = 300 \, \rm K$, to $\sim 10^{-7}$ at $T = 1000 \, \rm K$. Thus, if a star-forming cloud is heated to $\sim 1000 \rm \, K$, Ly$\alpha$ feedback could suddenly be dampened by a factor $\sim \textrm{few} - 100$ in clouds with $\taucl \sim 10^{11}$ (Fig. \ref{M_F p_d effect}), provided there is at least a residual abundance of H$_2$ ($f_{\rm H_2} \gtrsim 10^{-3}$). 

\subsubsection{Neglected processes that contribute to the destruction probability}
\label{neglected processes for pd}

There are other processes that can promote transitions between $2p$ and $2s$, or that contribute to the destruction probability more generally, that we have neglected. This is partially in the interest of keeping the subgrid model as simple as possible, and also because we think it is likely that they are of secondary importance compared to the processes that we do include. We briefly comment on some of the neglected processes:
\begin{itemize}
    \item \textit{Absorption of Ly$\alpha$ by other neighbouring lines}: Besides the H$_2$ lines, there are also other lines near Ly$\alpha$. One of particular interest is the $^1$S$_0$ -- $^1$P$_{1}^{\rm o}$ $1217.65 ~ \textrm{Å}$ line of \OI \citep{Forsman1973, NIST_ASD}. Very preliminary estimates by ourselves indicate that this could potentially suppress Ly$\alpha$ feedback by a factor $\lesssim \textrm{few}$ in some dense metal-enriched star-forming clouds. However, a more comprehensive study would need to couple the relevant level populations of \OI to the Ly$\alpha$ and continuum backgrounds, as well as consider oxygen chemistry in star-forming clouds \citep[e.g.][]{Omukai2005, Hollenbach2009}, making it considerably more complicated. Even if included, suppression of Ly$\alpha$ feedback by \OI line absorption is likely to be of minor importance compared to dust absorption. \\
    
    \item \textit{Maser amplification of the CMB}: \cite{Dijkstra2016} has noted that since Ly$\alpha$ will pump the $2p$ levels, we will get a $2s_{1/2}-2p_{3/2}$ population inversion. This in turn can lead to a maser effect, where the CMB is exponentially amplified around $\lambda = 3 ~ \rm cm$. It is conceivable that this could boost $2p \rightarrow 2s$ transitions enough and suppress Ly$\alpha$ feedback. Preliminary estimates by ourselves, however, suggest that (i) the effect is only potentially significant in very dense star-forming clouds where a massive star cluster can provide a strong enough Ly$\alpha$ background, and (ii) the effect is strongly self-regulated, such that any boost in $2p \rightarrow 2s$ by this mechanism will suppress $J$, and hence the  $2p \rightarrow 2s$ rate. A detailed analysis would need to couple the $2p$ and $2s$ levels to the Ly$\alpha$ and 3-cm backgrounds self-consistently to find the equilibrium rate of $2p \rightarrow 2s$. \\

    \item \textit{Other processes}: Besides the previously mentioned processes, we have also neglected: (i) collisional ionization and excitation/de-excitation (to $n \neq 2$), (ii) photoionization, and (iii) $\textrm{H}(2s)+\textrm{H}(2s) \rightleftharpoons \textrm{H}(2p)+\textrm{H}(2p)$. The rates of (i) are negligible in cold clouds compared to the other collisional processes included in this paper  \citep[e.g.][]{Omukai2001, McKee2008}, and (ii) can be neglected since the rate is negligible in self-shielded clouds, and because another Ly$\alpha$ photon is likely to be produced following recombination \citep{Dijkstra2006, McKee2008}. While the rate coefficient of (iii) is large, $\sim 10^{-6} ~ \rm cm^{3} ~ s^{-1}$ \citep{Forrey2000}, the level population in $n = 2$ is many orders of magnitude smaller than $n = 1$ in cold gas ($T < 10^4\,\rm K$). As a result, (iii) can be safely neglected.
\end{itemize}
Future work should revisit these processes with more detailed modelling, but we suspect that this will only lead to second-order corrections to the estimated $p_{\rm d}$ in this paper. 

\subsubsection{Summary of the estimated Ly$\alpha$ destruction probability}

\begin{figure*}
\includegraphics[width=0.95\textwidth]{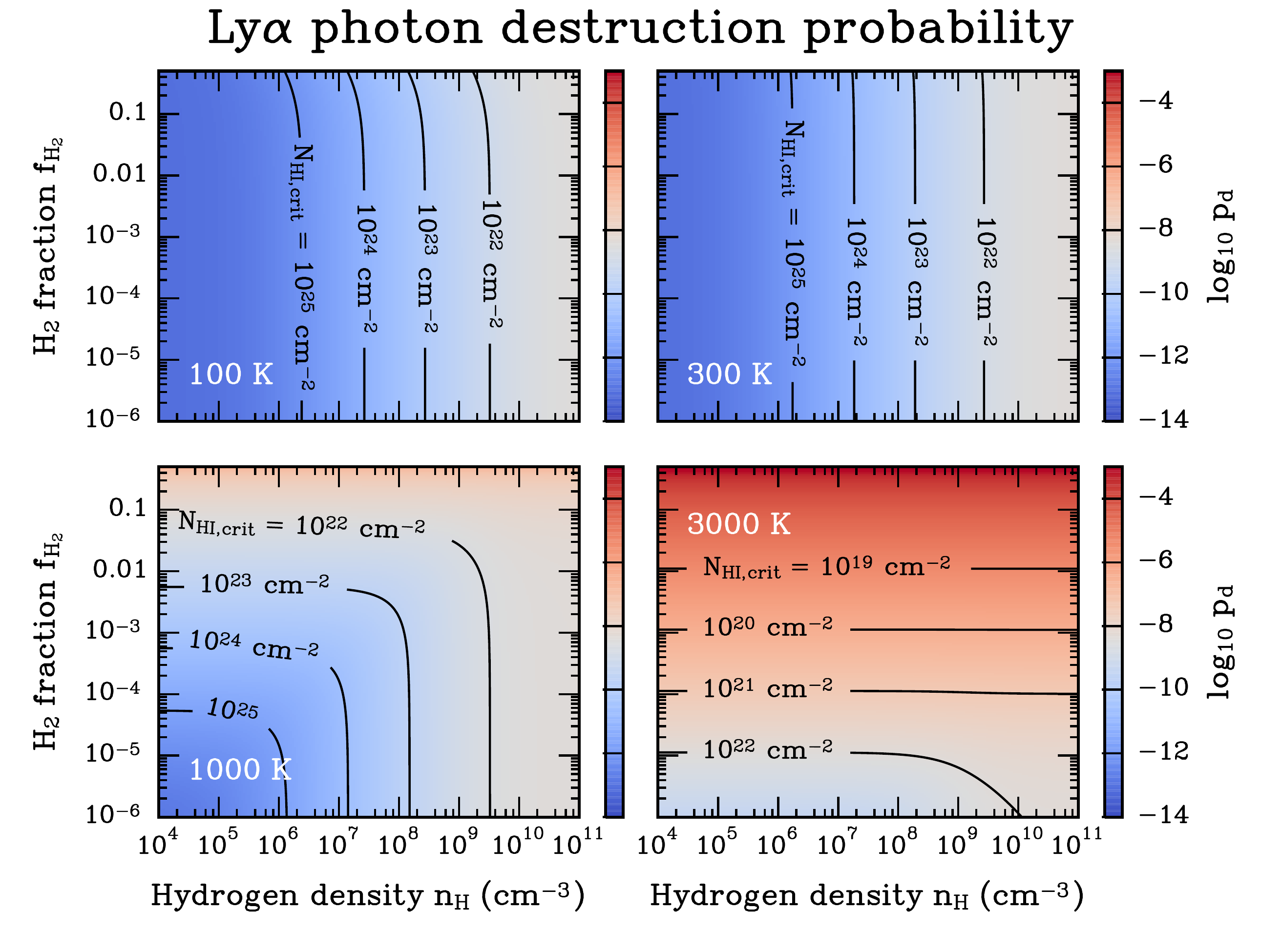}
\caption{ Colour plots of the estimated Ly$\alpha$ destruction probability, $p_{\rm d}$, as a function of the hydrogen density $n_{\rm H} = n_{\rm HI} + 2 n_{\rm H_2}$, and H$_2$ fraction $f_{\rm H_2} = n_{\rm H_2}/n_{\rm H}$, for different gas temperatures. The contours show where an \HI column density $N_{\rm HI} = N_{\rm HI,crit}$ is sufficient to yield $\mathcal{P} = 10$, corresponding to moderate suppression of Ly$\alpha$ radiation pressure (see Fig. \ref{M_F p_d effect}). In all panels, we have assumed a redshift $z = 15$ to compute the CMB-stimulated transition rates (Eqs. \ref{Gamm CMB 2p2s}--\ref{Gamm CMB 2s2p}), though this specific choice has a minor impact on the plotted $p_{\rm d}$. \textbf{Upper left panel}: The destruction probability for $T = 100 \, \rm K$. In this case, $p_{\rm d}$ is mainly determined by collisional processes. \textbf{Upper right panel}: The destruction probability for $T = 300 \, \rm K$, with results similar to $T = 100 \, \rm K$. \textbf{Lower left panel}: The destruction probability for $T = 1000 \, \rm K$. Here, H$_2$ line absorption becomes important, with a dominant impact at lower densities. \textbf{Lower right panel}: The destruction probability for $T = 3000 \, \rm K$. In this case, H$_2$ line absorption has a dominant impact on $p_{\rm d}$ across most of the parameter space, being capable of suppressing Ly$\alpha$ feedback even in clouds with low \HI column density.}
\label{pd_colorplot}
\end{figure*}

Here we summarize our final estimate of the total Ly$\alpha$ destruction probability $p_{\rm d} = p_{\rm d,HI} + p_{\rm d,H_2}$:\footnote{We provide a \textsc{Python} implementation of $p_{\rm d}$ in the module \texttt{LyaDestruction.py}, available at \url{https://github.com/olofnebrin/Lyman-alpha-feedback}.}
\begin{align}
    p_{\rm d} ~&\simeq~ \frac{\mathcal{R}_{2p,2s}}{(A_{\rm Ly \alpha}/A_{2 \gamma})( A_{2 \gamma} + \mathcal{R}_{2s,2p}) + \mathcal{R}_{2p,2s}} \label{pd final expression} \\ ~&+~ \min \left[ \dfrac{33 \, e^{-14540/T} + 13 \, e^{-15564/T}}{46 + 10^3 \, e^{-(10^3/T)^{3/2}}} \, , \, 1.3 \times 10^{-3} \right]  \dfrac{n_{\rm H_2}}{n_{\rm HI}} \nonumber \, ,
\end{align}
with $\mathcal{R}_{2p,2s}$ and $\mathcal{R}_{2s,2p}$ given by Eqs.~(\ref{R2p2s})--(\ref{R2s2p}), respectively, using $\mathcal{C}_{2s,2p} = 3\mathcal{C}_{2p,2s}$ from Eqs.~(\ref{HeRate})--(\ref{C_2s2p summary}). For reference, in Fig. \ref{pd_colorplot} we plot Eq.~(\ref{pd final expression}) for a few temperatures ($100\,\rm K \leq T \leq 3000\,\rm K$), as a function of the hydrogen number density $n_{\rm H}$, and the H$_2$ fraction $f_{\rm H_2}$. Also shown are contours, along which the required \HI column density for $\mathcal{P} = 10$ (moderate Ly$\alpha$ feedback suppression, see Fig. \ref{M_F p_d effect}) is $N_{\rm HI,crit}$.

Thus, for example, we see from the upper left panel that in a gas cloud with \HI column density $N_{\rm HI} = 10^{24} \, \rm cm^{-2}$ ($\Sigma_{\rm HI} = 8000 ~ \Msun \rm ~ pc^{-2}$) and temperature $T = 100 \, \rm K$, Ly$\alpha$ feedback can be dampened for densities $n_{\rm H} \gtrsim 3 \times 10^{7} \, \rm cm^{-3}$. While extreme by the standards of the low redshift Universe, such conditions are predicted to be more common in early dwarf galaxies at Cosmic Dawn \citep[e.g.][]{Kimm2016_GC, Nebrin2022}. At higher temperatures ($T \gtrsim 1000 \, \rm K$), $p_{\rm d}$ is mainly determined by H$_2$ line absorption, which can suppress Ly$\alpha$ feedback in clouds at significantly lower gas densities and \HI column densities.

Finally, we recall that we saw in Sec. \ref{turbulent cloud solution section} that, in a turbulent cloud with density fluctuations, one should replace $p_{\rm d}$ by $\langle p_{\rm d} \rangle_{m} \equiv \langle \alpha_0 p_{\rm d} \rangle/\langle \alpha_0 \rangle$, where $\alpha_0 \equiv n_{\rm HI} \sigma_0$. Since $p_{\rm d, HI}$ is approximately proportional to the gas density, and $p_{\rm d,H_2} \propto n_{\rm H_2}/n_{\rm HI}$, one can estimate $\langle p_{\rm d,HI} \rangle_{m}$ by making the replacement $n_{\rm HI} \rightarrow \langle n_{\rm HI} \rangle [1 + (b_s \mathcal{M})^2]$, and similarly for \HeI and H$_2$, in all collisional rates.\footnote{The factor $1 + (b_s \mathcal{M})^2$ is simply the clumping factor. From Table \ref{turbulent density fluctuations} and Eq.~(\ref{turbulence variance}) we find that $\langle \rho^2 \rangle/\langle \rho \rangle^2 = e^{\sigma_s^2} = 1 + (b_s \mathcal{M})^2$, which gives the quoted result. } And to get $\langle p_{\rm d,H_2} \rangle_{m}$ one can simply replace $n_{\rm H_2}/n_{\rm HI}$ by $\langle n_{\rm H_2}\rangle/\langle n_{\rm HI}\rangle$.

\subsection{Continuum absorption of Lyman-\texorpdfstring{\boldmath$\alpha$}{alpha} photons}
\label{Lya absorption processes}

\begin{table}
\caption{Predicted Ly$\alpha$ dust absorption cross-section per unit dust mass ($\kappa_{\rm d,abs}$), and dust absorption cross-section per H nucleon ($\sigma_{\rm d,abs,H} = \kappa_{\rm d,abs} \mathscr{D}m_{\rm H}/X$), for a few dust models. The latter is normalized to the Solar value, assuming a Solar dust-to-gas ratio $\mathscr{D}_{\odot} = 1/162$ \citep{Remy2014}, and a hydrogen mass fraction $X = 1-Y = 0.753$.}
\begin{tabular}{l l l}
\hline
\hline
Dust model & $\kappa_{\rm d,abs}$ & $\sigma_{\rm d,abs,H}$  \\
 & ($10^4$ cm$^{2}$ g$^{-1}$) & ($10^{-21}$ cm$^{2}$ H$^{-1}$)  \\

\hline
\hline

\\
\vspace{3 pt}

\cite{Hensley2023} & $6.873$ & $0.941 \, \mathscr{D}/\mathscr{D}_{\odot}$ \\
(Astrodust + PAH) & &  \\ \\

\vspace{3 pt}

\cite{WeingartnerDraine2001}$^\dagger$ & $6.854$ & $0.938 \, \mathscr{D}/\mathscr{D}_{\odot}$  \\

(Milky Way, $R_V = 3.1$) & &  \\ \\

\vspace{3 pt}

\cite{WeingartnerDraine2001}$^\dagger$ & $4.425$ &  $0.606 \, \mathscr{D}/\mathscr{D}_{\odot}$ \\

 (Milky Way, $R_V = 4.0$) & &  \\ \\

\vspace{3 pt}

\cite{WeingartnerDraine2001}$^\dagger$ & $3.182$ & $0.436 \, \mathscr{D}/\mathscr{D}_{\odot}$ \\
 (Milky Way, $R_V = 5.5$) & &  \\ \\

 \vspace{3 pt}

\cite{WeingartnerDraine2001} & $7.251$ & $0.993 \, \mathscr{D}/\mathscr{D}_{\odot}$ \\
 (LMC average) & &  \\ \\

  \vspace{3 pt}

\cite{WeingartnerDraine2001} & $7.181$ & $0.983 \, \mathscr{D}/\mathscr{D}_{\odot}$ \\
 (SMC bar) & &  \\ \\
 
\hline
\hline
\end{tabular}
\vspace{1 pt}\\
$^\dagger$: As updated by \citet{Li2001} and \citet{Draine2003_AnnRev, Draine2003albedo}.
\label{Dust models}
\end{table}

\begin{figure}
    \includegraphics[trim={0.1cm 0.1cm 0.1cm 0.1cm},clip,width = \columnwidth]{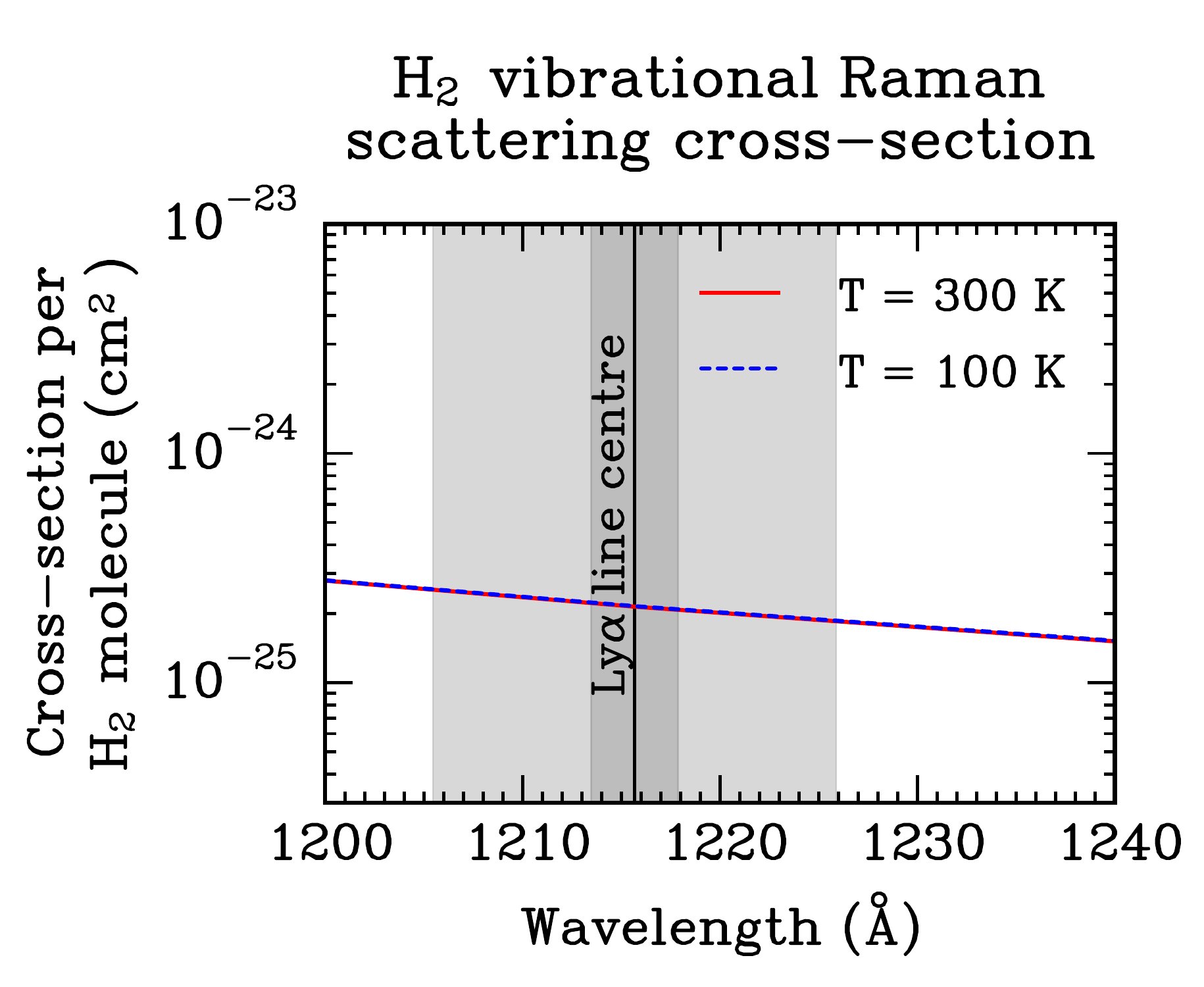}
    \caption{The total H$_2$ vibrational Raman scattering cross-section from \citet{Ford1973} and \citet{Ford1973_2} as a function of wavelength. The dark (light) gray region shows the characteristic Ly$\alpha$ line width $\sim 2(a_{\rm v}\taucl)^{1/3} \Delta \lambda_{\rm D}$ for $T = 200\,\rm K$ and $\taucl = 10^9$ ($\taucl = 10^{11}$). The red solid (blue dashed) line shows the cross-section for $T = 300\,\rm K$ ($T = 100\,\rm K$). The cross-section is insensitive to the gas temperature \citep[or more generally, the initial rotational quantum number, see][]{Ford1973}. The wavelength dependence over the Ly$\alpha$ line width is also fairly weak, and can be neglected to first order.
}
    \label{Raman cross-sec}
\end{figure}

We studied the general impact of continuum absorption on the Ly$\alpha$ force multiplier in Sec. \ref{Sec. M_F continuum absorption}. Here we will discuss some processes that contribute to the continuum absorption opacity. The most well-studied process is dust absorption. The cloud dust absorption optical depth is $\tau_{\rm d,abs} = \sigma_{\rm d,H} N_{\rm H}$, where $\sigma_{\rm d,H}$ is the dust absorption cross-section per hydrogen nucleon, and $N_{\rm H} = N_{\rm HI} + N_{\rm HII} + 2N_{\rm H_2}$ is the total hydrogen column density of the cloud.

Since the dust absorption cross-section depends on the assumed dust grain size distribution and composition, different dust models generally predict different values of $\sigma_{\rm d,H}$ for a given dust abundance. In Table \ref{Dust models} we list a few dust models and their predictions for $\sigma_{\rm d,H}$. While there is some variation, they all give $\sigma_{\rm d,H} \sim 10^{-21} \, (\mathscr{D}/\mathscr{D}_{\odot}) \, \rm cm^{2} \, H^{-1}$ to within a factor of $\sim 2$, where $\mathscr{D}/\mathscr{D}_{\odot}$ is the dust-to-gas ratio in Solar units. The resulting contribution to the cloud continuum absorption optical depth from dust is
\begin{equation}
    \tau_{\rm d,abs} = 100 \, \left( \dfrac{\sigma_{\rm d,H}}{10^{-21} \, \rm cm^{2}\, H^{-1}} \right) \left( \dfrac{N_{\rm H}}{10^{23} \, \rm cm^{-2}}\right) \, \label{tau dust}.
\end{equation}

In dust-free gas containing H$_2$, Ly$\alpha$ photons can also undergo Raman scattering, which is an inelastic scattering process. Vibrational Raman scattering by H$_2$ will shift the Ly$\alpha$ photons to wavelengths $\geq 1275 ~ \textrm{Å}$ \citep{Dalgarno1962, Ford1973, Ford1973_2, Oklopcic2016}, after which they can freely escape the cloud. Because of this, it can be treated as a continuum absorption process, similar to dust absorption. The impact of this process on Ly$\alpha$ feedback has not been explored before to our knowledge. 

In Fig. \ref{Raman cross-sec} we plot the vibrational Raman scattering cross-section per H$_2$ molecule. There is a weak wavelength dependence over the typical Ly$\alpha$ line width. However, this variation is less than a factor of $2$, so for simplicity we ignore this and adopt the cross-section at Ly$\alpha$ line centre, $2.1 \times 10^{-25} \, \rm cm^{2} \, H_{2}^{-1}$. Thus, the contribution to the continuum absorption optical depth from Raman scattering becomes
\begin{equation}
    \tau_{\rm H_2, abs} = 2.1 \times 10^{-4} \left( \dfrac{N_{\rm H_2}}{10^{21} \, \rm cm^{-2}}\right) \, . \label{tau H2}
\end{equation}
Note that even though $\tau_{\rm H_2, abs}$ is typically small, the relevant factor for Ly$\alpha$ feedback suppression, i.e. $(a_{\rm v} \taucl)^{1/3} \tau_{\rm H_2, abs}$, can still exceed unity. Because of this, we infer from Fig. \ref{M_F dust} that Raman scattering can potentially suppress Ly$\alpha$ feedback by up to a factor $\sim \rm few$ in dense and chemically pristine clouds where Pop III stars form. 

We neglect other continuum absorption processes that in principle could further suppress Ly$\alpha$ feedback, like photoionization of excited \HI and \HeI, and photodetachment of H$^{-}$. As noted in Sec. \ref{neglected processes for pd}, the populations of excited \HI (and \HeI) are extremely small in dense and cool star-forming clouds. Furthermore, the predicted abundance of H$^{-}$ is also exceedingly small \citep[e.g.][]{Dijkstra2016, Nebrin2023cooling}. For these reasons, and to not overcomplicate the modelling, we only take into account dust absorption and Raman scattering by H$_2$ in our treatment of the continuum opacity. We therefore estimate the total continuum absorption optical depth of the cloud to be:
\begin{equation}
    \tau_{\rm c, abs} = \tau_{\rm d,abs} +  \tau_{\rm H_2,abs}\, ,
\end{equation}
with $\tau_{\rm d,abs}$ and $\tau_{\rm H_2,abs}$ given by Eqs.~(\ref{tau dust}) and (\ref{tau H2}) above.

\section{Discussion}
\label{Discussion section}

\begin{figure*}
\includegraphics[width=0.9\textwidth]{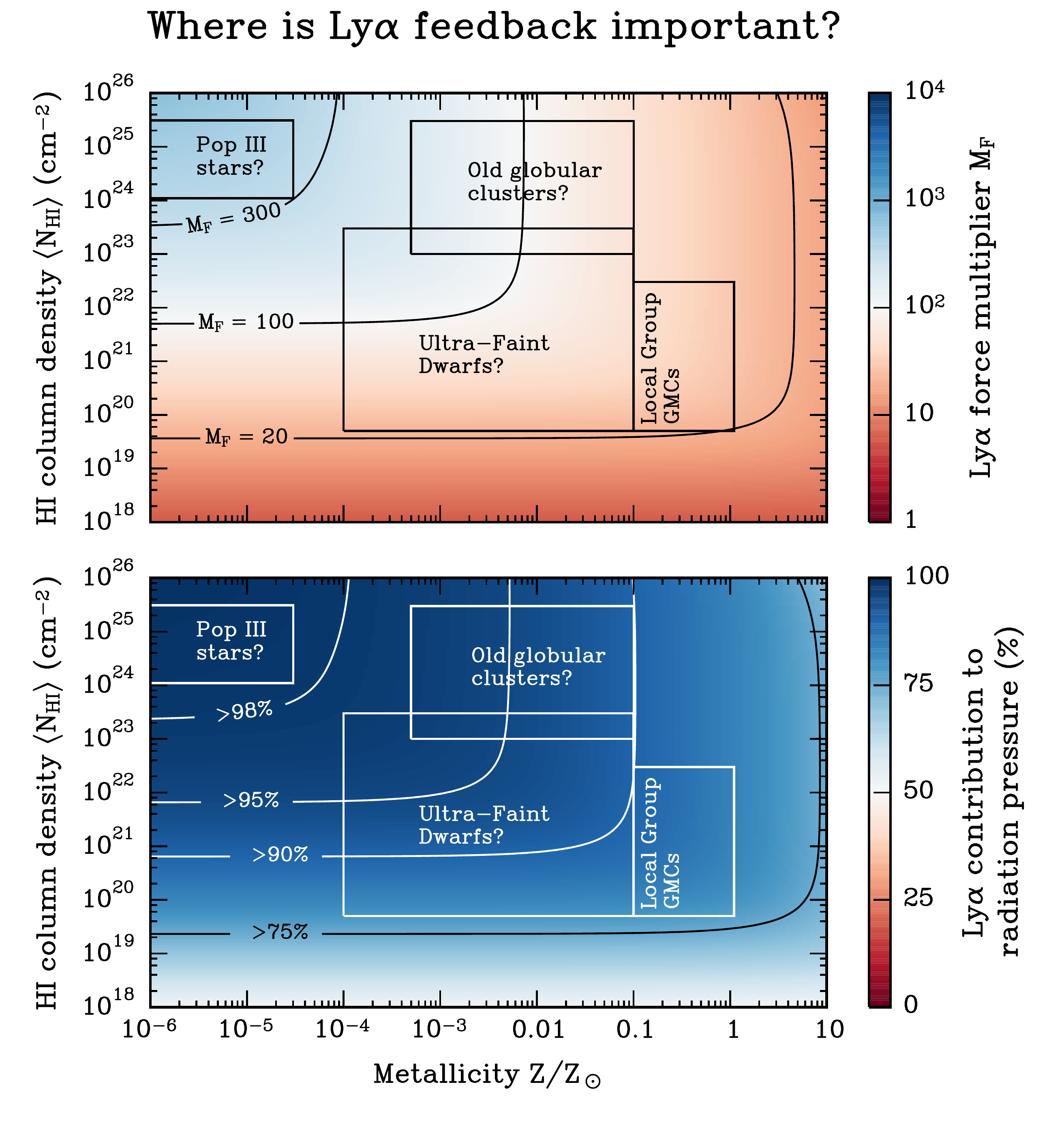}
\caption{Colour and contour plots mapping the importance of Ly$\alpha$ feedback in different regions of parameter space. \textbf{Upper panel}: The Ly$\alpha$ force multiplier as a function of gas metallicity $Z/Z_{\odot}$ and average cloud \HI column density $\langle N_{\rm HI} \rangle$. Several conservative assumptions have been made to produce this plot, including a relatively high Mach number ($\mathcal{M} = 10$) and gas density ($n_{\rm HI} = 10^6 ~ \rm cm^{-3}$), all of which dampen Ly$\alpha$ feedback. See the main text for further details on these assumptions. Also shown are the approximate regions of parameter space corresponding to objects of interest, such as GMCs in the Milky Way and other Local Group galaxies, old globular clusters, Ultra-Faint Dwarf galaxies, and Pop III stars in minihaloes. The \HI column densities and metallicities of local GMCs have some observational constraints \citep[e.g.][]{Blitz2007, Lee2015, Seifried2022}, along with underlying theoretical understanding \citep[e.g.][]{Bialy2016}. For the other classes of objects, there are are no tight observational constraints on the \HI column densities in their birth clouds, so we have assumed a wide range of values loosely based on their observed masses and sizes \citep[e.g.][]{Baumgardt2018, Simon2019, Adamo2024}, and theoretically predicted gas column densities \citep[e.g.][]{Ricotti2016, Kimm2016_GC, Jaura2022, Nebrin2022, Sugimura2023_PopIII, Kang2024}, and H$_2$ fractions at low metallicities \citep[e.g.][]{Nebrin2023cooling, Sugimura2023_PopIII, Sugimura2024}. Observational constraints on the metallicity of stars in globular clusters and Ultra-Faint Dwarfs exist \citep[e.g.][]{Harris2010,Martin2022, Fu2023_UFD}, but the metallicity threshold for Pop III star formation is uncertain \citep[e.g.][]{Sharda2022_IMF}. \textbf{Lower panel}: The corresponding relative importance of Ly$\alpha$ radiation pressure, in terms of momentum input rate, compared to the maximum input rate from direct radiation pressure ($L_{\rm bol}/c$). It is evident that for all objects of interest -- even at high metallicities -- Ly$\alpha$ radiation pressure dominates. The same conservative assumptions as in the upper panel were made here. Additionally, we stress that we have \textit{overestimated} the direct radiation pressure in low-metallicity (dust-poor) environments. Thus, Ly$\alpha$ feedback may have an even greater relative importance than shown here. }
\label{discussion_overview}
\end{figure*}

In the previous sections we have computed the expected Ly$\alpha$ force multiplier $M_{\rm F}$ using a novel analytical solution, verified by MCRT simulations. We further discussed important processes related to Ly$\alpha$ radiation pressure. Here we discuss some preliminary implications for the role of Ly$\alpha$ radiation pressure in different environments. To guide our discussion below, in Fig. \ref{discussion_overview} we plot an overview of where Ly$\alpha$ feedback is expected to be important. In the upper panel we plot the Ly$\alpha$ force multiplier (using Eq.~\ref{M_F fit turbulence, final}) as a function of the gas metallicity ($Z/Z_{\odot}$) and average cloud \HI column density, $\langle N_{\HI} \rangle \equiv \langle \alpha_0 \rangle R_{\rm cl}$. In the lower panel we compare the momentum input from Ly$\alpha$ radiation pressure to direct radiation pressure, to study the relative importance of the former. In making these plots we have made several assumption, most of which are conservative with respect to the importance of Ly$\alpha$ feedback: 
\begin{itemize}
    \item \textit{Dust-to-gas ratio}: We have assumed that the dust-to-gas ratio is proportional to the gas metallicity, i.e. $\mathscr{D}/\mathscr{D}_{\odot} = Z/Z_{\odot}$. This is a common assumption in galaxy formation simulations \citep[e.g.][]{Kimm2017minihaloes,Agertz2020, Hopkins2023}, but we note that there is observational evidence that this may overestimate $\mathscr{D}/\mathscr{D}_{\odot}$ when $Z/Z_{\odot} \lesssim 0.1$ \citep{Remy2014}. As a result, in Fig. \ref{discussion_overview} we are likely \textit{overestimating} the suppression of Ly$\alpha$ feedback by dust absorption for $Z/Z_{\odot} \lesssim 0.1$. We adopt the dust model of \cite{Hensley2023}, as given in Table \ref{Dust models}. The results are insensitive to this choice. 
    
    \item \textit{Turbulence \& gas temperature}: We have assumed a fixed Mach number $\mathcal{M} = 10$, and a high driving parameter $b_{\rm s} = 0.9$ (i.e. mostly compressive turbulence). Both are reasonable values, but are on the higher end of what is observed in the Milky Way and other Local Group galaxies \citep[e.g.][]{Orkisz2017, Syed2020, Wang2020,Sharda2022_turbulenceobs}. Furthermore, simulations and models of high-redshift minihaloes often find $\mathcal{M} \lesssim 2$ \citep{Greif2011_movingmesh, Safranek2012, Wise2019, Nebrin2022, Sugimura2023_PopIII}. As a result, we may be \textit{overestimating} the suppression of Ly$\alpha$ feedback by Ly$\alpha$ leakage through low-density channels. We also assume a fixed temperature $T = 100 ~ \rm K$, which may slightly underestimate $M_{\rm F}$ in colder star-forming clouds. 
    
    \item \textit{Gas density and Ly$\alpha$ destruction}: To estimate the Ly$\alpha$ destruction probability (Eq.~\ref{pd final expression}), we have assumed an average \HI density $\langle n_{\rm HI} \rangle = 10^6 ~ \rm cm^{-3}$, and \HeI density $\langle n_{\rm HeI} \rangle = 0.082 \langle n_{\rm HI} \rangle$. These values are fairly typical of what is expected for dense high-redshift environments where the first Pop II galaxies and star clusters form \citep{Kimm2016_GC, Nebrin2022, Adamo2024}. For Pop III star formation in minihaloes, however, this is likely an underestimate, as discussed below in more detail (Sec. \ref{First stars implications}). As for local GMCs, this is on the higher end of what is observed, so we are likely overestimating the impact of Ly$\alpha$ destruction in local star-forming clouds.
    
    \item \textit{Velocity gradients}: We have assumed a fixed cloud-scale velocity $\lvert \Dot{R}_{\rm cl} \rvert/b = 10$. This would correspond to a cloud undergoing large-scale collapse or dispersal (e.g. by feedback) at a rate $\simeq 13 ~ \rm km \, s^{-1}$. For clouds that are in virial equilibrium, this would underestimate $M_{\rm F}$ by a factor $\sim 2$ (see Fig. \ref{M_f velocity gradients plot}).
    
    \item \textit{Comparison to direct radiation pressure}: We compare the momentum input from Ly$\alpha$ radiation pressure to feedback from direct radiation pressure. As noted in the introduction of this paper, the latter has been implemented in several recent galaxy and star formation simulations, and known to be important for star and galaxy formation \citep[e.g.][]{Fall2010, Wise2012radpressure, Grudic2018, Hopkins2020}. The momentum input per unit stellar mass from Ly$\alpha$ feedback is estimated as $(L_{\rm Ly\alpha}/M_{\star}c) M_{\rm F}$ (Eq.~\ref{Lya luminosity}), using the ionizing output from a young Pop II star cluster, as given in Table \ref{Qdot}. We have estimated the momentum input per unit stellar mass from direct radiation pressure as $\Psi_{\rm bol}/c$, where $\Psi_{\rm bol} \simeq 800 ~ \Lsun ~ \Msun^{-1}$ is the bolometric light-to-mass ratio of a young Pop II star cluster \citep{Hopkins2023}. In dust-poor environments, this overestimates the importance of direct radiation pressure, since most of the UV and optical photons will escape the cloud. As a result, we are \textit{underestimating} the relative importance of Ly$\alpha$ radiation pressure at low $Z/Z_{\odot}$.
\end{itemize}
In the plots, we also assume an effective point source ($\Bar{\tau}_{\rm s} \rightarrow 0$).\footnote{As discussed in Sec. \ref{spatial source distribution section}, if Ly$\alpha$ photons are produced in an \HII region, we expect it to behave as a point source in optical depth space, i.e. $\Bar{\tau}_{\rm s} \rightarrow 0$, even if the \HII region is of comparable physical size to the cloud. } Despite several conservative assumptions, we see from Fig. \ref{discussion_overview} that Ly$\alpha$ feedback is predicted to dominate over direct radiation pressure by a factor $\sim 5 - 100$, depending mainly on the metallicity. As a result, Ly$\alpha$ feedback cannot be neglected in realistic star and galaxy formation simulations and models. The same conclusion has been reached in several other recent works \citep[][]{Smith2017, Abe2018, Kimm2018, Tomaselli2021, Nebrin2022, Kapoor2023, Thomson2024}. These previous studies, however, relied on more idealized assumptions (e.g. uniform clouds, no velocity gradients, no Ly$\alpha$ destruction). In this paper we have carefully studied the impact of including additional effects that could dampen Ly$\alpha$ feedback. While these effects are significant, we find that Ly$\alpha$ feedback prevails as a very strong, if not dominant, early feedback process. Below we discuss the implications for the first galaxies and stars in more detail.

\subsection{Implications for the first galaxies}
\label{first galaxies implications}

The first galaxies are expected to have formed in the smallest low-mass haloes able to to host cool gas, and retain metals from the first stars \citep[e.g.][]{ Nebrin2023cooling}. Because they formed in very low-mass haloes ($M_{\rm vir} \sim 10^6 - 10^9 \, \Msun$), they are particularly susceptible to bursty stellar feedback. Candidate `fossils' of these galaxies are observed around the Milky Way, and known as Ultra-Faint Dwarf (UFD) galaxies \citep{Simon2019}. UFDs have a wide range of stellar masses $\sim 10 - 10^5 \, \Msun$, and half-light radii $\sim \textrm{few} - \textrm{few} \times 100 ~ \, \rm pc$, and \textit{average} metallicities $-3 \lesssim \langle [\textrm{Fe/H}] \rangle \lesssim -2$ \citep[e.g.][]{Simon2019, Fu2023_UFD, Smith2024_UFD}.\footnote{The individual stars in a given UFD typically span a wide range of metallicities, $-4 \lesssim [\textrm{Fe/H}] \lesssim -1$ \citep[e.g.][]{Fu2023_UFD}. For this reason, and to be conservative regarding Ly$\alpha$ feedback, we have shown a wider range of metallicities in Fig. \ref{discussion_overview}. } Most, or a significant fraction, of their stars formed at redshifts $z \gtrsim 6$ \citep[e.g.][]{Brown2014, Simon2021, McQuinn2024}, consistent with quenching of star formation in low-mass haloes by reionization feedback \citep[e.g.][]{Ricotti2005, Jeon2017, Wheeler2019, Benitez2020, Nebrin2023cooling}. 

Several recent high-resolution galaxy formation simulations have studied the formation of UFDs \citep{Ricotti2016, Jeon2017, Wheeler2019, Agertz2020, Applebaum2021, Gutcke2022}. None of these simulations include Ly$\alpha$ feedback. In Fig. \ref{discussion_overview} we plot the expected region in parameter space of the star-forming clouds in which UFDs formed. It is seen that Ly$\alpha$ feedback is predicted to be strong in UFDs, contributing $\gtrsim 75-97\%$ of the total radiation pressure, even under our conservative assumptions. 

Furthermore, since UFDs are likely to have formed as gravitationally unbound star clusters \citep{Ricotti2016, Nebrin2022, Garcia2023}, this suggests that their birth clouds were sensitive to disruption by strong feedback. As a result, Ly$\alpha$ feedback may have an important effect on the predicted properties of UFDs. This was indeed found in \cite{Nebrin2022}, where starbursts in low-mass haloes were modelled semi-analytically in detail. Several feedback processes were implemented, including photoionization feedback, stellar winds, direct and indirect radiation pressure on dust, and Ly$\alpha$ feedback \citep[following][]{Tomaselli2021}. Ly$\alpha$ feedback was found to be important in reproducing the abundance and properties of UFDs, especially at the low-mass end.\footnote{The effect of turning off Ly$\alpha$ feedback in this model is the overproduction of very compact ($< 1 \, \rm pc$), gravitationally bound star clusters in low-mass haloes, and too few UFD-like galaxies. This leads to the (wrong) prediction that a significant fraction of local globular clusters formed in low-mass haloes at Cosmic Dawn. In contrast, with Ly$\alpha$ feedback, such bound clusters are still predicted to have formed, with properties similar to those recently found by \cite{Adamo2024}, but at lower abundance. The star clusters that become unbound following gas expulsion by Ly$\alpha$ feedback instead form low-mass galaxies, with properties consistent with UFDs around the Milky Way. } More detailed modelling and simulations should revisit the effect of Ly$\alpha$ feedback on UFD formation. Until this is done, great caution should be taken in interpreting simulation results as bona fide $\Lambda$CDM predictions, since they are likely missing a significant fraction, if not majority, of the momentum input from early stellar feedback.\footnote{There are, of course, additional uncertainties related to generating realistic UFD galaxies in simulations. For instance, the UFDs in the FIRE-2 simulations of \cite{Wheeler2019} have too low metallicity, which the authors suggest may indicate the importance of enrichment by Pop III stars. }

\subsection{Implications for the first star clusters}

\begin{figure}
    \includegraphics[trim={0.0cm 0.0cm 0.0cm 0.1cm},clip,width = \columnwidth]{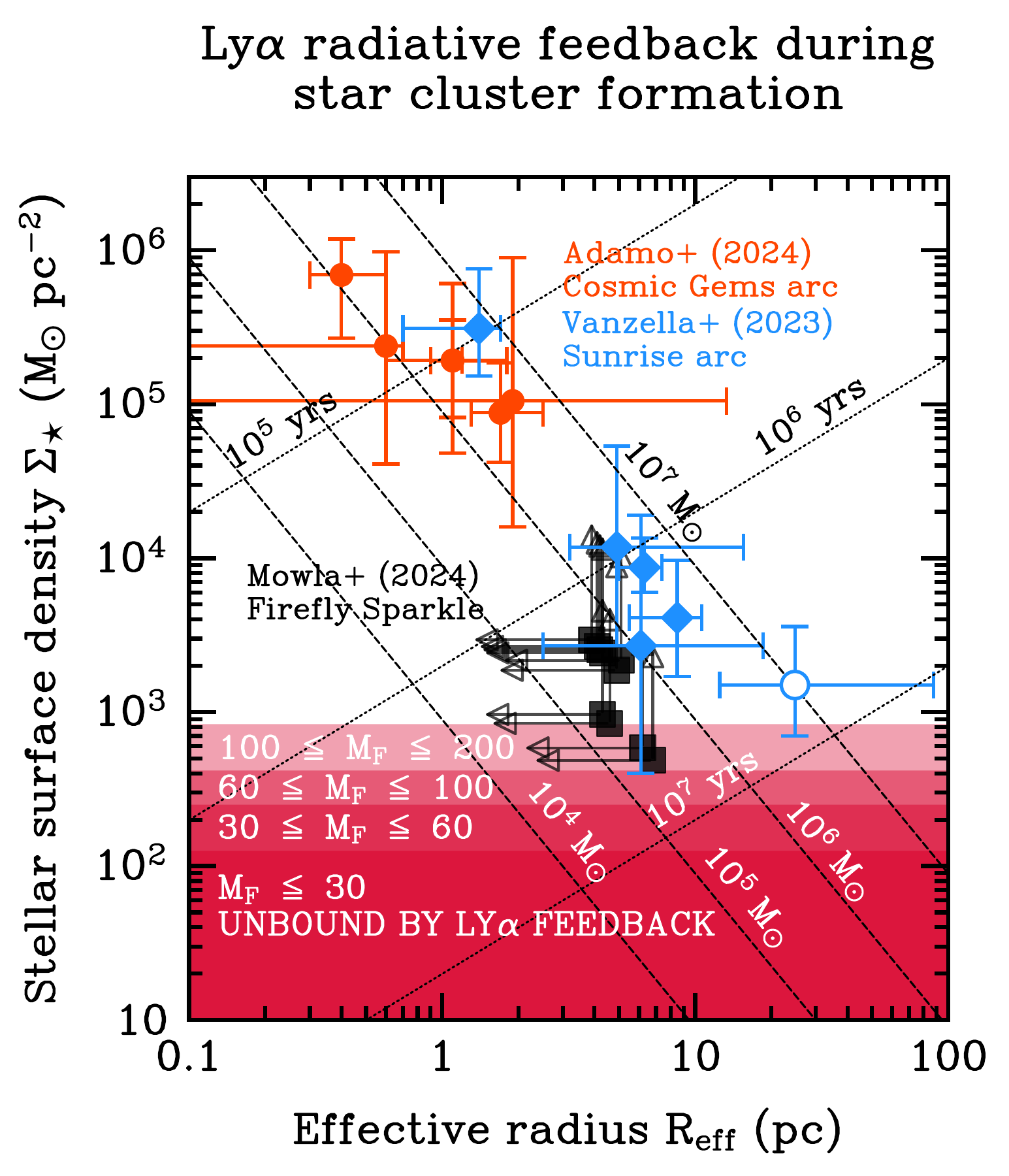}
    \caption{A plot of the stellar surface density $\Sigma_{\star}$ and effective radii of recently detected high-redshift ($z = 6 - 10$), metal-poor ($Z/Z_{\odot} \sim 0.01 - 0.1$) star clusters \citep[][]{Vanzella2023, Adamo2024, Mowla2024}. Most of the star clusters are gravitationally bound (filled symbols), with the cluster 4b from \citet{Vanzella2023} showing evidence of being unbound (blue unfilled circle). Gravitationally bound young star clusters in the red region are excluded by Ly$\alpha$ feedback, although the constraint depends on the value of $M_{\rm F}$. The detected star clusters are consistent with the typical values $30 \lesssim M_{\rm F} \lesssim 200$ expected for star clusters of similar metallicity (see Fig. \ref{discussion_overview} and the associated discussion). Note that the star clusters of \citet{Mowla2024} only have lower limits on $\Sigma_{\star}$, since the clusters are not resolved. Dashed and dotted lines show lines of constant mass and crossing time, $t_{\rm cross} \equiv 10 \, (R_{\rm eff}^3 /GM_{\star})^{1/2}$, respectively. 
}
    \label{StarClusters}
\end{figure}

If stellar feedback only expels a fraction $\lesssim 90\%$ of the initial gas reservoir in a star-forming cloud, the stars can remain gravitationally bound in clusters following gas expulsion \citep[e.g.][]{Hills1980, Adams2000, Baumgardt2007, Li2019, Fukushima2021, Farias2024}. Thus, the cluster formation efficiency is tightly related to the star formation efficiency (SFE). Analytical estimates and detailed simulations also find that the SFE in turn is mainly set by the initial cloud surface density $\Sigma_{\rm cl}$, and the IMF-averaged momentum injection rate per unit stellar mass from stellar feedback, $\langle \Dot{p}/m_{\star} \rangle$ \citep[e.g.][]{Fall2010, Raskutti2016, Grudic2018, Li2019, Fukushima2021, Menon2024}. In particular, stars are expected to form up until $\langle \Dot{p}/m_{\star} \rangle$ exceeds the gravitational binding force per unit mass of the cloud, $\sim G \Sigma_{\rm cl}$, after which the cloud is disrupted. Retaining factors of order unity, this can be shown to yield a maximum SFE $f_{\star}$ of \citep[e.g.][]{Fukushima2021}:
\begin{equation}
    \dfrac{f_{\star}}{1 - f_{\star}^2} \simeq \dfrac{\Sigma_{\rm cl}}{\Sigma_{\rm crit}}  \, , \quad \Sigma_{\rm crit} \equiv \dfrac{2}{\pi G} \left\langle \dfrac{\Dot{p}}{m_{\star}} \right\rangle \, .
\end{equation}
The SFE, therefore, saturates near unity for $\Sigma_{\rm cl} \gg \Sigma_{\rm crit}$. If efficient star cluster formation requires $f_{\star} \gtrsim 0.1$, we can derive a lower bound on the \textit{stellar} surface density $\Sigma_{\star} = f_{\star} \Sigma_{\rm cl}$ of a bound cluster at birth:
\begin{equation}
    \Sigma_{\star} \gtrsim 0.01 \Sigma_{\rm crit} =  \dfrac{0.02}{\pi G} \left\langle \dfrac{\Dot{p}}{m_{\star}} \right\rangle \;\;\; \textrm{(Efficient cluster formation)} . \label{critical cluster surf dens general}
\end{equation}
Direct radiation pressure and stellar winds only contribute $\langle \Dot{p}/m_{\star} \rangle  \sim 10^3\, \Lsun/\Msun c$ each, at most \citep[e.g.][]{Grudic2018, Hopkins2018_FIRE2}. This yields the fairly weak constraint $\Sigma_{\star} \gtrsim 60 ~ \Msun \, \rm pc^{-2}$ at $Z/Z_{\odot} \sim 1$ for newly born, gravitationally bound star clusters. The constraint from these processes becomes weaker still at lower metallicities. Ly$\alpha$ feedback, however, gives significantly stronger constraints. It contributes $\langle \Dot{p}_{\rm Ly\alpha}/m_{\star} \rangle = (L_{\rm Ly\alpha}/M_{\star}c) M_{\rm F}$, and so using Eqs.~(\ref{Lya luminosity}) and (\ref{critical cluster surf dens general}):
\begin{align}
    \Sigma_{\star} &\gtrsim~ 410 \, (1-f_{\rm esc,LyC}) \left( \dfrac{f_{\rm rec,Ly\alpha}}{2/3} \right) \label{Stellar surf density limit Lya feedback} \\ &\times~ \left( \dfrac{\Dot{Q}_{\rm LyC}}{5 \times 10^{46}~ \textrm{s}^{-1} ~ \Msun^{-1}} \right)  \left( \dfrac{M_{\rm F}}{100} \right) \, \rm \Msun ~ \textrm{pc}^{-2} \nonumber \, ,
\end{align}
where we have normalized to standard Pop II values, and a fairly typical value for the expected force multiplier (see Fig. \ref{discussion_overview}). In Fig. \ref{StarClusters} we compare the above prediction to recently discovered star clusters at redshifts $z \simeq 6-10$ \citep{Vanzella2023, Adamo2024, Mowla2024}. All the gravitationally bound star clusters are consistent with lying above the limit set by Ly$\alpha$ feedback. The star cluster 4b of \cite{Vanzella2023}, which shows evidence of being gravitationally unbound,\footnote{It has $\Pi \equiv \textrm{(Age)/(Crossing time)} = 0.2_{-0.1}^{+0.4}$, which is below unity, the expected dividing line between bound and unbound clusters \citep{Gieles2011}.} is situated near the limit set by Eq.~(\ref{Stellar surf density limit Lya feedback}), either for $M_{\rm F} \sim 200$, which is a reasonable value under less conservative assumptions, or for $M_{\rm F} \sim 100$ but a slightly larger assumed critical SFE for the formation of bound clusters.\footnote{For example, if turbulence and velocity gradients are negligible, we would predict $M_{\rm F} \simeq 220$ from Eq.~(\ref{M_F fit turbulence, final}). Furthermore, the fraction of stars that end up in bound star clusters is not a step function of the SFE. Rather, it goes from nearly zero at $10\%$ to near unity at $\sim 30\%$ \citep{Adams2000, Li2019, Fukushima2021}. Thus, we expect to see some unbound clusters with stellar surface densities up to a factor few above the conservative limit in Eq.~(\ref{Stellar surf density limit Lya feedback}), which had adopted a critical SFE for bound star cluster formation of $10\%$. } 

It is also likely that the implementation of Ly$\alpha$ feedback in simulations and models could alter the predicted abundance of globular clusters (GCs). Indeed, this was found in the simulations of \cite{Kimm2018}, who found that Ly$\alpha$ feedback reduced the number of star clusters formed in a metal-poor dwarf galaxy by a factor $\sim 5$. Similarly, \cite{Nebrin2022} found that the abundace of bound star clusters formed in low-mass haloes at Cosmic Dawn is sensitive to Ly$\alpha$ feedback. \cite{Abe2018} had argued that the apparent metallicity floor of GCs, $\textrm{[Fe/H]} \gtrsim -2.5$, could be explained by Ly$\alpha$ feedback becoming increasingly strong at lower metallicities, preventing the formation of bound star clusters. Indeed, we have also found that Ly$\alpha$ feedback put constraints on the formation of bound star clusters. 

However, we also note that these earlier studies had various shortcomings when it comes to estimating $M_{\rm F}$. For example, velocity gradients and turbulence were not taken into account in these estimates of $M_{\rm F}$. Furthermore, as discussed in Sec. \ref{Force multiplier}, \cite{Kimm2018} systematically underestimated $M_{\rm F}$ for uniform, static clouds by a factor $\sim 2-3$. \cite{Abe2018}, on the other hand, overestimated the suppression of $M_{\rm F}$ by dust absorption. As shown in this paper, and by \cite{Kimm2018} and \cite{Tomaselli2021}, there is no sharp transition in $M_{\rm F}$ at some specific metallicity. This conclusion is also consistent with more recent detections of GCs below the apparent metallicity floor \citep{Larsen2020, Martin2022, Weisz2023}, showing that star formation can be very efficient --- i.e. not limited by Ly$\alpha$ feedback --- in \textit{some} clouds down to at least $\textrm{[Fe/H]} \simeq -3.4$ \citep{Martin2022}. 

Additional observational constraints on the properties of high-$z$ young star clusters, coupled with more detailed theoretical predictions taking into account the processes studied in this paper, could help further constrain the strength and role of Ly$\alpha$ feedback during star cluster formation in the early Universe.

\subsection{Implications for the first stars}
\label{First stars implications}

The very first stars (Pop III stars) formed in zero-metallicity environments, where there is no dust on which H$_2$ can form. Because of this, and the extremely high predicted gas densities, large \HI column densities are guaranteed to be found. As a result, several authors have argued that Ly$\alpha$ feedback could be important in regulating formation and growth of Pop III protostars \citep{Doroshkevich1976, McKee2008, Jaura2022, Nebrin2022}. \cite{McKee2008} generalized, in an approximate manner, the analytical slab solution found by \cite{Neufeld1990} to study Ly$\alpha$ feedback from Pop III protostars. These authors found that Ly$\alpha$ feedback could potentially clear gas in the polar directions, after which Ly$\alpha$ photons would easily escape, and Ly$\alpha$ feedback would be suppressed. 

More recently, \cite{Jaura2022} studied Pop III star formation in high-resolution simulations. These authors found that the \HII regions remained trapped in the very dense gas on extremely small scales ($\lesssim \rm AU$). To facilitate \HII region breakout, Ly$\alpha$ feedback was suggested as a key missing ingredient in current simulations. Similarly, \cite{Nebrin2022} studied starbursts in low-mass haloes using an early version of the (semi-)analytical model \textsc{Anaxagoras}, and also found that Ly$\alpha$ radiation pressure is capable of regulating the total number of Pop III stars formed in minihaloes. 

However, these studies have several limitations. For instance, \cite{Nebrin2022} used the solution for the force multiplier by \cite{Tomaselli2021}, which did not include, e.g., velocity gradients, Ly$\alpha$ destruction by $2p \rightarrow 2s$ transitions and Raman scattering, and atomic recoil. \cite{McKee2008} did consider some of these effects, but mainly relying on order-of-magnitude scaling relations, and not including the rate of $2p \rightarrow 2s$ transitions in collisions with \HI, which we have found to dominate over collisions with protons and electrons in most cases (see Fig. \ref{RelativeCErates}). 

There are two separate basic questions we can ask about the role of Ly$\alpha$ feedback during Pop III star formation:
\begin{enumerate}
    \item Is Ly$\alpha$ feedback important compared to other feedback processes?
    \item If the answer is \textit{yes} to (i), is it strong enough to regulate Pop III star formation?
\end{enumerate}
In Fig. \ref{discussion_overview} it is evident that Ly$\alpha$ feedback dominates over direct radiation pressure. However, in Fig. \ref{discussion_overview} we had assumed an average \HI density $\langle n_{\rm HI} \rangle = 10^6 \, \rm cm^{-3}$ for all parameter space. In contrast, simulations and models of Pop III star formation in minihaloes find extremely dense discs, with temperature $T \sim 1000 \, \rm K$ and (total) gas number densities $n \sim 10^8 - 10^{11} \, \rm cm^{-3}$ \citep[e.g.][]{Jaura2022, Sugimura2023_PopIII}. As long as not all the gas is converted into H$_2$, this would imply \HI densities of a similar magnitude. With higher gas densities and temperatures like these, we estimate that the Ly$\alpha$ destruction probability is $p_{\rm d} \sim 4 \times 10^{-9} - 10^{-8}$ for $f_{\rm H_2} \lesssim 0.01$, in which case the destruction probability is mainly determined by $2p \rightarrow 2s$ transitions in collisions with \HI (see Figs. \ref{RelativeCErates} and \ref{pd_colorplot}).

If the \HI column density around Pop III protostars is $N_{\rm HI} \sim 10^{25} \, \rm cm^{-2}$, as estimated by \cite{Jaura2022}, we expect Ly$\alpha$ destruction to suppress $M_{\rm F}$ by a factor $\sim \textrm{few} \times 10$ (see Fig. \ref{M_F p_d effect}). Velocity gradients near the protostar, anisotropic Ly$\alpha$ escape, turbulence, and Raman scattering off H$_2$, could further suppress $M_{\rm F}$ by a factor $\sim \textrm{few}$. All in all, suppression of this magnitude could render Ly$\alpha$ feedback too weak to help \HII regions break out from the protostellar discs around Pop III stars, in contrast to the suggestion of \cite{Jaura2022}. This conclusion, however, is sensitive to the assumed Ly$\alpha$ output -- if Pop III stars grow more massive, their increased ionizing and Ly$\alpha$ output (Table \ref{Qdot}) could, at least partially, counteract the suppression of $M_{\rm F}$ by Ly$\alpha$ destruction. 

The key point here is that the presence of large \HI column densities and no dust, by itself, does not to ensure that Ly$\alpha$ feedback will be efficient.\footnote{A similar point has been made by \cite{Smith2019} regarding the importance of Ly$\alpha$ feedback, and by \cite{krumholz2018resolution} regarding radiation pressure on dust. In both cases, gas expulsion can only occur if feedback is strong enough to overcome the gravitational binding force of the cloud. A feedback process may be dominant, but still, in some circumstances, be too weak to have a significant dynamical effect on the gas. } The answers to questions (i) and (ii) above may very well be \textit{yes} and \textit{no}, respectively. More detailed analytical and numerical modelling should revisit the role of Ly$\alpha$ feedback during Pop III star formation, taking into account all major processes that effect Ly$\alpha$ radiation pressure. 

\section{Summary \& conclusion}
\label{Summary and conclusion}

Recent studies have emphasized the importance, and even dominance, of Ly$\alpha$ radiation pressure feedback during star and galaxy formation, especially in high-redshift and dust-poor environments \citep[e.g.][]{Smith2017, Kimm2018, Kimm2019, Tomaselli2021, Nebrin2022, Kapoor2023, Thomson2024}. In contrast, recent \textit{JWST} results indicate surprisingly efficient star formation in high-redshift galaxies. As simulations and models of the first stars and galaxies start to incorporate Ly$\alpha$ feedback in the pursuit of greater realism, there is an urgent need to understand under what conditions Ly$\alpha$ feedback is expected to be strong, or risk underpredicting the star formation efficiency.

In this paper, we have addressed Ly$\alpha$ feedback using a multifaceted approach. We have:
\begin{enumerate}
    \item \textit{Derived a novel analytical Ly$\alpha$ radiative transfer (RT) solution}: To study Ly$\alpha$ feedback in detail, we have greatly generalized previous analytical Ly$\alpha$ RT solutions for spherical clouds to incorporate the effects of continuum absorption, finite Ly$\alpha$ destruction probability (e.g. by $2p \rightarrow 2s$ transitions), gas velocity gradients, atomic recoil, and low-density channels from ISM turbulence. We have verified the solution for uniform clouds against extensive Monte Carlo Ly$\alpha$ RT (MCRT) experiments with \textsc{colt} \citep{Smith2015}, showing excellent agreement. This solution, and the framework used to, e.g., handle ISM turbulence, has broader applications beyond Ly$\alpha$ feedback, such as predicting and understanding observed Ly$\alpha$ spectra. We explore the effects of velocity gradients in greater detail in a companion study \citep{Smith2025}, and also plan to generalize analytical solutions further to study other effects (e.g. different cloud geometries).
    \\
    
    \item \textit{Studied suppression mechanisms for Ly$\alpha$ feedback}: We have conducted a comprehensive study of processes that can suppress Ly$\alpha$ feedback. We have identified and studied several processes that can be particularly important in star-forming clouds. These include (1) continuum absorption by dust and Raman scattering off H$_2$, (2) $2p \rightarrow 2s$ transitions in collisions with \HI, \HeI, and H$_2$, (3) velocity gradients, (4) ISM turbulence, and (5) atomic recoil. In the case of (2) we perform, to our knowledge, the first \textit{ab initio} calculation of the rate for \HI at relevant temperatures (Sec. \ref{Subsection CE/CDE rates} and Appendix \ref{Appendix H(2s) cross section}), and show it to be very important in regulating Ly$\alpha$ feedback around Pop III stars. The various suppression mechanisms covered in this paper are shown to be capable of suppressing Ly$\alpha$ feedback by a factor $\sim \textrm{few} - \textrm{few} \times 10$, depending on the precise conditions in the star-forming cloud. 
    \\

    \item \textit{Resolved the discrepancy between analytical and MCRT results}: We resolve an earlier unexplained factor $\sim 2-3$ discrepancy between analytical and MCRT estimates for the Ly$\alpha$ force multiplier $M_{\rm F}$ \citep{Kimm2018, Lao2020, Tomaselli2021}. Instead of missing physics (e.g. recoil), as speculated by \cite{Tomaselli2021}, we found that numerical errors in earlier MCRT experiments by \cite{Kimm2018} caused the discrepancy. We find that our analytical solution is in excellent agreement with our MCRT results. As a result, despite finding Ly$\alpha$ feedback to be very important, \cite{Kimm2018} may have systematically underestimated it, highlighting the urgency of implementing this feedback process in upcoming simulations. \\
    
    \item \textit{Developed an accurate analytical fit to the Ly$\alpha$ force multiplier}: We derived a computationally efficient, accurate, and easily implemented fit to the Ly$\alpha$ force multiplier $M_{\rm F}$, covering all major suppression mechanisms. This fit can be incorporated into a subgrid model for Ly$\alpha$ feedback in both star and galaxy formation simulations. Such a subgrid model greatly improves upon earlier modelling that lacked several processes considered in this paper.
    
\end{enumerate}
We also discuss some preliminary implications of our results:
\begin{itemize}
    \item \textit{Ly$\alpha$ feedback is dynamically important}: Despite considering a multitude of suppression mechanisms of Ly$\alpha$ feedback, we find that it prevails as an important, if not dominant, early feedback mechanism. This conclusion holds true for a wide range of metallicities and gas conditions. As a result, Ly$\alpha$ feedback cannot anymore be ignored, and should be implemented in star and galaxy formation simulations and models, even with approximate prescriptions until more accurate treatments can be widely adopted. \\

    \item \textit{Implications for the first galaxies}: Ly$\alpha$ feedback is predicted to be very strong, if not dominant, in the first faint dwarf galaxies that formed in low-mass haloes at Cosmic Dawn (Sec. \ref{first galaxies implications}). Ly$\alpha$ feedback is therefore likely to have a significant impact on the predicted properties and abundance of Ultra-Faint Dwarf galaxies in high-resolution galaxy formation simulations. This conclusion is consistent with earlier findings from semi-analytical modelling, which was less comprehensive with respect to Ly$\alpha$ feedback \citep{Nebrin2022}. This could also have significant implications for testing $\Lambda$CDM using the faintest satellite galaxies of the Milky Way \citep{Nadler2020}.   \\

    \item \textit{Implications for the first star clusters}: Ly$\alpha$ feedback is likely a dominant early feedback process during the formation of star clusters, as found by several recent authors \citep{Abe2018, Kimm2018, Nebrin2022}. We find that strong Ly$\alpha$ feedback imposes a lower limit on the stellar surface densities of young, gravitationally bound star clusters. We compare this constraint to recently discovered star clusters at redshifts $z \simeq 6-10$ \citep[][see Fig. \ref{StarClusters}]{Vanzella2023, Adamo2024, Mowla2024}, and find them to be consistent with the expected strong Ly$\alpha$ feedback in their birth clouds ($30 \lesssim M_{\rm F} \lesssim 200$).   \\

    \item \textit{Implications for the first stars}:  Ly$\alpha$ feedback is likely to dominate over direct (photoionizing) radiation pressure from Pop III protostars. However, whether it is sufficiently strong to aid in the breakout of \HII regions around these stars, as suggested by \cite{Jaura2022}, remains unclear. This is primarily due to the process of $2p \rightarrow 2s$ transitions in collisions with \HI, that we have identified and studied in this paper. We find that this process can suppress Ly$\alpha$ feedback by a factor $\sim \textrm{few} \times 10$, potentially making it too weak to expel gas from the extremely dense protostellar discs around Pop III protostars. However, this conclusion depends to some extent on the ionizing output, and hence masses, of Pop III stars. Regardless of its ultimate importance, Ly$\alpha$ feedback should be implemented in simulations of the first stars for physical completeness, as it is significantly stronger than the already implemented feedback from direct radiation pressure \citep[e.g.][]{Jaura2022, Latif2022}.
\end{itemize}

The present study highlights that Ly$\alpha$ feedback constitutes a rich astrophysical problem, where details of both resonant line RT and atomic/molecular physics can have a strong impact on one of the most dominant stellar feedback processes at high redshifts. While we believe that we have made significant strides towards realistic modelling of Ly$\alpha$ feedback, our work has also opened several new avenues for future exploration. Among other things, these include further studies of Ly$\alpha$ radiation pressure in clouds with complicated geometries, investigating the potential impact of Ly$\alpha$ destruction by neighbouring lines (e.g. in \OI), more accurate calculations of cross-sections at low temperatures for key processes (e.g. $2s \rightarrow 2p$ collisional excitation by \HeI and H$_2$), and studying the impact of deuterium scattering of Ly$\alpha$ \citep{Dijkstra2006, RASCAS}. State-of-the-art numerical and analytical methods will continue to address these questions and help us better understand how star formation is regulated in the early Universe. We therefore urge the simulation community to prioritize numerical implementations and benchmarks of Ly$\alpha$ radiation pressure.

\section*{Acknowledgements}

We thank Todd Thompson, Peter Laursen, Matthew Hayes, Tobias Buck, Angela Adamo, Andrea Ferrara, Volker Bromm, Dimitri Arramy, Mike Grudić, Taysun Kimm, and Pavel Kroupa for helpful comments, suggestions, and discussions during this project. O.N. acknowledges support
from the Swedish Research Council grant 2020-04691. J.H. and Å.L. acknowledge support from the project “Probing charge- and mass-transfer reactions on the atomic
level,” funded by the Knut and Alice Wallenberg Foundation
(2018.0028).

\textit{Software}: To evaluate, e.g., the series solutions for $M_{\rm F}$ and spectra, and to produce the figures in this paper, we have made use of \textsc{Scipy} \citep{Virtanen2020_Scipy}, \textsc{numpy} \citep{Harris2020_Numpy}, \textsc{Julia} \citep{Julia2017}, and \textsc{matplotlib} \citep{Hunter2007_matplotlib}. O.N. used a combination of \textsc{Krita} and \textsc{matplotlib} for the schematic sketches.

\section*{Data Availability}
A simple \textsc{Python} implementation of the fit to $M_{\rm F}$ (Eq.~\ref{M_F fit turbulence, final}), along with MCRT data, and the calculated cross-section for $\textrm{H}(2s)+\textrm{H}(1s) \rightarrow \textrm{H}(2p)+\textrm{H}(1s)$ (Fig. \ref{C2s2p_atomicH}), is available at the Github repository associated with this project: \url{https://github.com/olofnebrin/Lyman-alpha-feedback}. 





\bibliography{bibfile}

\begin{thebibliography}{}
\makeatletter
\relax
\def\mn@urlcharsother{\let\do\@makeother \do\$\do\&\do\#\do\^\do\_\do\%\do\~}
\def\mn@doi{\begingroup\mn@urlcharsother \@ifnextchar [ {\mn@doi@} {\mn@doi@[]}}
\def\mn@doi@[#1]#2{\def\@tempa{#1}\ifx\@tempa\@empty \href {http://dx.doi.org/#2} {doi:#2}\else \href {http://dx.doi.org/#2} {#1}\fi \endgroup}
\def\mn@eprint#1#2{\mn@eprint@#1:#2::\@nil}
\def\mn@eprint@arXiv#1{\href {http://arxiv.org/abs/#1} {{\tt arXiv:#1}}}
\def\mn@eprint@dblp#1{\href {http://dblp.uni-trier.de/rec/bibtex/#1.xml} {dblp:#1}}
\def\mn@eprint@#1:#2:#3:#4\@nil{\def\@tempa {#1}\def\@tempb {#2}\def\@tempc {#3}\ifx \@tempc \@empty \let \@tempc \@tempb \let \@tempb \@tempa \fi \ifx \@tempb \@empty \def\@tempb {arXiv}\fi \@ifundefined {mn@eprint@\@tempb}{\@tempb:\@tempc}{\expandafter \expandafter \csname mn@eprint@\@tempb\endcsname \expandafter{\@tempc}}}

\bibitem[\protect\citeauthoryear{{Abe} \& {Yajima}}{{Abe} \& {Yajima}}{2018}]{Abe2018}
{Abe} M.,  {Yajima} H.,  2018, \mn@doi [\mnras] {10.1093/mnrasl/sly018}, \href {https://ui.adsabs.harvard.edu/abs/2018MNRAS.475L.130A} {475, L130}

\bibitem[\protect\citeauthoryear{Abramovich \& Indelman}{Abramovich \& Indelman}{1995}]{Abramovich1995}
Abramovich B.,  Indelman P.,  1995, Journal of Physics A: Mathematical and General, 28, 693

\bibitem[\protect\citeauthoryear{{Adamo} et~al.,}{{Adamo} et~al.}{2024}]{Adamo2024}
{Adamo} A.,  et~al., 2024, \mn@doi [\nat] {10.1038/s41586-024-07703-7}, \href {https://ui.adsabs.harvard.edu/abs/2024Natur.632..513A} {632, 513}

\bibitem[\protect\citeauthoryear{{Adams}}{{Adams}}{1971}]{Adams1971}
{Adams} T.~F.,  1971, \mn@doi [\apj] {10.1086/151111}, \href {https://ui.adsabs.harvard.edu/abs/1971ApJ...168..575A} {168, 575}

\bibitem[\protect\citeauthoryear{{Adams}}{{Adams}}{1972}]{Adams1972}
{Adams} T.~F.,  1972, \mn@doi [\apj] {10.1086/151503}, \href {https://ui.adsabs.harvard.edu/abs/1972ApJ...174..439A} {174, 439}

\bibitem[\protect\citeauthoryear{{Adams}}{{Adams}}{1975}]{Adams1975}
{Adams} T.~F.,  1975, \mn@doi [\apj] {10.1086/153891}, \href {https://ui.adsabs.harvard.edu/abs/1975ApJ...201..350A} {201, 350}

\bibitem[\protect\citeauthoryear{{Adams}}{{Adams}}{2000}]{Adams2000}
{Adams} F.~C.,  2000, \mn@doi [\apj] {10.1086/317052}, \href {https://ui.adsabs.harvard.edu/abs/2000ApJ...542..964A} {542, 964}

\bibitem[\protect\citeauthoryear{{Agertz} et~al.,}{{Agertz} et~al.}{2020}]{Agertz2020}
{Agertz} O.,  et~al., 2020, \mn@doi [\mnras] {10.1093/mnras/stz3053}, \href {https://ui.adsabs.harvard.edu/abs/2020MNRAS.491.1656A} {491, 1656}

\bibitem[\protect\citeauthoryear{{Aggarwal}, {Owada}  \& {Igarashi}}{{Aggarwal} et~al.}{2018}]{Aggarwal2018}
{Aggarwal} K.~M.,  {Owada} R.,   {Igarashi} A.,  2018, \mn@doi [Atoms] {10.3390/atoms6030037}, \href {https://ui.adsabs.harvard.edu/abs/2018Atoms...6...37A} {6, 37}

\bibitem[\protect\citeauthoryear{{Akshaya}, {Murthy}, {Ravichandran}, {Henry}  \& {Overduin}}{{Akshaya} et~al.}{2019}]{Akshaya2019}
{Akshaya} M.~S.,  {Murthy} J.,  {Ravichandran} S.,  {Henry} R.~C.,   {Overduin} J.,  2019, \mn@doi [\mnras] {10.1093/mnras/stz2186}, \href {https://ui.adsabs.harvard.edu/abs/2019MNRAS.489.1120A} {489, 1120}

\bibitem[\protect\citeauthoryear{{Almada Monter} \& {Gronke}}{{Almada Monter} \& {Gronke}}{2024}]{Monter2024}
{Almada Monter} S.,  {Gronke} M.,  2024, \mn@doi [\mnras] {10.1093/mnrasl/slae074}, \href {https://ui.adsabs.harvard.edu/abs/2024MNRAS.534L...7A} {534, L7}

\bibitem[\protect\citeauthoryear{{Appel}, {Burkhart}, {Semenov}, {Federrath}  \& {Rosen}}{{Appel} et~al.}{2022}]{Appel2022}
{Appel} S.~M.,  {Burkhart} B.,  {Semenov} V.~A.,  {Federrath} C.,   {Rosen} A.~L.,  2022, \mn@doi [\apj] {10.3847/1538-4357/ac4be3}, \href {https://ui.adsabs.harvard.edu/abs/2022ApJ...927...75A} {927, 75}

\bibitem[\protect\citeauthoryear{{Applebaum}, {Brooks}, {Christensen}, {Munshi}, {Quinn}, {Shen}  \& {Tremmel}}{{Applebaum} et~al.}{2021}]{Applebaum2021}
{Applebaum} E.,  {Brooks} A.~M.,  {Christensen} C.~R.,  {Munshi} F.,  {Quinn} T.~R.,  {Shen} S.,   {Tremmel} M.,  2021, \mn@doi [\apj] {10.3847/1538-4357/abcafa}, \href {https://ui.adsabs.harvard.edu/abs/2021ApJ...906...96A} {906, 96}

\bibitem[\protect\citeauthoryear{{Arfken}, {Weber}  \& {Harris}}{{Arfken} et~al.}{2013}]{Arfken2013}
{Arfken} G.~B.,  {Weber} H.~J.,   {Harris} F.~E.,  2013, {Mathematical Methods for Physicists 7th ed}.
{Academic Press}

\bibitem[\protect\citeauthoryear{{Badnell}, {Guzm{\'a}n}, {Brodie}, {Williams}, {van Hoof}, {Chatzikos}  \& {Ferland}}{{Badnell} et~al.}{2021}]{Badnell2021}
{Badnell} N.~R.,  {Guzm{\'a}n} F.,  {Brodie} S.,  {Williams} R.~J.~R.,  {van Hoof} P.~A.~M.,  {Chatzikos} M.,   {Ferland} G.~J.,  2021, \mn@doi [\mnras] {10.1093/mnras/stab2266}, \href {https://ui.adsabs.harvard.edu/abs/2021MNRAS.507.2922B} {507, 2922}

\bibitem[\protect\citeauthoryear{{Baumgardt} \& {Hilker}}{{Baumgardt} \& {Hilker}}{2018}]{Baumgardt2018}
{Baumgardt} H.,  {Hilker} M.,  2018, \mn@doi [\mnras] {10.1093/mnras/sty1057}, \href {https://ui.adsabs.harvard.edu/abs/2018MNRAS.478.1520B} {478, 1520}

\bibitem[\protect\citeauthoryear{{Baumgardt} \& {Kroupa}}{{Baumgardt} \& {Kroupa}}{2007}]{Baumgardt2007}
{Baumgardt} H.,  {Kroupa} P.,  2007, \mn@doi [\mnras] {10.1111/j.1365-2966.2007.12209.x}, \href {https://ui.adsabs.harvard.edu/abs/2007MNRAS.380.1589B} {380, 1589}

\bibitem[\protect\citeauthoryear{{Begelman}}{{Begelman}}{2006}]{Begelman2006}
{Begelman} M.~C.,  2006, \mn@doi [\apj] {10.1086/503093}, \href {https://ui.adsabs.harvard.edu/abs/2006ApJ...643.1065B} {643, 1065}

\bibitem[\protect\citeauthoryear{{Behrens}, {Dijkstra}  \& {Niemeyer}}{{Behrens} et~al.}{2014}]{Behrens2014}
{Behrens} C.,  {Dijkstra} M.,   {Niemeyer} J.~C.,  2014, \mn@doi [\aap] {10.1051/0004-6361/201322949}, \href {https://ui.adsabs.harvard.edu/abs/2014A&A...563A..77B} {563, A77}

\bibitem[\protect\citeauthoryear{Belyaev}{Belyaev}{2015}]{belyaev2015}
Belyaev A.~K.,  2015, \mn@doi [Phys. Rev. A] {10.1103/PhysRevA.91.062709}, 91, 062709

\bibitem[\protect\citeauthoryear{Belyaev, Egorova, Grosser  \& Menzel}{Belyaev et~al.}{2001}]{belyaev2001}
Belyaev A.~K.,  Egorova D.,  Grosser J.,   Menzel T.,  2001, \mn@doi [Phys. Rev. A] {10.1103/PhysRevA.64.052701}, 64, 052701

\bibitem[\protect\citeauthoryear{{Benitez-Llambay} \& {Frenk}}{{Benitez-Llambay} \& {Frenk}}{2020}]{Benitez2020}
{Benitez-Llambay} A.,  {Frenk} C.,  2020, \mn@doi [\mnras] {10.1093/mnras/staa2698}, \href {https://ui.adsabs.harvard.edu/abs/2020MNRAS.498.4887B} {498, 4887}

\bibitem[\protect\citeauthoryear{Bezanson, Edelman, Karpinski  \& Shah}{Bezanson et~al.}{2017}]{Julia2017}
Bezanson J.,  Edelman A.,  Karpinski S.,   Shah V.~B.,  2017, \mn@doi [SIAM {R}eview] {10.1137/141000671}, 59, 65

\bibitem[\protect\citeauthoryear{{Bialy} \& {Sternberg}}{{Bialy} \& {Sternberg}}{2016}]{Bialy2016}
{Bialy} S.,  {Sternberg} A.,  2016, \mn@doi [\apj] {10.3847/0004-637X/822/2/83}, \href {https://ui.adsabs.harvard.edu/abs/2016ApJ...822...83B} {822, 83}

\bibitem[\protect\citeauthoryear{{Bithell}}{{Bithell}}{1990}]{Birthell1990}
{Bithell} M.,  1990, \mnras, \href {https://ui.adsabs.harvard.edu/abs/1990MNRAS.244..738B} {244, 738}

\bibitem[\protect\citeauthoryear{{Blitz}, {Fukui}, {Kawamura}, {Leroy}, {Mizuno}  \& {Rosolowsky}}{{Blitz} et~al.}{2007}]{Blitz2007}
{Blitz} L.,  {Fukui} Y.,  {Kawamura} A.,  {Leroy} A.,  {Mizuno} N.,   {Rosolowsky} E.,  2007, in {Reipurth} B.,  {Jewitt} D.,   {Keil} K.,  eds, Protostars and Planets V. p.~81 (\mn@eprint {arXiv} {astro-ph/0602600}), \mn@doi{10.48550/arXiv.astro-ph/0602600}

\bibitem[\protect\citeauthoryear{{Bonilha}, {Ferch}, {Salpeter}, {Slater}  \& {Noerdlinger}}{{Bonilha} et~al.}{1979}]{Bonilha1979}
{Bonilha} J.~R.~M.,  {Ferch} R.,  {Salpeter} E.~E.,  {Slater} G.,   {Noerdlinger} P.~D.,  1979, \mn@doi [\apj] {10.1086/157426}, \href {https://ui.adsabs.harvard.edu/abs/1979ApJ...233..649B} {233, 649}

\bibitem[\protect\citeauthoryear{{Boylan-Kolchin}}{{Boylan-Kolchin}}{2023}]{BoylanKolchin2023}
{Boylan-Kolchin} M.,  2023, \mn@doi [Nature Astronomy] {10.1038/s41550-023-01937-7}, \href {https://ui.adsabs.harvard.edu/abs/2023NatAs...7..731B} {7, 731}

\bibitem[\protect\citeauthoryear{{Braun} \& {Dekel}}{{Braun} \& {Dekel}}{1989}]{Braun1989}
{Braun} E.,  {Dekel} A.,  1989, \mn@doi [\apj] {10.1086/167878}, \href {https://ui.adsabs.harvard.edu/abs/1989ApJ...345...31B} {345, 31}

\bibitem[\protect\citeauthoryear{{Briegleb}}{{Briegleb}}{1992}]{NCAR1992}
{Briegleb} B.~P.,  1992, \mn@doi [\jgr] {10.1029/92JD00291}, \href {https://ui.adsabs.harvard.edu/abs/1992JGR....97.7603B} {97, 7603}

\bibitem[\protect\citeauthoryear{{Bromm} \& {Yoshida}}{{Bromm} \& {Yoshida}}{2011}]{Bromm2011review}
{Bromm} V.,  {Yoshida} N.,  2011, \mn@doi [\araa] {10.1146/annurev-astro-081710-102608}, \href {https://ui.adsabs.harvard.edu/abs/2011ARA&A..49..373B} {49, 373}

\bibitem[\protect\citeauthoryear{{Brown} et~al.,}{{Brown} et~al.}{2014}]{Brown2014}
{Brown} T.~M.,  et~al., 2014, \mn@doi [\apj] {10.1088/0004-637X/796/2/91}, \href {https://ui.adsabs.harvard.edu/abs/2014ApJ...796...91B} {796, 91}

\bibitem[\protect\citeauthoryear{{Buchler}}{{Buchler}}{1983}]{Buchler1983}
{Buchler} J.~R.,  1983, \mn@doi [\jqsrt] {10.1016/0022-4073(83)90102-4}, \href {https://ui.adsabs.harvard.edu/abs/1983JQSRT..30..395B} {30, 395}

\bibitem[\protect\citeauthoryear{{Buck}, {Pfrommer}, {Girichidis}  \& {Corobean}}{{Buck} et~al.}{2022}]{Buck2022}
{Buck} T.,  {Pfrommer} C.,  {Girichidis} P.,   {Corobean} B.,  2022, \mn@doi [\mnras] {10.1093/mnras/stac952}, \href {https://ui.adsabs.harvard.edu/abs/2022MNRAS.513.1414B} {513, 1414}

\bibitem[\protect\citeauthoryear{{Bullock} \& {Boylan-Kolchin}}{{Bullock} \& {Boylan-Kolchin}}{2017}]{Bullock2017}
{Bullock} J.~S.,  {Boylan-Kolchin} M.,  2017, \mn@doi [\araa] {10.1146/annurev-astro-091916-055313}, \href {https://ui.adsabs.harvard.edu/abs/2017ARA&A..55..343B} {55, 343}

\bibitem[\protect\citeauthoryear{{Burkhart}, {Lee}, {Murray}  \& {Stanimirovi{\'c}}}{{Burkhart} et~al.}{2015}]{Burkhart2015}
{Burkhart} B.,  {Lee} M.-Y.,  {Murray} C.~E.,   {Stanimirovi{\'c}} S.,  2015, \mn@doi [\apjl] {10.1088/2041-8205/811/2/L28}, \href {https://ui.adsabs.harvard.edu/abs/2015ApJ...811L..28B} {811, L28}

\bibitem[\protect\citeauthoryear{{Burkhart}, {Stalpes}  \& {Collins}}{{Burkhart} et~al.}{2017}]{Burkhart2017_PLT}
{Burkhart} B.,  {Stalpes} K.,   {Collins} D.~C.,  2017, \mn@doi [\apjl] {10.3847/2041-8213/834/1/L1}, \href {https://ui.adsabs.harvard.edu/abs/2017ApJ...834L...1B} {834, L1}

\bibitem[\protect\citeauthoryear{{Byron} \& {Gersten}}{{Byron} \& {Gersten}}{1971}]{Byron1971}
{Byron} F.~W.,  {Gersten} J.~I.,  1971, \mn@doi [\pra] {10.1103/PhysRevA.3.620}, \href {https://ui.adsabs.harvard.edu/abs/1971PhRvA...3..620B} {3, 620}

\bibitem[\protect\citeauthoryear{{Caselli}, {Walmsley}, {Terzieva}  \& {Herbst}}{{Caselli} et~al.}{1998}]{Caselli1998}
{Caselli} P.,  {Walmsley} C.~M.,  {Terzieva} R.,   {Herbst} E.,  1998, \mn@doi [\apj] {10.1086/305624}, \href {https://ui.adsabs.harvard.edu/abs/1998ApJ...499..234C} {499, 234}

\bibitem[\protect\citeauthoryear{{Castor}}{{Castor}}{2004}]{Castor2004}
{Castor} J.~I.,  2004, {Radiation Hydrodynamics}.
{Cambridge University Press}

\bibitem[\protect\citeauthoryear{{Cazaux} \& {Spaans}}{{Cazaux} \& {Spaans}}{2009}]{Cazaux2009}
{Cazaux} S.,  {Spaans} M.,  2009, \mn@doi [\aap] {10.1051/0004-6361:200811302}, \href {https://ui.adsabs.harvard.edu/abs/2009A&A...496..365C} {496, 365}

\bibitem[\protect\citeauthoryear{{Chemerynska} et~al.,}{{Chemerynska} et~al.}{2024}]{Chemerynska2024}
{Chemerynska} I.,  et~al., 2024, \mn@doi [\mnras] {10.1093/mnras/stae1260}, \href {https://ui.adsabs.harvard.edu/abs/2024MNRAS.531.2615C} {531, 2615}

\bibitem[\protect\citeauthoryear{{Chevance} et~al.,}{{Chevance} et~al.}{2022}]{Chevance2022}
{Chevance} M.,  et~al., 2022, \mn@doi [\mnras] {10.1093/mnras/stab2938}, \href {https://ui.adsabs.harvard.edu/abs/2022MNRAS.509..272C} {509, 272}

\bibitem[\protect\citeauthoryear{{Chevance}, {Krumholz}, {McLeod}, {Ostriker}, {Rosolowsky}  \& {Sternberg}}{{Chevance} et~al.}{2023}]{Chevance2023}
{Chevance} M.,  {Krumholz} M.~R.,  {McLeod} A.~F.,  {Ostriker} E.~C.,  {Rosolowsky} E.~W.,   {Sternberg} A.,  2023, in {Inutsuka} S.,  {Aikawa} Y.,  {Muto} T.,  {Tomida} K.,   {Tamura} M.,  eds,  Astronomical Society of the Pacific Conference Series Vol. 534, Protostars and Planets VII. p.~1 (\mn@eprint {arXiv} {2203.09570}), \mn@doi{10.48550/arXiv.2203.09570}

\bibitem[\protect\citeauthoryear{{Chibisov}}{{Chibisov}}{1969}]{Chibisov1969}
{Chibisov} M.~I.,  1969, Optics and Spectroscopy, \href {https://ui.adsabs.harvard.edu/abs/1969OptSp..27....4C} {27, 4}

\bibitem[\protect\citeauthoryear{{Chiu} \& {Draine}}{{Chiu} \& {Draine}}{1998}]{Chiu1998}
{Chiu} W.~A.,  {Draine} B.~T.,  1998, \mn@doi [arXiv e-prints] {10.48550/arXiv.astro-ph/9803209}, \href {https://ui.adsabs.harvard.edu/abs/1998astro.ph..3209C} {pp astro--ph/9803209}

\bibitem[\protect\citeauthoryear{{Chluba} \& {Sunyaev}}{{Chluba} \& {Sunyaev}}{2008}]{Chluba2008}
{Chluba} J.,  {Sunyaev} R.~A.,  2008, \mn@doi [\aap] {10.1051/0004-6361:20077921}, \href {https://ui.adsabs.harvard.edu/abs/2008A&A...480..629C} {480, 629}

\bibitem[\protect\citeauthoryear{{Chuzhoy} \& {Shapiro}}{{Chuzhoy} \& {Shapiro}}{2006}]{Chuzhoy2006}
{Chuzhoy} L.,  {Shapiro} P.~R.,  2006, \mn@doi [\apj] {10.1086/507670}, \href {https://ui.adsabs.harvard.edu/abs/2006ApJ...651....1C} {651, 1}

\bibitem[\protect\citeauthoryear{{Cox}}{{Cox}}{1985}]{Cox1985}
{Cox} D.~P.,  1985, \mn@doi [\apj] {10.1086/162812}, \href {https://ui.adsabs.harvard.edu/abs/1985ApJ...288..465C} {288, 465}

\bibitem[\protect\citeauthoryear{Dagan}{Dagan}{1993}]{Dagan1993}
Dagan G.,  1993, Transport in Porous Media, 12, 279

\bibitem[\protect\citeauthoryear{{Dale}, {Ercolano}  \& {Bonnell}}{{Dale} et~al.}{2012}]{Dale2012}
{Dale} J.~E.,  {Ercolano} B.,   {Bonnell} I.~A.,  2012, \mn@doi [\mnras] {10.1111/j.1365-2966.2012.21205.x}, \href {https://ui.adsabs.harvard.edu/abs/2012MNRAS.424..377D} {424, 377}

\bibitem[\protect\citeauthoryear{{Dale}, {Ercolano}  \& {Bonnell}}{{Dale} et~al.}{2013}]{Dale2013}
{Dale} J.~E.,  {Ercolano} B.,   {Bonnell} I.~A.,  2013, \mn@doi [\mnras] {10.1093/mnras/sts592}, \href {https://ui.adsabs.harvard.edu/abs/2013MNRAS.430..234D} {430, 234}

\bibitem[\protect\citeauthoryear{{Dalgarno} \& {Williams}}{{Dalgarno} \& {Williams}}{1962}]{Dalgarno1962}
{Dalgarno} A.,  {Williams} D.~A.,  1962, \mn@doi [\mnras] {10.1093/mnras/124.4.313}, \href {https://ui.adsabs.harvard.edu/abs/1962MNRAS.124..313D} {124, 313}

\bibitem[\protect\citeauthoryear{{Davis} \& {Marshak}}{{Davis} \& {Marshak}}{2001}]{Davis2001}
{Davis} A.~B.,  {Marshak} A.,  2001, \mn@doi [Nuclear Science and Engineering] {10.13182/NSE01-A2190}, \href {https://ui.adsabs.harvard.edu/abs/2001NSE...137..251D} {137, 251}

\bibitem[\protect\citeauthoryear{De~Wit}{De~Wit}{1995}]{DeWit1995}
De~Wit A.,  1995, Physics of Fluids, 7, 2553

\bibitem[\protect\citeauthoryear{{Dekel}, {Sarkar}, {Birnboim}, {Mandelker}  \& {Li}}{{Dekel} et~al.}{2023}]{Dekel2023}
{Dekel} A.,  {Sarkar} K.~C.,  {Birnboim} Y.,  {Mandelker} N.,   {Li} Z.,  2023, \mn@doi [\mnras] {10.1093/mnras/stad1557}, \href {https://ui.adsabs.harvard.edu/abs/2023MNRAS.523.3201D} {523, 3201}

\bibitem[\protect\citeauthoryear{{Deng}, {Li}, {Liu}, {Kannan}, {Smith}  \& {Bryan}}{{Deng} et~al.}{2024}]{Deng2024}
{Deng} Y.,  {Li} H.,  {Liu} B.,  {Kannan} R.,  {Smith} A.,   {Bryan} G.~L.,  2024, \mn@doi [\aap] {10.1051/0004-6361/202450699}, \href {https://ui.adsabs.harvard.edu/abs/2024A&A...691A.231D} {691, A231}

\bibitem[\protect\citeauthoryear{{Dennison}, {Turner}  \& {Minter}}{{Dennison} et~al.}{2005}]{Dennison2005}
{Dennison} B.,  {Turner} B.~E.,   {Minter} A.~H.,  2005, \mn@doi [\apj] {10.1086/462402}, \href {https://ui.adsabs.harvard.edu/abs/2005ApJ...633..309D} {633, 309}

\bibitem[\protect\citeauthoryear{{Dijkstra}}{{Dijkstra}}{2014}]{Dijkstra2014}
{Dijkstra} M.,  2014, \mn@doi [\pasa] {10.1017/pasa.2014.33}, \href {https://ui.adsabs.harvard.edu/abs/2014PASA...31...40D} {31, e040}

\bibitem[\protect\citeauthoryear{{Dijkstra} \& {Loeb}}{{Dijkstra} \& {Loeb}}{2008}]{Dijkstra2008}
{Dijkstra} M.,  {Loeb} A.,  2008, \mn@doi [\mnras] {10.1111/j.1365-2966.2008.13920.x}, \href {https://ui.adsabs.harvard.edu/abs/2008MNRAS.391..457D} {391, 457}

\bibitem[\protect\citeauthoryear{{Dijkstra}, {Haiman}  \& {Spaans}}{{Dijkstra} et~al.}{2006}]{Dijkstra2006}
{Dijkstra} M.,  {Haiman} Z.,   {Spaans} M.,  2006, \mn@doi [\apj] {10.1086/506243}, \href {https://ui.adsabs.harvard.edu/abs/2006ApJ...649...14D} {649, 14}

\bibitem[\protect\citeauthoryear{{Dijkstra}, {Sethi}  \& {Loeb}}{{Dijkstra} et~al.}{2016}]{Dijkstra2016}
{Dijkstra} M.,  {Sethi} S.,   {Loeb} A.,  2016, \mn@doi [\apj] {10.3847/0004-637X/820/1/10}, \href {https://ui.adsabs.harvard.edu/abs/2016ApJ...820...10D} {820, 10}

\bibitem[\protect\citeauthoryear{{Doroshkevich} \& {Kolesnik}}{{Doroshkevich} \& {Kolesnik}}{1976}]{Doroshkevich1976}
{Doroshkevich} A.~G.,  {Kolesnik} I.~G.,  1976, \sovast, \href {https://ui.adsabs.harvard.edu/abs/1976SvA....20....4D} {20, 4}

\bibitem[\protect\citeauthoryear{{Dose} \& {Hett}}{{Dose} \& {Hett}}{1974}]{Dose1974}
{Dose} V.,  {Hett} W.,  1974, \mn@doi [Journal of Physics B Atomic Molecular Physics] {10.1088/0022-3700/7/3/023}, \href {https://ui.adsabs.harvard.edu/abs/1974JPhB....7L..79D} {7, L79}

\bibitem[\protect\citeauthoryear{{Dose}, {Hett}, {Olson}, {Pradel}, {Roussel}, {Schlachter}  \& {Spiess}}{{Dose} et~al.}{1975}]{Dose1975}
{Dose} V.,  {Hett} W.,  {Olson} R.~E.,  {Pradel} P.,  {Roussel} F.,  {Schlachter} A.~S.,   {Spiess} G.,  1975, \mn@doi [\pra] {10.1103/PhysRevA.12.1261}, \href {https://ui.adsabs.harvard.edu/abs/1975PhRvA..12.1261D} {12, 1261}

\bibitem[\protect\citeauthoryear{{Draine}}{{Draine}}{2003a}]{Draine2003_AnnRev}
{Draine} B.~T.,  2003a, \mn@doi [\araa] {10.1146/annurev.astro.41.011802.094840}, \href {https://ui.adsabs.harvard.edu/abs/2003ARA&A..41..241D} {41, 241}

\bibitem[\protect\citeauthoryear{{Draine}}{{Draine}}{2003b}]{Draine2003albedo}
{Draine} B.~T.,  2003b, \mn@doi [\apj] {10.1086/379118}, \href {https://ui.adsabs.harvard.edu/abs/2003ApJ...598.1017D} {598, 1017}

\bibitem[\protect\citeauthoryear{{Draine}}{{Draine}}{2011a}]{Draine2011}
{Draine} B.~T.,  2011a, {Physics of the Interstellar and Intergalactic Medium}.
{Princeton University Press}

\bibitem[\protect\citeauthoryear{{Draine}}{{Draine}}{2011b}]{Draine2011_radpressure}
{Draine} B.~T.,  2011b, \mn@doi [\apj] {10.1088/0004-637X/732/2/100}, \href {https://ui.adsabs.harvard.edu/abs/2011ApJ...732..100D} {732, 100}

\bibitem[\protect\citeauthoryear{{Dykaar} \& {Kitanidis}}{{Dykaar} \& {Kitanidis}}{1992}]{Dykaar1992}
{Dykaar} B.~B.,  {Kitanidis} P.~K.,  1992, \mn@doi [Water Resources Research] {10.1029/91WR03083}, \href {https://ui.adsabs.harvard.edu/abs/1992WRR....28.1167D} {28, 1167}

\bibitem[\protect\citeauthoryear{{Fall}, {Krumholz}  \& {Matzner}}{{Fall} et~al.}{2010}]{Fall2010}
{Fall} S.~M.,  {Krumholz} M.~R.,   {Matzner} C.~D.,  2010, \mn@doi [\apjl] {10.1088/2041-8205/710/2/L142}, \href {https://ui.adsabs.harvard.edu/abs/2010ApJ...710L.142F} {710, L142}

\bibitem[\protect\citeauthoryear{{Farias}, {Offner}, {Grudi{\'c}}, {Guszejnov}  \& {Rosen}}{{Farias} et~al.}{2024}]{Farias2024}
{Farias} J.~P.,  {Offner} S. S.~R.,  {Grudi{\'c}} M.~Y.,  {Guszejnov} D.,   {Rosen} A.~L.,  2024, \mn@doi [\mnras] {10.1093/mnras/stad3609}, \href {https://ui.adsabs.harvard.edu/abs/2024MNRAS.527.6732F} {527, 6732}

\bibitem[\protect\citeauthoryear{{Faucher-Gigu{\`e}re}}{{Faucher-Gigu{\`e}re}}{2018}]{Faucher2018_bursty}
{Faucher-Gigu{\`e}re} C.-A.,  2018, \mn@doi [\mnras] {10.1093/mnras/stx2595}, \href {https://ui.adsabs.harvard.edu/abs/2018MNRAS.473.3717F} {473, 3717}

\bibitem[\protect\citeauthoryear{{Faucher-Gigu{\`e}re}, {Quataert}  \& {Hopkins}}{{Faucher-Gigu{\`e}re} et~al.}{2013}]{Faucher2013_feedbackregulated}
{Faucher-Gigu{\`e}re} C.-A.,  {Quataert} E.,   {Hopkins} P.~F.,  2013, \mn@doi [\mnras] {10.1093/mnras/stt866}, \href {https://ui.adsabs.harvard.edu/abs/2013MNRAS.433.1970F} {433, 1970}

\bibitem[\protect\citeauthoryear{{Federrath}}{{Federrath}}{2013}]{Federrath2013}
{Federrath} C.,  2013, \mn@doi [\mnras] {10.1093/mnras/stt1644}, \href {https://ui.adsabs.harvard.edu/abs/2013MNRAS.436.1245F} {436, 1245}

\bibitem[\protect\citeauthoryear{{Federrath}, {Klessen}  \& {Schmidt}}{{Federrath} et~al.}{2008}]{Federrath2008}
{Federrath} C.,  {Klessen} R.~S.,   {Schmidt} W.,  2008, \mn@doi [\apjl] {10.1086/595280}, \href {https://ui.adsabs.harvard.edu/abs/2008ApJ...688L..79F} {688, L79}

\bibitem[\protect\citeauthoryear{{Federrath}, {Roman-Duval}, {Klessen}, {Schmidt}  \& {Mac Low}}{{Federrath} et~al.}{2010}]{Federrath2010}
{Federrath} C.,  {Roman-Duval} J.,  {Klessen} R.~S.,  {Schmidt} W.,   {Mac Low} M.~M.,  2010, \mn@doi [\aap] {10.1051/0004-6361/200912437}, \href {https://ui.adsabs.harvard.edu/abs/2010A&A...512A..81F} {512, A81}

\bibitem[\protect\citeauthoryear{{Field}}{{Field}}{1958}]{Field1958}
{Field} G.~B.,  1958, \mn@doi [Proceedings of the IRE] {10.1109/JRPROC.1958.286741}, \href {https://ui.adsabs.harvard.edu/abs/1958PIRE...46..240F} {46, 240}

\bibitem[\protect\citeauthoryear{{Field}}{{Field}}{1959}]{Field1959}
{Field} G.~B.,  1959, \mn@doi [\apj] {10.1086/146654}, \href {https://ui.adsabs.harvard.edu/abs/1959ApJ...129..551F} {129, 551}

\bibitem[\protect\citeauthoryear{{Flannery}}{{Flannery}}{1971}]{Flannery1971}
{Flannery} M.~R.,  1971, \mn@doi [Physica] {10.1016/0031-8914(71)90099-1}, \href {https://ui.adsabs.harvard.edu/abs/1971Phy....53...28F} {53, 28}

\bibitem[\protect\citeauthoryear{{Ford} \& {Browne}}{{Ford} \& {Browne}}{1973a}]{Ford1973}
{Ford} A.~L.,  {Browne} J.~C.,  1973a, \mn@doi [Atomic Data] {10.1016/S0092-640X(73)80011-7}, \href {https://ui.adsabs.harvard.edu/abs/1973AD......5..305F} {5, 305}

\bibitem[\protect\citeauthoryear{Ford \& Browne}{Ford \& Browne}{1973b}]{Ford1973_2}
Ford A.~L.,  Browne J.~C.,  1973b, \mn@doi [Phys. Rev. A] {10.1103/PhysRevA.7.418}, 7, 418

\bibitem[\protect\citeauthoryear{{Forero-Romero}, {Yepes}, {Gottl{\"o}ber}, {Knollmann}, {Cuesta}  \& {Prada}}{{Forero-Romero} et~al.}{2011}]{ForeroRomero2011}
{Forero-Romero} J.~E.,  {Yepes} G.,  {Gottl{\"o}ber} S.,  {Knollmann} S.~R.,  {Cuesta} A.~J.,   {Prada} F.,  2011, \mn@doi [\mnras] {10.1111/j.1365-2966.2011.18983.x}, \href {https://ui.adsabs.harvard.edu/abs/2011MNRAS.415.3666F} {415, 3666}

\bibitem[\protect\citeauthoryear{{Forrey}, {C{\^o}t{\'e}}, {Dalgarno}, {Jonsell}, {Saenz}  \& {Froelich}}{{Forrey} et~al.}{2000}]{Forrey2000}
{Forrey} R.~C.,  {C{\^o}t{\'e}} R.,  {Dalgarno} A.,  {Jonsell} S.,  {Saenz} A.,   {Froelich} P.,  2000, \mn@doi [\prl] {10.1103/PhysRevLett.85.4245}, \href {https://ui.adsabs.harvard.edu/abs/2000PhRvL..85.4245F} {85, 4245}

\bibitem[\protect\citeauthoryear{Forsman \& Clark}{Forsman \& Clark}{1973}]{Forsman1973}
Forsman E.~N.,  Clark K.~C.,  1973, \mn@doi [Phys. Rev. A] {10.1103/PhysRevA.7.1203}, 7, 1203

\bibitem[\protect\citeauthoryear{{Fu} et~al.,}{{Fu} et~al.}{2023}]{Fu2023_UFD}
{Fu} S.~W.,  et~al., 2023, \mn@doi [\apj] {10.3847/1538-4357/ad0030}, \href {https://ui.adsabs.harvard.edu/abs/2023ApJ...958..167F} {958, 167}

\bibitem[\protect\citeauthoryear{{Fukushima} \& {Yajima}}{{Fukushima} \& {Yajima}}{2021}]{Fukushima2021}
{Fukushima} H.,  {Yajima} H.,  2021, \mn@doi [\mnras] {10.1093/mnras/stab2099}, \href {https://ui.adsabs.harvard.edu/abs/2021MNRAS.506.5512F} {506, 5512}

\bibitem[\protect\citeauthoryear{{Fukushima}, {Yajima}  \& {Omukai}}{{Fukushima} et~al.}{2018}]{Fukushima2018}
{Fukushima} H.,  {Yajima} H.,   {Omukai} K.,  2018, \mn@doi [\mnras] {10.1093/mnras/sty799}, \href {https://ui.adsabs.harvard.edu/abs/2018MNRAS.477.1071F} {477, 1071}

\bibitem[\protect\citeauthoryear{{Furlanetto} \& {Pritchard}}{{Furlanetto} \& {Pritchard}}{2006}]{Furlanetto2006_FokkerPlanck}
{Furlanetto} S.~R.,  {Pritchard} J.~R.,  2006, \mn@doi [\mnras] {10.1111/j.1365-2966.2006.10899.x}, \href {https://ui.adsabs.harvard.edu/abs/2006MNRAS.372.1093F} {372, 1093}

\bibitem[\protect\citeauthoryear{{Gao}, {Yoshida}, {Abel}, {Frenk}, {Jenkins}  \& {Springel}}{{Gao} et~al.}{2007}]{Gao2007}
{Gao} L.,  {Yoshida} N.,  {Abel} T.,  {Frenk} C.~S.,  {Jenkins} A.,   {Springel} V.,  2007, \mn@doi [\mnras] {10.1111/j.1365-2966.2007.11814.x}, \href {https://ui.adsabs.harvard.edu/abs/2007MNRAS.378..449G} {378, 449}

\bibitem[\protect\citeauthoryear{{Garavito-Camargo}, {Forero-Romero}  \& {Dijkstra}}{{Garavito-Camargo} et~al.}{2014}]{Garavito2014}
{Garavito-Camargo} J.~N.,  {Forero-Romero} J.~E.,   {Dijkstra} M.,  2014, \mn@doi [\apj] {10.1088/0004-637X/795/2/120}, \href {https://ui.adsabs.harvard.edu/abs/2014ApJ...795..120G} {795, 120}

\bibitem[\protect\citeauthoryear{{Garcia}, {Ricotti}, {Sugimura}  \& {Park}}{{Garcia} et~al.}{2023}]{Garcia2023}
{Garcia} F. A.~B.,  {Ricotti} M.,  {Sugimura} K.,   {Park} J.,  2023, \mn@doi [\mnras] {10.1093/mnras/stad1092}, \href {https://ui.adsabs.harvard.edu/abs/2023MNRAS.522.2495G} {522, 2495}

\bibitem[\protect\citeauthoryear{{Ge} \& {Wise}}{{Ge} \& {Wise}}{2017}]{Ge2017}
{Ge} Q.,  {Wise} J.~H.,  2017, \mn@doi [\mnras] {10.1093/mnras/stx2074}, \href {https://ui.adsabs.harvard.edu/abs/2017MNRAS.472.2773G} {472, 2773}

\bibitem[\protect\citeauthoryear{{George}}{{George}}{1973}]{George1973}
{George} D.,  1973, in {Remy-Battiau} L.,  {Vreux} J.~M.,   {Menzel} D.~H.,  eds,  Liege International Astrophysical Colloquia Vol. 18, Liege International Astrophysical Colloquia. pp 431--436

\bibitem[\protect\citeauthoryear{{Gerrard} et~al.,}{{Gerrard} et~al.}{2023}]{Gerrard2023}
{Gerrard} I.~A.,  et~al., 2023, \mn@doi [\mnras] {10.1093/mnras/stad2718}, \href {https://ui.adsabs.harvard.edu/abs/2023MNRAS.526..982G} {526, 982}

\bibitem[\protect\citeauthoryear{{Gieles} \& {Portegies Zwart}}{{Gieles} \& {Portegies Zwart}}{2011}]{Gieles2011}
{Gieles} M.,  {Portegies Zwart} S.~F.,  2011, \mn@doi [\mnras] {10.1111/j.1745-3933.2010.00967.x}, \href {https://ui.adsabs.harvard.edu/abs/2011MNRAS.410L...6G} {410, L6}

\bibitem[\protect\citeauthoryear{{Gordon}}{{Gordon}}{2004}]{Gordon2004}
{Gordon} K.~D.,  2004, in {Witt} A.~N.,  {Clayton} G.~C.,   {Draine} B.~T.,  eds,  Astronomical Society of the Pacific Conference Series Vol. 309, Astrophysics of Dust. p.~77 (\mn@eprint {arXiv} {astro-ph/0309709}), \mn@doi{10.48550/arXiv.astro-ph/0309709}

\bibitem[\protect\citeauthoryear{{Grachev}}{{Grachev}}{1989}]{Grachev1989}
{Grachev} S.~I.,  1989, Astrofizika, \href {https://ui.adsabs.harvard.edu/abs/1989Afz....30..347G} {30, 347}

\bibitem[\protect\citeauthoryear{{Greif}, {Springel}, {White}, {Glover}, {Clark}, {Smith}, {Klessen}  \& {Bromm}}{{Greif} et~al.}{2011}]{Greif2011_movingmesh}
{Greif} T.~H.,  {Springel} V.,  {White} S. D.~M.,  {Glover} S. C.~O.,  {Clark} P.~C.,  {Smith} R.~J.,  {Klessen} R.~S.,   {Bromm} V.,  2011, \mn@doi [\apj] {10.1088/0004-637X/737/2/75}, \href {https://ui.adsabs.harvard.edu/abs/2011ApJ...737...75G} {737, 75}

\bibitem[\protect\citeauthoryear{Grosser, Menzel  \& Belyaev}{Grosser et~al.}{1999}]{grosser1999}
Grosser J.,  Menzel T.,   Belyaev A.~K.,  1999, \mn@doi [Phys. Rev. A] {10.1103/PhysRevA.59.1309}, 59, 1309

\bibitem[\protect\citeauthoryear{{Grudi{\'c}}, {Hopkins}, {Faucher-Gigu{\`e}re}, {Quataert}, {Murray}  \& {Kere{\v{s}}}}{{Grudi{\'c}} et~al.}{2018}]{Grudic2018}
{Grudi{\'c}} M.~Y.,  {Hopkins} P.~F.,  {Faucher-Gigu{\`e}re} C.-A.,  {Quataert} E.,  {Murray} N.,   {Kere{\v{s}}} D.,  2018, \mn@doi [\mnras] {10.1093/mnras/sty035}, \href {https://ui.adsabs.harvard.edu/abs/2018MNRAS.475.3511G} {475, 3511}

\bibitem[\protect\citeauthoryear{{Grudi{\'c}}, {Hopkins}, {Lee}, {Murray}, {Faucher-Gigu{\`e}re}  \& {Johnson}}{{Grudi{\'c}} et~al.}{2019}]{Grudic2019_SFE}
{Grudi{\'c}} M.~Y.,  {Hopkins} P.~F.,  {Lee} E.~J.,  {Murray} N.,  {Faucher-Gigu{\`e}re} C.-A.,   {Johnson} L.~C.,  2019, \mn@doi [\mnras] {10.1093/mnras/stz1758}, \href {https://ui.adsabs.harvard.edu/abs/2019MNRAS.488.1501G} {488, 1501}

\bibitem[\protect\citeauthoryear{{Grudi{\'c}}, {Guszejnov}, {Offner}, {Rosen}, {Raju}, {Faucher-Gigu{\`e}re}  \& {Hopkins}}{{Grudi{\'c}} et~al.}{2022}]{Grudic2022_Starforge}
{Grudi{\'c}} M.~Y.,  {Guszejnov} D.,  {Offner} S. S.~R.,  {Rosen} A.~L.,  {Raju} A.~N.,  {Faucher-Gigu{\`e}re} C.-A.,   {Hopkins} P.~F.,  2022, \mn@doi [\mnras] {10.1093/mnras/stac526}, \href {https://ui.adsabs.harvard.edu/abs/2022MNRAS.512..216G} {512, 216}

\bibitem[\protect\citeauthoryear{{Guia}, {Pacucci}  \& {Kocevski}}{{Guia} et~al.}{2024}]{Audric2024}
{Guia} C.~A.,  {Pacucci} F.,   {Kocevski} D.~D.,  2024, \mn@doi [Research Notes of the American Astronomical Society] {10.3847/2515-5172/ad7262}, \href {https://ui.adsabs.harvard.edu/abs/2024RNAAS...8..207G} {8, 207}

\bibitem[\protect\citeauthoryear{{Gurian}, {Jeong}  \& {Liu}}{{Gurian} et~al.}{2024}]{Gurian2024}
{Gurian} J.,  {Jeong} D.,   {Liu} B.,  2024, \mn@doi [\apj] {10.3847/1538-4357/ad1e5b}, \href {https://ui.adsabs.harvard.edu/abs/2024ApJ...963...33G} {963, 33}

\bibitem[\protect\citeauthoryear{{Gutcke}, {Pfrommer}, {Bryan}, {Pakmor}, {Springel}  \& {Naab}}{{Gutcke} et~al.}{2022}]{Gutcke2022}
{Gutcke} T.~A.,  {Pfrommer} C.,  {Bryan} G.~L.,  {Pakmor} R.,  {Springel} V.,   {Naab} T.,  2022, \mn@doi [\apj] {10.3847/1538-4357/aca1b4}, \href {https://ui.adsabs.harvard.edu/abs/2022ApJ...941..120G} {941, 120}

\bibitem[\protect\citeauthoryear{{Guzm{\'a}n}, {Badnell}, {Williams}, {van Hoof}, {Chatzikos}  \& {Ferland}}{{Guzm{\'a}n} et~al.}{2016}]{Guzman2016}
{Guzm{\'a}n} F.,  {Badnell} N.~R.,  {Williams} R.~J.~R.,  {van Hoof} P.~A.~M.,  {Chatzikos} M.,   {Ferland} G.~J.,  2016, \mn@doi [\mnras] {10.1093/mnras/stw893}, \href {https://ui.adsabs.harvard.edu/abs/2016MNRAS.459.3498G} {459, 3498}

\bibitem[\protect\citeauthoryear{{Guzm{\'a}n}, {Badnell}, {Williams}, {van Hoof}, {Chatzikos}  \& {Ferland}}{{Guzm{\'a}n} et~al.}{2017a}]{Guzman2017_PSM}
{Guzm{\'a}n} F.,  {Badnell} N.~R.,  {Williams} R.~J.~R.,  {van Hoof} P.~A.~M.,  {Chatzikos} M.,   {Ferland} G.~J.,  2017a, \mn@doi [\mnras] {10.1093/mnras/stw2304}, \href {https://ui.adsabs.harvard.edu/abs/2017MNRAS.464..312G} {464, 312}

\bibitem[\protect\citeauthoryear{{Guzm{\'a}n}, {Badnell}, {Chatzikos}, {van Hoof}, {Williams}  \& {Ferland}}{{Guzm{\'a}n} et~al.}{2017b}]{Guzman2017}
{Guzm{\'a}n} F.,  {Badnell} N.~R.,  {Chatzikos} M.,  {van Hoof} P.~A.~M.,  {Williams} R.~J.~R.,   {Ferland} G.~J.,  2017b, \mn@doi [\mnras] {10.1093/mnras/stx269}, \href {https://ui.adsabs.harvard.edu/abs/2017MNRAS.467.3944G} {467, 3944}

\bibitem[\protect\citeauthoryear{{Han}, {Kimm}, {Katz}, {Devriendt}  \& {Slyz}}{{Han} et~al.}{2022}]{Han2022}
{Han} D.,  {Kimm} T.,  {Katz} H.,  {Devriendt} J.,   {Slyz} A.,  2022, \mn@doi [\apj] {10.3847/1538-4357/ac7ff3}, \href {https://ui.adsabs.harvard.edu/abs/2022ApJ...935...53H} {935, 53}

\bibitem[\protect\citeauthoryear{{Hanasoge}, {Gizon}  \& {Bal}}{{Hanasoge} et~al.}{2013}]{Hanasoge2013}
{Hanasoge} S.~M.,  {Gizon} L.,   {Bal} G.,  2013, \mn@doi [\apj] {10.1088/0004-637X/773/2/101}, \href {https://ui.adsabs.harvard.edu/abs/2013ApJ...773..101H} {773, 101}

\bibitem[\protect\citeauthoryear{{Harrington}}{{Harrington}}{1973}]{Harrington1973}
{Harrington} J.~P.,  1973, \mn@doi [\mnras] {10.1093/mnras/162.1.43}, \href {https://ui.adsabs.harvard.edu/abs/1973MNRAS.162...43H} {162, 43}

\bibitem[\protect\citeauthoryear{{Harris}}{{Harris}}{2010}]{Harris2010}
{Harris} W.~E.,  2010, \mn@doi [arXiv e-prints] {10.48550/arXiv.1012.3224}, \href {https://ui.adsabs.harvard.edu/abs/2010arXiv1012.3224H} {p. arXiv:1012.3224}

\bibitem[\protect\citeauthoryear{{Harris} et~al.,}{{Harris} et~al.}{2020}]{Harris2020_Numpy}
{Harris} C.~R.,  et~al., 2020, \mn@doi [\nat] {10.1038/s41586-020-2649-2}, \href {https://ui.adsabs.harvard.edu/abs/2020Natur.585..357H} {585, 357}

\bibitem[\protect\citeauthoryear{{Henney} \& {Arthur}}{{Henney} \& {Arthur}}{1998}]{Henney1998}
{Henney} W.~J.,  {Arthur} S.~J.,  1998, \mn@doi [\aj] {10.1086/300433}, \href {https://ui.adsabs.harvard.edu/abs/1998AJ....116..322H} {116, 322}

\bibitem[\protect\citeauthoryear{{Hensley} \& {Draine}}{{Hensley} \& {Draine}}{2023}]{Hensley2023}
{Hensley} B.~S.,  {Draine} B.~T.,  2023, \mn@doi [\apj] {10.3847/1538-4357/acc4c2}, \href {https://ui.adsabs.harvard.edu/abs/2023ApJ...948...55H} {948, 55}

\bibitem[\protect\citeauthoryear{{Henyey} \& {Greenstein}}{{Henyey} \& {Greenstein}}{1941}]{Henyey1941}
{Henyey} L.~G.,  {Greenstein} J.~L.,  1941, \mn@doi [\apj] {10.1086/144246}, \href {https://ui.adsabs.harvard.edu/abs/1941ApJ....93...70H} {93, 70}

\bibitem[\protect\citeauthoryear{{Heyer} \& {Dame}}{{Heyer} \& {Dame}}{2015}]{Heyer2015}
{Heyer} M.,  {Dame} T.~M.,  2015, \mn@doi [\araa] {10.1146/annurev-astro-082214-122324}, \href {https://ui.adsabs.harvard.edu/abs/2015ARA&A..53..583H} {53, 583}

\bibitem[\protect\citeauthoryear{{Higgins} \& {Meiksin}}{{Higgins} \& {Meiksin}}{2012}]{Higgins2012}
{Higgins} J.,  {Meiksin} A.,  2012, \mn@doi [\mnras] {10.1111/j.1365-2966.2012.21917.x}, \href {https://ui.adsabs.harvard.edu/abs/2012MNRAS.426.2380H} {426, 2380}

\bibitem[\protect\citeauthoryear{{Hillier} \& {Dessart}}{{Hillier} \& {Dessart}}{2012}]{Hillier2012}
{Hillier} D.~J.,  {Dessart} L.,  2012, \mn@doi [\mnras] {10.1111/j.1365-2966.2012.21192.x}, \href {https://ui.adsabs.harvard.edu/abs/2012MNRAS.424..252H} {424, 252}

\bibitem[\protect\citeauthoryear{{Hills}}{{Hills}}{1980}]{Hills1980}
{Hills} J.~G.,  1980, \mn@doi [\apj] {10.1086/157703}, \href {https://ui.adsabs.harvard.edu/abs/1980ApJ...235..986H} {235, 986}

\bibitem[\protect\citeauthoryear{{Hirano} \& {Yoshida}}{{Hirano} \& {Yoshida}}{2024}]{Hirano2024}
{Hirano} S.,  {Yoshida} N.,  2024, \mn@doi [\apj] {10.3847/1538-4357/ad22e0}, \href {https://ui.adsabs.harvard.edu/abs/2024ApJ...963....2H} {963, 2}

\bibitem[\protect\citeauthoryear{{Hirano}, {Zhu}, {Yoshida}, {Spergel}  \& {Yorke}}{{Hirano} et~al.}{2015}]{Hirano2015_bluetilt}
{Hirano} S.,  {Zhu} N.,  {Yoshida} N.,  {Spergel} D.,   {Yorke} H.~W.,  2015, \mn@doi [\apj] {10.1088/0004-637X/814/1/18}, \href {https://ui.adsabs.harvard.edu/abs/2015ApJ...814...18H} {814, 18}

\bibitem[\protect\citeauthoryear{{Hirashita} \& {Inoue}}{{Hirashita} \& {Inoue}}{2019}]{Hirashita2019_radiationpressuredust}
{Hirashita} H.,  {Inoue} A.~K.,  2019, \mn@doi [\mnras] {10.1093/mnras/stz1348}, \href {https://ui.adsabs.harvard.edu/abs/2019MNRAS.487..961H} {487, 961}

\bibitem[\protect\citeauthoryear{{Hirata}}{{Hirata}}{2006}]{Hirata2006}
{Hirata} C.~M.,  2006, \mn@doi [\mnras] {10.1111/j.1365-2966.2005.09949.x}, \href {https://ui.adsabs.harvard.edu/abs/2006MNRAS.367..259H} {367, 259}

\bibitem[\protect\citeauthoryear{{Hollenbach} \& {Tielens}}{{Hollenbach} \& {Tielens}}{1999}]{Hollenbach1999}
{Hollenbach} D.~J.,  {Tielens} A.~G.~G.~M.,  1999, \mn@doi [Reviews of Modern Physics] {10.1103/RevModPhys.71.173}, \href {https://ui.adsabs.harvard.edu/abs/1999RvMP...71..173H} {71, 173}

\bibitem[\protect\citeauthoryear{{Hollenbach}, {Kaufman}, {Bergin}  \& {Melnick}}{{Hollenbach} et~al.}{2009}]{Hollenbach2009}
{Hollenbach} D.,  {Kaufman} M.~J.,  {Bergin} E.~A.,   {Melnick} G.~J.,  2009, \mn@doi [\apj] {10.1088/0004-637X/690/2/1497}, \href {https://ui.adsabs.harvard.edu/abs/2009ApJ...690.1497H} {690, 1497}

\bibitem[\protect\citeauthoryear{{Holmes}}{{Holmes}}{2013}]{Holmes2013_Perturbation}
{Holmes} M.,  2013, Introduction to Perturbation Methods.
Texts in Applied Mathematics, Springer New York

\bibitem[\protect\citeauthoryear{{Hopkins}}{{Hopkins}}{2012}]{Hopkins2012_GMCs}
{Hopkins} P.~F.,  2012, \mn@doi [\mnras] {10.1111/j.1365-2966.2012.20730.x}, \href {https://ui.adsabs.harvard.edu/abs/2012MNRAS.423.2016H} {423, 2016}

\bibitem[\protect\citeauthoryear{{Hopkins} et~al.,}{{Hopkins} et~al.}{2018}]{Hopkins2018_FIRE2}
{Hopkins} P.~F.,  et~al., 2018, \mn@doi [\mnras] {10.1093/mnras/sty1690}, \href {https://ui.adsabs.harvard.edu/abs/2018MNRAS.480..800H} {480, 800}

\bibitem[\protect\citeauthoryear{{Hopkins}, {Grudi{\'c}}, {Wetzel}, {Kere{\v{s}}}, {Faucher-Gigu{\`e}re}, {Ma}, {Murray}  \& {Butcher}}{{Hopkins} et~al.}{2020}]{Hopkins2020}
{Hopkins} P.~F.,  {Grudi{\'c}} M.~Y.,  {Wetzel} A.,  {Kere{\v{s}}} D.,  {Faucher-Gigu{\`e}re} C.-A.,  {Ma} X.,  {Murray} N.,   {Butcher} N.,  2020, \mn@doi [\mnras] {10.1093/mnras/stz3129}, \href {https://ui.adsabs.harvard.edu/abs/2020MNRAS.491.3702H} {491, 3702}

\bibitem[\protect\citeauthoryear{{Hopkins} et~al.,}{{Hopkins} et~al.}{2023}]{Hopkins2023}
{Hopkins} P.~F.,  et~al., 2023, \mn@doi [\mnras] {10.1093/mnras/stac3489}, \href {https://ui.adsabs.harvard.edu/abs/2023MNRAS.519.3154H} {519, 3154}

\bibitem[\protect\citeauthoryear{H\"ornquist, Hedvall, Larson  \& Orel}{H\"ornquist et~al.}{2022}]{hornquist2022}
H\"ornquist J.,  Hedvall P.,  Larson A.,   Orel A.~E.,  2022, \mn@doi [Phys. Rev. A] {10.1103/PhysRevA.106.062821}, 106, 062821

\bibitem[\protect\citeauthoryear{H\"ornquist, Hedvall, Orel  \& Larson}{H\"ornquist et~al.}{2023}]{hornquist2023}
H\"ornquist J.,  Hedvall P.,  Orel A.~E.,   Larson A.,  2023, \mn@doi [Phys. Rev. A] {10.1103/PhysRevA.108.052811}, 108, 052811

\bibitem[\protect\citeauthoryear{{Hummer}}{{Hummer}}{1962}]{Hummer1962}
{Hummer} D.~G.,  1962, \mn@doi [\mnras] {10.1093/mnras/125.1.21}, \href {https://ui.adsabs.harvard.edu/abs/1962MNRAS.125...21H} {125, 21}

\bibitem[\protect\citeauthoryear{{Hunter}}{{Hunter}}{2007}]{Hunter2007_matplotlib}
{Hunter} J.~D.,  2007, \mn@doi [Computing in Science and Engineering] {10.1109/MCSE.2007.55}, \href {https://ui.adsabs.harvard.edu/abs/2007CSE.....9...90H} {9, 90}

\bibitem[\protect\citeauthoryear{{Ito} \& {Omukai}}{{Ito} \& {Omukai}}{2024}]{Ito2024}
{Ito} M.,  {Omukai} K.,  2024, \mn@doi [\pasj] {10.1093/pasj/psae054}, \href {https://ui.adsabs.harvard.edu/abs/2024PASJ..tmp...63I} {}

\bibitem[\protect\citeauthoryear{Janev, Langer, Post  \& Evans}{Janev et~al.}{1987}]{Janev1987}
Janev R.~K.,  Langer W.~D.,  Post D.~E.,   Evans K.,  1987, Elementary Processes in Hydrogen-Helium Plasmas: Cross Sections and Reaction Rate Coefficients.
Springer Berlin, Heidelberg, \mn@doi{https://doi.org/10.1007/978-3-642-71935-6}

\bibitem[\protect\citeauthoryear{{Jankovic}, {Maghrebi}, {Fiori}  \& {Dagan}}{{Jankovic} et~al.}{2017}]{Jankovic2017}
{Jankovic} I.,  {Maghrebi} M.,  {Fiori} A.,   {Dagan} G.,  2017, \mn@doi [Advances in Water Resources] {10.1016/j.advwatres.2016.10.024}, \href {https://ui.adsabs.harvard.edu/abs/2017AdWR..100..199J} {100, 199}

\bibitem[\protect\citeauthoryear{{Jaura}, {Glover}, {Wollenberg}, {Klessen}, {Geen}  \& {Haemmerl{\'e}}}{{Jaura} et~al.}{2022}]{Jaura2022}
{Jaura} O.,  {Glover} S. C.~O.,  {Wollenberg} K. M.~J.,  {Klessen} R.~S.,  {Geen} S.,   {Haemmerl{\'e}} L.,  2022, \mn@doi [\mnras] {10.1093/mnras/stac487}, \href {https://ui.adsabs.harvard.edu/abs/2022MNRAS.512..116J} {512, 116}

\bibitem[\protect\citeauthoryear{{Jeffreson}, {Krumholz}, {Fujimoto}, {Armillotta}, {Keller}, {Chevance}  \& {Kruijssen}}{{Jeffreson} et~al.}{2021}]{Jeffreson2021}
{Jeffreson} S. M.~R.,  {Krumholz} M.~R.,  {Fujimoto} Y.,  {Armillotta} L.,  {Keller} B.~W.,  {Chevance} M.,   {Kruijssen} J.~M.~D.,  2021, \mn@doi [\mnras] {10.1093/mnras/stab1536}, \href {https://ui.adsabs.harvard.edu/abs/2021MNRAS.505.3470J} {505, 3470}

\bibitem[\protect\citeauthoryear{{Jeon}, {Besla}  \& {Bromm}}{{Jeon} et~al.}{2017}]{Jeon2017}
{Jeon} M.,  {Besla} G.,   {Bromm} V.,  2017, \mn@doi [\apj] {10.3847/1538-4357/aa8c80}, \href {https://ui.adsabs.harvard.edu/abs/2017ApJ...848...85J} {848, 85}

\bibitem[\protect\citeauthoryear{{Johnson}}{{Johnson}}{1973}]{johnson1973}
{Johnson} B.~R.,  1973, \mn@doi [Journal of Computational Physics] {10.1016/0021-9991(73)90049-1}, \href {https://ui.adsabs.harvard.edu/abs/1973JCoPh..13..445J} {13, 445}

\bibitem[\protect\citeauthoryear{{Johnson} \& {Dijkstra}}{{Johnson} \& {Dijkstra}}{2017}]{Johnson2017}
{Johnson} J.~L.,  {Dijkstra} M.,  2017, \mn@doi [\aap] {10.1051/0004-6361/201630010}, \href {https://ui.adsabs.harvard.edu/abs/2017A&A...601A.138J} {601, A138}

\bibitem[\protect\citeauthoryear{{Joseph}, {Wiscombe}  \& {Weinman}}{{Joseph} et~al.}{1976}]{Joseph1976}
{Joseph} J.~H.,  {Wiscombe} W.~J.,   {Weinman} J.~A.,  1976, \mn@doi [Journal of Atmospheric Sciences] {10.1175/1520-0469(1976)033<2452:TDEAFR>2.0.CO;2}, \href {https://ui.adsabs.harvard.edu/abs/1976JAtS...33.2452J} {33, 2452}

\bibitem[\protect\citeauthoryear{{Kainulainen} \& {Tan}}{{Kainulainen} \& {Tan}}{2013}]{Kainulainen2013}
{Kainulainen} J.,  {Tan} J.~C.,  2013, \mn@doi [\aap] {10.1051/0004-6361/201219526}, \href {https://ui.adsabs.harvard.edu/abs/2013A&A...549A..53K} {549, A53}

\bibitem[\protect\citeauthoryear{{Kakiichi} \& {Gronke}}{{Kakiichi} \& {Gronke}}{2021}]{Kakiichi2021}
{Kakiichi} K.,  {Gronke} M.,  2021, \mn@doi [\apj] {10.3847/1538-4357/abc2d9}, \href {https://ui.adsabs.harvard.edu/abs/2021ApJ...908...30K} {908, 30}

\bibitem[\protect\citeauthoryear{{Kang}, {Kimm}, {Han}, {Katz}, {Devriendt}, {Slyz}  \& {Teyssier}}{{Kang} et~al.}{2024}]{Kang2024}
{Kang} C.,  {Kimm} T.,  {Han} D.,  {Katz} H.,  {Devriendt} J.,  {Slyz} A.,   {Teyssier} R.,  2024, \mn@doi [arXiv e-prints] {10.48550/arXiv.2407.12090}, \href {https://ui.adsabs.harvard.edu/abs/2024arXiv240712090K} {p. arXiv:2407.12090}

\bibitem[\protect\citeauthoryear{{Kapoor} et~al.,}{{Kapoor} et~al.}{2023}]{Kapoor2023}
{Kapoor} A.~U.,  et~al., 2023, \mn@doi [\mnras] {10.1093/mnras/stad2977}, \href {https://ui.adsabs.harvard.edu/abs/2023MNRAS.526.3871K} {526, 3871}

\bibitem[\protect\citeauthoryear{{Kim}, {Kim}  \& {Ostriker}}{{Kim} et~al.}{2018}]{Kim2018_RadPressure}
{Kim} J.-G.,  {Kim} W.-T.,   {Ostriker} E.~C.,  2018, \mn@doi [\apj] {10.3847/1538-4357/aabe27}, \href {https://ui.adsabs.harvard.edu/abs/2018ApJ...859...68K} {859, 68}

\bibitem[\protect\citeauthoryear{{Kimm}, {Cen}, {Rosdahl}  \& {Yi}}{{Kimm} et~al.}{2016}]{Kimm2016_GC}
{Kimm} T.,  {Cen} R.,  {Rosdahl} J.,   {Yi} S.~K.,  2016, \mn@doi [\apj] {10.3847/0004-637X/823/1/52}, \href {https://ui.adsabs.harvard.edu/abs/2016ApJ...823...52K} {823, 52}

\bibitem[\protect\citeauthoryear{{Kimm}, {Katz}, {Haehnelt}, {Rosdahl}, {Devriendt}  \& {Slyz}}{{Kimm} et~al.}{2017}]{Kimm2017minihaloes}
{Kimm} T.,  {Katz} H.,  {Haehnelt} M.,  {Rosdahl} J.,  {Devriendt} J.,   {Slyz} A.,  2017, \mn@doi [\mnras] {10.1093/mnras/stx052}, \href {https://ui.adsabs.harvard.edu/abs/2017MNRAS.466.4826K} {466, 4826}

\bibitem[\protect\citeauthoryear{{Kimm}, {Haehnelt}, {Blaizot}, {Katz}, {Michel-Dansac}, {Garel}, {Rosdahl}  \& {Teyssier}}{{Kimm} et~al.}{2018}]{Kimm2018}
{Kimm} T.,  {Haehnelt} M.,  {Blaizot} J.,  {Katz} H.,  {Michel-Dansac} L.,  {Garel} T.,  {Rosdahl} J.,   {Teyssier} R.,  2018, \mn@doi [\mnras] {10.1093/mnras/sty126}, \href {https://ui.adsabs.harvard.edu/abs/2018MNRAS.475.4617K} {475, 4617}

\bibitem[\protect\citeauthoryear{{Kimm}, {Blaizot}, {Garel}, {Michel-Dansac}, {Katz}, {Rosdahl}, {Verhamme}  \& {Haehnelt}}{{Kimm} et~al.}{2019}]{Kimm2019}
{Kimm} T.,  {Blaizot} J.,  {Garel} T.,  {Michel-Dansac} L.,  {Katz} H.,  {Rosdahl} J.,  {Verhamme} A.,   {Haehnelt} M.,  2019, \mn@doi [\mnras] {10.1093/mnras/stz989}, \href {https://ui.adsabs.harvard.edu/abs/2019MNRAS.486.2215K} {486, 2215}

\bibitem[\protect\citeauthoryear{{Kimm}, {Bieri}, {Geen}, {Rosdahl}, {Blaizot}, {Michel-Dansac}  \& {Garel}}{{Kimm} et~al.}{2022}]{Kimm2022}
{Kimm} T.,  {Bieri} R.,  {Geen} S.,  {Rosdahl} J.,  {Blaizot} J.,  {Michel-Dansac} L.,   {Garel} T.,  2022, \mn@doi [\apjs] {10.3847/1538-4365/ac426d}, \href {https://ui.adsabs.harvard.edu/abs/2022ApJS..259...21K} {259, 21}

\bibitem[\protect\citeauthoryear{{Klessen} \& {Glover}}{{Klessen} \& {Glover}}{2023}]{Klessen2023}
{Klessen} R.~S.,  {Glover} S. C.~O.,  2023, \mn@doi [\araa] {10.1146/annurev-astro-071221-053453}, \href {https://ui.adsabs.harvard.edu/abs/2023ARA&A..61...65K} {61, 65}

\bibitem[\protect\citeauthoryear{{Kokubo}}{{Kokubo}}{2024}]{Kokubo2024}
{Kokubo} M.,  2024, \mn@doi [\mnras] {10.1093/mnras/stae515}, \href {https://ui.adsabs.harvard.edu/abs/2024MNRAS.529.2131K} {529, 2131}

\bibitem[\protect\citeauthoryear{Kozlov}{Kozlov}{1993}]{Kozlow1993_centrallimit}
Kozlov S.,  1993, in Flow in Porous Media: Proceedings of the Oberwolfach Conference, June 21--27, 1992. pp 117--127

\bibitem[\protect\citeauthoryear{Kramida, {Yu.~Ralchenko}, Reader  \& {and NIST ASD Team}}{Kramida et~al.}{2023}]{NIST_ASD}
Kramida A.,  {Yu.~Ralchenko} Reader J.,   {and NIST ASD Team} 2023, {NIST Atomic Spectra Database (ver. 5.11), [Online]. Available: {\tt{https://physics.nist.gov/asd}} [2024, May 4]. National Institute of Standards and Technology, Gaithersburg, MD.}

\bibitem[\protect\citeauthoryear{{Kroupa}}{{Kroupa}}{2001}]{Kroupa2001}
{Kroupa} P.,  2001, \mn@doi [\mnras] {10.1046/j.1365-8711.2001.04022.x}, \href {https://ui.adsabs.harvard.edu/abs/2001MNRAS.322..231K} {322, 231}

\bibitem[\protect\citeauthoryear{{Krti{\v{c}}ka} \& {Kub{\'a}t}}{{Krti{\v{c}}ka} \& {Kub{\'a}t}}{2006}]{Krtivcka2006_winds}
{Krti{\v{c}}ka} J.,  {Kub{\'a}t} J.,  2006, \mn@doi [\aap] {10.1051/0004-6361:20053289}, \href {https://ui.adsabs.harvard.edu/abs/2006A&A...446.1039K} {446, 1039}

\bibitem[\protect\citeauthoryear{{Kruijssen} et~al.,}{{Kruijssen} et~al.}{2019}]{Kruijssen2019}
{Kruijssen} J.~M.~D.,  et~al., 2019, \mn@doi [\nat] {10.1038/s41586-019-1194-3}, \href {https://ui.adsabs.harvard.edu/abs/2019Natur.569..519K} {569, 519}

\bibitem[\protect\citeauthoryear{{Krumholz}}{{Krumholz}}{2014}]{Krumholz2014review}
{Krumholz} M.~R.,  2014, \mn@doi [\physrep] {10.1016/j.physrep.2014.02.001}, \href {https://ui.adsabs.harvard.edu/abs/2014PhR...539...49K} {539, 49}

\bibitem[\protect\citeauthoryear{Krumholz}{Krumholz}{2018}]{krumholz2018resolution}
Krumholz M.~R.,  2018, Monthly Notices of the Royal Astronomical Society, 480, 3468

\bibitem[\protect\citeauthoryear{{Krumholz} \& {Matzner}}{{Krumholz} \& {Matzner}}{2009}]{Krumholz2009}
{Krumholz} M.~R.,  {Matzner} C.~D.,  2009, \mn@doi [\apj] {10.1088/0004-637X/703/2/1352}, \href {https://ui.adsabs.harvard.edu/abs/2009ApJ...703.1352K} {703, 1352}

\bibitem[\protect\citeauthoryear{{Krumholz} \& {Thompson}}{{Krumholz} \& {Thompson}}{2012}]{Krumholz2012}
{Krumholz} M.~R.,  {Thompson} T.~A.,  2012, \mn@doi [\apj] {10.1088/0004-637X/760/2/155}, \href {https://ui.adsabs.harvard.edu/abs/2012ApJ...760..155K} {760, 155}

\bibitem[\protect\citeauthoryear{{Krumholz} \& {Thompson}}{{Krumholz} \& {Thompson}}{2013}]{Krumholz2013}
{Krumholz} M.~R.,  {Thompson} T.~A.,  2013, \mn@doi [\mnras] {10.1093/mnras/stt1174}, \href {https://ui.adsabs.harvard.edu/abs/2013MNRAS.434.2329K} {434, 2329}

\bibitem[\protect\citeauthoryear{{Labb{\'e}} et~al.,}{{Labb{\'e}} et~al.}{2023}]{Labbe2023}
{Labb{\'e}} I.,  et~al., 2023, \mn@doi [\nat] {10.1038/s41586-023-05786-2}, \href {https://ui.adsabs.harvard.edu/abs/2023Natur.616..266L} {616, 266}

\bibitem[\protect\citeauthoryear{{Lah{\'e}n}, {Naab}, {Johansson}, {Elmegreen}, {Hu}, {Walch}, {Steinwandel}  \& {Moster}}{{Lah{\'e}n} et~al.}{2020}]{Lahen2020}
{Lah{\'e}n} N.,  {Naab} T.,  {Johansson} P.~H.,  {Elmegreen} B.,  {Hu} C.-Y.,  {Walch} S.,  {Steinwandel} U.~P.,   {Moster} B.~P.,  2020, \mn@doi [\apj] {10.3847/1538-4357/ab7190}, \href {https://ui.adsabs.harvard.edu/abs/2020ApJ...891....2L} {891, 2}

\bibitem[\protect\citeauthoryear{{Landau} \& {Lifshitz}}{{Landau} \& {Lifshitz}}{1960}]{Landau1960}
{Landau} L.~D.,  {Lifshitz} E.~M.,  1960, {Electrodynamics of continuous media}.
Pergmanon Press, Oxford

\bibitem[\protect\citeauthoryear{{Lane}, {Grudi{\'c}}, {Guszejnov}, {Offner}, {Faucher-Gigu{\`e}re}  \& {Rosen}}{{Lane} et~al.}{2022}]{Lane2022}
{Lane} H.~B.,  {Grudi{\'c}} M.~Y.,  {Guszejnov} D.,  {Offner} S. S.~R.,  {Faucher-Gigu{\`e}re} C.-A.,   {Rosen} A.~L.,  2022, \mn@doi [\mnras] {10.1093/mnras/stab3739}, \href {https://ui.adsabs.harvard.edu/abs/2022MNRAS.510.4767L} {510, 4767}

\bibitem[\protect\citeauthoryear{{Lao} \& {Smith}}{{Lao} \& {Smith}}{2020}]{Lao2020}
{Lao} B.-X.,  {Smith} A.,  2020, \mn@doi [\mnras] {10.1093/mnras/staa2198}, \href {https://ui.adsabs.harvard.edu/abs/2020MNRAS.497.3925L} {497, 3925}

\bibitem[\protect\citeauthoryear{{Larsen}, {Romanowsky}, {Brodie}  \& {Wasserman}}{{Larsen} et~al.}{2020}]{Larsen2020}
{Larsen} S.~S.,  {Romanowsky} A.~J.,  {Brodie} J.~P.,   {Wasserman} A.,  2020, \mn@doi [Science] {10.1126/science.abb1970}, \href {https://ui.adsabs.harvard.edu/abs/2020Sci...370..970L} {370, 970}

\bibitem[\protect\citeauthoryear{{Latif}, {Zaroubi}  \& {Spaans}}{{Latif} et~al.}{2011}]{Latif2011}
{Latif} M.~A.,  {Zaroubi} S.,   {Spaans} M.,  2011, \mn@doi [\mnras] {10.1111/j.1365-2966.2010.17796.x}, \href {https://ui.adsabs.harvard.edu/abs/2011MNRAS.411.1659L} {411, 1659}

\bibitem[\protect\citeauthoryear{{Latif}, {Whalen}  \& {Khochfar}}{{Latif} et~al.}{2022}]{Latif2022}
{Latif} M.~A.,  {Whalen} D.,   {Khochfar} S.,  2022, \mn@doi [\apj] {10.3847/1538-4357/ac3916}, \href {https://ui.adsabs.harvard.edu/abs/2022ApJ...925...28L} {925, 28}

\bibitem[\protect\citeauthoryear{{Laursen}}{{Laursen}}{2010}]{Laursen2010}
{Laursen} P.,  2010, PhD thesis, Niels Bohr Institute for Astronomy, Physics and Geophysics

\bibitem[\protect\citeauthoryear{{Laursen}, {Sommer-Larsen}  \& {Andersen}}{{Laursen} et~al.}{2009}]{Laursen2009}
{Laursen} P.,  {Sommer-Larsen} J.,   {Andersen} A.~C.,  2009, \mn@doi [\apj] {10.1088/0004-637X/704/2/1640}, \href {https://ui.adsabs.harvard.edu/abs/2009ApJ...704.1640L} {704, 1640}

\bibitem[\protect\citeauthoryear{{Lee}, {Stanimirovi{\'c}}, {Murray}, {Heiles}  \& {Miller}}{{Lee} et~al.}{2015}]{Lee2015}
{Lee} M.-Y.,  {Stanimirovi{\'c}} S.,  {Murray} C.~E.,  {Heiles} C.,   {Miller} J.,  2015, \mn@doi [\apj] {10.1088/0004-637X/809/1/56}, \href {https://ui.adsabs.harvard.edu/abs/2015ApJ...809...56L} {809, 56}

\bibitem[\protect\citeauthoryear{{Lee}, {Miville-Desch{\^e}nes}  \& {Murray}}{{Lee} et~al.}{2016}]{Lee2016}
{Lee} E.~J.,  {Miville-Desch{\^e}nes} M.-A.,   {Murray} N.~W.,  2016, \mn@doi [\apj] {10.3847/1538-4357/833/2/229}, \href {https://ui.adsabs.harvard.edu/abs/2016ApJ...833..229L} {833, 229}

\bibitem[\protect\citeauthoryear{{Li} \& {Draine}}{{Li} \& {Draine}}{2001}]{Li2001}
{Li} A.,  {Draine} B.~T.,  2001, \mn@doi [\apj] {10.1086/323147}, \href {https://ui.adsabs.harvard.edu/abs/2001ApJ...554..778L} {554, 778}

\bibitem[\protect\citeauthoryear{{Li}, {Vogelsberger}, {Marinacci}  \& {Gnedin}}{{Li} et~al.}{2019}]{Li2019}
{Li} H.,  {Vogelsberger} M.,  {Marinacci} F.,   {Gnedin} O.~Y.,  2019, \mn@doi [\mnras] {10.1093/mnras/stz1271}, \href {https://ui.adsabs.harvard.edu/abs/2019MNRAS.487..364L} {487, 364}

\bibitem[\protect\citeauthoryear{{Loeb} \& {Rybicki}}{{Loeb} \& {Rybicki}}{1999}]{Loeb1999}
{Loeb} A.,  {Rybicki} G.~B.,  1999, \mn@doi [\apj] {10.1086/307844}, \href {https://ui.adsabs.harvard.edu/abs/1999ApJ...524..527L} {524, 527}

\bibitem[\protect\citeauthoryear{{Lupu}, {France}  \& {McCandliss}}{{Lupu} et~al.}{2006}]{Lupu2006}
{Lupu} R.~E.,  {France} K.,   {McCandliss} S.~R.,  2006, \mn@doi [\apj] {10.1086/503603}, \href {https://ui.adsabs.harvard.edu/abs/2006ApJ...644..981L} {644, 981}

\bibitem[\protect\citeauthoryear{{Ma}, {Hopkins}, {Kasen}, {Quataert}, {Faucher-Gigu{\`e}re}, {Kere{\v{s}}}, {Murray}  \& {Strom}}{{Ma} et~al.}{2016}]{Ma2016}
{Ma} X.,  {Hopkins} P.~F.,  {Kasen} D.,  {Quataert} E.,  {Faucher-Gigu{\`e}re} C.-A.,  {Kere{\v{s}}} D.,  {Murray} N.,   {Strom} A.,  2016, \mn@doi [\mnras] {10.1093/mnras/stw941}, \href {https://ui.adsabs.harvard.edu/abs/2016MNRAS.459.3614M} {459, 3614}

\bibitem[\protect\citeauthoryear{{Martin} et~al.,}{{Martin} et~al.}{2022}]{Martin2022}
{Martin} N.~F.,  et~al., 2022, \mn@doi [\nat] {10.1038/s41586-021-04162-2}, \href {https://ui.adsabs.harvard.edu/abs/2022Natur.601...45M} {601, 45}

\bibitem[\protect\citeauthoryear{{Matheron}}{{Matheron}}{1967}]{Matheron1967}
{Matheron} G.,  1967, {El{\'e}ments pour une th{\'e}orie des milieux poreux}.
{Masson, Paris}

\bibitem[\protect\citeauthoryear{{McClellan}, {Davis}  \& {Arras}}{{McClellan} et~al.}{2022}]{McClellan2022}
{McClellan} B.~C.,  {Davis} S.~W.,   {Arras} P.,  2022, \mn@doi [\apj] {10.3847/1538-4357/ac7724}, \href {https://ui.adsabs.harvard.edu/abs/2022ApJ...934...37M} {934, 37}

\bibitem[\protect\citeauthoryear{{McKee} \& {Tan}}{{McKee} \& {Tan}}{2008}]{McKee2008}
{McKee} C.~F.,  {Tan} J.~C.,  2008, \mn@doi [\apj] {10.1086/587434}, \href {https://ui.adsabs.harvard.edu/abs/2008ApJ...681..771M} {681, 771}

\bibitem[\protect\citeauthoryear{{McQuinn}, {Mao}, {Tollerud}, {Cohen}, {Shih}, {Buckley}  \& {Dolphin}}{{McQuinn} et~al.}{2024}]{McQuinn2024}
{McQuinn} K. B.~W.,  {Mao} Y.-Y.,  {Tollerud} E.~J.,  {Cohen} R.~E.,  {Shih} D.,  {Buckley} M.~R.,   {Dolphin} A.~E.,  2024, \mn@doi [\apj] {10.3847/1538-4357/ad429b}, \href {https://ui.adsabs.harvard.edu/abs/2024ApJ...967..161M} {967, 161}

\bibitem[\protect\citeauthoryear{{Mead} \& {Truhlar}}{{Mead} \& {Truhlar}}{1982}]{mead1982}
{Mead} C.~A.,  {Truhlar} D.~G.,  1982, \mn@doi [\jcp] {10.1063/1.443853}, \href {https://ui.adsabs.harvard.edu/abs/1982JChPh..77.6090M} {77, 6090}

\bibitem[\protect\citeauthoryear{{Meiksin}}{{Meiksin}}{2006}]{Meiksin2006}
{Meiksin} A.,  2006, \mn@doi [\mnras] {10.1111/j.1365-2966.2006.10632.x}, \href {https://ui.adsabs.harvard.edu/abs/2006MNRAS.370.2025M} {370, 2025}

\bibitem[\protect\citeauthoryear{{Menager}, {Barth{\'e}lemy}, {Koskinen}, {Lilensten}, {Ehrenreich}  \& {Parkinson}}{{Menager} et~al.}{2013}]{Menager2013}
{Menager} H.,  {Barth{\'e}lemy} M.,  {Koskinen} T.,  {Lilensten} J.,  {Ehrenreich} D.,   {Parkinson} C.~D.,  2013, \mn@doi [\icarus] {10.1016/j.icarus.2013.02.028}, \href {https://ui.adsabs.harvard.edu/abs/2013Icar..226.1709M} {226, 1709}

\bibitem[\protect\citeauthoryear{{Menon}, {Federrath}  \& {Kuiper}}{{Menon} et~al.}{2020}]{Menon2020_HIIexpansionturbulence}
{Menon} S.~H.,  {Federrath} C.,   {Kuiper} R.,  2020, \mn@doi [\mnras] {10.1093/mnras/staa580}, \href {https://ui.adsabs.harvard.edu/abs/2020MNRAS.493.4643M} {493, 4643}

\bibitem[\protect\citeauthoryear{{Menon}, {Federrath}  \& {Krumholz}}{{Menon} et~al.}{2022}]{Menon2022_IR}
{Menon} S.~H.,  {Federrath} C.,   {Krumholz} M.~R.,  2022, \mn@doi [\mnras] {10.1093/mnras/stac2702}, \href {https://ui.adsabs.harvard.edu/abs/2022MNRAS.517.1313M} {517, 1313}

\bibitem[\protect\citeauthoryear{{Menon}, {Federrath}  \& {Krumholz}}{{Menon} et~al.}{2023}]{Menon2023_directRP}
{Menon} S.~H.,  {Federrath} C.,   {Krumholz} M.~R.,  2023, \mn@doi [\mnras] {10.1093/mnras/stad856}, \href {https://ui.adsabs.harvard.edu/abs/2023MNRAS.521.5160M} {521, 5160}

\bibitem[\protect\citeauthoryear{{Menon}, {Lancaster}, {Burkhart}, {Somerville}, {Dekel}  \& {Krumholz}}{{Menon} et~al.}{2024}]{Menon2024}
{Menon} S.~H.,  {Lancaster} L.,  {Burkhart} B.,  {Somerville} R.~S.,  {Dekel} A.,   {Krumholz} M.~R.,  2024, \mn@doi [\apjl] {10.3847/2041-8213/ad462d}, \href {https://ui.adsabs.harvard.edu/abs/2024ApJ...967L..28M} {967, L28}

\bibitem[\protect\citeauthoryear{{Michel-Dansac}, {Blaizot}, {Garel}, {Verhamme}, {Kimm}  \& {Trebitsch}}{{Michel-Dansac} et~al.}{2020}]{RASCAS}
{Michel-Dansac} L.,  {Blaizot} J.,  {Garel} T.,  {Verhamme} A.,  {Kimm} T.,   {Trebitsch} M.,  2020, \mn@doi [\aap] {10.1051/0004-6361/201834961}, \href {https://ui.adsabs.harvard.edu/abs/2020A&A...635A.154M} {635, A154}

\bibitem[\protect\citeauthoryear{{Mihalas} \& {Auer}}{{Mihalas} \& {Auer}}{2001}]{Mihalas2001}
{Mihalas} D.,  {Auer} L.,  2001, \mn@doi [\jqsrt] {10.1016/S0022-4073(01)00013-9}, \href {https://ui.adsabs.harvard.edu/abs/2001JQSRT..71...61M} {71, 61}

\bibitem[\protect\citeauthoryear{{Mihalas}, {Kunasz}  \& {Hummer}}{{Mihalas} et~al.}{1975}]{Mihalas1975}
{Mihalas} D.,  {Kunasz} P.~B.,   {Hummer} D.~G.,  1975, \mn@doi [\apj] {10.1086/153996}, \href {https://ui.adsabs.harvard.edu/abs/1975ApJ...202..465M} {202, 465}

\bibitem[\protect\citeauthoryear{{Milosavljevi{\'c}}, {Bromm}, {Couch}  \& {Oh}}{{Milosavljevi{\'c}} et~al.}{2009}]{Milosavljevic2009}
{Milosavljevi{\'c}} M.,  {Bromm} V.,  {Couch} S.~M.,   {Oh} S.~P.,  2009, \mn@doi [\apj] {10.1088/0004-637X/698/1/766}, \href {https://ui.adsabs.harvard.edu/abs/2009ApJ...698..766M} {698, 766}

\bibitem[\protect\citeauthoryear{{Mortlock}}{{Mortlock}}{2016}]{Mortlock2016}
{Mortlock} D.,  2016, in {Mesinger} A.,  ed.,  Astrophysics and Space Science Library Vol. 423, Understanding the Epoch of Cosmic Reionization: Challenges and Progress. p.~187 (\mn@eprint {arXiv} {1511.01107}), \mn@doi{10.1007/978-3-319-21957-8_7}

\bibitem[\protect\citeauthoryear{{Mowla} et~al.,}{{Mowla} et~al.}{2024}]{Mowla2024}
{Mowla} L.,  et~al., 2024, \mn@doi [arXiv e-prints] {10.48550/arXiv.2402.08696}, \href {https://ui.adsabs.harvard.edu/abs/2024arXiv240208696M} {p. arXiv:2402.08696}

\bibitem[\protect\citeauthoryear{{Munirov} \& {Kaurov}}{{Munirov} \& {Kaurov}}{2023}]{Munirov2023}
{Munirov} V.~R.,  {Kaurov} A.~A.,  2023, \mn@doi [\mnras] {10.1093/mnras/stad1165}, \href {https://ui.adsabs.harvard.edu/abs/2023MNRAS.522.2747M} {522, 2747}

\bibitem[\protect\citeauthoryear{{Nadler} et~al.,}{{Nadler} et~al.}{2020}]{Nadler2020}
{Nadler} E.~O.,  et~al., 2020, \mn@doi [\apj] {10.3847/1538-4357/ab846a}, \href {https://ui.adsabs.harvard.edu/abs/2020ApJ...893...48N} {893, 48}

\bibitem[\protect\citeauthoryear{Nebrin}{Nebrin}{2022}]{Nebrin2022}
Nebrin O.,  2022, Master's thesis, Stockholm University, Department of Astronomy, \url {https://urn.kb.se/resolve?urn=urn:nbn:se:su:diva-205625}

\bibitem[\protect\citeauthoryear{{Nebrin}}{{Nebrin}}{2023}]{Nebrin2023}
{Nebrin} O.,  2023, \mn@doi [Research Notes of the American Astronomical Society] {10.3847/2515-5172/acd37a}, \href {https://ui.adsabs.harvard.edu/abs/2023RNAAS...7...90N} {7, 90}

\bibitem[\protect\citeauthoryear{{Nebrin}, {Giri}  \& {Mellema}}{{Nebrin} et~al.}{2023}]{Nebrin2023cooling}
{Nebrin} O.,  {Giri} S.~K.,   {Mellema} G.,  2023, \mn@doi [\mnras] {10.1093/mnras/stad1852}, \href {https://ui.adsabs.harvard.edu/abs/2023MNRAS.524.2290N} {524, 2290}

\bibitem[\protect\citeauthoryear{{Neufeld}}{{Neufeld}}{1990}]{Neufeld1990}
{Neufeld} D.~A.,  1990, \mn@doi [\apj] {10.1086/168375}, \href {https://ui.adsabs.harvard.edu/abs/1990ApJ...350..216N} {350, 216}

\bibitem[\protect\citeauthoryear{Neuman \& Orr}{Neuman \& Orr}{1993}]{Neuman1993}
Neuman S.~P.,  Orr S.,  1993, Water resources research, 29, 341

\bibitem[\protect\citeauthoryear{Noetinger}{Noetinger}{1994}]{Noetinger1994}
Noetinger B.,  1994, Transport in porous media, 15, 99

\bibitem[\protect\citeauthoryear{{Oh} \& {Haiman}}{{Oh} \& {Haiman}}{2002}]{Oh2002}
{Oh} S.~P.,  {Haiman} Z.,  2002, \mn@doi [\apj] {10.1086/339393}, \href {https://ui.adsabs.harvard.edu/abs/2002ApJ...569..558O} {569, 558}

\bibitem[\protect\citeauthoryear{{Oklop{\v{c}}i{\'c}}, {Hirata}  \& {Heng}}{{Oklop{\v{c}}i{\'c}} et~al.}{2016}]{Oklopcic2016}
{Oklop{\v{c}}i{\'c}} A.,  {Hirata} C.~M.,   {Heng} K.,  2016, \mn@doi [\apj] {10.3847/0004-637X/832/1/30}, \href {https://ui.adsabs.harvard.edu/abs/2016ApJ...832...30O} {832, 30}

\bibitem[\protect\citeauthoryear{{Olivier}, {Lopez}, {Rosen}, {Nayak}, {Reiter}, {Krumholz}  \& {Bolatto}}{{Olivier} et~al.}{2021}]{Olivier2021}
{Olivier} G.~M.,  {Lopez} L.~A.,  {Rosen} A.~L.,  {Nayak} O.,  {Reiter} M.,  {Krumholz} M.~R.,   {Bolatto} A.~D.,  2021, \mn@doi [\apj] {10.3847/1538-4357/abd24a}, \href {https://ui.adsabs.harvard.edu/abs/2021ApJ...908...68O} {908, 68}

\bibitem[\protect\citeauthoryear{{Omukai}}{{Omukai}}{2001}]{Omukai2001}
{Omukai} K.,  2001, \mn@doi [\apj] {10.1086/318296}, \href {https://ui.adsabs.harvard.edu/abs/2001ApJ...546..635O} {546, 635}

\bibitem[\protect\citeauthoryear{{Omukai} \& {Inutsuka}}{{Omukai} \& {Inutsuka}}{2002}]{Omukai2002}
{Omukai} K.,  {Inutsuka} S.-i.,  2002, \mn@doi [\mnras] {10.1046/j.1365-8711.2002.05276.x}, \href {https://ui.adsabs.harvard.edu/abs/2002MNRAS.332...59O} {332, 59}

\bibitem[\protect\citeauthoryear{{Omukai}, {Tsuribe}, {Schneider}  \& {Ferrara}}{{Omukai} et~al.}{2005}]{Omukai2005}
{Omukai} K.,  {Tsuribe} T.,  {Schneider} R.,   {Ferrara} A.,  2005, \mn@doi [\apj] {10.1086/429955}, \href {https://ui.adsabs.harvard.edu/abs/2005ApJ...626..627O} {626, 627}

\bibitem[\protect\citeauthoryear{{Orkisz} et~al.,}{{Orkisz} et~al.}{2017}]{Orkisz2017}
{Orkisz} J.~H.,  et~al., 2017, \mn@doi [\aap] {10.1051/0004-6361/201629220}, \href {https://ui.adsabs.harvard.edu/abs/2017A&A...599A..99O} {599, A99}

\bibitem[\protect\citeauthoryear{{Pan}, {Padoan}  \& {Nordlund}}{{Pan} et~al.}{2018}]{Pan2018}
{Pan} L.,  {Padoan} P.,   {Nordlund} {\r{A}}.,  2018, \mn@doi [\apjl] {10.3847/2041-8213/aae57c}, \href {https://ui.adsabs.harvard.edu/abs/2018ApJ...866L..17P} {866, L17}

\bibitem[\protect\citeauthoryear{{Passot} \& {V{\'a}zquez-Semadeni}}{{Passot} \& {V{\'a}zquez-Semadeni}}{1998}]{Passot1998}
{Passot} T.,  {V{\'a}zquez-Semadeni} E.,  1998, \mn@doi [\pre] {10.1103/PhysRevE.58.4501}, \href {https://ui.adsabs.harvard.edu/abs/1998PhRvE..58.4501P} {58, 4501}

\bibitem[\protect\citeauthoryear{{Peebles}}{{Peebles}}{1993}]{Peebles1993}
{Peebles} P.~J.~E.,  1993, {Principles of Physical Cosmology}.
{Princeton University Press}, \mn@doi{10.1515/9780691206721}

\bibitem[\protect\citeauthoryear{{Pengelly} \& {Seaton}}{{Pengelly} \& {Seaton}}{1964}]{Pengelly1964}
{Pengelly} R.~M.,  {Seaton} M.~J.,  1964, \mn@doi [\mnras] {10.1093/mnras/127.2.165}, \href {https://ui.adsabs.harvard.edu/abs/1964MNRAS.127..165P} {127, 165}

\bibitem[\protect\citeauthoryear{{Pineda} et~al.,}{{Pineda} et~al.}{2024}]{Pineda2024}
{Pineda} J.~E.,  et~al., 2024, \mn@doi [\aap] {10.1051/0004-6361/202347997}, \href {https://ui.adsabs.harvard.edu/abs/2024A&A...686A.162P} {686, A162}

\bibitem[\protect\citeauthoryear{{Prole}, {Schauer}, {Clark}, {Glover}, {Priestley}  \& {Klessen}}{{Prole} et~al.}{2023}]{Prole2023}
{Prole} L.~R.,  {Schauer} A. T.~P.,  {Clark} P.~C.,  {Glover} S. C.~O.,  {Priestley} F.~D.,   {Klessen} R.~S.,  2023, \mn@doi [\mnras] {10.1093/mnras/stad188}, \href {https://ui.adsabs.harvard.edu/abs/2023MNRAS.520.2081P} {520, 2081}

\bibitem[\protect\citeauthoryear{{Purcell}}{{Purcell}}{1952}]{Purcell1952}
{Purcell} E.~M.,  1952, \mn@doi [\apj] {10.1086/145637}, \href {https://ui.adsabs.harvard.edu/abs/1952ApJ...116..457P} {116, 457}

\bibitem[\protect\citeauthoryear{{Raskutti}, {Ostriker}  \& {Skinner}}{{Raskutti} et~al.}{2016}]{Raskutti2016}
{Raskutti} S.,  {Ostriker} E.~C.,   {Skinner} M.~A.,  2016, \mn@doi [\apj] {10.3847/0004-637X/829/2/130}, \href {https://ui.adsabs.harvard.edu/abs/2016ApJ...829..130R} {829, 130}

\bibitem[\protect\citeauthoryear{{R{\'e}my-Ruyer} et~al.,}{{R{\'e}my-Ruyer} et~al.}{2014}]{Remy2014}
{R{\'e}my-Ruyer} A.,  et~al., 2014, \mn@doi [\aap] {10.1051/0004-6361/201322803}, \href {https://ui.adsabs.harvard.edu/abs/2014A&A...563A..31R} {563, A31}

\bibitem[\protect\citeauthoryear{Renard \& De~Marsily}{Renard \& De~Marsily}{1997}]{Renard1997}
Renard P.,  De~Marsily G.,  1997, Advances in water resources, 20, 253

\bibitem[\protect\citeauthoryear{{Ricotti} \& {Gnedin}}{{Ricotti} \& {Gnedin}}{2005}]{Ricotti2005}
{Ricotti} M.,  {Gnedin} N.~Y.,  2005, \mn@doi [\apj] {10.1086/431415}, \href {https://ui.adsabs.harvard.edu/abs/2005ApJ...629..259R} {629, 259}

\bibitem[\protect\citeauthoryear{{Ricotti}, {Parry}  \& {Gnedin}}{{Ricotti} et~al.}{2016}]{Ricotti2016}
{Ricotti} M.,  {Parry} O.~H.,   {Gnedin} N.~Y.,  2016, \mn@doi [\apj] {10.3847/0004-637X/831/2/204}, \href {https://ui.adsabs.harvard.edu/abs/2016ApJ...831..204R} {831, 204}

\bibitem[\protect\citeauthoryear{{Rosdahl} et~al.,}{{Rosdahl} et~al.}{2018}]{Rosdahl2018}
{Rosdahl} J.,  et~al., 2018, \mn@doi [\mnras] {10.1093/mnras/sty1655}, \href {https://ui.adsabs.harvard.edu/abs/2018MNRAS.479..994R} {479, 994}

\bibitem[\protect\citeauthoryear{Ryan, Czuchlewski  \& McCusker}{Ryan et~al.}{1977}]{Ryan1977}
Ryan S.~R.,  Czuchlewski S.~J.,   McCusker M.~V.,  1977, \mn@doi [Phys. Rev. A] {10.1103/PhysRevA.16.1892}, 16, 1892

\bibitem[\protect\citeauthoryear{{Rybicki}}{{Rybicki}}{2006}]{Rybicki2006}
{Rybicki} G.~B.,  2006, \mn@doi [\apj] {10.1086/505327}, \href {https://ui.adsabs.harvard.edu/abs/2006ApJ...647..709R} {647, 709}

\bibitem[\protect\citeauthoryear{{Rybicki} \& {Lightman}}{{Rybicki} \& {Lightman}}{1986}]{Rybicki1986}
{Rybicki} G.~B.,  {Lightman} A.~P.,  1986, {Radiative Processes in Astrophysics}.
{Wiley}

\bibitem[\protect\citeauthoryear{{Rybicki} \& {dell'Antonio}}{{Rybicki} \& {dell'Antonio}}{1994}]{Rybicki1994}
{Rybicki} G.~B.,  {dell'Antonio} I.~P.,  1994, \mn@doi [\apj] {10.1086/174170}, \href {https://ui.adsabs.harvard.edu/abs/1994ApJ...427..603R} {427, 603}

\bibitem[\protect\citeauthoryear{{Safranek-Shrader}, {Agarwal}, {Federrath}, {Dubey}, {Milosavljevi{\'c}}  \& {Bromm}}{{Safranek-Shrader} et~al.}{2012}]{Safranek2012}
{Safranek-Shrader} C.,  {Agarwal} M.,  {Federrath} C.,  {Dubey} A.,  {Milosavljevi{\'c}} M.,   {Bromm} V.,  2012, \mn@doi [\mnras] {10.1111/j.1365-2966.2012.21852.x}, \href {https://ui.adsabs.harvard.edu/abs/2012MNRAS.426.1159S} {426, 1159}

\bibitem[\protect\citeauthoryear{{Safranek-Shrader}, {Milosavljevi{\'c}}  \& {Bromm}}{{Safranek-Shrader} et~al.}{2014}]{Safranek2014}
{Safranek-Shrader} C.,  {Milosavljevi{\'c}} M.,   {Bromm} V.,  2014, \mn@doi [\mnras] {10.1093/mnras/stt2307}, \href {https://ui.adsabs.harvard.edu/abs/2014MNRAS.438.1669S} {438, 1669}

\bibitem[\protect\citeauthoryear{Sanchez-Vila, Guadagnini  \& Carrera}{Sanchez-Vila et~al.}{2006}]{Sanchez2006}
Sanchez-Vila X.,  Guadagnini A.,   Carrera J.,  2006, Reviews of Geophysics, 44

\bibitem[\protect\citeauthoryear{{Schneider} et~al.,}{{Schneider} et~al.}{2022}]{Schneider2022_turb}
{Schneider} N.,  et~al., 2022, \mn@doi [\aap] {10.1051/0004-6361/202039610}, \href {https://ui.adsabs.harvard.edu/abs/2022A&A...666A.165S} {666, A165}

\bibitem[\protect\citeauthoryear{{Seaton}}{{Seaton}}{1955}]{Seaton1955}
{Seaton} M.~J.,  1955, \mn@doi [Proceedings of the Physical Society A] {10.1088/0370-1298/68/6/301}, \href {https://ui.adsabs.harvard.edu/abs/1955PPSA...68..457S} {68, 457}

\bibitem[\protect\citeauthoryear{{Seifried}, {Beuther}, {Walch}, {Syed}, {Soler}, {Girichidis}  \& {W{\"u}nsch}}{{Seifried} et~al.}{2022}]{Seifried2022}
{Seifried} D.,  {Beuther} H.,  {Walch} S.,  {Syed} J.,  {Soler} J.~D.,  {Girichidis} P.,   {W{\"u}nsch} R.,  2022, \mn@doi [\mnras] {10.1093/mnras/stac607}, \href {https://ui.adsabs.harvard.edu/abs/2022MNRAS.512.4765S} {512, 4765}

\bibitem[\protect\citeauthoryear{{Seon} \& {Kim}}{{Seon} \& {Kim}}{2020}]{Seon2020}
{Seon} K.-i.,  {Kim} C.-G.,  2020, \mn@doi [\apjs] {10.3847/1538-4365/aba2d6}, \href {https://ui.adsabs.harvard.edu/abs/2020ApJS..250....9S} {250, 9}

\bibitem[\protect\citeauthoryear{{Sharda} \& {Krumholz}}{{Sharda} \& {Krumholz}}{2022}]{Sharda2022_IMF}
{Sharda} P.,  {Krumholz} M.~R.,  2022, \mn@doi [\mnras] {10.1093/mnras/stab2921}, \href {https://ui.adsabs.harvard.edu/abs/2022MNRAS.509.1959S} {509, 1959}

\bibitem[\protect\citeauthoryear{{Sharda} \& {Menon}}{{Sharda} \& {Menon}}{2024}]{Sharda2024}
{Sharda} P.,  {Menon} S.~H.,  2024, arXiv e-prints, \href {https://ui.adsabs.harvard.edu/abs/2024arXiv240518265S} {p. arXiv:2405.18265}

\bibitem[\protect\citeauthoryear{{Sharda} et~al.,}{{Sharda} et~al.}{2022}]{Sharda2022_turbulenceobs}
{Sharda} P.,  et~al., 2022, \mn@doi [\mnras] {10.1093/mnras/stab3048}, \href {https://ui.adsabs.harvard.edu/abs/2022MNRAS.509.2180S} {509, 2180}

\bibitem[\protect\citeauthoryear{{Shull}}{{Shull}}{1978}]{Shull1978}
{Shull} J.~M.,  1978, \mn@doi [\apj] {10.1086/156433}, \href {https://ui.adsabs.harvard.edu/abs/1978ApJ...224..841S} {224, 841}

\bibitem[\protect\citeauthoryear{{Simon}}{{Simon}}{2019}]{Simon2019}
{Simon} J.~D.,  2019, \mn@doi [\araa] {10.1146/annurev-astro-091918-104453}, \href {https://ui.adsabs.harvard.edu/abs/2019ARA&A..57..375S} {57, 375}

\bibitem[\protect\citeauthoryear{{Simon} et~al.,}{{Simon} et~al.}{2021}]{Simon2021}
{Simon} J.~D.,  et~al., 2021, \mn@doi [\apj] {10.3847/1538-4357/abd31b}, \href {https://ui.adsabs.harvard.edu/abs/2021ApJ...908...18S} {908, 18}

\bibitem[\protect\citeauthoryear{{Skinner} \& {Ostriker}}{{Skinner} \& {Ostriker}}{2015}]{Skinner2015}
{Skinner} M.~A.,  {Ostriker} E.~C.,  2015, \mn@doi [\apj] {10.1088/0004-637X/809/2/187}, \href {https://ui.adsabs.harvard.edu/abs/2015ApJ...809..187S} {809, 187}

\bibitem[\protect\citeauthoryear{Slocomb, Miller  \& Schaefer}{Slocomb et~al.}{1971}]{Slocomb1971}
Slocomb C.~A.,  Miller W.~H.,   Schaefer H.~F. I.,  1971, \mn@doi [The Journal of Chemical Physics] {10.1063/1.1676163}, 55, 926

\bibitem[\protect\citeauthoryear{{Smith}}{{Smith}}{2014}]{Smith2014_winds}
{Smith} N.,  2014, \mn@doi [\araa] {10.1146/annurev-astro-081913-040025}, \href {https://ui.adsabs.harvard.edu/abs/2014ARA&A..52..487S} {52, 487}

\bibitem[\protect\citeauthoryear{{Smith}, {Safranek-Shrader}, {Bromm}  \& {Milosavljevi{\'c}}}{{Smith} et~al.}{2015}]{Smith2015}
{Smith} A.,  {Safranek-Shrader} C.,  {Bromm} V.,   {Milosavljevi{\'c}} M.,  2015, \mn@doi [\mnras] {10.1093/mnras/stv565}, \href {https://ui.adsabs.harvard.edu/abs/2015MNRAS.449.4336S} {449, 4336}

\bibitem[\protect\citeauthoryear{{Smith}, {Bromm}  \& {Loeb}}{{Smith} et~al.}{2016}]{Smith2016}
{Smith} A.,  {Bromm} V.,   {Loeb} A.,  2016, \mn@doi [\mnras] {10.1093/mnras/stw1129}, \href {https://ui.adsabs.harvard.edu/abs/2016MNRAS.460.3143S} {460, 3143}

\bibitem[\protect\citeauthoryear{{Smith}, {Bromm}  \& {Loeb}}{{Smith} et~al.}{2017a}]{Smith2017}
{Smith} A.,  {Bromm} V.,   {Loeb} A.,  2017a, \mn@doi [\mnras] {10.1093/mnras/stw2591}, \href {https://ui.adsabs.harvard.edu/abs/2017MNRAS.464.2963S} {464, 2963}

\bibitem[\protect\citeauthoryear{{Smith}, {Becerra}, {Bromm}  \& {Hernquist}}{{Smith} et~al.}{2017b}]{Smith2017_DCBH}
{Smith} A.,  {Becerra} F.,  {Bromm} V.,   {Hernquist} L.,  2017b, \mn@doi [\mnras] {10.1093/mnras/stx1993}, \href {https://ui.adsabs.harvard.edu/abs/2017MNRAS.472..205S} {472, 205}

\bibitem[\protect\citeauthoryear{{Smith}, {Tsang}, {Bromm}  \& {Milosavljevi{\'c}}}{{Smith} et~al.}{2018}]{Smith2018_discretediffusion}
{Smith} A.,  {Tsang} B. T.~H.,  {Bromm} V.,   {Milosavljevi{\'c}} M.,  2018, \mn@doi [\mnras] {10.1093/mnras/sty1509}, \href {https://ui.adsabs.harvard.edu/abs/2018MNRAS.479.2065S} {479, 2065}

\bibitem[\protect\citeauthoryear{{Smith}, {Ma}, {Bromm}, {Finkelstein}, {Hopkins}, {Faucher-Gigu{\`e}re}  \& {Kere{\v{s}}}}{{Smith} et~al.}{2019}]{Smith2019}
{Smith} A.,  {Ma} X.,  {Bromm} V.,  {Finkelstein} S.~L.,  {Hopkins} P.~F.,  {Faucher-Gigu{\`e}re} C.-A.,   {Kere{\v{s}}} D.,  2019, \mn@doi [\mnras] {10.1093/mnras/sty3483}, \href {https://ui.adsabs.harvard.edu/abs/2019MNRAS.484...39S} {484, 39}

\bibitem[\protect\citeauthoryear{{Smith} et~al.,}{{Smith} et~al.}{2024}]{Smith2024_UFD}
{Smith} S. E.~T.,  et~al., 2024, \mn@doi [\apj] {10.3847/1538-4357/ad0d9f}, \href {https://ui.adsabs.harvard.edu/abs/2024ApJ...961...92S} {961, 92}

\bibitem[\protect\citeauthoryear{{Smith}, {Lorinc}, {Nebrin}  \& {Lao}}{{Smith} et~al.}{2025}]{Smith2025}
{Smith} A.,  {Lorinc} K.,  {Nebrin} O.,   {Lao} B.-X.,  2025, arXiv e-prints, \href {https://ui.adsabs.harvard.edu/abs/2025arXiv250101928S} {p. arXiv:2501.01928}

\bibitem[\protect\citeauthoryear{{Spaans} \& {Silk}}{{Spaans} \& {Silk}}{2006}]{Spaans2006}
{Spaans} M.,  {Silk} J.,  2006, \mn@doi [\apj] {10.1086/508444}, \href {https://ui.adsabs.harvard.edu/abs/2006ApJ...652..902S} {652, 902}

\bibitem[\protect\citeauthoryear{{Spitzer} \& {Greenstein}}{{Spitzer} \& {Greenstein}}{1951}]{Spitzer1951}
{Spitzer} Lyman J.,  {Greenstein} J.~L.,  1951, \mn@doi [\apj] {10.1086/145480}, \href {https://ui.adsabs.harvard.edu/abs/1951ApJ...114..407S} {114, 407}

\bibitem[\protect\citeauthoryear{{Stacy}, {Greif}  \& {Bromm}}{{Stacy} et~al.}{2012}]{Stacy2012}
{Stacy} A.,  {Greif} T.~H.,   {Bromm} V.,  2012, \mn@doi [\mnras] {10.1111/j.1365-2966.2012.20605.x}, \href {https://ui.adsabs.harvard.edu/abs/2012MNRAS.422..290S} {422, 290}

\bibitem[\protect\citeauthoryear{{Stacy}, {Bromm}  \& {Lee}}{{Stacy} et~al.}{2016}]{Stacy2016}
{Stacy} A.,  {Bromm} V.,   {Lee} A.~T.,  2016, \mn@doi [\mnras] {10.1093/mnras/stw1728}, \href {https://ui.adsabs.harvard.edu/abs/2016MNRAS.462.1307S} {462, 1307}

\bibitem[\protect\citeauthoryear{{Stanway} \& {Eldridge}}{{Stanway} \& {Eldridge}}{2019}]{Stanway2019}
{Stanway} E.~R.,  {Eldridge} J.~J.,  2019, \mn@doi [\aap] {10.1051/0004-6361/201834359}, \href {https://ui.adsabs.harvard.edu/abs/2019A&A...621A.105S} {621, A105}

\bibitem[\protect\citeauthoryear{{Stenrup}, {Larson}  \& {Elander}}{{Stenrup} et~al.}{2009}]{stenrup2009}
{Stenrup} M.,  {Larson} {\r{A}}.,   {Elander} N.,  2009, \mn@doi [\pra] {10.1103/PhysRevA.79.012713}, \href {https://ui.adsabs.harvard.edu/abs/2009PhRvA..79a2713S} {79, 012713}

\bibitem[\protect\citeauthoryear{{Sternberg}, {Gurman}  \& {Bialy}}{{Sternberg} et~al.}{2021}]{Sternberg2021}
{Sternberg} A.,  {Gurman} A.,   {Bialy} S.,  2021, \mn@doi [\apj] {10.3847/1538-4357/ac167b}, \href {https://ui.adsabs.harvard.edu/abs/2021ApJ...920...83S} {920, 83}

\bibitem[\protect\citeauthoryear{{Struensee} \& {Cohen}}{{Struensee} \& {Cohen}}{1988}]{Struensee1988}
{Struensee} M.~C.,  {Cohen} J.~S.,  1988, \mn@doi [\pra] {10.1103/PhysRevA.38.3377}, \href {https://ui.adsabs.harvard.edu/abs/1988PhRvA..38.3377S} {38, 3377}

\bibitem[\protect\citeauthoryear{{Sugimura}, {Matsumoto}, {Hosokawa}, {Hirano}  \& {Omukai}}{{Sugimura} et~al.}{2023}]{Sugimura2023_PopIII}
{Sugimura} K.,  {Matsumoto} T.,  {Hosokawa} T.,  {Hirano} S.,   {Omukai} K.,  2023, \mn@doi [\apj] {10.3847/1538-4357/ad02fc}, \href {https://ui.adsabs.harvard.edu/abs/2023ApJ...959...17S} {959, 17}

\bibitem[\protect\citeauthoryear{{Sugimura}, {Ricotti}, {Park}, {Garcia}  \& {Yajima}}{{Sugimura} et~al.}{2024}]{Sugimura2024}
{Sugimura} K.,  {Ricotti} M.,  {Park} J.,  {Garcia} F. A.~B.,   {Yajima} H.,  2024, \mn@doi [\apj] {10.3847/1538-4357/ad499a}, \href {https://ui.adsabs.harvard.edu/abs/2024ApJ...970...14S} {970, 14}

\bibitem[\protect\citeauthoryear{{Syed} et~al.,}{{Syed} et~al.}{2020}]{Syed2020}
{Syed} J.,  et~al., 2020, \mn@doi [\aap] {10.1051/0004-6361/202038449}, \href {https://ui.adsabs.harvard.edu/abs/2020A&A...642A..68S} {642, A68}

\bibitem[\protect\citeauthoryear{{Tasitsiomi}}{{Tasitsiomi}}{2006}]{Tasitsiomi2006}
{Tasitsiomi} A.,  2006, \mn@doi [\apj] {10.1086/505682}, \href {https://ui.adsabs.harvard.edu/abs/2006ApJ...648..762T} {648, 762}

\bibitem[\protect\citeauthoryear{{Tegmark}, {Silk}, {Rees}, {Blanchard}, {Abel}  \& {Palla}}{{Tegmark} et~al.}{1997}]{Tegmark1997}
{Tegmark} M.,  {Silk} J.,  {Rees} M.~J.,  {Blanchard} A.,  {Abel} T.,   {Palla} F.,  1997, \mn@doi [\apj] {10.1086/303434}, \href {https://ui.adsabs.harvard.edu/abs/1997ApJ...474....1T} {474, 1}

\bibitem[\protect\citeauthoryear{{Terazawa}, {Ukai}, {Kouchi}, {Kameta}, {Hatano}  \& {Tanaka}}{{Terazawa} et~al.}{1993}]{Terazawa1993}
{Terazawa} N.,  {Ukai} M.,  {Kouchi} N.,  {Kameta} K.,  {Hatano} Y.,   {Tanaka} K.,  1993, \mn@doi [\jcp] {10.1063/1.465333}, \href {https://ui.adsabs.harvard.edu/abs/1993JChPh..99.1637T} {99, 1637}

\bibitem[\protect\citeauthoryear{Thompson \& Heckman}{Thompson \& Heckman}{2024}]{Thomson2024}
Thompson T.~A.,  Heckman T.~M.,  2024, Annual Review of Astronomy and Astrophysics, 62

\bibitem[\protect\citeauthoryear{{Thompson} \& {Krumholz}}{{Thompson} \& {Krumholz}}{2016}]{Thompson2016}
{Thompson} T.~A.,  {Krumholz} M.~R.,  2016, \mn@doi [\mnras] {10.1093/mnras/stv2331}, \href {https://ui.adsabs.harvard.edu/abs/2016MNRAS.455..334T} {455, 334}

\bibitem[\protect\citeauthoryear{{Tomaselli} \& {Ferrara}}{{Tomaselli} \& {Ferrara}}{2021}]{Tomaselli2021}
{Tomaselli} G.~M.,  {Ferrara} A.,  2021, \mn@doi [\mnras] {10.1093/mnras/stab876}, \href {https://ui.adsabs.harvard.edu/abs/2021MNRAS.504...89T} {504, 89}

\bibitem[\protect\citeauthoryear{{Unno}}{{Unno}}{1955}]{Unno1955}
{Unno} W.,  1955, \pasj, \href {https://ui.adsabs.harvard.edu/abs/1955PASJ....7...81U} {7, 81}

\bibitem[\protect\citeauthoryear{{Vanzella} et~al.,}{{Vanzella} et~al.}{2023}]{Vanzella2023}
{Vanzella} E.,  et~al., 2023, \mn@doi [\apj] {10.3847/1538-4357/acb59a}, \href {https://ui.adsabs.harvard.edu/abs/2023ApJ...945...53V} {945, 53}

\bibitem[\protect\citeauthoryear{{Vassilev}, {Perales}, {Miniatura}, {Robert}, {Reinhardt}, {Vecchiocattivi}  \& {Baudon}}{{Vassilev} et~al.}{1990}]{Vassilev1990}
{Vassilev} G.,  {Perales} F.,  {Miniatura} C.,  {Robert} J.,  {Reinhardt} J.,  {Vecchiocattivi} F.,   {Baudon} J.,  1990, \mn@doi [Zeitschrift fur Physik D Atoms Molecules Clusters] {10.1007/BF01437663}, \href {https://ui.adsabs.harvard.edu/abs/1990ZPhyD..17..101V} {17, 101}

\bibitem[\protect\citeauthoryear{{Virtanen} et~al.,}{{Virtanen} et~al.}{2020}]{Virtanen2020_Scipy}
{Virtanen} P.,  et~al., 2020, \mn@doi [Nature Methods] {10.1038/s41592-019-0686-2}, \href {https://ui.adsabs.harvard.edu/abs/2020NatMe..17..261V} {17, 261}

\bibitem[\protect\citeauthoryear{Vrinceanu \& Flannery}{Vrinceanu \& Flannery}{2001}]{Vrinceaunu2001}
Vrinceanu D.,  Flannery M.~R.,  2001, \mn@doi [Phys. Rev. A] {10.1103/PhysRevA.63.032701}, 63, 032701

\bibitem[\protect\citeauthoryear{{Wang} et~al.,}{{Wang} et~al.}{2020}]{Wang2020}
{Wang} Y.,  et~al., 2020, \mn@doi [\aap] {10.1051/0004-6361/201935866}, \href {https://ui.adsabs.harvard.edu/abs/2020A&A...634A.139W} {634, A139}

\bibitem[\protect\citeauthoryear{{Weingartner} \& {Draine}}{{Weingartner} \& {Draine}}{2001}]{WeingartnerDraine2001}
{Weingartner} J.~C.,  {Draine} B.~T.,  2001, \mn@doi [\apj] {10.1086/318651}, \href {https://ui.adsabs.harvard.edu/abs/2001ApJ...548..296W} {548, 296}

\bibitem[\protect\citeauthoryear{{Weisz}, {Savino}  \& {Dolphin}}{{Weisz} et~al.}{2023}]{Weisz2023}
{Weisz} D.~R.,  {Savino} A.,   {Dolphin} A.~E.,  2023, \mn@doi [\apj] {10.3847/1538-4357/acc328}, \href {https://ui.adsabs.harvard.edu/abs/2023ApJ...948...50W} {948, 50}

\bibitem[\protect\citeauthoryear{{Wheeler} et~al.,}{{Wheeler} et~al.}{2019}]{Wheeler2019}
{Wheeler} C.,  et~al., 2019, \mn@doi [\mnras] {10.1093/mnras/stz2887}, \href {https://ui.adsabs.harvard.edu/abs/2019MNRAS.490.4447W} {490, 4447}

\bibitem[\protect\citeauthoryear{{Wiscombe} \& {Joseph}}{{Wiscombe} \& {Joseph}}{1977}]{Wiscombe1977}
{Wiscombe} W.~J.,  {Joseph} J.~H.,  1977, \mn@doi [\icarus] {10.1016/0019-1035(77)90008-2}, \href {https://ui.adsabs.harvard.edu/abs/1977Icar...32..362W} {32, 362}

\bibitem[\protect\citeauthoryear{{Wise}, {Abel}, {Turk}, {Norman}  \& {Smith}}{{Wise} et~al.}{2012}]{Wise2012radpressure}
{Wise} J.~H.,  {Abel} T.,  {Turk} M.~J.,  {Norman} M.~L.,   {Smith} B.~D.,  2012, \mn@doi [\mnras] {10.1111/j.1365-2966.2012.21809.x}, \href {https://ui.adsabs.harvard.edu/abs/2012MNRAS.427..311W} {427, 311}

\bibitem[\protect\citeauthoryear{{Wise}, {Regan}, {O'Shea}, {Norman}, {Downes}  \& {Xu}}{{Wise} et~al.}{2019}]{Wise2019}
{Wise} J.~H.,  {Regan} J.~A.,  {O'Shea} B.~W.,  {Norman} M.~L.,  {Downes} T.~P.,   {Xu} H.,  2019, \mn@doi [\nat] {10.1038/s41586-019-0873-4}, \href {https://ui.adsabs.harvard.edu/abs/2019Natur.566...85W} {566, 85}

\bibitem[\protect\citeauthoryear{{Wolcott-Green} \& {Haiman}}{{Wolcott-Green} \& {Haiman}}{2019}]{Wolcott2019}
{Wolcott-Green} J.,  {Haiman} Z.,  2019, \mn@doi [\mnras] {10.1093/mnras/sty3280}, \href {https://ui.adsabs.harvard.edu/abs/2019MNRAS.484.2467W} {484, 2467}

\bibitem[\protect\citeauthoryear{{Wouthuysen}}{{Wouthuysen}}{1952}]{Wouthuysen1952}
{Wouthuysen} S.~A.,  1952, \mn@doi [\aj] {10.1086/106661}, \href {https://ui.adsabs.harvard.edu/abs/1952AJ.....57R..31W} {57, 31}

\bibitem[\protect\citeauthoryear{{Xiao} et~al.,}{{Xiao} et~al.}{2024}]{Xiao2024}
{Xiao} M.,  et~al., 2024, \mn@doi [\nat] {10.1038/s41586-024-08094-5}, \href {https://ui.adsabs.harvard.edu/abs/2024Natur.635..311X} {635, 311}

\bibitem[\protect\citeauthoryear{{Yajima} \& {Khochfar}}{{Yajima} \& {Khochfar}}{2014}]{Yajima2014}
{Yajima} H.,  {Khochfar} S.,  2014, \mn@doi [\mnras] {10.1093/mnras/stu505}, \href {https://ui.adsabs.harvard.edu/abs/2014MNRAS.441..769Y} {441, 769}

\bibitem[\protect\citeauthoryear{{Yuan}, {Martin-Alvarez}, {Haehnelt}, {Garel}  \& {Sijacki}}{{Yuan} et~al.}{2024}]{Yuan2024}
{Yuan} Y.,  {Martin-Alvarez} S.,  {Haehnelt} M.~G.,  {Garel} T.,   {Sijacki} D.,  2024, \mn@doi [\mnras] {10.1093/mnras/stae1606}, \href {https://ui.adsabs.harvard.edu/abs/2024MNRAS.532.3643Y} {532, 3643}

\bibitem[\protect\citeauthoryear{{Zheng} \& {Wallace}}{{Zheng} \& {Wallace}}{2014}]{Zheng2014}
{Zheng} Z.,  {Wallace} J.,  2014, \mn@doi [\apj] {10.1088/0004-637X/794/2/116}, \href {https://ui.adsabs.harvard.edu/abs/2014ApJ...794..116Z} {794, 116}

\bibitem[\protect\citeauthoryear{{Zwillinger}}{{Zwillinger}}{1992}]{Zwillinger1992_Diffeqs}
{Zwillinger} D.,  1992, {Handbook of differential equations}.
{Academic Press}

\makeatother
\end{thebibliography}



\appendix
\onecolumn

\section{Lyman-\texorpdfstring{\boldmath$\alpha$}{alpha} Radiative transfer}

In this Appendix we derive the Ly$\alpha$ radiative transfer equation (Appendix \ref{AppendixRTeq}), solve for the frequency dependence (Appendix \ref{AppendixRTfreq}), and derive some quantities of interest, including the emergent spectrum (Appendix \ref{AppendixSpectrum}), the Ly$\alpha$ escape fraction (Appendix \ref{Escape fraction appendix}), and the Ly$\alpha$ radiation pressure (Appendix \ref{AppendixRadPressure}).

\subsection{The Ly\texorpdfstring{\boldmath$\alpha$}{alpha} radiative transfer equation}
\label{AppendixRTeq}

We now give a derivation of the Ly$\alpha$ radiative transfer (RT) equation, valid in the optically thick regime, and allowing for Ly$\alpha$ destruction, (non-relativistic) gas motion, and atomic recoil. We start with the RT equation for the Ly$\alpha$ intensity $I(\nu, \boldsymbol{r}, \boldsymbol{\hat{n}}, t)$ in the comoving frame \citep[e.g.][]{Mihalas1975, Loeb1999, Castor2004}:
\begin{align}
    \dfrac{1}{c} \dfrac{\partial I}{\partial t} + \boldsymbol{\hat{n}} \boldsymbol{\cdot} \boldsymbol{\nabla} I - \dfrac{1}{c} \boldsymbol{\hat{n}} \boldsymbol{\cdot} \boldsymbol{\nabla} \boldsymbol{u}\boldsymbol{\cdot} \boldsymbol{\hat{n}} \nu \dfrac{\partial I}{\partial \nu} ~&=~ j_{\rm s} - (\alpha + \alpha_{\rm c}) I + \omega \alpha_{\rm c} \int_{4\pi} \dfrac{\textrm{d}\Omega'}{4 \pi} I(x, \boldsymbol{r}, \boldsymbol{\hat{n}}', t) \Phi_{\rm c} (\boldsymbol{\hat{n}}, \boldsymbol{\hat{n}}') \nonumber \\ ~&+~ \alpha_0 \int_{4\pi} \dfrac{\textrm{d}\Omega'}{4 \pi} \int_{0}^{\infty} \textrm{d}x' \hspace{1 pt}  I(x', \boldsymbol{r}, \boldsymbol{\hat{n}}', t) R_{\rm II}(x, x',\boldsymbol{\hat{n}}, \boldsymbol{\hat{n}}') \hspace{1 pt} \label{A1_LyaGeneralEq} ,
\end{align}
where the term proportional to $\partial I/\partial \nu$ describes the Doppler shift due to the gas velocity $\boldsymbol{u}$.\footnote{We have ignored aberration and advection terms here, which is valid when we are dealing with a resonant line and $\lvert \boldsymbol{u} \rvert/c \ll 1$ \citep{Mihalas1975}. See e.g. \cite{Castor2004} for the more general case where these extra terms are included.} As for the other terms, $j_{\rm s}(x, \boldsymbol{r})$ is the (assumed isotropic) emissivity (in erg s$^{-1}$ cm$^{-3}$ Hz$^{-1}$ sr$^{-1}$), and $\alpha(\nu, \boldsymbol{r}) = n_{\rm HI} \sigma_{0} \mathcal{H}(x) \equiv \alpha_0 \mathcal{H}(x)$ is the Ly$\alpha$ extinction coefficient (in cm$^{-1}$). We have also taken into account continuum absorption/scattering, with extinction coefficient $\alpha_{\rm c}$, single-scattering albedo $\omega$, and scattering phase function $\Phi_{\rm c}(\boldsymbol{\hat{n}}, \boldsymbol{\hat{n}}')$. This allow us to consider e.g. absorption/scattering by dust, or Raman scattering by H$_2$ \citep[e.g.][]{Dalgarno1962, Oklopcic2016}. To model potentially anisotropic scattering we can use the $\delta$-Eddington approximation \citep{Joseph1976,Wiscombe1977}, often employed in climate models to model anisotropic scattering in the atmosphere \cite[e.g.][]{NCAR1992}. Its application in astrophysics will be discussed in greater detail in a separate paper on radiation pressure feedback on dust (Nebrin, in prep.). In the $\delta$-Eddington approximation the `true' phase function \citep[e.g.][]{Henyey1941, Draine2003albedo} is replaced by an approximation:
\begin{equation}
    \Phi_{\rm c, \delta-Edd}(\boldsymbol{\hat{n}}, \boldsymbol{\hat{n}}') = f \delta_{\rm D}(1 - \boldsymbol{\hat{n}} \boldsymbol{\cdot} \boldsymbol{\hat{n}}') + (1-f)(1 + 3 g' \boldsymbol{\hat{n}} \boldsymbol{\cdot} \boldsymbol{\hat{n}}') \hspace{1 pt}.
\end{equation}
Here it is assumed that some fraction $f$ of the light is scattered in the forward direction, and the remaining fraction $1-f$ is scattered in a diffuse manner. The parameters $f$ and $g'$ are fixed to reproduce the scattering asymmetry\footnote{A scattering asymmetry $g = 1$, $g = -1$, and $g = 0$ would correspond to forward, backward, and isotropic scattering, respectively. Observational constraints and dust models typically give $g \sim 0.7$ around Ly$\alpha$ line centre \citep{Draine2003albedo, Gordon2004, Akshaya2019}.} $g \equiv \langle  \boldsymbol{\hat{n}} \boldsymbol{\cdot} \boldsymbol{\hat{n}}' \rangle$ of the true phase function, which yields the following consistency condition:
\begin{equation}
    f+(1-f)g' = g \hspace{1 pt}.
\end{equation}
For later reference, we have:
\begin{align}
    \int_{4\pi} \dfrac{\textrm{d}\Omega'}{4 \pi} I(x, \boldsymbol{r}, \boldsymbol{\hat{n}}', t) \Phi_{\rm c, \delta-Edd} (\boldsymbol{\hat{n}}, \boldsymbol{\hat{n}}')  ~&=~ fI(x, \boldsymbol{r}, \boldsymbol{\hat{n}}, t) + (1-f) J(x, \boldsymbol{r}, t) + 3(1-f)g'  \boldsymbol{\hat{n}} \boldsymbol{\cdot} \boldsymbol{H}(x, \boldsymbol{r}, t) \hspace{1 pt} , \label{zeroth moment of continuum scattering phase function} \\  \int_{4\pi} \dfrac{\textrm{d}\Omega}{4\pi} \int_{4\pi} \dfrac{\textrm{d}\Omega'}{4 \pi} I(x, \boldsymbol{r}, \boldsymbol{\hat{n}}', t) \Phi_{\rm c, \delta-Edd} (\boldsymbol{\hat{n}}, \boldsymbol{\hat{n}}') ~&=~ J(x, \boldsymbol{r}, t) \hspace{1 pt} \label{first moment of continuum scattering phase function}, \\ \int_{4\pi} \dfrac{\textrm{d}\Omega}{4\pi} \hspace{1 pt} \boldsymbol{\hat{n}} \int_{4\pi} \dfrac{\textrm{d}\Omega'}{4 \pi} I(x, \boldsymbol{r}, \boldsymbol{\hat{n}}', t) \Phi_{\rm c, \delta-Edd} (\boldsymbol{\hat{n}}, \boldsymbol{\hat{n}}') ~&=~ f \boldsymbol{H} (x, \boldsymbol{r}, t) + (1-f)g'  \boldsymbol{H} (x, \boldsymbol{r}, t) \hspace{1 pt} \label{second moment of continuum scattering phase function}, 
\end{align}
where $J$ is the mean specific intensity (zeroth moment of $I$), and $\boldsymbol{H}$ is the first moment of $I$, given in Eq.~(\ref{A1moments}) below. 

Finally, $R_{\rm II}(x, x',\boldsymbol{\hat{n}}, \boldsymbol{\hat{n}}')$ in Eq.~(\ref{A1_LyaGeneralEq}) is the Ly$\alpha$ redistribution function, which is proportional to the probability that a photon absorbed by H$(1s)$ at frequency $x'$ and direction $\boldsymbol{\hat{n}}'$ will be re-emitted with a new frequency $x$ and direction $\boldsymbol{\hat{n}}$ \citep[for explicit expressions, see e.g.][]{Hummer1962, Smith2018_discretediffusion}. The problem can be made analytically tractable using a Fokker-Planck approach \citep[e.g.][]{Unno1955, Harrington1973, Rybicki1994, Chuzhoy2006, Meiksin2006, Rybicki2006, McClellan2022}. In this approach one can Taylor expand $I(x', \boldsymbol{r}, \boldsymbol{\hat{n}}', t)$ in the integral in Eq.~(\ref{A1_LyaGeneralEq}) to second order in $\Delta x \equiv x' - x$. If one evaluate the resulting integrals using the frequency moments of $R_{\rm II}(x, x',\boldsymbol{\hat{n}}, \boldsymbol{\hat{n}}')$, one can show that \citep[see e.g.][]{Rybicki1994, Chuzhoy2006, Rybicki2006, McClellan2022}:
\begin{equation}
    \int_{4\pi} \dfrac{\textrm{d}\Omega'}{4 \pi}  \int_{0}^{\infty} \textrm{d}x' \hspace{1 pt}  I(x', \boldsymbol{r}, \boldsymbol{\hat{n}}', t) R_{\rm II}(x, x',\boldsymbol{\hat{n}}, \boldsymbol{\hat{n}}') \simeq (1-p_{\rm d}) \mathcal{H}J+  (1-p_{\rm d}) \dfrac{1}{2} \dfrac{\partial}{\partial x} \left( \mathcal{H} \dfrac{\partial J}{\partial x} + 2 \Bar{x} \mathcal{H} J \right) \hspace{1 pt} \label{expansion of redistribution function}, 
\end{equation}
where $p_{\rm d}$ is the probability that the Ly$\alpha$ photon is not re-emitted (i.e. the destruction probability), $\mathcal{H}(x)$ is the Voigt profile (Eq.~\ref{Voigt definition}), and $\Bar{x}$ is the recoil parameter. The recoil parameter is equal to the average change in $x$ due to recoil, i.e. $\Bar{x} \equiv \langle \Delta x \rangle_{\rm recoil}$, and given by:
\begin{equation}
    \Bar{x} = \dfrac{h \Delta \nu_{\rm D}}{2 k_{\rm B} T} = 2.54 \times 10^{-3} \left( \dfrac{T}{100\,\rm K} \right)^{-1/2} \hspace{1 pt}.
\end{equation}
The implication of recoil is that the Ly$\alpha$ photon is slowly losing energy to the gas, and thereby pushed to the red side of the spectrum (note that $x'$ is the frequency before the scattering, so $\Delta x > 0$ corresponds to the photon losing energy). We note that in writing Eq.~(\ref{expansion of redistribution function}) we have neglected additional small corrections from hyperfine transitions \citep{Chuzhoy2006}, and assumed that the Ly$\alpha$ intensity is nearly isotropic so that anisotropic corrections in the Fokker-Planck expansion are small \citep{Rybicki1994, McClellan2022}. 
Using Eqs.~(\ref{expansion of redistribution function}) and (\ref{zeroth moment of continuum scattering phase function}) in Eq.~(\ref{A1_LyaGeneralEq}) then gives us 
\begin{align}
    \dfrac{1}{c} \dfrac{\partial I}{\partial t} + \boldsymbol{\hat{n}} \boldsymbol{\cdot} \boldsymbol{\nabla} I - \dfrac{1}{c} \boldsymbol{\hat{n}} \boldsymbol{\cdot} \boldsymbol{\nabla} \boldsymbol{u}\boldsymbol{\cdot} \boldsymbol{\hat{n}} \nu \dfrac{\partial I}{\partial \nu} ~&=~ j_{\rm s} - (\alpha + \alpha_{\rm c}) I + \omega  \alpha_{\rm c} fI + \omega \alpha_{\rm c} (1-f) J + 3 \omega \alpha_{\rm c} (1-f) g' \boldsymbol{\hat{n}} \boldsymbol{\cdot} \boldsymbol{H} \nonumber \\ ~&+~ \alpha_0 (1 - p_{\rm d}) \left[\mathcal{H} J +  \dfrac{1}{2} \dfrac{\partial}{\partial x} \left( \mathcal{H} \dfrac{\partial J}{\partial x} +  2 \Bar{x} \mathcal{H} J\right) \right] \hspace{1 pt} \label{A1_LyaGeneralEq2} .
\end{align}
Next we want a single equation for the mean specific intensity $J$ (the zeroth moment of $I$). We define the usual zeroth, first, second, and third angular moments of the intensity:
\begin{align}
    J(x, \boldsymbol{r}, t) &\equiv  \int_{4\pi} \dfrac{\textrm{d}\Omega}{4 \pi}  \hspace{1 pt} I(x, \boldsymbol{r}, \boldsymbol{\hat{n}}, t) \hspace{1 pt} , \hspace{10 pt} \boldsymbol{H}(x, \boldsymbol{r}, t) \equiv \int_{4\pi} \dfrac{\textrm{d}\Omega}{4 \pi} \hspace{1 pt} I(x, \boldsymbol{r}, \boldsymbol{\hat{n}}, t) \hspace{1 pt} \boldsymbol{\hat{n}} \hspace{1 pt} , \nonumber \\ \mathbfss{K}(x, \boldsymbol{r}, t) &\equiv \int_{4\pi} \dfrac{\textrm{d}\Omega}{4 \pi} \hspace{1 pt} I(x, \boldsymbol{r}, \boldsymbol{\hat{n}}, t) \hspace{1 pt} \boldsymbol{\hat{n}} \otimes \boldsymbol{\hat{n}} \hspace{1 pt} , \hspace{10 pt} \mathbfss{L}(x, \boldsymbol{r}, t) \equiv \int_{4\pi} \dfrac{\textrm{d}\Omega}{4 \pi} \hspace{1 pt} I(x, \boldsymbol{r}, \boldsymbol{\hat{n}}, t) \hspace{1 pt} \boldsymbol{\hat{n}} \otimes \boldsymbol{\hat{n}} \otimes \boldsymbol{\hat{n}} \hspace{1 pt} \label{A1moments}.
\end{align}
Here $\boldsymbol{H}$ is related to the specific flux via $\boldsymbol{H} = \boldsymbol{F}/4 \pi$. Taking the zeroth and first angular moments of Eq.~(\ref{A1_LyaGeneralEq2}), along with using Eqs.~(\ref{zeroth moment of continuum scattering phase function}), (\ref{first moment of continuum scattering phase function}), and (\ref{second moment of continuum scattering phase function}), then yields
\begin{align}
      \dfrac{1}{c} \dfrac{\partial J}{\partial t} + \boldsymbol{\nabla} \boldsymbol{\cdot} \boldsymbol{H} - \dfrac{\nu}{c }  \dfrac{\partial  \mathbfss{K}}{\partial \nu} \boldsymbol{:} \boldsymbol{\nabla u}  ~&=~ j_{\rm s} - \alpha_{\rm c}(1-\omega)J - p_{\rm d} \alpha J + \alpha_0 (1 - p_{\rm d})  \dfrac{1}{2} \dfrac{\partial}{\partial x} \left( \mathcal{H} \dfrac{\partial J}{\partial x} +  2 \Bar{x} \mathcal{H} J \right)  \hspace{1 pt} , \\
      \dfrac{1}{c} \dfrac{\partial \boldsymbol{H}}{\partial t} + \boldsymbol{\nabla} \boldsymbol{\cdot} \mathbfss{K} - \dfrac{\nu}{c }  \dfrac{\partial \mathbfss{L}}{\partial \nu} \boldsymbol{:} \boldsymbol{\nabla u}  ~&=~  - [\alpha + \alpha_{\rm c}(1 - \omega g)] \boldsymbol{H} \hspace{1 pt} \label{div K equation},
\end{align}
where $\mathbfss{K} \boldsymbol{:} \boldsymbol{\nabla u} \equiv \mathbfss{K}_{ij} \partial_{i} u_{j}$, and $\mathbfss{L} \boldsymbol{:} \boldsymbol{\nabla u} \equiv \mathbfss{L}_{ijk} \partial_{i} u_{j}$, with summation over repeated indices. To proceed we will make four approximations: 
\begin{enumerate}
    \item \textbf{Time-independence}: We will assume steady-state conditions, so that $\partial J/\partial t = 0$ and $\partial \boldsymbol{H}/\partial t = 0$. This is a good approximation for our purposes, since the Ly$\alpha$ trapping time $t_{\rm trap} \sim (a_{\rm v} \taucl)^{1/3} (R_{\rm cl}/c)$ is typically much shorter than characteristic time-scales of star formation (e.g. the free-fall time-scale). Future work could relax this assumption, following \cite{McClellan2022}. \\
    
    \item \textbf{Eddington approximation}: We will --- as alluded to earlier --- assume the Eddington approximation, in which the intensity is assumed to be nearly isotropic in the comoving frame. This yields the closure relation $\mathbfss{K} = \mathbfss{1} \hspace{1 pt} J/3$, where $\mathbfss{1}$ is the identity matrix. The Eddington approximation is expected to be valid for the very optically thick clouds of interest to us \citep[for numerical tests see e.g.][]{Dijkstra2008, Smith2017}. Furthermore, we had implicitly assumed the Eddington approximation when deriving the Fokker-Planck approximation --- abandoning this assumption would introduce additional terms that would become increasingly important as the intensity becomes more anisotropic \citep{Rybicki1994}. \\
    
    \item \textbf{Continuum extinction coefficient}: We will assume that the continuum extinction coefficient is small compared to the Ly$\alpha$ extinction coefficient, so that $\alpha \gg \alpha_{\rm c}(1-\omega g)$ in Eq.~(\ref{div K equation}) for all frequencies of interest, $\lvert x \rvert \lesssim 2 (a_{\rm v}\taucl)^{1/3}$. This assumption was also made by \cite{Neufeld1990} and \cite{Tomaselli2021}, and is expected to be valid for all reasonable dust abundances. In terms of the dust absorption cross-section per H nucleon, this condition becomes $\sigma_{\rm d,abs,H} \ll \sigma_0 a_{\rm v}/4 \sqrt{\pi} (a_{\rm v} \taucl)^{2/3}$, where we have used $(1-\omega g) \simeq (1-\omega)$. From the dust models in Table \ref{Dust models}, we then get the condition on the dust-to-gas ratio: 
    \begin{equation}
        \mathscr{D}/\mathscr{D}_{\odot} \ll 3 \, \left( \dfrac{\taucl}{10^{10}} \right)^{-2/3} \left( \dfrac{T}{100 \, \rm K} \right)^{-2/3} \, .
    \end{equation}
    In reality, since dust absorption makes the Ly$\alpha$ line more narrow (see Fig. \ref{spectrum}), we would get a less strict condition. It follows that the approximation made here, $\alpha \gg \alpha_{\rm c}(1-\omega g)$, is valid for optically thick clouds with dust-to-gas ratios up to at least $\sim \mathscr{D}_{\odot}$.
    \\

    \item \textbf{Neglected term}: Finally, we will neglect the term $\propto \partial \mathbfss{L}/\partial \nu$ in Eq.~(\ref{div K equation}) to render the problem analytically tractable. This term has been argued to be unimportant in most radiation hydrodynamical contexts by \cite{Buchler1983} \citep[see also][]{Castor2004}, although it is sometimes retained \citep[e.g.][]{Hillier2012}. For Ly$\alpha$ RT, we can check when it is expected to become important with the following order of magnitude estimate. First, we note that $\mathbfss{L} \boldsymbol{: \nabla u}$ is of order $\sim \boldsymbol{H} \Dot{R}_{\rm cl}/R_{\rm cl}$ for a cloud experiencing radial expansion or contraction at rate $\Dot{R}_{\rm cl}$.\footnote{This can be seen either by expanding the term, as done by \cite{Buchler1983} (see his eq. 38), or by starting from the assumption of Hubble-like expansion/contraction, $u(r) \propto r$, and working out the corresponding term \citep[e.g.][]{Mihalas1975, Hillier2012}.} If this term indeed \textit{is} negligible, we would get Fick's law from Eq.~(\ref{div K equation}), i.e., $\boldsymbol{H} = -\frac{1}{3} (\boldsymbol{\nabla}J)/\alpha$. Thus, at least up until this term becomes important we have, up to a sign:
    \begin{equation}
        \dfrac{\nu}{c }  \dfrac{\partial \mathbfss{L}}{\partial \nu} \boldsymbol{:} \boldsymbol{\nabla u} \sim \dfrac{\nu_{\rm Ly\alpha}}{3c} \dfrac{\Dot{R}_{\rm cl}}{R_{\rm cl}} \dfrac{\partial }{\partial \nu} \left( \dfrac{1}{\alpha} \boldsymbol{\nabla} J \right) = \dfrac{\Dot{R}_{\rm cl}}{3b \taucl} \dfrac{\partial }{\partial x} \left( \dfrac{1}{\mathcal{H}} \boldsymbol{\nabla} J \right) \simeq \dfrac{\Dot{R}_{\rm cl}}{3b }  \left( \dfrac{2 \sqrt{\pi} x}{a_{\rm v}\taucl} \boldsymbol{\nabla}J + \dfrac{\sqrt{\pi}x^2}{a_{\rm v}\taucl} \dfrac{\partial }{\partial x} \boldsymbol{\nabla} J \right) \sim 2 \left( \dfrac{\Dot{R}_{\rm cl}}{b } \right) \left( \dfrac{ x}{a_{\rm v}\taucl}\right) \boldsymbol{\nabla}J \, .
    \end{equation}
    Here we have used $\Delta \nu_{\rm D} = (b/c) \nu_{\rm Ly\alpha}$, and the wing limit for $\mathcal{H}$ (Eq.~\ref{VoigtApprox}), and in the last expression, we have crudely estimated that $\partial/\partial x (\boldsymbol{\nabla}J) \sim (\boldsymbol{\nabla}J)/x$ (up to a sign). Thus, this term will be negligible compared to $\alpha \boldsymbol{H} \simeq - \frac{1}{3}\boldsymbol{\nabla}J$ as long as
    \begin{equation}
        \dfrac{\lvert \Dot{R}_{\rm cl} \rvert}{b} \ll \dfrac{1}{6} \left(\dfrac{\lvert x \rvert}{a \taucl} \right)^{-1} \hspace{1 pt} .
    \end{equation}
    Since most of the Ly$\alpha$ intensity is expected to be concentrated around $\lvert x \rvert \lesssim 2 (a_{\rm v} \taucl)^{1/3}$ (see Fig. \ref{spectrum}), we arrive at the condition:
    \begin{equation}
        \dfrac{\lvert \Dot{R}_{\rm cl} \rvert}{b} \ll \dfrac{1}{12} (a_{\rm v} \taucl)^{2/3} \sim 500 \, \left( \dfrac{T}{100 \, \rm K} \right)^{-1/3} \left( \dfrac{\taucl}{10^8} \right)^{2/3} \hspace{1 pt} , \hspace{10 pt} \textrm{or equivalently:}  \hspace{10 pt} \lvert \Dot{R}_{\rm cl} \rvert \ll 600 \, \left( \dfrac{T}{100 \, \rm K} \right)^{1/6} \left( \dfrac{\taucl}{10^8} \right)^{2/3} \hspace{1 pt} \rm km \, s^{-1} \, .
    \end{equation}
    If this condition is satisfied we expect that we can safely drop the $\partial \mathbfss{L} / \partial \nu$ term in Eq.~(\ref{div K equation}). For dense and bound star-forming clouds, this condition is expected to hold. Even Ly$\alpha$ feedback itself is predicted to drive winds with $\Dot{R}_{\rm cl} \sim 10 - 200 \, \rm km \, s^{-1}$ \citep{Dijkstra2008, Smith2016, Smith2017}.\footnote{Feedback from, e.g., SNe can also drive fast winds. However, in this case, we expect the gas to be heated and largely ionized to the extent that, together with the velocity gradients, Ly$\alpha$ feedback plays at best a subdominant role. Thus, even if our Ly$\alpha$ RT solution becomes somewhat less accurate in this regime, it would have little impact on the predicted outflows in galaxy formation simulations.}

\end{enumerate}
We now apply these four approximations. From the Eddington approximation we have $\mathbfss{K} \boldsymbol{: \nabla u} = \frac{1}{3} J \, \delta_{ij} \partial_i u_j = \frac{1}{3} J \, \boldsymbol{\nabla \cdot u}$, and $\boldsymbol{\nabla} 
\boldsymbol{\cdot} \mathbfss{K} =  \frac{1}{3} \boldsymbol{\nabla}J$. Together with the other approximations we arrive at
\begin{align}
     \boldsymbol{\nabla} \boldsymbol{\cdot} \boldsymbol{H} - \dfrac{1}{3 b} (\boldsymbol{\nabla \cdot u}) \, \dfrac{\partial J}{\partial x} ~&=~ j_{\rm s} - \alpha_{\rm c} (1-\omega)J - p_{\rm d} \alpha J + \alpha_0 (1 - p_{\rm d})  \dfrac{1}{2} \dfrac{\partial}{\partial x} \left( \mathcal{H} \dfrac{\partial J}{\partial x} +  2 \Bar{x} \mathcal{H} J \right)  \hspace{1 pt} , \\
     \dfrac{1}{3} \boldsymbol{\nabla} J  ~&=~  - \alpha \boldsymbol{H} \hspace{1 pt} ,
\end{align}
where we have simplified the Doppler term using $\nu/c \simeq \nu_{\rm Ly\alpha}/c$, written the frequency derivative with respect to $x$ instead of $\nu$, and used $\Delta \nu_{\rm D} = (b/c) \nu_{\rm Ly \alpha}$. These can be combined into a single equation for $J$. After a little algebra we get:
\begin{equation}
     \dfrac{1}{\alpha_0} \boldsymbol{\nabla} \boldsymbol{\cdot} \left( \dfrac{1}{\alpha_0}  \boldsymbol{\nabla} J  \right) + \dfrac{(\boldsymbol{\nabla \cdot u})}{b \alpha_0 } \mathcal{H} \dfrac{\partial J}{\partial x} = - \dfrac{3 \mathcal{H} j_{\rm s}}{\alpha_0} + 3 (p_{\rm d} \mathcal{H}^2 + \epsilon \mathcal{H}) J - (1 - p_{\rm d})  \dfrac{3}{2} \mathcal{H} \dfrac{\partial}{\partial x} \left(  \mathcal{H} \dfrac{\partial J}{\partial x} +  2 \Bar{x} \mathcal{H} J \right)  \hspace{1 pt} \label{Appendix Lya RT equation final general result},
\end{equation}
where we have introduced the continuum absorption parameter $\epsilon$:
\begin{equation}
    \epsilon \equiv \dfrac{\alpha_{\rm c}(1-\omega)}{\alpha_0} \hspace{1 pt} .
\end{equation}
Eq.~(\ref{Appendix Lya RT equation final general result}) is generally valid as long as the previously mentioned approximations hold -- even for highly inhomogeneous clouds.\footnote{As long as the density does not vary on scales smaller than or comparable to the Ly$\alpha$ mean free path. For large density fluctuations on smaller scales, one could use a slightly modified closure relation following \cite{Begelman2006}. However, in typical ISM conditions, the Eddington approximation suffices. } In the special case of a uniform spherical cloud ($\alpha_0 = \rm const.$) of radius $R_{\rm cl}$, and Hubble-like radial expansion/contraction, $\boldsymbol{u} = (\Dot{R}_{\rm cl}/R_{\rm cl})r \boldsymbol{\hat{r}}$, Eq.~(\ref{Appendix Lya RT equation final general result}) is simplified further with the following replacements:
\begin{equation}
    \dfrac{1}{\alpha_0} \boldsymbol{\nabla} \boldsymbol{\cdot} \left( \dfrac{1}{\alpha_0}  \boldsymbol{\nabla} J  \right) \rightarrow \dfrac{\partial ^2 J}{\partial \tau^2} + \dfrac{2}{\tau} \dfrac{\partial J}{\partial \tau} \, , \hspace{15 pt} \dfrac{(\boldsymbol{\nabla \cdot u})}{b \alpha_0 } \mathcal{H} \dfrac{\partial J}{\partial x} \rightarrow \dfrac{3\Dot{R}_{\rm cl}}{b \taucl } \mathcal{H} \dfrac{\partial J}{\partial x} \, ,  
\end{equation}
where $\tau = n_{\rm HI} \sigma_0 r$ and $\taucl =  n_{\rm HI} \sigma_0 R_{\rm cl}$ are the optical depth to radius $r$ and the cloud edge $R_{\rm cl}$, respectively. We also have $\epsilon = \tau_{\rm c,abs}/\taucl$, where $\tau_{\rm c,abs} = \alpha_{\rm c} (1 - \omega) R_{\rm cl}$ is the cloud continuum absorption optical depth. With these replacements in Eq.~(\ref{Appendix Lya RT equation final general result}), and using $p_{\rm d} \ll 1$, we arrive at Eq.~(\ref{radtransfereq}) in the main text for a spherical uniform cloud, undergoing Hubble-like expansion/contraction. Before concluding, we also note two immediate consequences of Eq.~(\ref{Appendix Lya RT equation final general result}). First, there is no dependence on the continuum scattering asymmetry $g$, and secondly, the Doppler term depends on the \textit{divergence} of the velocity field. Both are consistent with Monte Carlo results, which find that the Ly$\alpha$ escape fraction has no noticeable dependence on $g$ \citep{Laursen2009}, and that bulk rotation (giving $\boldsymbol{\nabla \cdot u}=0$) has no significant impact on Ly$\alpha$ scattering and escape up to at least $300 \rm \, km \, s^{-1}$ \citep{Garavito2014}.

\subsection{Solution for the frequency dependence of the Ly\texorpdfstring{\boldmath$\alpha$}{alpha} mean intensity}
\label{AppendixRTfreq}

\begin{figure*}
\includegraphics[width=0.9\textwidth]{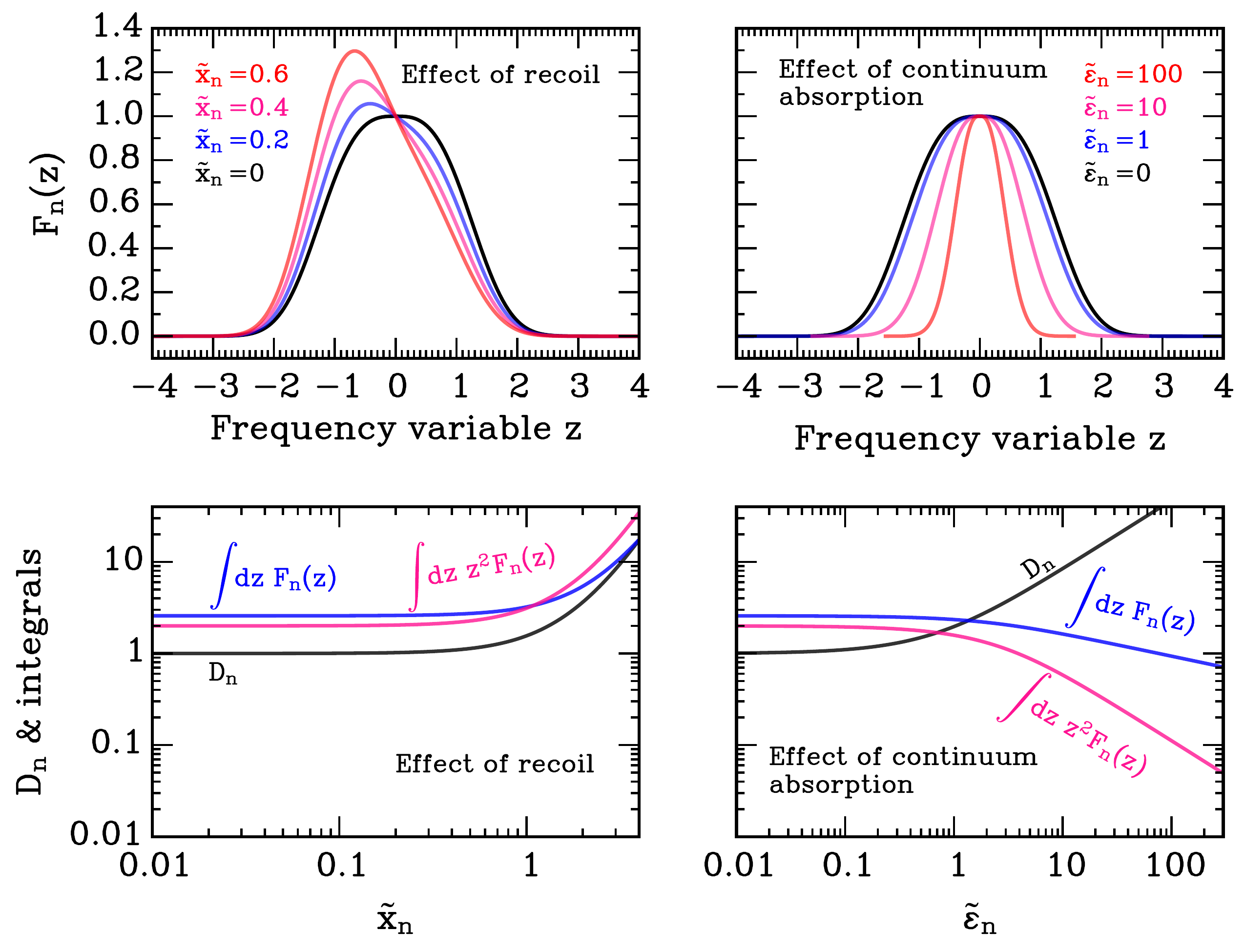}
\caption{ Plot of the series solution $\mathcal{F}_n(z)$ and related quantities. \textbf{Upper left panel}: The series solution $\mathcal{F}_n (z)$ for different values of the dimensionless recoil parameter $\Tilde{x}_n$, keeping other parameters zero ($\Tilde{\epsilon}_n = \Tilde{\gamma}_n = 0$). Recoil is seen to produce a red-tilted spectrum, as expected. \textbf{Upper right panel}: Same as the upper left panel, but for the dimensionless continuum absorption parameter $\Tilde{\epsilon}_n$. \textbf{Lower left panel}: Plots of $\mathcal{D}_n$ (black line), $\int_{-\infty}^{\infty} \textrm{d}z \hspace{1 pt} \mathcal{F}_n(z)$ (blue line), and $\int_{-\infty}^{\infty} \textrm{d}z \hspace{1 pt} z^2 \mathcal{F}_n(z)$ (pink line), as a function of the dimensionless recoil parameter $\Tilde{x}$, keeping other parameters zero. \textbf{Lower right panel}: Same as the lower left panel, but for continuum absorption.}
\label{F_n series plot}
\end{figure*}
In the main text, we saw that the Ly$\alpha$ mean specific intensity could be written as $J(x,\tau) = \sum_n T_n(\tau) f_n(x)$, with the frequency dependence $f_{n}$ being the solution to the following differential equation (Eq.~\ref{freqeq2}):
\begin{equation}
    \dfrac{\textrm{d}^2 f_n}{\textrm{d}y^2} + \gamma \dfrac{\textrm{d} f_n}{\textrm{d}y} + \sqrt{6} \Bar{x} \dfrac{\textrm{d}}{\textrm{d}y} \left( \mathcal{H} f_n \right) -  (\lambda_n^2 + 3\epsilon \mathcal{H} )f_n =  \left[ \sqrt{6 \pi} p_{\rm d} f_n - \mathcal{I}_{\textrm{s},n} \right] \delta_{\rm D}(y) \hspace{1 pt} \label{AppendixFrequencyEq}.
\end{equation}
Here we solve Eq.~(\ref{AppendixFrequencyEq}). In the absence of recoil and dust absorption (i.e. $\Bar{x} = \epsilon = 0$), Eq.~(\ref{AppendixFrequencyEq}) can be solved by a simple exponential Ansatz (see Eq.~\ref{Solution f_n no recoil or dust}). With dust and/or recoil, the equation becomes harder to solve, as there are no known closed-form solutions, and a WKB approximation is inapplicable for a wide range of parameters \citep[][]{Tomaselli2021}. Instead we can proceed using a power series Ansatz, following \cite{Neufeld1990} who considered the case with dust absorption (but not recoil and cloud expansion/contraction). Eq.~(\ref{AppendixFrequencyEq}) is more straightforward to solve in terms of $x$. Using the wing limit $\mathcal{H} \simeq a_{\rm v}/\sqrt{\pi}x^2$ gives the following equation away from line centre (i.e. $x \neq 0$),
\begin{equation}
    \dfrac{\textrm{d}^2f_n}{\textrm{d}x^2} + \left( 2 \Bar{x} + \gamma \sqrt{\dfrac{2 \pi}{3}} \dfrac{x^2}{a_{\rm v}} - \dfrac{2}{x} \right)\dfrac{\textrm{d}f_n}{\textrm{d}x} - \left( \dfrac{2 \lambda_n^2 \pi x^4}{3 a_{\rm v}^2} + \dfrac{2 \sqrt{\pi} \epsilon x^2}{ a_{\rm v}} + \dfrac{4\Bar{x}}{x} \right) f_n = 0 \hspace{1 pt} .
\end{equation}
To reduce clutter we can make a variable change, $x \equiv z \beta_n$, with $\beta_n = (3 a_{\rm v}^2/ 2\pi \lambda_n^2)^{1/6}$. We then get
\begin{equation}
    \dfrac{\textrm{d}^2f_n}{\textrm{d}z^2} + \left( \Tilde{x}_n + \Tilde{\gamma}_n z^2 - \dfrac{2}{z} \right)\dfrac{\textrm{d}f_n}{\textrm{d}z} - \left( z^4 + \Tilde{\epsilon}_n z^2 + \dfrac{2\Tilde{x}_n}{z} \right) f_n \equiv \mathcal{L}[f_n(z)] = 0 \hspace{1 pt} , \hspace{5 pt}  \textrm{where} ~ \Tilde{x}_n \equiv 2 \beta_n \Bar{x} \hspace{1 pt} , \hspace{2 pt} \Tilde{\gamma}_n \equiv \gamma \sqrt{2\pi/ 3} \beta_n^3 /a_{\rm v} \hspace{1 pt} , \hspace{2 pt} \Tilde{\epsilon}_n \equiv 2 \sqrt{\pi} \beta_n^4 \epsilon/a_{\rm v}  \hspace{1 pt} \label{diffeq f_n(z)}.
\end{equation}
To avoid clutter in the following derivation of the solution to Eq.~(\ref{diffeq f_n(z)}), we will temporarily drop the subscript $n$ for $\Tilde{x}_n$, $\Tilde{\gamma}_n$, and $\Tilde{\epsilon}_n$ in this Appendix. Thus, for example, $\Tilde{x}$ should be understood to mean $\Tilde{x}_n$ below. 

Eq.~(\ref{diffeq f_n(z)}) is a linear differential equation with a regular singular point at $z = 0$. It can be solved using Frobenius method \citep[e.g.][]{Arfken2013}. The indicial equation for Eq.~(\ref{diffeq f_n(z)}) is $r(r-1) -2r = 0$, which has the two roots $r = 3$ and $r = 0$. Since the roots differ by an integer we will have two linearly independent solutions $S(z)$ and $P(z)$ of the form
\begin{equation}
    S(z) = \sum_{k=0}^{\infty} s_k z^{k+3} \hspace{1 pt} , \hspace{5 pt} P(z) = \mathcal{A} S(z) \ln z + \sum_{k=0}^{\infty} p_k z^{k} \hspace{1 pt} ,
\end{equation}
for some constant $\mathcal{A}$ to be determined (we show it to be zero below). Plugging in a power series of the form $\sum_{k=0}^{\infty} c_k z^{r+k}$ (to represent $S$ and the power series part of $P$) in Eq.~(\ref{diffeq f_n(z)}) yields 
\begin{align}
    \mathcal{L} \sum_{k=0}^{\infty} c_k z^{r+k} ~&=~ \sum_{k = 0}^{\infty} (r+k)(r+k-1) c_k z^{r+k-2} + (\Tilde{x} +\Tilde{\gamma})\sum_{k = 0}^{\infty} (r+k) c_k z^{r+k-1} - 2 \sum_{k = 0}^{\infty} (r+k) c_k z^{r+k-2} \nonumber \\ ~&-~ \sum_{k = 0}^{\infty} c_k z^{r+k+4} - \Tilde{\epsilon} \sum_{k = 0}^{\infty} c_k z^{r+k+2} - 2 \Tilde{x} \sum_{k = 0}^{\infty} c_k z^{r+k-1} \nonumber \\ ~&=~ \sum_{k = 0}^{\infty} (r+k)(r+k-3)c_k z^{r+k-2} + \sum_{k = 0}^{\infty}[\Tilde{x}(r+k-2) + \Tilde{\gamma}(r+k)] c_k z^{r+k-1} - \sum_{k = 0}^{\infty} c_k z^{r+k+4} - \Tilde{\epsilon} \sum_{k = 0}^{\infty} c_k z^{r+k+2}  \hspace{1 pt} .
\end{align}
Expanding, collecting terms, and shifting indices gives us
\begin{align}
    \mathcal{L} \sum_{k=0}^{\infty} c_k z^{r+k} ~&=~ \underbrace{[r(r-1) - 2r]}_{=~0}c_{0} z^{r-2} + [(r+1)(r-2) c_1 + \Tilde{x}(r-2) c_0] z^{r-1} + [(r+2)(r-1)c_2 + \Tilde{x}(r-1)c_1 ] z^r \nonumber \\ ~&+~  [(r+3)r c_3 + \Tilde{x}r c_2 + \Tilde{\gamma} r c_0] z^{r+1} + [(r+4)(r+1)c_4 + \Tilde{x}(r+1)c_3 + \Tilde{\gamma}(r+1) c_1 - \Tilde{\epsilon} c_0 ]z^{r+2} \nonumber \\ ~&+~ [(r+5)(r+2) c_5 + \Tilde{x}(r+2)c_4 + \Tilde{\gamma}(r+2) c_2 - \Tilde{\epsilon} c_1 ]z^{r+3}
    \nonumber \\ ~&+~ \sum_{k = 0}^{\infty} \{ (r+k+6)(r+k+3)c_{k+6} + \Tilde{x}(r+k+3)  c_{k+5} + \Tilde{\gamma}(r+k+3) c_{k+3} - c_k - \Tilde{\epsilon} c_{k+2} \}z^{r+k+4} \hspace{1 pt} . \label{L of c_k z^r+k}
\end{align}
We can now derive the coefficients for $S(z)$, which had $r = 3$. Since $\mathcal{L} [S(z)] = 0$ we find, after fixing $s_0 = 1$:
\begin{align}
    s_0 &= 1 \hspace{1 pt}, \hspace{8 pt} s_1 = - \dfrac{1}{4} \Tilde{x} \hspace{1 pt}, \hspace{8 pt} s_2 = \dfrac{1}{5\times 4} \Tilde{x}^2 \hspace{1 pt}, \hspace{8 pt}  s_3 = - \dfrac{1}{6 \times 5 \times 4} \Tilde{x}^3 - \dfrac{1}{6} \Tilde{\gamma}  \hspace{1 pt}, \hspace{8 pt} s_4 = - \dfrac{1}{7}\Tilde{x} s_3 -\dfrac{1}{7} \Tilde{\gamma} s_1 + \dfrac{1}{7 \times 4} \Tilde{\epsilon} \hspace{1 pt} , \nonumber \\ s_5 &= - \dfrac{1}{8} \Tilde{x} s_4 - \dfrac{1}{8} \Tilde{\gamma} s_2 + \dfrac{1}{8 \times 5} \Tilde{\epsilon} s_1 \hspace{1 pt} , \hspace{8 pt} s_{k+6} = - \dfrac{\Tilde{x} s_{k+5}}{(k+9)} - \dfrac{\Tilde{\gamma} s_{k+3}}{(k+9)}  + \dfrac{s_k + \Tilde{\epsilon} s_{k+2}}{(k+9)(k+6)} \hspace{5 pt} \textrm{for }k \geq 0 \hspace{1 pt}.  \label{s_k coeffs}
\end{align}
If we ignore recoil and consider a static cloud (i.e. $\Tilde{x} = \Tilde{\gamma} = 0$) we recover the series solution found by \cite{Neufeld1990} (see his eq. B9). Next we consider $P(z)$. We first note that
\begin{align}
    \mathcal{L}[S(z) \ln z] ~&=~ \underbrace{\mathcal{L}[S(z)]}_{= ~0}  \ln z  + \dfrac{2}{z} \dfrac{\textrm{d}S}{\textrm{d}z} - \dfrac{S}{z^2} + \left( \Tilde{x} + \Tilde{\gamma}z^2 - \dfrac{2}{z} \right) \dfrac{S}{z}  =   \sum_{k=0}^{\infty} [(2k+3)s_kz^{k+1} + \Tilde{x} s_k z^{k+2} + \Tilde{\gamma} s_k z^{k+4}] \hspace{1 pt} \label{L(S lnz)}.
\end{align}
Using Eqs.~(\ref{L(S lnz)}) and (\ref{L of c_k z^r+k}) with $r=0$ then yields
\begin{align}
    \mathcal{L}[P(z)] = 0 ~&=~ \mathcal{A} (3 s_0 z + \Tilde{x} s_0 z^2 + \Tilde{\gamma} s_0 z^4) + \mathcal{A} \sum_{k=1}^{\infty} [(2k+3)s_kz^{k+1} + \Tilde{x} s_k z^{k+2} + \Tilde{\gamma} s_k z^{k+4}] \nonumber \\ &+~ (-2p_1 - 2\Tilde{x}p_0) z^{-1} + (-2p_2 - \Tilde{x}p_1) + 0 \times z + (4 p_4 + \Tilde{x}p_3 + \Tilde{\gamma}p_1 - \Tilde{\epsilon} p_0) z^2 + (10 p_5 + 2\Tilde{x}p_4 + 2\Tilde{\gamma}p_2 - \Tilde{\epsilon} p_1) z^5 \nonumber \\ ~&+~ \sum_{k=0}^{\infty} [(k+6)(k+3)p_{k+6} + (k+3) \Tilde{x} p_{k+5} + (k+3)\Tilde{\gamma}p_{k+3} - \Tilde{\epsilon} p_{k+2} - p_k] z^{k+4}
\end{align}
We see from the $\mathcal{O}(z)$ coefficients that $\mathcal{A} = 0$, and so there is no logarithmic contribution in the solution $P(z)$. Setting each coefficient of $z^k$ to zero then yields (after fixing $p_0 = 1$, $p_3 = 0$):
\begin{align}
    p_0 &= 1 \hspace{1 pt} , \hspace{8 pt} p_1 = - \Tilde{x} \hspace{1 pt} , \hspace{8 pt} p_2 = \dfrac{1}{2}\Tilde{x}^2  \hspace{1 pt} , \hspace{8 pt} p_3 = 0 \hspace{1 pt} , \hspace{8 pt} p_4 = \dfrac{1}{4} \Tilde{\gamma} \Tilde{x} +  \dfrac{1}{4} \Tilde{\epsilon}  \hspace{1 pt} , \hspace{8 pt} p_5 = \dfrac{1}{10} \Tilde{\epsilon} p_1 - \dfrac{1}{5} \Tilde{\gamma} p_2 - \dfrac{1}{10} \Tilde{x} p_4 \hspace{1 pt} , \nonumber \\ p_{k+6} &= \dfrac{p_k + \Tilde{\epsilon} p_{k+2}}{(k+6)(k+3)} - \dfrac{\Tilde{x} p_{k+5} + \Tilde{\gamma} p_{k+3}}{(k+6)} \hspace{5 pt} \textrm{for }k \geq 0 \hspace{1 pt}.\label{p_k coeffs}
\end{align}
In summary so far, we have found that the general solution for $f_n(z)$ is
\begin{equation}
    f_n(z) = \mathcal{C}_1 P(z) + \mathcal{C}_2 S(z) = \mathcal{C}_1 \sum_{k=0}^{\infty} p_k z^k + \mathcal{C}_2 \sum_{k=0}^{\infty} s_k z^{k+3} \hspace{1 pt} ,
\end{equation}
where the coefficients $p_k$ and $s_k$ are given by Eqs.~(\ref{p_k coeffs}) and (\ref{s_k coeffs}), respectively, and where $\mathcal{C}_1$ and $\mathcal{C}_2$ are constants. To get the physically relevant solution we impose the boundary condition $f_n(\lvert z \rvert = \infty) = 0$ (no intensity very far from line centre). This fixes one of the constants, and we can therefore write the solution away from line centre as
\begin{equation}
    f_n(z>0) = \mathcal{C} [P(z) - S(z) \mathcal{R}_+] \hspace{1 pt} , \hspace{5 pt}  f_n(z<0) = \mathcal{C} [P(z) - S(z) \mathcal{R}_-] \hspace{1 pt} ,\hspace{5 pt} \textrm{where } \mathcal{R}_{\pm} \equiv \lim_{z \rightarrow \pm \infty} \dfrac{P(z)}{S(z)} \hspace{1 pt} \label{f_n solution before C determined}.
\end{equation}
The final constant $\mathcal{C}$ can be determined from the jump in derivative imposed from the source term on the right-hand side of Eq.~(\ref{AppendixFrequencyEq}). Integrating Eq.  (\ref{AppendixFrequencyEq}) from $y = -y_{\rm max}$ to $y= y_{\rm max}$ and taking the limit $y_{\rm max} \rightarrow 0$ yields for the left-hand side:
\begin{align}
    \lim_{y_{\rm max} \rightarrow 0} \left( \dfrac{\textrm{d}f_n}{\textrm{d}y}\bigg|_{y_{\rm max}} + \sqrt{6} \Bar{x}\mathcal{H} f_n \big|_{y_{\rm max}} - \dfrac{\textrm{d}f_n}{\textrm{d}y} \bigg|_{-y_{\rm max}} - \sqrt{6} \Bar{x}\mathcal{H} f_n \big|_{-y_{\rm max}} \right) ~&=~  \dfrac{a_{\rm v}}{\sqrt{\pi}\beta_n^2} \lim_{z_{\rm max} \rightarrow 0} \dfrac{1}{z_{\rm max}^2} \sqrt{\dfrac{3}{2}} \dfrac{1}{\beta_n}\left( \dfrac{\textrm{d}f_n}{\textrm{d}z}\bigg|_{z_{\rm max}} - \dfrac{\textrm{d}f_n}{\textrm{d}z} \bigg|_{-z_{\rm max}} \right) \nonumber \\ ~&+~  \dfrac{a_{\rm v}}{\sqrt{\pi}\beta_n^2} \lim_{z_{\rm max} \rightarrow 0} \sqrt{\dfrac{3}{2}}  \dfrac{\Tilde{x}}{\beta_n z_{\rm max}^2} \left( f_n\big|_{z_{\rm max}} - f_n\big|_{-z_{\rm max}} \right) \hspace{1 pt} .
\end{align}
Expanding the result using Eq.~(\ref{f_n solution before C determined}), with the coefficients from Eqs.~(\ref{p_k coeffs}) and (\ref{s_k coeffs}), then gives us after some simplification:
\begin{align}
    \lim_{y_{\rm max} \rightarrow 0} \left( \dfrac{\textrm{d}f_n}{\textrm{d}y}\bigg|_{y_{\rm max}} + \sqrt{6} \Bar{x}\mathcal{H} f_n \big|_{y_{\rm max}} - \dfrac{\textrm{d}f_n}{\textrm{d}y} \bigg|_{-y_{\rm max}} - \sqrt{6} \Bar{x}\mathcal{H} f_n \big|_{-y_{\rm max}} \right) ~&=~ - 3 \lambda_n (\mathcal{R}_+ - \mathcal{R}_-) \mathcal{C} \equiv -2 \lambda_n \mathcal{D}_n \mathcal{C}  \hspace{1 pt} \label{limit of derivative LHS}.
\end{align}
In the last step we have introduced $\mathcal{D}_n \equiv (3/2)(\mathcal{R}_+ - \mathcal{R}_-)$, which is equal to unity for the special case of no dust, expansion/contraction, or recoil -- i.e. the special case considered by several earlier authors \citep[e.g.][]{Dijkstra2006, Tasitsiomi2006, Lao2020, Tomaselli2021, McClellan2022}. More generally, $\mathcal{D}_n = \mathcal{D}_n(\Tilde{x}_n. \Tilde{\gamma}_n, \Tilde{\epsilon}_n)$ (where we have now restored the subscript $n$ for the dimensionless parameters $\Tilde{x}_n$, $\Tilde{\gamma}_n$, and $\Tilde{\epsilon}_n$). Equating Eq.~(\ref{limit of derivative LHS}) to the integral over $y$ of the right-hand side of Eq.~(\ref{AppendixFrequencyEq}), solving for $\mathcal{C}$, and then using Eq.~(\ref{f_n solution before C determined}) gives us the following solution:
\begin{equation}
    f_n(z) = \dfrac{\mathcal{I}_{\textrm{s},n} \taucl}{2(\mathcal{P} + n \pi \mathcal{D}_n)} \hspace{1 pt} \mathcal{F}_n(z) \hspace{1 pt} , \hspace{5 pt} \textrm{where } \mathcal{F}_n(z) = \begin{cases}
        P(z) - S(z) \mathcal{R}_+ ~~ \textrm{for } z > 0 \\ P(z) - S(z) \mathcal{R}_- ~~ \textrm{for } z < 0 
    \end{cases} , \label{f_n series solution final}
\end{equation}
where, as usual, $\mathcal{P} \equiv \sqrt{6 \pi} p_{\rm d} \taucl/2$, and we have used $\lambda_n \simeq n\pi/\taucl$. In Fig. \ref{F_n series plot} we plot $\mathcal{F}_n(z)$, $\mathcal{D}_n$, and relevant integrals of $\mathcal{F}_n(z)$, for different values of the recoil and continuum absorption parameters $\Tilde{x}_n$ and $\Tilde{\epsilon}_n$, respectively. The series solution for velocity gradients can be written in closed form, given in Eq.~(\ref{Solution f_n no recoil or dust}) in the main text.

\subsection{Emergent Ly\texorpdfstring{\boldmath$\alpha$}{alpha} spectrum}
\label{AppendixSpectrum}

The emergent Ly$\alpha$ spectrum at $r = R_{\rm cl}$ is $J(x,\taucl)$. Naively, one would predict no escaping emission at all from Eq.~(\ref{series solution}), which would be wrong. The issue has to do with the approximation that the eigenvalues are exactly $\lambda_n = n \pi/\taucl$, which is only an approximate solution of Eq.~(\ref{BCcloudedge}). To get the correct result at the cloud edge one could look at the first-order corrections to the eigenvalues, as done by \cite{Dijkstra2006} and \cite{Seon2020}. Another approach to get the same result, which we take here, is to use the fact that the boundary condition at the cloud edge yields $J = \sqrt{3} H$ (which was used to derive the eigenvalues in the first place). Using $H = - (1/3 \mathcal{H}) (\partial J/\partial\tau)$ (from the Eddington approximation) and $J$ from Eq.~(\ref{series solution}) then yields:
\begin{align}
    J(x,\taucl) &=~ -\dfrac{1}{\sqrt{3} \mathcal{H}(x) \taucl} \dfrac{\partial J}{\partial \Bar{\tau}}\bigg|_{\Bar{\tau} = 1} \nonumber \\ &\simeq~ -\dfrac{\sqrt{18} L_{\rm Ly\alpha}}{(4\pi)^2  \taucl \Delta \nu_{\rm D} R_{\rm cl}^2 \mathcal{H}(x)} \sum_{n=1}^{\infty} \dfrac{ [\sin(n\pi \Bar{\tau}_{\rm s}) - n\pi \Bar{\tau}_{\rm s} \cos( n\pi \Bar{\tau}_{\rm s})] }{\Bar{\tau}_{\rm s}^3 (n \pi) (\mathcal{P} + n\pi \mathcal{D}_n)} \hspace{1 pt}  (-1)^n \mathcal{F}_n(z) \hspace{1 pt} \label{EmergentSpectrum General} ,
\end{align}
where we have used $\partial/\partial \Bar{\tau} [\sin(n \pi \Bar{\tau})/\Bar{\tau}] = n\pi (-1)^n$ at $\Bar{\tau} = 1$. In Fig. \ref{spectrum} we compare predicted spectra using Eq.~(\ref{EmergentSpectrum General}) to MCRT results for the same setups using \textsc{colt}.

\subsection{The Ly\texorpdfstring{\boldmath$\alpha$}{alpha} escape fraction}
\label{Escape fraction appendix}

The  Ly$\alpha$ escape fraction $f_{\rm esc, Ly\alpha}$ from a spherical uniform cloud is given by $f_{\rm esc, Ly\alpha} = 4\pi R_{\rm cl}^2 \Delta \nu_{\rm D} \int_{-\infty}^{\infty} \textrm{d}x \hspace{1 pt} F(x, \taucl)/L_{\rm Ly\alpha}$, where $F(x, \taucl) = 4\pi H(x, \taucl)$ is the specific flux at the cloud edge. From the boundary condition at the cloud edge we have $F(x,\taucl) = 4 \pi J(x, \taucl)/\sqrt{3}$. Thus, with Eq.~(\ref{EmergentSpectrum General}) for the emergent spectrum, the Ly$\alpha$ escape fraction becomes:
\begin{equation}
    f_{\rm esc, Ly\alpha} =  -\dfrac{\sqrt{6}}{ \taucl} \sum_{n=1}^{\infty}  \dfrac{ [\sin(n\pi \Bar{\tau}_{\rm s}) - n\pi \Bar{\tau}_{\rm s} \cos( n\pi \Bar{\tau}_{\rm s})] }{\Bar{\tau}_{\rm s}^3 (n \pi) (\mathcal{P} + n \pi \mathcal{D}_n)} (-1)^n \int_{-\infty}^{\infty} \dfrac{\textrm{d}x}{\mathcal{H}(x)} \mathcal{F}_n(z)  \hspace{1 pt} .
\end{equation}
In the general case where recoil and dust are included, it is more convenient to convert the integral into an integral over $z$. Using the wing limit $\mathcal{H} \simeq a_{\rm v}/\sqrt{\pi}x^2$, and $\textrm{d}x = \beta_n \textrm{d}z$, where $\beta_n = (3/2\pi^3)^{1/6} (a_{\rm v} \taucl)^{1/3} n^{-1/3}$, yields 
\begin{equation}
    f_{\rm esc, Ly\alpha} =  -3 \sum_{n=1}^{\infty}  \dfrac{ [\sin(n\pi \Bar{\tau}_{\rm s}) - n\pi \Bar{\tau}_{\rm s} \cos( n\pi \Bar{\tau}_{\rm s})] }{\Bar{\tau}_{\rm s}^3 (n \pi)^2 (\mathcal{P} + n \pi \mathcal{D}_n)} (-1)^n \int_{-\infty}^{\infty} \textrm{d}z \hspace{1 pt} z^2 \mathcal{F}_n(z)  \hspace{1 pt} \label{Lya escape fraction appendix}.
\end{equation}
For a given source distribution in a static cloud, and assuming $p_{\rm d} = 0$, the Ly$\alpha$ escape fraction $f_{\rm esc, Ly\alpha}$ is only a function of the dimensionless variable $\Tilde{\epsilon}_n$. In Fig. \ref{fesc Lya prediction vs Monte Carlo} we plot the predicted Ly$\alpha$ escape fraction from dusty clouds using Eq (\ref{Lya escape fraction appendix}) and compare it to MCRT results from \textsc{colt}.  

\subsection{Ly\texorpdfstring{\boldmath$\alpha$}{alpha} radiation hydrodynamics}
\label{AppendixRadPressure}

In this Appendix we give a short derivation of the Ly$\alpha$ radiation pressure force on gas. The momentum equation for the gas in the lab frame is given by \citep[e.g.][]{Mihalas2001, Castor2004, Smith2017}:
\begin{align}
\dfrac{\partial }{\partial t} (\rho \boldsymbol{u}) + \boldsymbol{\nabla} \boldsymbol{\cdot} (\rho \boldsymbol{u} \otimes \boldsymbol{u}) ~&=~ - \boldsymbol{\nabla}P - \rho \boldsymbol{\nabla}\Phi + \boldsymbol{\mathcal{G}}_{\rm LF} - \dfrac{\boldsymbol{u}}{c} \mathcal{G}_{\rm LF}^{0} \, , \label{Appendix gas equation 1}
\end{align}
where $\boldsymbol{u}$ is the gas velocity, $P$ the gas pressure, and $\Phi$ the gravitational potential. As for the two last terms on the right, $\boldsymbol{\mathcal{G}}_{\rm LF}$ and $c \mathcal{G}_{\rm LF}^{0}$ are the rates of momentum and energy deposition from radiation, as computed in the lab frame (LF). They are related to the corresponding comoving frame quantities $\boldsymbol{\mathcal{G}}$ and $\mathcal{G}^0$ via a Lorentz transformation \citep{Mihalas2001}:
\begin{equation}
    \boldsymbol{\mathcal{G}}_{\rm LF} = \boldsymbol{\mathcal{G}} + \gamma \dfrac{\boldsymbol{u}}{c} \left( \mathcal{G}^0 + \dfrac{\gamma}{\gamma + 1} \dfrac{\boldsymbol{u}}{c} \boldsymbol{\cdot} \boldsymbol{\mathcal{G}} \right) \, , \hspace{10 pt} \mathcal{G}_{\rm LF}^0 = \gamma \left( \mathcal{G}^0 + \dfrac{\boldsymbol{u}}{c} \boldsymbol{\cdot} \boldsymbol{\mathcal{G}} \right) \, ,
\end{equation}
where $\gamma \equiv 1/\sqrt{1 - \lvert \boldsymbol{u} \rvert^2/c^2}$ is the Lorentz factor. To first order in $\boldsymbol{u}/c$, Eq.~(\ref{Appendix gas equation 1}) then simplifies to 
\begin{align}
\dfrac{\partial }{\partial t} (\rho \boldsymbol{u}) + \boldsymbol{\nabla} \boldsymbol{\cdot} (\rho \boldsymbol{u} \otimes \boldsymbol{u}) ~&=~ - \boldsymbol{\nabla}P - \rho \boldsymbol{\nabla}\Phi + \boldsymbol{\mathcal{G}} \, ,
\end{align}
with $\boldsymbol{\mathcal{G}}$, the Ly$\alpha$ radiation pressure force per unit volume, given by $-4\pi/c$ times the right-hand side of Eq.~(\ref{div K equation}), integrated over frequency:
\begin{equation}
    \boldsymbol{\mathcal{G}} = \dfrac{1}{c} \int_{0}^{\infty} \textrm{d}\nu \, [\alpha + \alpha_{\rm c}(1 - \omega g)] \boldsymbol{F} \simeq \dfrac{1}{c} \int_{0}^{\infty} \textrm{d}\nu \, n_{\rm HI} \sigma_0 \mathcal{H} \boldsymbol{F} \hspace{1 pt} \label{Lya force per unit volume},
\end{equation}
where in the last step we have used $\alpha \gg \alpha_{\rm c}(1 - \omega g)$ for the Ly$\alpha$ line, as shown in Appendix \ref{AppendixRTeq}, and $\alpha = n_{\rm HI} \sigma_0 \mathcal{H}$. The acceleration due to Ly$\alpha$ radiation pressure is simply $\boldsymbol{a}_{\rm Ly\alpha} = \boldsymbol{\mathcal{G}}/\rho$, or: 
\begin{equation}
    \boldsymbol{a}_{\rm Ly\alpha} = \dfrac{1}{c \rho} \int_{0}^{\infty} \textrm{d}\nu \, n_{\rm HI} \sigma_0 \mathcal{H} \boldsymbol{F} \simeq - \dfrac{4\pi}{3c \rho} \int_{0}^{\infty} \textrm{d}\nu \, \boldsymbol{\nabla} J \, ,
\end{equation}
where the last step follows from the Eddington approximation (see Appendix \ref{AppendixRTeq}).

\section{The cross-section for quenching by collisions with H(1s)}
\label{Appendix H(2s) cross section}

In this Appendix we outline how the cross-section for $\textrm{H}(2s)+\textrm{H}(1s) \rightarrow \textrm{H}(2p)+\textrm{H}(1s)$ was calculated. New structure calculations of the adiabatic potential energy curves and non-adiabatic couplings of $^3\Sigma^+_g$ states associated with the $n=2$ and $n=3$ asymptotic limits are performed. The multi reference configuration interaction (MRCI) method is used with the basis set described in \cite{stenrup2009}. Molecular orbitals are generated using a state-averaged complete active space self-consistent field (CASSCF) calculation. The active space both in MRCI and CASSCF contains (7$\sigma_g$,3$\pi_g$,1$\delta_g$,7$\sigma_u$,3$\pi_u$,1$\delta_u$) and 5 states of $^3\Sigma^+_g$ are computed.

The method used to calculate the $\textrm{H}(2s)+\textrm{H}(1s) \rightarrow \textrm{H}(2p)+\textrm{H}(1s)$ cross-section has been described in detail in previous papers \citep[see e.g.][]{stenrup2009,hornquist2022}, and we will therefore only repeat the main steps here. For convenience, atomic units are used in this Appendix. The molecular wave function is expanded using adiabatic electronic states, $\ket{\phi_i}$, as
\begin{align}
\Psi(\boldsymbol{R}) = \dfrac{1}{R}\sum_{i,l,m}Y_{lm}(\hat{R})F_{il}(R)\ket{\phi_i} \hspace{1 pt},
\end{align}
where $R$ and $\hat{R}$ are the radial and angular components of the internuclear coordinate $\boldsymbol{R}$, respectively. The radial nuclear wave functions are denoted by $F_{il}(R)$ and the functions $Y_{lm}(\hat{R})$ are spherical harmonics, where $l$ denotes the orbital angular momentum quantum number. The radial nuclear wave functions are solutions to the adiabatic coupled Schrödinger equation,
\begin{align}
&\bigg[-\dfrac{1}{2\mu}\dfrac{ \textrm{d}^2}{\textrm{d} R^2}+V^{\rm ad}_i(R)+\dfrac{l(l+1)}{2\mu R^2}-E\bigg]F_{il}(R) - \dfrac{1}{\mu}\sum_j\bigg[2f_{ij}(R)\dfrac{\textrm{d}}{\textrm{d}R}+g_{ij}(R)\bigg]F_{jl}(R)=0 \hspace{1 pt},
\end{align}
where $\mu$ is the reduced mass of the nuclei, and where
\begin{align}
f_{ij}(R) &= \bra{\phi_i}\dfrac{\textrm{d}}{\textrm{d}R}\ket{\phi_j} \hspace{1 pt},\\
g_{ij}(R) &= \bra{\phi_i}\dfrac{\textrm{d}^2}{\textrm{d}R^2}\ket{\phi_j} \hspace{1 pt},
\end{align}
are the first and second derivative non-adiabatic coupling elements, respectively. We have here neglected the angular components of the non-adiabatic couplings. 

The adiabatic coupled Schrödinger equation is then transformed to a strict diabatic representation. The adiabatic-to-diabatic transformation matrix is obtained by numerically solving the equation \citep[][]{mead1982}
\begin{align}
\dfrac{\textrm{d}}{\textrm{d} R}\mathbfss{T}+\mathbfss{f}\mathbfss{T}=\mathbfss{0} \hspace{1 pt},
\end{align}
where $\mathbfss{f}$ is the anti-symmetric matrix containing the first derivative non-adiabatic coupling elements. The non-adiabatic couplings are assumed to be zero at large internuclear distances, and the adiabatic-to-diabatic transformation matrix is therefore equal to the identity matrix asymptotically. The diabatic potential matrix is obtained by the similarity transformation $\mathbfss{V}^{\rm d}=\mathbfss{T}^T\mathbfss{V}^{\rm ad}\mathbfss{T}$. The diabatic nuclear wave functions, which we denote by $\tilde{F}_{il}(R)$, are solutions to the diabatic coupled Schrödinger equation
\begin{align}
\bigg(-\dfrac{1}{2\mu}\dfrac{\textrm{d}^2}{\textrm{d}R^2} + \dfrac{l(l+1)}{2\mu R^2} -E\bigg)\tilde{F}_{il}(R) + \sum_jV_{ij}^{\rm d}(R)\tilde{F}_{jl}(R) = 0 \hspace{1 pt}.
\end{align}
This equation is solved numerically by using Johnson's log-derivative method \citep[][]{johnson1973}, in which the equation is transformed into the matrix Riccati equation for the logarithmic derivative of the nuclear wave function. The matrix Riccati equation is then numerically integrated. From the logarithmic derivative, the scattering matrix $S_{ij,l}$ can be extracted by matching the solution with the appropriate boundary conditions of the open or closed covalent or ionic channels. The cross-section going from state $j$ to state $i$ is then calculated from the open-open partition of the scattering matrix by 
\begin{align}
\sigma_{ij}(E) = \frac{\pi}{k_j^2}\sum_{l=0}^\infty(2l+1)|S_{ij,l}-\delta_{ij}|^2 \hspace{1 pt}.\label{cross_section}
\end{align}
For each partial wave, the matrix Riccati equation is numerically integrated from $0.6 \hspace{1 pt} a_0$ to $50 \hspace{1 pt} a_0$ using an integration step size of $0.005 \hspace{1 pt} a_0$. The total cross-section is obtained by summing the contributions from the partial waves. When the ratios of the partial cross-sections and the accumulated cross-sections remain less than $10^{-5}$ for 25 terms in succession, the summation of Eq.~\eqref{cross_section} is terminated.

In the present formulation, we use a molecular description of the electronic states. The excited states correlating with $n\geq 2$ are degenerate in the asymptotic region and are linear combinations of atomic states. In order to know which potential energy curve that corresponds to $\mathrm{H}(1s)+\mathrm{H}(2s)$ or $\mathrm{H}(1s)+\mathrm{H}(2p)$ in the asymptotic region, these linear combinations need to be determined. This can be achieved by using the method proposed by \cite{belyaev2015}. The molecular electronic wave function is thus written as a linear combination of the atomic states
\begin{align}
\mathbf{\Phi}_{\rm mol} = \mathbfss{B} \mathbf{\Phi}_{\rm at} \hspace{1 pt},\label{mol-at}
\end{align}
where $\mathbfss{B}$ is an orthogonal matrix. In order to determine the matrix $\mathbfss{B}$, one can utilize the \textit{ab initio} calculated first derivative non-adiabatic couplings. It can be shown \citep[see e.g.][]{grosser1999,belyaev2001} that the following relation holds for the asymptotic values of the non-adiabatic couplings written in an atomic basis:
\begin{align}
\bra{\phi^{\rm at}_i}\dfrac{\textrm{d}}{\textrm{d}R}\ket{\phi^{\rm at}_j}_\infty = \gamma[V_i(\infty) - V_j(\infty)]\bra{\phi^{\rm at}_i}z\ket{\phi^{\rm at}_j} \hspace{1 pt},\label{atomcouplings}
\end{align}
where $\bra{\phi^{\rm at}_i}z\ket{\phi^{\rm at}_j}$ are atomic transition dipole moment that can be calculated analytically using the eigenstates of atomic hydrogen. In the present case, $\gamma = 1/2$. The $B$ matrix is parameterized by blocks of $n\times n$ orthogonal matrices, which is possible since the molecular states are linear combinations of atomic states with the same $n$. In the present case, we neglect any mixing between the $n=4$ states. Therefore, we parameterize the $B$ matrix using block matrices composed of $2\times2$ and $3\times3$ rotation matrices while the rest of the $B$ matrix contains ones along the diagonal and zeroes otherwise. By using Eq.~\eqref{mol-at}, one can show that
\begin{align}
\overline{\tau}^{\rm mol}_{ij} = \sum_{k,l} B_{ik}\overline{\tau}^{\rm at}_{kl}B_{jl} \hspace{1 pt},
\end{align}
where $\overline{\tau}^{\rm mol}_{ij}$ are the asymptotic values of the non-adiabatic couplings in a molecular basis and $\overline{\tau}^{\rm at}_{ij}$ the same quantities expressed in an atomic basis. This set of equations can be further simplified by using the selection rules for the transition dipole moments. To obtain the relevant elements of the $B$ matrix, the set of equations was minimized using least squares. The optimization was performed separately for each electronic symmetry. For the $2\times 2$ block of $\mathbfss{B}$, which is written as
\begin{align}
\mathbfss{B}_{2\times 2} =
\begin{pmatrix}
\alpha & -\beta \\
\beta & \alpha
\end{pmatrix} \hspace{1 pt},
\end{align}
the values obtained were $\alpha = 0.9992$ and $\beta=0.0410$ for $^1\Sigma^+_g$, $\alpha =-0.1122$ and $\beta=-0.9937$ for $^1\Sigma^+_u$ and $\alpha =-0.0022$ and $\beta=0.9999$ for $^3\Sigma^+_g$. A comparison of the \textit{ab initio} calculated non-adiabatic couplings of $^3\Sigma^+_u$ symmetry with the couplings calculated using Eq.~\eqref{atomcouplings} shows that there is no mixing among the $n=2$ states in that symmetry and the corresponding $B$ matrix will therefore be equal to the identity matrix. Once the $B$ matrix has been determined, the cross-section can be calculated in the usual way using the transformed scattering matrix $\overline{\mathbfss{S}}$, which is defined as
\begin{align}
\overline{\mathbfss{S}} = \mathbfss{B}^T \mathbfss{S} \mathbfss{B}  \hspace{1 pt}.
\end{align}
However, since the mixing between the $n=2$ states is weak (i.e. they correspond almost entirely to either $2s$ or $2p$), we approximate the $S$ matrix by neglecting the smallest component of $\mathbfss{B}_{2\times 2}$ for each symmetry, which should be sufficient for the present purpose.

\label{lastpage}
\end{document}